\documentclass[12pt]{article}
\usepackage{amsmath}
\usepackage{amsfonts}
\usepackage{amssymb}
\usepackage{srcltx}
\usepackage{hyperref}
\usepackage{imakeidx}
\usepackage[utf8]{inputenc}
\usepackage[T1]{fontenc}

\def\oi{b_{\infty }}
\def\di{d_{\infty}}

\numberwithin{equation}{section}

\newtheorem{theorem}{Theorem}[section]
\newtheorem{proposition}[theorem]{Proposition}
\newtheorem{lemma}[theorem]{Lemma}
\newtheorem{corollary}[theorem]{Corollary}

\newtheorem{conjecture}[theorem]{Conjecture}
\newtheorem{remark}[theorem]{Remark}
\newtheorem{example}[theorem]{Example}

\makeindex
\begin{document}
\noindent
{\Large {\bf Multicomponent KP type hierarchies and their reductions, associated to conjugacy classes of Weyl groups of  classical Lie algebras }}
\vskip 9 mm
\begin{center}
\begin{minipage}[t]{70mm}
{\bf Victor Kac}\\
\\
Department of Mathematics,\\
Massachusetts Institute of Technology,\\
Cambridge, Massachusetts 02139, U.S.A\\
e-mail: kac@math.mit.edu\\
\end{minipage}\qquad
\begin{minipage}[t]{70mm} 
{\bf Johan van de Leur}\\
\\
Mathematical Institute,\\
Utrecht University,\\
P.O. Box 80010, 3508 TA Utrecht,\\
The Netherlands\\
e-mail: J.W.vandeLeur@uu.nl
\end{minipage}
\end{center}
\begin{abstract}
This,    to a large extent,
expository paper, describes the theory of multicomponent 
hierarchies of evolution equations of XKP type, where 
X=A, B, C or D, and AKP=KP, and their
reductions, associated to the conjugacy classes of 
the Weyl groups of classical Lie algebras of type X. As usual,
the main tool is the multicomponent boson-fermion correspondence,
which leads to the corresponding tau-functions, wave functions, dressing operators
and Lax operators.
\end{abstract}
\tableofcontents
\pagebreak
\section{Introduction}
Let $V$ be a vector space over $\mathbb C$ with a symmetric  (resp. skewsymmetric) nondegenerate bilinear form $(\cdot,\cdot)$. Recall that the Clifford (resp.  Weyl) algebra $C\ell(V)$ (resp.  $W(V)$) is the factor algebra of the tensor algebra over $V$ by the 2-sided ideal, generated by the relations
\[
uv+vu=(u,v)\quad (\mbox{resp. } uv-vu=(u,v)),\quad u,v\in V.
\]
\indent The key construction of the paper is that of the {\it Casimir operator,} acting on $F\otimes F$, where $F$ is an irreducible module over $C\ell(V)$ (resp.  $W(V)$). 
For this we need two sets of vectors $\{ v^+_i\}_{i\in I}$ and $\{ v^-_i\}_{i\in I}$ in $V$ indexed by a set $I\subset \mathbb Q$,  invariant under sign change, satisfying the following properties:
\[
\begin{aligned}
(i)&\quad \{ v^+_i\}_{i\in I}\cup \{ v^-_i\}_{i\in I}\quad \mbox{span } V,\\
(ii)&\quad (v_i^+,v_j^-)=\delta_{i,-j}, \quad (v_i^\pm, v_i^\pm) =0\quad \mbox{if } i\ne 0, \quad v^+_0=v^-_0:=v_0.
\end{aligned}
\]
\indent  Let $\epsilon=1$ (resp. $0$) if $0\in I$ (resp.  $\not\in I$). The module $F$ is the irreducible $C\ell(V)$- (resp. $W(V)$-) module,  called the spin (resp. Weyl) module,  admitting a non-zero vector $|0\rangle$ (vacuum vector), such that
\begin{equation}
\label{intro1.1} 
v_i^+|0\rangle=v^-_i|0\rangle =0 \quad\mbox{if } i>0,\quad v_0|0\rangle=\frac{\epsilon}{\sqrt 2}|0\rangle.
\end{equation}
Define the Casimir operator $S$ by the following formula
\begin{equation}
\label{intro1.2}
S=\epsilon v_0\otimes v_0 +\sum_{i\in I \backslash \{0\}} v_i^+\otimes v_{-i}^-.
\end{equation}
 Due to \eqref{intro1.1}, we have 
\begin{equation}
\label{intro1.3}
S(|0\rangle\otimes |0\rangle)=\frac{\epsilon}2 |0\rangle\otimes |0\rangle.
\end{equation}
\indent 
Let $\mathfrak g$ be the Lie subalgebra of $C\ell(V)$ (resp. $W(V)$), spanned by the elements  of the form  $v^+v^-$,  where $v^\pm\in V^\pm := {\rm span}_{i\in I} \{ v_i^\pm\}$.
In the case of $C\ell(V)$ we have also the corresponding group $G$,  generated by the elements of the form $1+v^+v^-$, where $v^\pm \in V^\pm$ and span$\, \{ v^+,v^-\}$ is an isotropic subspace of $V$ (its inverse being $1-v^+v^-$).
It is straightforward to check that the operator $S$ on $F\otimes F$ commutes with the action of $\mathfrak g$ by derivations of the tensor product, hence $S$ commutes with the diagonal action of $G$. It follows from \eqref{intro1.3} that all elements $\tau$  from the orbit $G\cdot|0\rangle$ satisfy the equation
\begin{equation}
\label{intro1.4}
S(\tau\otimes\tau)=\frac{\epsilon}2 \tau\otimes\tau.
\end{equation}
Equation\eqref{intro1.4} is called the {\it generalized KP equation} in {\it fermionic picture}.

It is easy to prove (along the lines of \cite{KvdLB}, Theorem 1.2) that,   conversely,  a non-zero $\tau\in F$, satisfying equation \eqref{intro1.4}, lies on the orbit $G\cdot |0\rangle$.

The most important examples of the above setup are the Clifford algebras ~$C\ell_x$, $x=a,\ b,$ or $d$, and the Weyl algebra,  which is,  for the sake of uniformity, denoted by $C\ell_c$, constructed as follows. We denote the corresponding $C\ell_x$-module $F$ by $F_x$.
\\
\ 
\\
{$\mathbf {C\ell_a}$.  $I=\frac12+\mathbb Z$; $\psi^+_i:=v^+_i$,  $\psi_i^-:=v^-_i$,  $i\in I$,  form a basis of $V$; the symmetric bilinear form is 
$(\psi_i^+ ,\psi_j^-)_a=\delta_{i,-j}$, $(\psi_i^\pm, \psi_j^\pm)_a=0$; the action on the vacuum vector is $\psi^\pm _i|0\rangle=0 $ for $i>0$.
The generating series
\[
\psi^\pm (z) =\sum_{j\in\frac12 +\mathbb Z}\psi^\pm_j z^{-j-\frac12},
\]
are called charged free fermions. The relations of $C\ell_a$ become
\[
[ \psi^+(z), \psi^-(w)]_+=\delta(z-w),\quad  [ \psi^\pm(z), \psi^\pm(w)]_+=0,
\]
where $\delta(z-w)=z^{-1}\sum_{n\in\mathbb Z}\left( \frac{w}{z}\right)^n$ is the formal delta function.
The equation \eqref{intro1.4}
becomes
\begin{equation}
\label{intro1.5}
{\rm Res}_{z=0}\psi^{+}(z)\tau\otimes\psi^-(z)\tau dz=0, \quad\tau\in F_a.
\end{equation}
It is called the {\it KP hierarchy} in the {\it fermionic picture.}

The Lie algebra $\mathfrak g$ is the Lie algebra $gl_\infty$ of matrices $(a_{i,j})_{i,j\in\mathbb Z}$ over $\mathbb C$ with finitely many non-zero entries via the identification
\[ E_{ij}=\psi^+_{-i+\frac12}\psi^-_{j-\frac12}, \quad i,j\in\mathbb Z.
\]
\
\\
$\mathbf {C\ell_b}$.  $I=\mathbb Z$, $\tilde\phi_i:=v_i^+=(-1)^iv^-_i$,  $i\in I$, 
 form a basis of $V$, the symmetric bilinear form $(\cdot,\cdot)_b$ is $(\tilde\phi_i,\tilde\phi_j)_b=(-1)^i\delta_{i,-j}$;
the action on the vacuum vector is $\tilde\phi_i|0\rangle=0$ if $i>0$, 
$\tilde\phi_0|0\rangle=\frac1{\sqrt 2}|0\rangle$.
The generating series
\[
\tilde\phi(z)=\sum_{j\in \mathbb Z}\tilde\phi_j z^{-j}
\]
is called the twisted neutral free fermion. The relations of $C\ell_b$ become
\begin{equation}
\label{comb}
[\tilde\phi(z),\tilde\phi(w)]_+=z\delta(z+w).
\end{equation}
The equation \eqref{intro1.4} becomes
\begin{equation}
\label{intro1.6b}
{\rm Res}_{z=0}\tilde\phi(z)\tau\otimes\tilde\phi(-z)\tau\frac{dz}z=\frac12 \tau \otimes\tau, \quad\tau\in F_b.
\end{equation}
It is called the {\it BKP hierarchy} in the {\it fermionic picture}.

The Lie algebra $\mathfrak g$ is the subalgebra of $gl_\infty$, consisting of matrices, skewadjoint with respect to the bilinear form $(\cdot,\cdot)_b$, which we denote by 
$so_{\infty,odd}$, via the identification 
\[
(-1)^jE_{-i,j}-(-1)^iE_{-j,i}=\tilde\phi_i\tilde\phi_j \quad \mbox{for } i>j.
\] 
Note,  that we use a a slightly different basis of the Clifford algebra $C\ell_b$ in this introduction, than in the other sections.  This means that we also have a different embedding in $gl_\infty$.
\ 
\\
$\mathbf {C\ell_d}$.
$I=\frac12+\mathbb Z$, $\phi_i:=v_i^+=v^-_i$,  $i\in I$, form a basis of $V$, the symmetric bilinear form  is $(\phi_i,\phi_j)_d=\delta_{i,-j}$, $i,j\in I$;
the action on the vacuum vector is $\phi_i|0\rangle=0$ if $i>0$.
The generating series
\[
\phi(z)=\sum_{j\in\frac12+ \mathbb Z}\phi_j z^{-j-\frac12}
\]
is called the neutral free fermion.  The relations of $C\ell_d$ become
\begin{equation}
\label{comd}
[\phi(z),\phi(w)]_+=\delta(z-w).
\end{equation}
The equation \eqref{intro1.4} becomes
\begin{equation}
\label{intro1.7}
{\rm Res}_{z=0}\phi(z)\tau\otimes\phi(z)\tau {dz}=0, \quad\tau\in F_d.
\end{equation}
It is called the {\it DKP hierarchy} in the {\it fermionic picture}.

The Lie algebra $\mathfrak g$ is the subalgebra of $gl_\infty$, consisting of matrices, skewadjoint with respect to the bilinear form $(\cdot,\cdot)_d$, which we denote by 
$so_{\infty,even}$, via the identification 
\[
E_{-i+\frac12,j+\frac12}-E_{-j+\frac12,i+\frac12}=\phi_i\phi_j \quad \mbox{for } i> j.
\]
\ 
\\
$\mathbf {C\ell_c}$.
$I=\frac12+\mathbb Z$, $b_i:= v_i^+=(-1)^{i+\frac12} v^-_i$,  $i\in I$, 
form a basis of $V$, the skewsymmetric bilinear form   is $(b_i,b_j)_c=(-1) ^{i-\frac12}\delta_{i,-j}$, $i,j\in I$;
the action on the vacuum vector is $b_i|0\rangle=0$ if $i>0$.
The generating series
\begin{equation}
\label{bz1}
b(z)=\sum_{j\in\frac12+ \mathbb Z}b_j z^{-j-\frac12}
\end{equation}
is called the neutral free symplectic boson. The relations of $C\ell_c$ become
\begin{equation}
\label{bz2}
[b(z),b(w)]_-=\delta(z+w).
\end{equation}
The equation \eqref{intro1.4} becomes
\begin{equation}
\label{intro1.6c}
{\rm Res}_{z=0}b(z)\tau\otimes b(-z)\tau {dz}=0,\quad \tau\in F_c.
\end{equation}
It is called the {\it CKP hierarchy} in the {\it fermionic picture}.

The Lie algebra $\mathfrak g$ is the subalgebra of $gl_\infty$,  consisting of matrices, skewadjoint with respect to the bilinear form $(\cdot,\cdot)_c$, which we denote by 
$sp_{\infty}$, via the identification 
\[
  (-1)^{j+1}E_{-i,j+1}+(-1)^{i+1}E_{-j,1+i}= b_{i+\frac12}b_{j+\frac12}
\quad \mbox{for } i\ge j.
\] 

Note that we use the subscripts $a$,$b$, $c$ and $d$ since the Lie algebra $\mathfrak g$  in these cases has Dynkin diagram $A_\infty$, $B_\infty$, $C_\infty$ and $D_\infty$,  respectively.
Recall that the equation \eqref{intro1.4} describes the group orbit of the vacuum vector under the group, corresponding to the Lie algebra $\mathfrak g$,
in the space $F_x$  for $x=a,b,d$.

The generating series $\psi^\pm(z)$, $\tilde\phi(z)$, $\phi(z)$ and $b(z)$ are quantum fields in the sense that, when applied to $v\in F_x$,  one obtains
an element from $F_x((z))$. An important operation on quantum fields $\alpha(z) $  and $\beta(z)$ is their normally ordered product, defined by
\begin{equation}
\label{intro1.9}
:\alpha(z)\beta(z):= \alpha(z)_-\beta(z)\pm \beta(z)\alpha_+(z),
\end{equation}
where $\alpha(z)=\alpha_-(z)+\alpha_+(z)$,   $\alpha_+(z)|0\rangle\in F[[z]]$,
$\alpha_-(z)|0\rangle\in z^{-1}F[[z^{-1}]]$ and we have in \eqref{intro1.9}  a minus sign  if $\alpha(z)$ and $\beta(z)$ are both fermions.

The simple formalism outlined above is applied to the theory of integrable evolution PDE's by using {\it bosonization}.
For example, the simplest,  1-component bosonization,   in cases $a$, $b$, $d$, $c$ is,  respectively,
\begin{equation}
\label{alphax}
\begin{aligned}
\alpha_a(z)&=:\psi^+(z)\psi^-(z):,\\
\alpha_b(z)&=\frac12:\tilde\phi(z)\tilde\phi(-z):,\\
\alpha_d(z)&= \frac12:\phi(z)\phi(-z):, \\
\alpha_c(z)&=\frac12 :b(z)b(-z):.
\end{aligned}
\end{equation}

In all cases bosonization is expressed in terms of vertex operators from string theory.
In order to introduce them we need the following notation, which will
be used throughout the paper:
\begin{equation}
\label{1.9a}
z\cdot t=\sum_{j=1}^\infty z^j t_j, \quad  z^{-1}\cdot\tilde  \partial_t=\sum_{j=1}^\infty z^{-j} \frac 1{j}\frac{\partial }{\partial t_j};
\end{equation}
also $z\circ t$ and $z^{-1}\circ \tilde  \partial_t$ will denote the expressions in \eqref{1.9a}, where the summation is taken over positive odd integers.

In  the simplest  $x=a$ case, $\alpha_a(z)=\sum_{n\in\mathbb Z}\alpha_n z^{-n-1}$, where
\begin{equation}
\label{intro1.10}
[\alpha_m,\alpha_n]=m\delta_{m,-n},\quad \alpha_j|0\rangle=0\quad\mbox{for }j\ge 0.
\end{equation}
Hence we have a Heisenberg Lie algebra, with respect to which $F_a$ decomposes in a direct sum of eigenspaces $F_a=\oplus_{m\in\mathbb Z}F^{(m)}$
of $\alpha_0$, so that each $F^{(m)}$ is an irreducible Heisenberg Lie algebra module with a non-zero  vector $|m\rangle$, such that $\alpha_j|m\rangle=\delta_{j,0} m|m\rangle$ for $j\ge 0$.
This allows one to construct an isomorphism
\begin{equation}
\label{intro1.11}
\sigma_1: F_a\xrightarrow{\sim}B_1=\mathbb C[q,q^{-1}, t_1,t_2,\ldots ],
\end{equation}
called the bosonization 
of type $a$, which is uniquely determined by the following properties:
\begin{equation}
\label{intro1.12}
\sigma_1(|m\rangle)=q^m, \ \sigma_1\alpha_0\sigma_1^{-1}=q\frac{\partial}{\partial q},\ \sigma_1\alpha_n\sigma_1^{-1}=\frac{\partial}{\partial t_n}\
\mbox{and }\sigma_1\alpha_{-n}\sigma_1^{-1}=n t_n, \ \mbox{for } n>0.
\end{equation}
Furthermore, 
since $[\alpha_k,  \psi^\pm(z)]=\pm z^k \psi^\pm(z) $,    we can identify the charged free fermions,
under the isomorphism $\sigma_1$, with vertex operators:
\begin{equation}
\label{intro1.13}
\sigma_1\psi^\pm (z)\sigma_1^{-1}=q^{\pm 1}z^{\pm q \frac{\partial}{\partial q}}
e^{\pm z\cdot t} e^{\mp z^{-1}\cdot\tilde  \partial_t}  .
\end{equation}
Recall that the isomorphism \eqref{intro1.11} is equivalent to the Jacobi triple product identity.
 Indeed, define energy$\, |0\rangle$=0  and energy$\, \psi_j^\pm=-j$, then 
\[
F_a=\bigoplus_{i\in\frac12\mathbb Z_{\ge 0}} F_i, \ \mbox{where }F_i =\{ w\in F|\, \mbox{energy}(w)=i\}.
\]
Let
\[
F_{i}^{(m)}:=  F_i\cap F^{(m)}   \ \mbox{and }\dim_{q,t}F_a:=\sum_{i,m}\dim(F_{i}^{(m)})q^it^m.
\]
Then, looking at the spin module, $F_a$,    we have 
\begin{equation}
\label{JTI1}
\dim_{q,t}F_a=\prod_{j\in \frac12+\mathbb Z_{\ge 0}}(1+tq^j)(1+t^{-1}q^j).
\end{equation}
Looking at $B_1$,  one finds
\begin{equation}
\label{JTI2}
\dim_{q,t}F_a=\sum_{m\in\mathbb Z} t^mq^{\frac{m^2}2}\prod_{j\in \mathbb Z_{> 0}}\frac1{1-q^j}.
\end{equation}
Thus, the  equality between \eqref{JTI1} and \eqref{JTI2},     is the famous Jacobi triple product identity.

Applying $\sigma_1$ to the  KP hierarchy in the fermionic picture  \eqref{intro1.5},  gives the famous KP hierarchy of bilinear equations on the tau-function  \cite{DJKM2}:
\begin{equation}
\label{ClasKP}
\begin{aligned}
{\rm Res}_{z=0}
e^{ z\cdot (t  -  \overline t) } 
e^{z^{-1}\cdot(\tilde  \partial_{\overline t }- \tilde  \partial_{ t })}
%
\tau ( t)\tau ({\overline t})dz
=0,\quad \tau(t)\in B_1.
\end{aligned}
\end{equation}
Here and futher $\bar t$ denotes another copy of $t=(t_1,t_2,\ldots)$.

The KP  hierarchy of equations on the tau-function, like \eqref{ClasKP}, can be rewritten in terms of Hirota bilinear equations, following \cite{S} and many other papers.  However, the most  important approach, introduced by Sato in \cite{S}, is by the Lax type equations via the wave functions and dressing operators.
To describe  this, consider the algebra of pseudo-differential operators
$\mathbb C[[t_1,t_2,\ldots]][[\partial^{-1}]]$,  where 
$\partial =\frac{\partial}{\partial t_1}$. 
Recall that for the  KP hierarchy the
{\it dressing operator } $P(t,\partial)$ for the tau-function $\tau(t)$ is the monic pseudo-differential operator,  whose symbol is
\begin{equation}
\label{dressP}
P(t,z)=\frac{e^{-z^{-1}\cdot \tilde\partial_t}\tau(t)}{\tau(t)}.
\end{equation}
The associated {\it Lax operator} $L(\partial)=L(t,\partial)$ is defined as the  "dressed" by $P(t,\partial) $ operator $\partial$:
\begin{equation}
\label{defL}
L(\partial)=P(\partial)\partial P(\partial)^{-1}.
\end{equation}
The equations \eqref{ClasKP} on the tau-function 
are  equivalent to the Sato-Wilson equations
\begin{equation}
\label{LaxSato}
L(\partial)P(\partial)=P(\partial)\partial,\quad
\frac{\partial P(\partial)}{\partial t_k}=
-(L(\partial)^k)_-P(\partial),\quad k=1,2,\ldots,
\end{equation}
where the subscript$\ _-$,  as usual, denotes the integral part of $L(\partial)^k$.
This is equivalent to the following {\it Lax-Sato equations} on $L(\partial )$:
\begin{equation}
\label{introlax}
\frac{\partial L(\partial)}{\partial t_k}=[(L(\partial)^k)_+,L(\partial)], 
\quad k=1,2,\ldots,
\end{equation}
where  the subscript$\ _+$,  as usual,  denotes the differential part of $L(\partial)^k$,   which are   equivalent to      \eqref{ClasKP}.

%
%

It turns out that the bosonization \eqref{intro1.11}-\eqref{intro1.13}   in the $x=a$ case
can be generalized to an $s$-component bosonization $\sigma_s$ for an arbitrary positive integer $s$ (see \cite{KLmult}).
For that one breaks the fields $\psi^\pm (z)=\sum_{k\in\frac12+\mathbb Z} \psi^\pm _k z^{-k-\frac12}$ in a sum of $s$ fields $\psi^{\pm a}(z)=\sum_{k\in\frac12+\mathbb Z} \psi^{\pm a} _k z^{-k-\frac12}$,  where $a=1,\ldots ,s$,
by letting
\begin{equation}
\label{relabela}
\psi^{\pm a}_j=\psi^\pm_{sj\pm \frac12(s-2a+1)}, \quad j\in\frac12+\mathbb Z,
\end{equation}
namely the modes $\psi^{\pm a}_j$ have indices $\equiv \pm \frac12(1-2a)\mod s$. Then one constructs $s$ bosonic fields
$\alpha^a(z)=:\psi^{+a}(z)\psi^{-a}(z):=\sum_{j\in\mathbb Z}\alpha^a_j z^{-j-1}$,  $a=1,\ldots,s$,
whose modes form a Heisenberg Lie algebra of "rank" $s$:
\begin{equation}
\label{intro1.14}
[\alpha^a_j, \alpha^b_k]=j \delta_{ab}\delta_{j,-k}, \quad \alpha_j^a|0\rangle=0\quad\mbox{for } j\ge 0.
\end{equation}
Like in  the 1-component case, this Heisenberg Lie algebra does not act irreducibly on $F=F_a$, so in order to get irreducibility one needs to 
introduce additional operators $Q_a$, $a=1,\ldots s$, on $F$,  analogous to the operator of multiplication by $q$. 
They are uniquely determined by the following properties:
\begin{equation}
\label{intro1.15}
Q_a|0\rangle=\psi^{+a}_{-\frac12}|0\rangle,\quad Q_a\psi^{\pm b}_k=(-1)^{1-\delta_{ab}}\psi^{\pm b}_{k\mp \delta_{ab}}Q_a.
\end{equation}
Then the $s$-component bosonization is a vector space isomorphism
\begin{equation}
\label{intro1.16}
\sigma_s:F\xrightarrow{\sim}B_s=\mathbb C[q_a, q_a^{-1}, t^{(a)}_1,t^{(a)}_2,\ldots | \, 1\le a\le s],
\end{equation}
uniquely determined by $\sigma_s(|0\rangle)=1$ and the following properties ($a=1,\ldots,s$):
\begin{equation}
\label{intro1.17}
\begin{aligned}
&\sigma_sQ_a\sigma_s^{-1}=(-1)^{\sum_{j=1}^{a-1} q_j\frac{\partial}{\partial q_j}}q_a,
\  \sigma_s\alpha_0^a\sigma_s^{-1}=q_a\frac{\partial}{\partial q_a},\\
&\sigma_s\alpha_j^a\sigma_s^{-1}=\frac{\partial}{\partial t_j^{(a)}}\ 
 \mbox{and }\sigma_s\alpha_{-j}^a\sigma_s^{-1}=jt^{(a)}_j,\ \mbox{for }j>0.
\end{aligned}
\end{equation}
As in the 1-component case,  under the isomorphism $\sigma_s$, one has the following identification
\begin{equation}
\label{intro1.18}
\sigma_s\psi^{\pm a} (z)\sigma_s^{-1}=(-1)^{\sum_{j=1}^{a-1}  q_j \frac{\partial}{\partial q_j}}q_a^{\pm 1}z^{\pm q_a \frac{\partial}{\partial q_a}}
e^{\pm z\cdot t^{(a)}}
e^{\mp z^{-1}\cdot\tilde  \partial_{t^{(a)}}}
\end{equation}
The  isomorphism \eqref{intro1.16}  is equivalent  to 
 the product of $s$ copies of \eqref{JTI1}, which is equal to the product of $s$ copies of \eqref{JTI2},  by the Jacobi triple product identity.
 
Note that the KP hierarchy in the fermionic picture \eqref{intro1.5} becomes
\begin{equation}
\label{intro1.19}
{\rm Res}_{z=0}\sum_{a=1}^s\psi^{+a}(z)\tau\otimes\psi^{-a}(z)\tau dz=0.
\end{equation}
Furthermore under the isomorphism $\sigma_s$,
we have
\[
\sigma_s:F^{(0)} \xrightarrow{\sim} B^{(0)}_s= \mathbb C[q^{\underline k}, t_1^{(a)},t_2^{(a)},\dots|\, 1\le a\le s,\ |\underline k|=0],
\]
where $\underline k=(k_1,\ldots,k_s)\in \mathbb Z^s$,
$q^{\underline k}=q_1^{k_1}q_2^{k_2}\cdots q_s^{k_s}
$
 and $ |\underline k|_j$ stands for $\sum_{i=1}^j k_i$,  and $|\underline k|=|\underline k|_s$,
so that for $\tau\in F^{(0)}$ we can write
$\sigma_s(\tau)=\sum_{|\underline k|=0}\tau_{\underline k}( t)q^{\underline k}\in B^{(0)}$,  
where $ t=(t_1^{(a)},t_2^{(a)},\ldots|\, 1\le a\le s)$.
Applying the  identification
$\sigma_s\otimes\sigma_s:F^{(0)}\otimes F^{(0)}\xrightarrow{\sim} B^{(0)}_s\otimes B^{(0)}_s$ to \eqref{intro1.19} and using \eqref{intro1.18}, we obtain the $s$-component {\it KP hierarchy} in the {\it bosonic picture} on the   tau-functions $\tau_{\underline k}( t)\in  B_s^{(0)}$:
\begin{equation}
\label{BosKP}
\begin{aligned}
{\rm Res}_{z=0}
\sum_{j=1}^s &(-1)^{|\underline k+\underline \ell |_{j-1}} z^{k_j-\ell_j-2+2\delta_{js}}
e^{ z\cdot (t^{(j)}-  \overline t^{(j)})} 
e^{z^{-1}\cdot(\tilde  \partial_{\overline t^{(j)}}- \tilde  \partial_{ t^{(j)}})}
%
\tau_{\underline k+\underline e_s-\underline e_j}( t)\tau_{\underline \ell+\underline e_j-\underline e_s}({\overline t})
=0,
\end{aligned}
\end{equation}
for  each pair $\underline k, \underline\ell\in\mathbb Z^s$, such that $|\underline k|=|\underline\ell|=0$.
Here $\underline e_j\in\mathbb Z^s$ are the standard basis vectors of $\mathbb Z^s$.
A special case for $s=1$ is the classical KP hierarchy \eqref{ClasKP} on the tau-function $\tau(t)\in \mathbb C[t_1,t_2,\ldots]$.

In the case of the $s$-component KP hierarchy,  the dressing operator $P(\underline k; t,\partial)$,  associated to the tau-function $\{ \tau_{\underline k}|\, |\underline k|=0\}$, is defined in a similar way,  except that $\partial= \sum_{j=1}^s \frac{\partial}{\partial t_1^{(j)}}$ and its symbol is now  a
$s\times s$ matrix $P(\underline k; t,z)$, with entries 
\begin{equation}
\label{dressing}
P_{ij}(\underline k; t,z)=\frac{
(-1)^{\eta_{ij}}
z^{\delta_{ij}-1}
e^{-z^{-1}\cdot \tilde\partial_{t^{(j)}}}
\tau_{\underline k+\underline e_i-\underline e_j}(t)
}
{\tau_{\underline k}(t)}, 
\quad i,j=1,\ldots,s,
\end{equation}
where 
\begin{equation}
\label{eta}
\eta_{ij}=0\ \mbox{ if }i\ge j\ \mbox{ and }=1\  \mbox{otherwise.}
\end{equation}
This  is a monic $s\times s$ matrix pseudo-differential operator. 
The (commuting) Lax operators are the dressed   by  $P(\underline k; t,\partial)$   operators  $I\partial$ and $E_{jj}$,    $j=1, \ldots, s$:
\[
L(\underline k; t,\partial)=P(\underline k; t,\partial)\partial P(\underline k; t,\partial)^{-1},
\quad
C^{(j)}(\underline k; t,\partial)=P(\underline k; t,\partial)E_{jj} P(\underline k; t,\partial)^{-1}.
\]
The equations \eqref{BosKP} on the tau-functions imply the following $s\times s$ matrix Lax-Sato  equations  on
$L=L(\underline k ,:t, \partial)$ and $C^{(j)}=C^{(j)}(\underline k ,:t, \partial)$   for each $\underline k$:
\begin{equation}
\label{1.36a}
\frac{\partial L }{\partial t_i^{(j)}}=[(L ^iC^{(j)})_+,L], 
\quad
\frac{\partial C^{(k)} }{\partial t_i^{(j)}}=[(L ^iC^{(j)})_+,C^{(k)}], \quad j,k=1,\ldots ,s, 
\quad i=1,2,\ldots.
\end{equation}
See Subsection \ref{SS7.1} for more details.
Note that for $s>1$ one needs the auxilary operators   $   C^{(j)}$ in order to have Lax equations \eqref{1.36a},  which are  equivalent to
\eqref{BosKP}, see \cite{KLmult}.
Note that the Lax operator $L$ (resp. the auxiliary opertors $C^{(k)}$) is an arbitrary $s\times s$ matrix pseudo-differential operator of the form $I_s\partial+\sum_{i\ge 1} A_i(t)\partial^{-i}$
(resp. $E_{kk}+\sum_{i\ge 1} C_i^{(k)}(t)\partial^{-i}$),  where  $A_i(t)$ and $C_i^{(k)}(t)$ are $s\times s$ matrices with entries in 
$
\mathbb C[[t_i^{(a)}| a=1,\ldots,s,\ i=1,2,\ldots]].
$

\
%
%

For    the  1-component $x=b$ case,
$\alpha_b(z)=\sum_{n\in\mathbb Z_{odd}}\alpha_n z^{-n-1}$, where
\begin{equation}
\label{intro1.10b}
[\alpha_m,\alpha_n]=\frac{m}2\delta_{m,-n},\quad \alpha_j|0\rangle=0\quad\mbox{for }j> 0.
\end{equation}
Hence we have a Heisenberg Lie algebra, with respect to which $F_b$ is irreducible.
This allows one to construct an isomorphism
\begin{equation}
\label{intro1.11b}
\sigma_1: F_b\xrightarrow{\sim}B_1=\mathbb C[t_1,t_3,\ldots ],
\end{equation}
called bosonization 
of type $b$, which is uniquely determined by the following properties:
\begin{equation}
\label{intro1.12b}
\sigma_1(|0\rangle)=1,  \ \sigma_1\alpha_n\sigma_1^{-1}=\frac{\partial}{\partial t_n}\
\mbox{and }
\sigma_1\alpha_{-n}\sigma_1^{-1}=\frac{n}2 t_n, \ \mbox{for } n>0.
\end{equation}
This isomorphism is equivalent to the following simple identity:
\begin{equation}
\label{qb}
\dim_q(F_b):=\sum_{j\in\mathbb Z_\ge0} \dim(F_j)q^j= \prod_{j\in \mathbb Z_{> 0}}(1+q^j)=
\prod_{j\in \mathbb Z_{> 0, odd}}\frac1{1-q^j},
\end{equation}
where the energy decomposition  $F=\oplus_j F_j$ is defined by energy$\, |0\rangle=0$,  energy$\,\tilde\phi_j=-j$.
Furthermore, 
since $[\alpha_k,  \tilde\phi(z)]=\pm z^k \tilde \phi(z) $,    we can identify the neutral twisted free fermion,
under the isomorphism $\sigma_1$,  with the vertex operator:
\begin{equation}
\label{intro1.13b}
\sigma_1\tilde \phi (z)\sigma_1^{-1}= \frac1{\sqrt 2}
e^{z\circ t} e^{ -2(z^{-1}\circ\tilde  \partial_t)}  .
\end{equation}
Applying
$\sigma_1$ turns equation \eqref{intro1.6b} into the (small)  BKP hierarchy \cite{DJKM3}, \cite{KvdLB}
(there also exists the so-called large BKP hierarchy, which was introduced in the Introduction of \cite{KvdLB}):
\begin{equation}
\label{ClasBKP}
{\rm Res}_{z=0}
e^{z\circ (t-\bar t)} e^{ 2(z^{-1}\circ(\tilde  \partial_{\bar t}-(\tilde  \partial_t))}\tau ( t)\tau ({\overline t})dz
= \tau ( t)\tau ({\overline t}).
\end{equation}

In the    $x=b$ case the dressing operator is,   see \cite{DJKM3} or \cite{SLumini}, 
\[
P(t,z)=\frac{e^{-2(z^{-1}\circ \tilde\partial_t)}\tau(t)}{\tau(t)}.
\]
Then  the Lax opertor $L$, defined again by \eqref{defL}, satisfies the same Lax-Sato equations \eqref{introlax}, but only for odd $k$.  In this case $L$ is a pseudo-differential  operator
of the form $L=\partial+\sum_{j>0} u_j(t)\partial^{-j}$, where $u_j(t)\in\mathbb C[[t_1,t_3, \ldots]]$,
satisfying the extra condition \cite{SLumini}:
\begin{equation}
\label{1.42a}
L(\partial)^*= -\partial^{-1}L(\partial)\partial.
\end{equation}
\\
%
%

For   the 1-component  $x=c$ case,
$\alpha_c(z)=\sum_{n\in\mathbb Z_{odd}}\alpha_n z^{-n-1}$,  where
\begin{equation}
\label{intro1.10c}
[\alpha_m,\alpha_n]=-\frac{m}2\delta_{m,-n},\quad \alpha_j|0\rangle=0\quad\mbox{for }j> 0.
\end{equation}
Since $F_c$ is not irreducible,  when restricted to this Heisenberg  Lie algebra,  we need to construct more  operators.
Since  
$
[\alpha_k , b(z)]=- z^kb(z)$,  
one finds that 
\begin{equation}
\label{bTheta}
b(z)=V_-(z)^{-1}\Theta(z)V_+(z),  
\end{equation}
where (cf.  \cite{LOS} or Subsection \ref{3.6.2} of this paper for details)
\begin{equation}
\label{introV}
V_-(z)=\exp\left(
-2 \sum_{k<0,\,\rm odd} \frac{\alpha_k}k z^{-k}
\right),\qquad V_+(z)=\exp\left(
-2 \sum_{k>0,\,\rm odd} \frac{\alpha_k}k {z^{-k}}
\right)
,
\end{equation}
and 
\[
\Theta(z)=\sum_{k\in\frac12 +\mathbb Z}\theta_kz^{-k-\frac12}:=V_-(z)^{-1}b(z)V_+(z)^{-1}.
\]
One has
\[
\theta(z)|0\rangle =V_-(z)^{-1}\sum_{i>0}b_{-i}z^{i-\frac12}|0\rangle\in F_c[[z]],
\] 
and 
\[
\Theta(y)\Theta(-z)+\Theta(-z)\Theta(y)=2
z^{\frac12}\partial_z(z^{\frac12}\delta(y-z)).
\]
We construct an isomorphism
\begin{equation}
\label{intro1.11c}
\sigma_1: F_c\xrightarrow{\sim}B_1=\mathbb C[t_1,t_3,\ldots, \xi_{\frac12}, \xi_{\frac32}, \ldots  ],
\end{equation}
where $\xi_j$ are anti-commuting variables,
called bosonization  of type $c$, which is uniquely determined by the following properties:
\begin{equation}
\label{intro1.13c}
\begin{aligned}
\sigma_1(|0\rangle)=1,\ 
\sigma_1 \alpha_{-j}\sigma_1^{-1} =\frac12 jt_j, 
\
\sigma_1 \alpha_{j}\sigma_1^{-1} =-\frac{\partial}{\partial t_j},\\
\sigma_1 \theta_{-i}\sigma_1^{-1}=
(-1)^{i+\frac12} 2i\xi_i,\ 
\sigma_1 \theta_{i}\sigma_1^{-1}=
\frac{\partial}{\partial \xi_i }
,
\end{aligned}
\end{equation}
and we find 
 \begin{equation}
\label{vertexc}
\begin{aligned}
&\sigma _1b(z)\sigma_1^{-1}=e^{ z\circ  t} 
 \sum_{i\in  \frac12+\mathbb Z_{\ge  0}}\left(
\frac{\partial}{\partial \xi_i}z^{- i-\frac12}
- 2i\xi_i (-z)^{i-\frac12}\right)
e^{2 (z^{-1}\circ \tilde  \partial_t) }     .
\end{aligned}
\end{equation}
The isomorphism \eqref{intro1.11c} is equivalent 
to the following obvious identity   for the Weyl  module $F_c$,  when comparing the generators of the   algebra $C\ell_c$ and the operators that appear in the vertex operator \eqref{vertexc}:
\begin{equation}
\label{Cqidentity}
\dim_q F_c=
\prod_{j\in \frac12+\mathbb Z_{\ge 0}} \frac1{1-q^j}=
 \prod_{j\in \mathbb Z_{\ge 1, odd}}\frac1{1-q^{j}}
\prod_{j\in\frac12+\mathbb Z_{\ge 0}}(1+q^j),
\end{equation}
where the energy decomposition is defined by energy$\,|0\rangle=0$,  energy$\, b_j=-j$,  hence  energy$\,\alpha_k=-k$,  and 
energy$\, \theta_j=-j$.

Equation \eqref{intro1.6c} turns under  $\sigma_1$  into \cite{LOS}:
\[
\begin{aligned}
{\rm Res}_{z=0} 
\sum_{i,j\in  \frac12+\mathbb Z_{\ge  0}}
e^{ z\circ ( t-\bar t)} &
e^{2 (z^{-1}\circ (\tilde  \partial_t-\tilde  \partial_{\bar t})}
\left(
\frac{\partial}{\partial \xi_i}z^{- i-\frac12}
- 2i\xi_i (-z)^{i-\frac12}\right) \times
\\
& \left(
\frac{\partial}{\partial \overline \xi_j}(-z)^{- j-\frac12}
- 2j\overline\xi_j(z)^{j-\frac12}\right) 
\tau(t,\xi)\tau(\overline t,\overline \xi)dz=0.
\end{aligned}
\]

To construct the dressing operator $P$ in the  1-component $x=c$ case,     we write, see \cite{LOS}, 
$$\tau(t,\xi)
=\tau_0(t)+ \sum_{m\in\frac12+\mathbb Z_{> 0} }\tau_{   m}( t)\xi_{\frac12}  \xi_m+\mbox{terms of degree $>2$ in the  $\xi_j$'s},
$$
and let
\[
P(t,z)= \frac1{\tau_0(t)}
e^{2(z^{-1}\circ\tilde\partial_{t})}
\left(\tau_0(t)-
\sum_{m\in\frac12+\mathbb Z_{> 0} }
\!\!\!\!
 \tau_{   m}( t)z^{-m-\frac12}
\right).
\]
Then $L$, defined by \eqref{defL}, satisfies the Lax-Sato equations \eqref{introlax}, but as in the $x=b$ case,   only for odd $k$.
Furthermore, $L$ satisfies the extra condition  (see  \cite{DJKMVI} or \cite {LOS})
$L^*=-L$.
\\
\
\\

%
%

For     the 1-component $x=d$ case, 
$\alpha_d(z)=\sum_{n\in\mathbb Z_{even}}\alpha_n z^{-n-1}$,   where
\begin{equation}
\label{intro1.10d}
[\alpha_m,\alpha_n]=\frac{m}2\delta_{m,-n},\quad \alpha_j|0\rangle=0\quad\mbox{for }j\ge 0.
\end{equation}
Hence we have a Heisenberg Lie algebra, 
with respect to which $F_d$ splits  
into the direct sum $F_d=\bigoplus_{i\in\mathbb Z} F^{(i)}$,  where $\alpha_0 f= i f$   for $f\in F^{(i)}$.
The Heisenberg Lie algebra module $F^{(i)}$ is irreducible and  has the highest weight vector  
\[
|i\rangle=
\begin{cases}
\phi_{2i+\frac32}\phi_{2i-\frac12}\cdots \phi_{-\frac52}\phi_{-\frac12}|0\rangle& i<0,\\
\phi_{-2i+\frac12}\phi_{-2i+\frac52}\cdots \phi_{-\frac72}\phi_{-\frac32}|0\rangle& i\ge 0,
\end{cases}
\]
and $\alpha_0|i\rangle=i|i\rangle$.
This allows one to construct an isomorphism
\begin{equation}
\label{intro1.11d}
\sigma_1: F_d\xrightarrow{\sim}B_1=\mathbb C[q,q^{-1}, t_1,t_2,\ldots ],
\end{equation}
called the bosonization 
of type $d$,   which is uniquely determined by the following properties:
\begin{equation}
\label{intro1.12d}
\sigma_1(|i\rangle)= q^{i},\ \sigma_1\alpha_0\sigma_1^{-1}=q\frac{\partial}{\partial q},\ \sigma_1\alpha_{2n}\sigma_1^{-1}=\frac{\partial}{\partial t_n}\
\mbox{and }\sigma_1\alpha_{-2n}\sigma_1^{-1}=n t_n, \ \mbox{for } n\in \mathbb Z_{>0}.
\end{equation}
Since $[\alpha_{2k},  \phi(z)]= -z^{2k} \phi(-z) $,     we cannot find a vertex operator formulation of the field $\phi(z)$.
Instead, we define the Heisenberg eigenfunctions
\begin{equation}
\label{intro1.d}
 \phi^\pm(z):=\frac{\phi(z)\mp \phi(-z)}2, \quad\mbox{so  that }[\alpha_{2k},\phi^\pm (z)]=\pm z^{2k}\phi^\pm (z).
\end{equation}
These fermionic fields  can be identified with the following vertex operators
\begin{equation}
\label{intro1.13d}
\sigma_1\phi^\pm (z)\sigma_1^{-1}=q^{\pm 1}z^{-\frac12\pm \frac12 \pm 2q \frac{\partial}{\partial q}}
e^{\pm z^2\cdot t} e^{\mp z^{-2}\cdot\tilde  \partial_t}
\end{equation}
With respect to $so_{\infty,even}$ the module $F$ splits into two irreducible modules,  namely
\[
F_d=F^{\overline 0}\oplus F^{\overline 1}=\bigoplus_{i\in 2\mathbb Z} F^{(i)} \oplus \bigoplus_{i\in 1+2\mathbb Z} F^{(i)},
\]
with a highest weight vector $|\overline 0\rangle= |0\rangle$,  respectively $|\overline 1\rangle=\phi_{-\frac12} |0\rangle= |-1\rangle$,
 such that $\alpha_j|\overline \epsilon \rangle=-\epsilon \delta_{j,0} |\overline \epsilon\rangle$ for $j\ge 0$,  where $\epsilon=0$ or $1$.

Thus the DKP hierarchy \eqref{intro1.7} turns into 
\begin{equation}
\label{intro1.7ddd}
{\rm Res}_{z=0}\left(\phi^{+}(z)\tau\otimes\phi^-(z)\tau +  \phi^{-}(z)\tau\otimes\phi^+(z)\tau \right)dz=0,
\end{equation}
with $\tau\in F^{\overline 0}$ (the same equation holds for $\tau\in F^{\overline 1}$).
Applying $\sigma_1$ to this equation gives
 the charged version of the DKP hierarchy \cite{JM}, \cite {KvdLB}:
\begin{equation}
\label{intro1.7d}
\begin{aligned}
{\rm Res}_{z=0} \Big(&
z^{n-m-2}
e^{ z\cdot (t  -  \overline t) } 
e^{z^{-1}\cdot(\tilde  \partial_{\overline t }- \tilde  \partial_{ t })}
%
\tau_{n-1}( t)\tau_{m+1} ({\overline t})+\\
&\quad +z^{m-n-2}
e^{ -z\cdot (t  -  \overline t) } 
e^{-z^{-1}\cdot(\tilde  \partial_{\overline t }- \tilde  \partial_{ t })}
%
\tau_{n+1} ( t)\tau_{m-1} ({\overline t})\Big)dz
=0,
\end{aligned}
\end{equation}
where $n$ and $m$ are odd and  $\tau(t)=\sum_{k\in2\mathbb Z}\tau_k(t) q^{k}$.  
Note that this holds for $ F^{\overline 0}$.  In the case of  $ F^{\overline 1}$,    $n$ and $m$  are even  and  $\tau(t)=\sum_{k\in 1+2\mathbb Z}\tau_k(t) q^{k}$.

The isomorphism \eqref{intro1.11d} is equivalent to the following version of the Jacobi triple product identity:
\[
\dim_{q,t}F_d=
\prod_{j\in \frac12+2\mathbb Z_{\ge 0}}(1+tq^{j+1})(1+t^{-1}q^j)=
\sum_{m\in\mathbb Z} t^mq^{\frac{2m^2+m}2}\prod_{j\in \mathbb Z_{> 0}}\frac1{1-q^{2j}},
\]
where again energy$\, |0\rangle=0$ and energy$\, \phi_j=-j$.  
The two irreducible $so_{\infty, even}$ representations correspond to 
\[
\sum_{m\in 2\mathbb Z} t^mq^{\frac{2m^2+m}2}\prod_{j\in \mathbb Z_{> 0}}\frac1{1-q^{2j}}\quad\mbox{and } 
\sum_{m\in 1+ 2\mathbb Z} t^mq^{\frac{2m^2+m}2}\prod_{j\in \mathbb Z_{> 0}}\frac1{1-q^{2j}}.
\]

Since we have the appearance  of two terms in  the   bilinear equation \eqref{intro1.7d},   this  hierarchy  can 
 be rewritten only in terms  of a $2\times 2$ matrix Lax-Sato equations (see the Introduction of \cite{KvdLB} for more details).  
Namely,  
for $\tau(t)=\sum_{j\in 2\mathbb Z}\tau_j q^j$,  define, for each $k\in\mathbb Z$,  a monic $2\times 2$ matrix pseudo-differential dressing  operator
$P(k;t,\partial)$ with the symbol
\[
P(k;t,z)=\frac1{\tau_k(t)}
\begin{pmatrix}
e^{-z^{-1}\cdot \tilde\partial_t}\tau_k(t)&
z^{-2}e^{(-z)^{-1}\cdot \tilde\partial_t}\tau_{k+2}(t)\\
z^{-2}e^{-z^{-1}\cdot \tilde\partial_t}\tau_{k-2}(t)&
e^{(-z)^{-1}\cdot \tilde\partial_t}\tau_{k}(t)
\end{pmatrix}, 
\]
and  the Lax opertor   by $L= L(k;t,\partial)=P(k;t,\partial) (E_{11}\partial-E_{22} \partial)P(k;t,\partial) ^{-1}$. 
One also needs an auxilary operator $D(\partial) = D(k;t,\partial)$,  defined by  $D(\partial)=P(\partial)(E_{11}-E_{22})
P(\partial)^{-1}$. Then these operators, for each $k$,  clearly commute and satisfy the following Lax-Sato equations, see Subsection \ref{section4}:
\[
\frac{\partial L}{\partial t_i }=[(L^iD) _+,L],\quad \frac{\partial D }{\partial t_i }=[(L^iD )_+,D ] .
\]
Furthermore, $L$ and $D$ satisfy the following additional conditions
\[
L(\partial)^*=(E_{11}-E_{22})L(\partial)(E_{11}-E_{22}),\quad D(\partial)^*=-(E_{11}-E_{22})D(\partial)(E_{11}-E_{22}).
\]

In order to obtain  a  $1\times 1$ matrix dressing operator,  one has to modify the bosonization,  see   \cite{You2} for this construction.  However,  that construction does not allow a reduction to an affine Lie algebra, therefore we will not describe this in the introduction.  It appears as a special case of  the construction in Subsection \ref{section4},  viz.  for  $r=0$ and $s=1$.
\\
\
\\

The multicomponent bosonizations in the $x=b$,    $x= d$ and $ x=c$ cases
are more involved.  They can be constructed for an arbitrary pair of non-negative integers $s$ and $r$, such that $r+s>0$.  We refer for their constructions  to Sections \ref{subsec4.1} and \ref{subsec4.2} and the construction of the corresponding 
$s,r$-component BKP,  DKP and CKP hierarchies to Section \ref{subsec4.3}.

The identity equivalent to the bosonizations in the $s,r$-component $b$ (and $d$ cases,)  is the product of
  $s$ copies of the Jacobi triple product identity \eqref{JTI1}-\eqref{JTI2} and $r$ copies of \eqref{qb},
while in the  the $c$ case,  the corresponding identity is the product of $r$ copies of \eqref{Cqidentity} together with $s$ copies of the following identity,  which, as far as we know, is new,  see Subsection \ref{3.6.2},
\begin{equation}
 \prod_{i\in\mathbb Z_{>0}}
\frac{1-q^i}{(1-tq^{i-\frac12})(1-t^{-1}q^{i-\frac12})}=
\sum_{j,k=0}^\infty \frac{q^{\frac{k}2}t^k} {(1-q)\cdots(1-q^k)}
q^{jk} \frac{q^{\frac{j}2}t^{-j} }{(1-q)\cdots(1-q^j)} 
.
\end{equation}

\ 
\\

In the papers \cite{DJKM1}, \cite{JM},
Date, Jimbo,  Kashiwara and Miwa initiated the study of reductions of KP type hierarchies   to affine Lie algebras. We generalize this to the multicomponent case, see Subsection \ref{subsec4.2}.

In the 1-component case,  one fixes a positive integer $N$,  and let $\omega=e^{\frac{2\pi \sqrt{-1}}N}$, if $ x=a,b,c$ and 
$\omega=e^{\frac{\pi \sqrt{-1}}N}$, if $x=d$.
Define the fields
\begin{equation}
\label{alphaxx}
\begin{aligned}
A_k(z)&=:\psi^+(z)\psi^-(z\omega^k):,\\
B_k(z)&=\frac12:\tilde\phi(z)\tilde\phi(-z\omega^k):,\\
C_k(z)&=\frac12 :b(z)b(-z\omega^k):,\\
D_k(z)&= \frac12:\phi(z)\phi(-z\omega^k):.
\end{aligned}
\end{equation}
Then clearly,  see \eqref{alphax},  $X_0(z)=\alpha_x(z)$,  for $X=A,B,C$ and  $D$ and the modes of the  fields $X_k(z)$ span a Lie algebra that contains the corresponding Heisenberg Lie algebra.

 In the $x=a$ case,  the modes of the fields $A_k(z)$,  $k=0,\ldots,  N-1$, span the affine Lie algebra  
$\hat{gl_N} $,  which is the central extension of the loop algebra $gl_N(\mathbb C [t,t^{-1}])$.
The diagonal action of an element $g\in \hat{gl_N}$,on  $F_a\otimes F_a$  commutes not only with the Casimir operator $S$,  but also  with the operators
\[
{\rm Res}_{z=0}z^{\ell N}\psi^{+}(z)\otimes\psi^-(z) dz,\quad \ell=1,2,\ldots,
\]
which means that an $N$-reduced tau-function $\tau$, in the fermionic picture,  also satisfies the equation
\[
{\rm Res}_{z=0}z^{\ell N}\psi^{+}(z)\tau\otimes\psi^-(z)\tau dz,\quad \ell=1,2,\ldots.
\]
This gives in the bosonic picture,  that  $\tau(t)$ satisfies  not only \eqref{ClasKP},  but  all the following equations 
\begin{equation}
\label{ClasKPred}
\begin{aligned}
{\rm Res}_{z=0}z^{\ell N}
e^{ z\cdot (t  -  \overline t) } 
e^{z^{-1}\cdot(\tilde  \partial_{\overline t }- \tilde  \partial_{ t })}
%
\tau ( t)\tau ({\overline t})dz
=0, \quad \ell=0,1 ,\ldots.
\end{aligned}
\end{equation} 
This is the $N$-KdV hierarchy,  the case $N=2$,  being  the classical  KdV hierarchy,    see \cite {DJKM2}.  
Note that the equation \eqref{ClasKPred} for $\ell=0$ is the KP hierarchy \eqref{ClasKP}.  It is well-known,  see e.g. \cite{DJKM1},
that this is equivalent to the fact that the tau-function satisfies
\begin{equation}
\label{taured}
\frac{\partial\tau(t)}{\partial t_{\ell N}}=c_\ell \tau(t), \quad \ell=1,2,\ldots.,\quad \mbox{with  }c_\ell\in\mathbb C.
\end{equation}
If  a KP tau-function $\tau(t)$ satisfies \eqref{taured}, then,  it is straightforward to check that, when differentiating \eqref{ClasKPred} for $\ell=0$ by  $t_{k N}$, we get  \eqref{ClasKPred} for $\ell=k$.  The converse is also true,  but not so straightforward,  and we refer to Subsection  \ref{SS7.1} for a proof.

Since the tau-function $\tau(t)$ satisfies \eqref{taured},  the dressing operator $P(t,\partial)$, whose symbol is given by 
 \eqref{dressP} , satisfies 
$\frac{\partial P(t,\partial)}{\partial t_{\ell N}}=0$. 
Thus the second equation of \eqref{LaxSato},   implies that $(L^{\ell N})_-=0$, which means that ${\cal L}:=L^N$ is a monic $N$-th order differential operator.  One has the folowing Lax-Sato equation for $\cal L$:
\begin{equation}
\label{LAXRED}
\frac{\partial{ \cal L}}{\partial t_k}=
[({\cal L}^{\frac{k}{N}})_+,{\cal{L}}],  \quad k=1,2,\ldots .
\end{equation}

In the $x=b$ case,  the modes of the generating fields  $B_k(z)$ (see \eqref{alphaxx}) span  the twisted affine Lie algebras $\hat{gl}_N^{(2)}$ if $N$ is odd, and $\hat{so}_N^{(2)}$ if $N$ is even.  
An  $N$-reduced BKP tau-function, satisfies 
\begin{equation}
\label{intro1.6red}
{\rm Res}_{z=0}z^{\ell N} \tilde\phi(z)\tau\otimes\tilde\phi(-z)\tau\frac{dz}z=\frac{\delta_{\ell,0}}2 \tau \otimes\tau,\quad \ell=0,1 ,\ldots.
\end{equation}
As in the $x=a$ case, this implies that $(L^{kN})_-=0$.   Thus,  the differential operator ${\cal L} =L^N$ satisfies
\eqref{LAXRED}.  It satisfies the extra condition  ${\cal L}^*=(-1)^N\partial^{-1}{\cal L}\partial$.

Recall that a BKP tau-function depends  only on $t_k$ for $k$ odd.  Hence, if $N$ is odd, it is again straightforward  to check that  Equation \eqref{taured}  for $\ell $ odd  implies \eqref{intro1.6red}   for this same $\ell$.

The reduction in the $x=c$ case gives the affine Lie algebra    $\hat{sp}_N $,  if  $N$ is even,   and $\hat{gl}_N^{(2)}$ if  $N$ is odd (cf.  \cite{DJKM1}).
An $N$-reduced CKP tau-function satisfies the equations (see Subsection \ref{SubsecC}):
\begin{equation}
\label{intro1.6cred}
{\rm Res}_{z=0}z^{\ell N}b(z)\tau\otimes b(-z)\tau {dz}=0, \quad \ell=0,1 ,\ldots.
\end{equation}
As in the $x=a$ and $b$ cases, we find that ${\cal L} :=L^N$ is again a differential operator that satisfies \eqref{LAXRED},  and the condition ${\cal L}^*=(-1)^N {\cal L}$.

When $x=d$,     the reduction gives $\hat{so}_{2N} $.
The tau-function satisfies the equations
\begin{equation}
\label{intro1.7dddd}
{\rm Res}_{z=0}z^{2\ell N}\left(\phi^{+}(z)\tau\otimes\phi^-(z)\tau +  \phi^{-}(z)\tau\otimes\phi^+(z)\tau \right)dz=0, \quad \ell=0,1 ,\ldots.
\end{equation}
As before,  ${\cal L}=L^{N}$    is an $N$-th order ($2\times2$-matrix) differential operator. One has the folowing Lax-Sato equations for $\cal L$ 
and $D$:
\begin{equation}
\label{LAXREDD}
\frac{\partial{ \cal L} }{\partial t_k}= [({\cal L}^{\frac{k}{N}} D)_+,{\cal L}],\quad
\frac{\partial  D}{\partial t_k}= [({\cal L}^{\frac{k}{N}}   D)_+,D], \quad k=1,2,\ldots. \
,
\end{equation}
where, as before, $D(\partial)=P(\partial)(E_{11}-E_{22})
P(\partial)^{-1}$. 
Note that when $N$ is even,  $LD$ is  also an  $N$-th root of $\cal L$.

The reductions in the multicomponent case   for $x=a$
(see e.g \cite{tKLA} and  \cite{KLmult})
are parametrized by the conjugacy classes of the Weyl group of $gl_N$ or $sl_N$,  which is the permutation group $S_N$.  Conjugacy classes of $S_N$ are in one to one correspondence with  partitions 
$\lambda=(\lambda_1,\lambda_2,\ldots,\lambda_s)$ of $N$,   
where the $\lambda_i$ are  the lengths of the   cycles in the cycle decomposition of a permutation from the conjugacy class.

Each partition $\lambda$ of $N$ in $s$ non-zero parts
gives a reduction of the $s$-component KP hierarchy, which we call the $\lambda$-KdV hierarchy.    For instance, when $N=2$, there are two partitions $(1,1)$ and $(2)$.  The former leads to the AKNS hierarchy,  while the latter to the KdV hierarchy.  

We can express the generating series of the affine Lie algebra $\hat{gl}_N$ in terms of the generating series
\[
:\psi^{+a} (\omega_a^kz^{\lambda_b} )\psi^{-b} (\omega_b^\ell z^{\lambda_a} ):, \quad 1\le a,b\le s,\quad 
0\le k<\lambda_a, \quad 0\le \ell<\lambda_b,
\]
where $\omega_a=e^{\frac{2\pi \sqrt{-1}}{\lambda_a}}$ for  $a=1,\dots,s$.  Each part $\lambda_a$ of the  partition $\lambda$
corresponds to a $\hat{gl}_{\lambda_a}$, i.e.  the modes of the fields 
$
:\psi^{+a} (\omega_a^kz^{\lambda_a})\psi^{-a} (\omega_a^\ell z^{\lambda_a}):$, 
with $0\le k,\ell<\lambda_a$), span $\hat{gl}_{\lambda_a}$.
A $ (\lambda_1,\lambda_2,\ldots,\lambda_s)$-KdV tau-function satisfies
  the equations
\begin{equation}
\label{intro1.19red}
{\rm Res}_{z=0}\sum_{a=1}^s z^{ p\lambda_a} \psi^{+a}(z)\tau\otimes\psi^{-a}(z)\tau dz=0,\quad p=0, 1,2,\ldots ,
\end{equation}
which for $p=0$ is equation \eqref{intro1.19}.
Hence,  after the $s$-component bosonization, we have for $p=0,1,2,\ldots$,
\begin{equation}
\begin{aligned}
{\rm Res}_{z=0}
\sum_{j=1}^s (-1)^{|\underline k+\underline \ell |_{j-1}} z^{p \lambda_j +k_j-\ell_j-2+2\delta_{js}}
e^{ z\cdot (t^{(j)}-  \overline t^{(j)})} 
e^{z^{-1}\cdot(\tilde  \partial_{\overline t^{(j)}}- \tilde  \partial_{ t^{(j)}})}
%
&
\tau_{\underline k+\underline e_s-\underline e_j}( t)\tau_{\underline \ell+\underline e_j-\underline e_s}({\overline t})
=0,
\label{BosKPred}
\end{aligned}
\end{equation}
of which $p=0$ equation is \eqref{BosKP}.
As in the 1-component case, this  reduction corresponds to taking a differential operator $\cal L$ instead of a pseudo-differential operator $L$.  However in general the differential  operator $\cal L$ is not the  power of the pseudo-differential operator $L$.  For a general partition $\lambda=(\lambda_1,\lambda_2,\ldots,\lambda_s)$,    the equation \eqref{BosKPred} for $p=1$ implies that  
\[
{\cal L}= \sum_{j=1}^s L^{\lambda_j} C^{(j)}
\]
is a differential operaror, see Subsection \ref{SS7.1} for more details.
This operator satisfies the  Lax-Sato equations
\[
\frac{\partial {\cal L} }{\partial t_i^{(j)}}=[({\cal L }^{\frac{i}{\lambda_j}}C^{(j)})_+,{\cal L}],\quad i=1,2,\ldots,
\]
and the auxilary operators  $C^{(k)}$ with $1\le k\le s$  satisfy
\[
\frac{\partial C^{(k)} }{\partial t_i^{(j)}}=[({\cal L }^{\frac{i}{\lambda_j}}C^{(j)})_+,C^{(k)}],\quad i=1,2,\ldots.
\]
As in the 1-component $N$-KdV case, the tau-function satisfies an additional constraint, viz.
\[
\sum_{a=1}^s\frac{\partial\tau_{\underline  k}(t)}
{\partial t_{\ell \lambda_a}^{(a)}}= c_\ell \tau_{\underline  k}(t), \quad\mbox{for }
\ell=1,2,\ldots, \  \mbox{with  }c_{\ell}\in\mathbb C,
\]
and $c_\ell$ is the same for all $\tau_{\underline k}(t)$.  See Subsection \ref{SS7.1} for more details.

For other classical Lie algebras,  $so_N$ and $sp_N$, the reductions of the corresponding multicomponent hierarchies correspond to conjugacy classes of their Weyl groups $W$ as well.

Recall that,  according to one of the definitions of $W$ for $so_{2n+1}$ and $sp_{2n}$ (resp.  $so_{2n}$), this group consits of all permutations of non-zero integers  from the interval $[-n,n]$,  such that $\pi(-j)=-\pi(j)$ (in addition the number of $j$, such that $j\pi(j)<0$, is a multiple of 4).  Each such permutation is a  product of disjoined positive and negative cycles; a cycle is called positive (resp. negative) if for $j$ appearing in it, $-j$ doesn't appear (resp.  does appear). Reversing the signs of all entries of a positve cycle,  we again obtain a positive cycle, called its twin.  If we do this for a negative cycle,  it  does not change. Note also that the length of a negative cycle is always even.

One associates to an element $w\in W$ a pair of partitions $(\lambda,\mu)$ as follows. The partition $\lambda=(\lambda_1,\ldots,\lambda_s)$ consists of the lengths of positive cycles, where only one of the twin cycles is counted. The partition $\mu=(\mu_1,\ldots,\mu_r)$ cosists of the halves of the lengths of negative cycles.

The construction of the reduced  hierarchies for the affine Lie algebra $\hat{so}_N$  is based on $s$ pairs of charged fermionic fields $\psi^{\pm a}(z)$,  $a=1,\ldots, s$, and $r$  twisted neutral fermionic fields $\tilde \phi^{s+b}(z)$,  $b=1,\ldots, r$.   For $\hat{so}_{2n+1} $,  there is an additional field $\sigma(z)$,  that we do not bosonize using vertex operators.   The reason for this is that the elements of the Heisenbeg Lie algebra,   corresponding to this field,  do not  lie in $\hat{so}_{2n+1} $. 

The simpler case is $\hat{so}_{2n}$,    i.e.  the $x=d$ case (see Subsection \ref{subs7.2} for more details).  This affine Lie algebra is a subalgebra of $d_\infty$ and the bilinear identities are,  see \eqref{SD2nfield}, 
\begin{equation}
\label{introSD2nfield}
\begin{aligned}
{\rm Res}_{z=0}& 
\left(\sum_{b=1}^s  z^{p\lambda_b}\left( \psi^{+b}(z) \otimes \psi^{-b}(z)
+   \psi^{-b}(z) \otimes \psi^{+b}(z)\right)+\right.
\\
&\qquad+\left.
\sum_{c=s+1}^{r+s}z^{2p\mu_c-1} \tilde\phi^c(z)\otimes \tilde\phi^c(-z)\right)(\tau\otimes \tau)dz=0,\quad \mbox{for }p=0,1,2,\ldots .
\end{aligned}
\end{equation} 
The equation for $p=0$ is the $s,r$-component DKP hierarchy,  while the equations for $p=1,2,\ldots$ give the reduction to $\hat{so}_{2n}$ 
associated  to the pair of partitions $(\lambda,\mu)$,   so that $\lambda_1+\cdots+\lambda_s+\mu_1+\cdots+\mu_r=n$.
Note that in this case $r$ is always even.  
The $s,r$-bosonization is the isomorphism 
\[
\sigma_{s,r}:F_d \xrightarrow{\sim} B_{s,r}= \mathbb C[q^{\underline k}, \theta_1,\ldots,\theta_{\frac{r}2},t_1^{(a)},t_2^{(a)},\dots, t_1^{(b)}, t_3^{(b)},\ldots\, |\, 1\le a\le s< b\le r+s],
\]
where $\underline k=(k_1,\ldots,k_s)\in \mathbb Z^s$,  and $\theta_j$ are anticommuting variables.
Applying $\sigma_{s,r}$ to this expression and using 
the vertex operators  \eqref{intro1.18} for the charged free fermionic fields,   
and (cf.
\eqref{intro1.13b})
\begin{equation}
\label{intro1.13bb}
\sigma_{s,r}\tilde \phi^{s+b} (z)\sigma_{s,r}^{-1}= R_b
e^{z\circ t^{(s+b)} }e^{ -2(z^{-1}\circ\tilde  \partial_{t^{(s+b)}})} ,
\quad 
R_b={\rm const} \left(
\frac{\partial}{\theta_{\left[
\frac{b}2\right]}}
-(-1)^b
\theta_{\left[\frac{b}2\right]}
\right)
\end{equation}
for the twisted neutral free fermionic fields,   we obain,  in a similar way as in the $x=a$ case,  the, rather complex,  bilinear identities 
\eqref{BosDKPred} on the (collection of) tau-function(s).   

In the $x=b$ case we can embed the  affine Lie algebra $\hat{so}_{2n+1}$ in both $b_\infty$ as well as $d_\infty$,  which gives different level 1 representations of this affine Lie algebra.  
The bilinear identity for the $d_\infty$ embedding is similar to the expression \eqref{introSD2nfield}, but one has to add the following term on the left-hand sides
$
\sum_j \sigma_j\tau\otimes \sigma_{p-j}\tau.
$
Then applying $\sigma_{s,r}$   gives the  bilinear identities 
\eqref{BosDBKPred}.  
Here the additional field $\sigma(z)$
is bosonized as  generating series of anti-commuting  variables $\sigma_{-j}= \xi_j$ and $\sigma_j=\frac{\partial}{\partial \xi_j}$,  $j>0$,
with  $j\in \mathbb Z$ if $r$ is odd,  and  $j\in \frac12+\mathbb Z$ if $r$ is even.  $\sigma_0$ is special,  it is the operator $R_{r+1}$.
 Hence,  in this case the tau-function (or rather the collection of tau-functions)  depend not only on $t_i^{(c)}$, for $c=1,\ldots ,r+s$, but also on these $\xi_j$.    
The embedding in $b_\infty$  leads to a similar bilinear identity as \eqref{BosDBKPred}.

The construction of the reduced hierarchy related to $\hat{so}_N$ in terms of these fields  suggests    a $(2s+r)\times (2s+r)$ matrix valued dressing operator.  However,  if $r>1$,   this dressing operator is not invertible.  We solve this  problem by considering  a certain principal minor,  and  put  both the anticommuting variables $\xi_j$ and  the times $t_k^{(s+b)}$,   for $b=1, \ldots, s$    or
for $b=1, \ldots, s-1$,  equal to zero    in the biliner identity.   This leads to two different approaches,  one with   a
$2s \times 2s $ matrix   dressing operator,  and the second one with a    $(2s+1)\times (2s+1)$ matrix   dressing operator. 
Since the second one is rather involved,   we shall only sketch the first one
 in this introduction,  for the second one  and more details on the first one,  we refer to Section \ref{section4} and  Subsections \ref{subs7.2}--\ref{subs7.4}. 

In the first approach,  in the $x=b$ and $d$ case,    we construct  a $2s\times 2s$ matrix dressing operator $P(t,\partial)$ in a similar way as in the $x=a$ case from the bilinear identity \eqref{introSD2nfield} for $p=0$
 (and the one in the $\hat{so}_{2n+1}$, where one has the extra field $\sigma(z)$), by only using the terms related to the 2s charged fields $\psi^{\pm a}(a)$,   $a=1,\ldots ,s$.
The corresponding Lax operator $L$   and the auxilary operators $D^j$ and $E^j$   are the  dressed by $P$ operators   $\sum_{j=1}^s (E_{jj}-E_{s+j,s+j})\partial$, and  $E_{jj}-E_{s+j,s+j}$,   $E_{jj}+E_{s+j,s+j}$, respectively.
The reduced Lax operator $\cal L$ is determined by the partition $\lambda$,  viz. 
$${\cal L}= \sum_{j=1}^s L^{\lambda_j}E^j  ,$$ that, together with the auxilary operators,  satisfy the Lax-Sato equations
\[ \frac{\partial \cal L}{\partial t^{(a)}_j}=[ ({\cal L}^{\frac{j}{\lambda_a}}D^a)_+,{\cal L}],\quad
\frac{\partial D^k}{\partial t^{(a)}_j}=[ ({\cal L}^{\frac{j}{\lambda_a}}D^a)_+,D^k],\quad
 \frac{\partial E^k}{\partial t^{(a)}_j}=[ ({\cal L}^{\frac{j}{\lambda_a}}D^a)_+,D^k]
. \]
If $\mu\ne \emptyset$,  $\cal L$ is no longer a differential operator,  but an operator of constrained KP type,
i.e. its integral part has a certain simple form (given below).
The partition $\mu$   determines the integral part of this operator. 
For $\hat{so}_{2n}$   we find
\[
{\cal L} ={\cal L} _++ \sum_{i=1}^r 
\sum_{n=0}^{2\mu_i} W^{n} E_{ii}\partial^{-1}
\tilde W^{2\mu_i-n}  ,
\]
where 
$W^j$  (resp. $\tilde W^j$) are certain $2s\times r$ (resp.  $r\times 2s$) matrix functions, depending on $t_i^{(a)}$, for  $a=1,\ldots ,s$,  $i=1,2,\ldots$.   Note that if $\mu=\emptyset$,  then there is no integral part of $\cal L$ and $\cal L$ is a differential operator.

For   $\hat{so}_{2n+1}$, 
one has the extra field $\sigma(z)$.  This field leads to some extra terms in the integral part of the operator,  see Subsections \ref{SecDred}--\ref{subs7.4} for details.

Since the Weyl group of the Lie algebra $sp_{2n}$ is the same as the Weyl group of $so_{2n+1}$,
the reduction to the affine Lie algebra  $\hat{sp}_{2n}$   depends,  as  in the $\hat{so}_{2n+1}$ case, on the decomposition of 
$n$ into two partitions  $\lambda=(\lambda_1,\lambda_2,\ldots,\lambda_s)$ for the positive cycles and
$\mu=(\mu_1,\mu_2,\ldots, \mu_r)$, for the negative cycles,   such that
$n=\lambda_1+\cdots +\lambda_s+\mu_1+\cdots+\mu_r$.  Each part of $\mu$ is related to a neutral symplectic bosonic field $b^{c}(z)$, $c=s+1,\ldots , s+r$,  see \eqref{bz1} and \eqref{bz2}.  Each part of $\lambda$    is related to a pair of {\it charged free symplectic bosons} $b^{\pm a}(z)$, $a=1,\ldots, s$. These fields satisfy
\[
[ b^{\pm a}(z), b^{\pm c}(w)]_-=0, \quad [b^{+a}(z),b^{-c}(w)]_-=\delta_{ac}\delta(z-w) \quad a,c=1,\ldots, s.
\]
The bilinear identities in this case are,    see \eqref{SC2},
\begin{equation}
\label{introSC2}
\begin{aligned}
{\rm Res}_{z=0}\big\{
\sum_{a=1}^s &z^{p\lambda_a} \left(b^{-a}(z)\tau \otimes b^{+a}(z)\tau- b^{+a}(z)\tau \otimes b^{-a}(z)\tau\right)+\\
+&\sum_{c=s+1}^{r+s}z^{2p\mu_{c-s}} b^c(z)\tau\otimes b^c (-z)\tau \big\} dz =0.
\end{aligned}
\end{equation}
Note that this is similar to equation \eqref{introSD2nfield},  but now with the fermionic fields $\psi^{\pm a}(z)$,
$\phi^{c}(z)$ replaced by the symplectic boson fields $b^{\pm a}(z)$,  $ b^{c}(z)$,  respectively.
Special cases,  viz.  $s=1$ and $r=0$  (resp.  $r=0$ and $s=1$) for only $p=0$ were studied in \cite{A}
(resp.  \cite{LOS}).
For the bosonization of these fields we refer to Subsection \ref{3.7.2}.  These fields depend on commuting  times $t_k^{(a)}$ for $1\le a\le s$ and $k=1,2,\ldots$,  and anticommuting variables $\xi^a_\ell$,  $\ell\in \mathbb Z$.  Again there is a whole collection of tau-functions,  which are connected via a bilinear identity.   As in the 1-component $x=c$ case,  we focus on $\tau_0$  and construct a $(2s+r)\times (2s+r)$ matrix monic dressing operator $P(t,\partial)$.  This time $P$ is invertible,  and we construct the Lax operator (see Subsection~\ref{section4}) $L$,  and the auxillary operators $C^b$ and $D^a$ by dressing the operators
\[
\left(\sum_{a=1}^s \left( E_{aa}- E_{s+a,s+a}\right)\partial +\sum_{c=s+1}^{s+r} E_{s+c,s+c}\right)\partial,\ \mbox{and }
\ E_{bb}, \ E_{aa}-E_{s+a,s+a},
\]
by $P$.
The reduced Lax operator $\cal L$ is the operator that one gets by dressing the operator
\[
 \sum_{a=1}^s \left( E_{aa}+(-1)^{\lambda_a} E_{s+a,s+a}\right)\partial^{\lambda_a}+\sum_{c=s+1}^{s+r} E_{s+c,s+c}\partial^{2\mu_{c-s}}. 
\]  
It  satisfies  the following Lax-Sato equations 
\[
\frac{\partial {\cal L}}{\partial t_k^a}= [({\cal L}^{\frac{k}{\lambda_a}}D^a)_+, {\cal L}]
,\quad
\frac{\partial {\cal L}}{\partial t_k^c}= [({\cal L}^{\frac{k}{2\mu_{c-s}}} C^{s+c})_+, {\cal L}],\quad a=1,\ldots s,\ c=s+1,\ldots, s+r,
\]
and the Lax equations for the auxillary operators are 
\[
\begin{aligned}
\frac{\partial D^b}{\partial t_k^a}= [({\cal L}^{\frac{k}{\lambda_a}}D^a)_+, D^b]
,\quad
\frac{\partial D^b}{\partial t_k^c}= [({\cal L}^{\frac{k}{2\mu_{c-s}}} C^{s+c})_+, D^b],\\
\frac{\partial C_{s+d}}{\partial t_k^a}= [({\cal L}^{\frac{k}{\lambda_a}}D^a)_+, C_{s+d}]
,\quad
\frac{\partial C_{s+d}}{\partial t_k^c}= [({\cal L}^{\frac{k}{2\mu_{c-s}}} C^{s+c})_+, C_{s+d}].
\end{aligned}
\]
The Lax operator $\cal L$ obeys the  constraint
\[
{\cal L}^*=M{\cal L}M,\quad\mbox{where } M=\sum_{a=1}^s(E_{a,s+a}+E_{s+a,a})+\sum_{b=1}^rE_{2s+b,2s+b}.
\]
Note that $\cal L=L^*$, if $\lambda=\emptyset$.
\\
\
\\
\

Below we  describe the contents of the paper.

In Section 2 we recall the description of  the classical finite-dimensional  Lie algebras,  the corresponding affine Lie algebras,   and the central extensions of the  infinite matrix Lie algebras denoted by  $a_\infty,b_\infty,c_\infty$, and   $d_\infty$.  By considering a vector space  isomorphism between $\mathbb C^\infty$ and $\mathbb C^n\otimes \mathbb C[t,t^{-1}]$,  we construct each  classical affine Lie algebra as  a subalgebra of $a_\infty,b_\infty,c_\infty$ or   $d_\infty$.     In Subsection 
\ref{twistaf} we establish a similar result for the twisted affine Lie algebras.
This Section is based on  \cite{Kacbook},  Chapter 7.

In Section 3   we define the Clifford (resp. Weyl) algebra and their corresponding spin (resp. Weyl) module $F_x$ for the  Lie algebras $x_\infty$ for $x=a,b, d$ (resp $c $) of level 1 (resp $-\frac12$).     The generating series  of the generators  of these  Clifford (resp. Weyl) algebras,   called the (free) fermionic (resp.  bosonic) fields,  admit a vertex operator construction.  Since the generating series of the natural bases of $x_\infty $ can be expressed as the normal ordered product of two such fields, we obtain a vertex operator realization for these  Lie  algebras,  called a bosonization of $F_x$.   

 In Subsection \ref{section 3.5}, and \ref{BRF}  we construct   multicomponent bosonizations of $F_x$.  This idea was introduced  by Date, Jimbo, Kashiwara and Miwa, in a series of papers \cite{DJKM1}-\cite{DJKMVI}  in the 1980's and further developed in \cite{KLmult} and \cite{KvdLB}.  The construction for $c_\infty$,    as far as we know,  is new.
Note that, while for $x=a$ there exists an $s$-component bosonization for each positive integer 
$s$, for $x=b,c$ and $d$ there exists an $(s,r)$-component bosonization for each pair of non-negative $s$ and $r$,
$s+r\ne 0$.

In Section 4 we use these bosonizations to define the hierarchies of equations of KP type  on the  
tau-functions $\tau\in F_x$, called the  KP, BKP, DKP (resp. CKP) hierarchy for $x=a,b,d$ (resp. $c$),
  in terms of the free fields  of the corresponding Clifford (resp. Weyl) algebra.  Using the 
multicomponent bosonizations, we obtain  Hirota bilinear equations  for the tau-function.  Using ideas of \cite{S},
\cite{DJKM1}-\cite{DJKMVI}, \cite{KLmult} and \cite{KvdLB},  we turn these equations into some Lax-Sato equations for certain pseudo-differential operators.  Again, the construction for $x=c$ is new.
There are various CKP hierarchies in the literature.  Our hierarchy coincides with    the one given in \cite{A} (resp.  \cite{LOS}) for $(s,r)=(1,0)$ (resp.  $=(0,1)$).  The CKP hierarchies described in \cite{KZ}, \cite{HHH} and
\cite{AHH} (see also \cite{KvdLC}) are different.   These  authors study a certain restriction of the KP hierarchy,  wich is related to the fixed poins under some involution.  This gives special solutions of the KP hierarchy wich are fixed under this involution.  The CKP tau-function is the square root of this special KP tau-function when all even flows are set to zero.  In particular the level of their $c_\infty$ representation is $1$,  while our (metaplectic) representation  of $c_\infty$ in $F_c$ has level $-\frac12$.

In Section 5    we begin the discussion of reductions of the hierarchies of KP type, which correspond to
restrictions of the representations in $F_x$ of the Lie algebras $x_\infty$ ($x=a,b,c,d$)
to their classical affine subalgebras $\hat{\mathfrak g}$, where $\mathfrak g$ is a classical Lie algebra of type $X$.  The simplest such reductions have been studied before in \cite{DJKM1}-\cite{DJKMVI}.
Our main observation  is that for any conjugacy class in the Weyl group of $\mathfrak g$
one can naturally associate such a reduction.
For this we use the following construction intoduced in \cite{KP112}
for any simple Lie algebra $\mathfrak g$. Let $\mathfrak h$ be its Cartan subalgebra,
$W$ its Weyl group. and $(\cdot|\cdot)$ a non-zero invariant bilinear form on $\mathfrak g$.
 Pick $w$ in a conjugacy class of $W$,  and let $\tilde w$ be a lift of $w$ such that  
 $\tilde w$ is conjugate in $G$ to an element $\sigma$ of order $N$ of
the form $\sigma= e^{2\pi\sqrt{-1}h_w}$,  where $h_w\in\mathfrak h$ 
is orthogonal to the fixed point set of $w$ in $\mathfrak h$.
Let $\mathfrak g=\bigoplus_{\overline j\in\mathbb Z/N\mathbb Z} \mathfrak g_{\overline j}$
be the $\mathbb Z/N\mathbb Z$ gradation of $\mathfrak g$, defined by the eigenspaces of $\sigma$,
 and let 
$
L(\mathfrak g, \sigma)=\bigoplus_{j\in\mathbb Z}t^j \mathfrak g_{j\mod N}
\subset \mathfrak g [t,t^{-1}]
$
be the corresponding equivariant loop algebra.
It is isomorphic to the loop algebra 
$ \mathfrak g [t,t^{-1}]$ via the map
$t^j g\mapsto g(j):=t^{-{\rm ad}\, h_w}(t^{\frac{j}{N} }g)$, 
for $g\in \mathfrak g_{j\mod N}$. 
Then the subalgebra $H_w={\rm span}\{ h(j)|\, h\in \mathfrak h,\ j\in\mathbb Z\}$
is a maximal abelian  subalgebra of $ \mathfrak g [t,t^{-1}]$.
Let $ Z_w$ be the centralizer of $H_w$ in the loop group $G(\mathbb C(t,t^{-1})$.
The main result of \cite{KP112} is that a level 1 representation of $\hat{\mathfrak g}$ is irreducible with respect to the central extensions $\hat H_w $ and $\hat{ Z}_w $,  
provided that $\mathfrak g$ is simply laced.  Since $\hat H_w$ is 
a Heisenberg Lie algebra, this leads to a "coordination" 
of the representation and to a vertex operator construction of 
$\hat{\mathfrak g}$,
associated to the conjugacy class of $w$.
In this section we recall the explicit form of this construction, which was given for $\mathfrak g$ of types $A$ and $D$ in \cite{tKLA} and \cite{tKLD}, and also for the twisted affine Lie algebras of this type in  \cite{tKLtwist}, and for  $\mathfrak g$ of type $B$ in \cite{tKLB}.
In this section we also give a vertex operator construction
of the metaplectic representation (of level $-\frac12$) of $\hat{\mathfrak g}$ for
$\mathfrak g$
of type $C$.
This construction seems to be new.

As a side application of the results of Section 5, we construct in 
Section 6
a map from the conjugacy classes of $W$ to the nilpotent orbits of $\mathfrak g$ for all classical Lie algebras $\mathfrak g$.
It coincides with Lusztig's map for typr $C$, but is different for types $B$ an $D$.

  In Section  7 we present a systematic description
of reductions of the KP,  DKP,  BKP and CKP hierarchies to the affine Lie algebras $\hat{\mathfrak g}$ of type $A$, $D$, $B$ and $C$,
associated to each conjugacy class of the Weyl group of $\mathfrak g$.  We obtain  the corresponding equations on the tau-functions,
construct the   associated dressing operators, Lax operators and auxilary operators. The
Lax operators  are matrix  differential operators for the Lie algebras of type $A$ and $C$; for the Lie algebras of type $B$ and $D$  they are  matrix  pseudo-differential operator  of constrained KP type.
We show that these Lax operators and auxilary operators
satisfy certain Lax-Sato equations, which  in the $x=a$ case are equivalent to the  bilinear equations on the tau-function.

Although some of the above constructions,   except for the case of type $c$,  have been known for many years,  a systematic description of the hierarchies related to these classical affine Lie algebras,  has not been given before.   (The  case of affine Lie algebras of type $a$ was briefly mentioned in \cite{KLmult}.)  

Finally,  in Section 8, we establish this construction for the twisted affine Lie algebras of type $A$ and $D$.
In both  Sections 7 and 8  some cases related to specific conjugacy classes were known,  but as far as we know,  no systematic description of all cases has been given before.


In conclusion of the introduction we would like to propose conjectures about
solutions of bi-Hamiltonian hierarchies of evolution PDE, associated to
Poisson vertex algebras $W(\mathfrak{g},f)$ (see e.g. \cite{DeSole}),
and tau-functions of the $w$-reduced XKP
hierarchies, where $\mathfrak g$ is a classical Lie algebra of type X=A, B, C,
D,
 and $w$ is a representative of a conjugacy class of the Weyl group $W$ of $\mathfrak g$.
\begin{conjecture}
Let $F$ be a map from the set of conjugacy classes of $W$ to nilpotent orbits of $\mathfrak g$,
constructed in Chapter \ref{S6}, and let 
$W(\mathfrak{g},F(w))$
be the classical affine $W$-algebra, associated to
$\mathfrak g$ and the orbit of $F(w)$. Then tau-functions of the $w$-reduced XKP hierarchy , constructed in Chapter \ref{S7},
give solutions of the bi-Hamiltonian hierarchy,
associated with the Poisson vertex algebra $W(\mathfrak{g},F(w))$.
\end{conjecture}

The validity  of this conjecture for the reduced AKP=KP hierarchies was demonstrated 
in \cite{Carpentier}.

A related conjecture for an arbitrary simply laced simple Lie algebra $\mathfrak g$ and its nilpotent element $f$ (rather its adjoint orbit) is as follows:
\begin{conjecture}
Solutions of  $W(\mathfrak{g},f)$-hierarchy can be obtained from the elements of the orbit of the highest weight vector of the $w$-realization 
of the basic representation of  $\hat{\mathfrak g}$, constructed in 
\cite{KP112},  where  $f$ is the image of $w$ under the Kazhdan-Lusztig map (see \cite{Yun}).
\end{conjecture}
Throughout the paper the base field is $\mathbb C$.

\section{Lie algebras 
}
This section more or less follows  \cite{Kacbook}, Chapter 7.
\subsection{Classical Lie algebras}

We denote by $gl_n$  the Lie algebra over $\mathbb C$, consisting of complex    $n\times n$-matrices with the bracket $[a,b]=ab-ba$. It has as basis the matrices $e_{ij}$,  that have a 1 on the $i,j$-th entry and zeros elsewhere.

Let $v_j$ with $1\le j\le n$ be the standard basis vector of $\mathbb C^{n}$. Consider the following symmetric bilinear form on $\mathbb C^{n}$:
\begin{equation}
\label{bilD}
(v_i,v_j)_{so_n}=\delta_{i+j,n+1}.
\end{equation}
The Lie algebra $so_n$ is the subalgebra of $gl_n$, consisting of elements $a\in gl_n$ that satisfy
\begin{equation}
\label{defso}
(a v,w)_{so_n}=-(v, aw)_{so_n}.
\end{equation}
The elements 
\begin{equation}
\label{2a}
e_{ij}-e_{n+1-j,n+1-i}, \qquad 1\le i,j\le n, 
\end{equation}
span $so_{n}$. Note, that the trace of such an element is zero, hence $so_{n}$ is   a subalgebra of $sl_{n}$. For later use, we introduce for $1\le i,j\le \frac{n}2$
\begin{equation}
\label{redefbasis bd}
\begin{aligned}
e_{ij}^{+-}&=
  e_{ ij}-  e_{n+1-j, n+1 -i},\\
 e_{ij}^{++}&=-e_{ji}^{++}= e_{ i,n+1-j}-e_{  j, n+1 -i},\\
 e_{ij}^{--}&=-e_{ji}^{--}= e_{n+1-i, j}- e_{ n+1 -j,   i}.
\end{aligned}
\end{equation}
and when $n$ is odd, also
\begin{equation}
\label{redefbasis b}
e_{i}^{+}= e_{ i, \frac{n+1}2}-  e_{\frac{n+1}2, n+1 -i},\quad 
e_{i}^{-}= e_{ n+1-i,  \frac{n+1}2 }-  e_{\frac{n+1}2, i}.
\end{equation}
Consider the following skewsymmetric bilinear form on $\mathbb C^{2n}$:
\begin{equation}
\label{bilC}
(v_i,v_j)_{sp_{2n}}=(-1)^i\delta_{i+j,2n+1},
\end{equation}
then the elements  $a\in gl_{2n}$ that satisfy condition \eqref{defso}, where we replace the symmetric bilinear form   $(\cdot,\cdot)_{so_{ n}}$ by the skewsymmetric bilinear form $(\cdot,\cdot)_{sp_{2n}}$, form the subalgebra $sp_{2n}$. The elements 
\begin{equation}
\label{2b}
(-1)^j e_{ij}-(-1)^ie_{2n+1-j,2n+1-i},  \qquad 1\le i,j\le 2n,
\end{equation}
span  $sp_{2n}$.
Again, $sp_{2n}$ is  a subalgebra of $sl_{2n}$.
We also introduce for $1\le i,j\le n$
\begin{equation}
\label{redefbasis c}
\begin{aligned}
e_{ij}^{+-}&=
  e_{ ij}- (-1)^{i+j}e_{2n+1-j, 2n+1 -i},\\
 e_{ij}^{++}&=(-1)^{i+j}e_{ji}^{++}= e_{ i,2n+1-j}- (-1)^{ i+j+1 }e_{  j, 2n+1 -i},\\
 e_{ij}^{--}&=(-1)^{ i+j}e_{ji}^{--}= e_{2n+1-i, j}- (-1)^{ i +j+1}e_{ 2n+1 -j,   i}.
\end{aligned}
\end{equation}

For a classical Lie algebra $\mathfrak g$, we choose the non-degenerate symmetric invariant  bilinear form as follows:
\begin{equation}
\label{2.5a}
(g|h)=\gamma_x \rm{Tr}\, gh,\quad g,h\in\mathfrak g,
\end{equation}
where $x=a$, $b$, $c$ or $d$, corresponding to the cases $\mathfrak g=gl_n$, $so_{2n+1}$, $sp_{2n}$ or $so_{2n}$, respectively, and 
\[
\gamma_x=1\quad\mbox{if } x=a\ \mbox{or }c,\ \mbox{and }\gamma_x=\frac12\ \mbox{if }x=b\ \mbox{or }d.
\]
This is the standard choice for which the square of the length of a long root is 2.
 
The standard {\it Cartan subalgebra} $\mathfrak h$ of a classical Lie algebra  $\mathfrak g$  consists of all the diagonal matrices in $\mathfrak g$. It has as orthonormal basis, with respect to the bilinear form \eqref{2.5a}, the following elements ($1\le i\le n$):
\begin{equation}
\label{hh}
\begin{aligned}
\mathfrak g=gl_n:&\quad \epsilon_i= e_{ii},\\
\mathfrak g=so_{2n+1} :&\quad \epsilon_i= e_{ii}- e_{2n+2-i,2n+2-i},\\
\mathfrak g= sp_{2n}:&\quad \epsilon_i= \frac1{\sqrt 2}(e_{ii}-e_{2n+1-i,2n+1-i}),\\
\mathfrak g=so_{2n} :&\quad \epsilon_i= e_{ii}-e_{2n+1-i,2n+1-i}.
\end{aligned}
\end{equation}
One has the following {\it root space decomposition} of $\mathfrak {g}$:
\[
\mathfrak g=\mathfrak h \bigoplus\bigoplus_{\alpha\in\Delta}
\mathfrak g_\alpha,\quad\mbox{where } 
\mathfrak g_\alpha=\{ g\in\mathfrak g\, |\, [h, g]=(\alpha,h)g\ \mbox{for }h\in\mathfrak h \}.
\]
Every root space $\mathfrak g_\alpha$ is 1-dimensional and one has the following {\it set of roots} $\Delta $:
\begin{equation}
\label{roots}
\begin{aligned}
gl_n:&\quad  \{\pm (\epsilon_i-\epsilon_j)\,  |\, 1\le i<j\le n \},\\
 so_{2n+1} : &\quad \{ \pm \epsilon_k,\ \pm (\epsilon_i-\epsilon_j),\ \pm (\epsilon+\epsilon_j)\, |\, 1\le k\le n,\ 1\le i<j\le n \},\\
 sp_{2n } :& \quad\{ \pm 2\epsilon_k,\ \pm (\epsilon_i-\epsilon_j),\ \pm (\epsilon+\epsilon_j)\, |\, 1\le k\le n,\ 1\le i<j\le n \},\\
 so_{2n}:&\quad  \{\pm (\epsilon_i-\epsilon_j),\ \pm (\epsilon+\epsilon_j)\, |\, 1\le i<j\le n \},
\end{aligned}
\end{equation}
and the following root spaces:
\begin{equation}
\label{rootspace}
\begin{aligned}
 gl_n:&\quad 
\mathfrak g_{\epsilon_i-\epsilon_j}=\mathbb C e_{ij};
\\
 so_{2n+1} :&\quad 
\mathfrak g_{\epsilon_i }=\mathbb C e^+_i,\  
\mathfrak g_{-\epsilon_i }=\mathbb C e^-_i,\ 
\mathfrak g_{\epsilon_i-\epsilon_j}=\mathbb C e_{ij}^{+-}, \
\mathfrak g_{\epsilon_i+\epsilon_j}=\mathbb C e_{ij}^{++},
\  
\mathfrak g_{-\epsilon_i-\epsilon_j}=\mathbb C e_{ij}^{--};
 \\
 sp_{2n} :&\quad 
 \mathfrak g_{\epsilon_i-\epsilon_j}=\mathbb C e_{ij}^{+-},\
\mathfrak g_{\epsilon_i+\epsilon_j}=\mathbb C e_{ij}^{++},
\ 
\mathfrak g_{-\epsilon_i-\epsilon_j}=\mathbb C e_{ij}^{--};
\\
 so_{2n} :&\quad 
\mathfrak g_{\epsilon_i-\epsilon_j}=\mathbb C e_{ij}^{+-}, \ 
\mathfrak g_{\epsilon_i+\epsilon_j}=\mathbb C e_{ij}^{++},\ 
\mathfrak g_{-\epsilon_i-\epsilon_j}=\mathbb C e_{ij}^{--}.
\end{aligned}
\end{equation}

Recall that 
  the Weyl group $W$ of $\mathfrak g$ is the subgroup of $GL(\mathfrak h)$  generated by all reflections $r_\alpha$,
$\alpha\in\Delta$ . The elements of $W$ acts on the  orthonormal basis,   given by \eqref{hh}, of the Cartan subalgebra, by permuting these basis vectors for $\mathfrak g=gl_n$. For $\mathfrak g=so_{2n+1}$ or $sp_{2n}$,  elements of the Weyl group   permute these basis vectors and  also change signs of an arbitrary number of them.   For $\mathfrak g=so_{2n}$, the sign changes are restricted only to an even number.  
Thus conjugacy classes of the Weyl group are parametrized by the cycle type for $\mathfrak g=gl_n$, and for the positive and negative cycle type for the other $\mathfrak g$, the number of negative cycles being even for $\mathfrak g=so_{2n}$.
\subsection{Loop algebras and affine Lie algebras}
Let $\mathfrak g$ be a finite-dimensional    Lie algebra. The associated   loop algebra is defined as the Lie algebra  $\tilde{\mathfrak g}=\mathfrak g \otimes \mathbb{C}[t,t^{-1}]$ with the bracket 
\[
[g\otimes t^k,h\otimes t^\ell ]= [g,h]\otimes t^{k+\ell}, \quad g,h\in\mathfrak g, \ k,\ell\in\mathbb Z.
\]
We will usually write $t^kg$ in place of $g\otimes t^k$.
Given  a non-degenerate symmetric invariant bilinear form $ (\cdot|\cdot)$ on $\mathfrak g$, one constructs a Lie algebra central extension  $\hat{\mathfrak g}=\tilde{\mathfrak g}\oplus \mathbb C K$  of $\tilde{\mathfrak g}$  by 1-dimensional center $\mathbb C K$,    with the bracket
\[
[  t^kg+\lambda K,  t^\ell h+\mu K]= t^{k+\ell} [g,h]+k\delta_{k,-\ell}(g|h)K.
\]
In the paper we shall consider $\mathfrak g$ to be one of the classical Lie algebras $gl_n$ (or $sl_n$), $sp_{2n}, so_{2n+1}$ and $so_{2n}$,
with  the bilinear form on $\mathfrak g$ is   chosen   as \eqref{2.5a}.
The Lie algebra $\hat{\mathfrak g}$ with this choice of bilinear form $(\cdot| \cdot)$ is called an {\it affine Lie algebra}.
\\
\ 
\\
We can twist this construction with an automorphism $\sigma$.   Let $\sigma$ be an automorphism of $\mathfrak g$ of finite order $N$, i.e.  $\sigma^N=id$, then this defines a 
$\mathbb Z_N=\mathbb Z/ N\mathbb Z$-gradation of $\mathfrak g=\bigoplus_{\overline j\in\mathbb Z_N}\mathfrak g_{\overline j}$,   defined by $\sigma(g_{\overline j})= e^{\frac{2\pi j}{N}}g_{\overline j}$ for $g_{\overline j}\in \mathfrak g_{\overline j}$.

Define the twisted affine Lie algebra  $ \hat L(\mathfrak g, \sigma)$ as the following  subalgebra of $\hat{\mathfrak g}$:
\begin{equation}
\label{2.4}
\hat L(\mathfrak g, \sigma)=\mathbb C K\oplus \bigoplus_{j\in\mathbb Z} t^j\mathfrak g_{j\,{\rm mod}\, N} .
\end{equation}
The Lie algebra $\hat{\mathfrak g}$ and $ \hat L(\mathfrak g,\sigma)$  are isomorphic for any inner automorphism $\sigma$ of $\mathfrak g$, but not for an outer automorphism (see \cite{Kacbook}, Chapter 8).  Let $J_n= \sum_{k=1}^n  e_{j,n+1-j}\in GL_n$ and $O$ be the matrix of transposition of $v_n$ and $v_{n+1}$. Define $\sigma \in {\rm Aut}\, \mathfrak g$ by
\begin{equation}
\label{out}
\begin{aligned}
&\sigma (g)=- {\rm Ad}\, J_n( g^t)= -J_n g^t J_n,\quad\mbox{ for }g\in\mathfrak g=gl_n\quad\mbox{and }\\
 &\sigma (g)={\rm Ad}\, O( g)=OgO,\quad \mbox{for } g\in\mathfrak g=so_{2n}.
\end{aligned}
\end{equation}
In both cases $\sigma$ is an outer automorphism of order 2,  hence we get the corresponding twisted affine Lie algebra $\hat {\mathfrak g}^{(2)}:=\hat L(\mathfrak g, \sigma)$,
which is not  isomorphic to $\hat{\mathfrak g}$.  

Note that if $\mathfrak g=gl_n$, $\sigma(e_{ij})= -e_{n+1-j, n+1-i}$,  thus 
the fixed point set $(gl_n)_{\overline 0}$ is equal to  exactly  $so_n$ as in \eqref{2a}, i.e.,
\begin{equation}
\label{gl-twist}
\begin{aligned}
e_{ij}-e_{n+1-j,n+1-i}\in &(gl_n)_{\overline 0}, \quad \mbox{and}\\
e_{ij}+e_{n+1-j,n+1-i}\in &(gl_n)_{\overline 1}, \quad 1\le i,j\le n.
\end{aligned}
\end{equation}
For $\mathfrak g=so_{2n}$,  $e_{ij}-e_{2n+1-j, 2n+1-i} \in (so_{2n})_{\overline 0}$ for  $1\le i,j\le 2n$ with both $i,j\ne n$ or $n+1$.
And 
\begin{equation}
\label{so-twist}
\begin{aligned}
e_{in}- e_{n+1,2n+1-i}+e_{i,n+1}-e_{n, 2n+1-i}&\in (so_{2n})_{\overline 0},\\
 e_{in}- e_{n+1,2n+1-i}-e_{i,n+1}+e_{n, 2n+1-i}&\in (so_{2n})_{\overline 1},\qquad 1\le i\le 2n,
\end{aligned}
\end{equation}
in particular $e_{n,n}-e_{n+1,n+1}\in (so_{2n})_{\overline 1}$.
Thus $(so_{2n})_{\overline 0}=so_{2n-1}$.

\subsection{The classical infinite matrix algebras}
\label{clasinf}
Let $\overline a_\infty$ be the Lie algebra consisting of infinite matrices $(a_{ij})_{i,j\in\mathbb Z}$ such that $a_{ij}\in\mathbb C$, $a_{ij}=0$ when $|i-j|>>0$. This Lie algebra has a non-trivial central extension $a_\infty=\overline a_\infty\oplus \mathbb C K$ with the bracket 
\begin{equation}
\label{a1}
[a+\lambda K,b+\mu K]= ab-ba+\omega( a,b)K,\quad a,b\in \overline a_\infty,
\end{equation}
where $\omega$ is a two-cocycle, described below.
 Let $E_{ij}$  be the matrix with a 1 on the $(i,j)$-th entry and zeros elsewhere, then $\omega$ is given by
\begin{equation}
\label{omega}
\omega (E_{ij},E_{ji})=-\omega(E_{ji},E_{ij})=1\quad\mbox{if }i\le 0, j>0,\quad\mbox{and } \omega (E_{ij},E_{k\ell})=0\ \mbox{otherwise,}
\end{equation}
which extends by bilinearity to the whole $\overline a_\infty$.
The Lie algebra  $  a_\infty$ acts naturally  on the space of column vectors $\mathbb C^\infty=\bigoplus_{j\in\mathbb Z} \mathbb C e_j$, by
\[
E_{ij} e_k=\delta_{jk} e_i, \quad K=0,
\] 
where $e_j$ is the column vector with a  1 on the $j$-th entry and zeros elsewhere.

We   define the other classical infinite matrix algebras $\overline b_\infty$, $\overline c_\infty$ and $\overline d_\infty$, by introducing a bilinear form on  $\mathbb C^\infty$ as follows (cf. \eqref{bilD} and
\eqref{defso})
\begin{equation}
\label{bil1}
(e_i, e_j)_b= \delta_{i,-j}, \quad (e_i,e_j)_c=(-1)^i\delta_{i,1-j} ,\quad\mbox{and } (e_i, e_j)_d=\delta_{i,1-j},\quad i,j\in\mathbb Z.
\end{equation}
Then for $x=b,c,d$
\[
\overline x_\infty= \{
 g\in\overline a_\infty\, |\, (gv,w)_x=-(v,gw)_x\ \mbox{for all }v,w\in\mathbb C^\infty
\},
\]
and its central extension is a subalgebra of $a_\infty$:
\[
x_\infty=\overline x_\infty\oplus\mathbb C K, 
\]
where the  bracket is given by \eqref{a1}.

In the $b$, $c$,   $d$  cases  the elements  ($i,j\in\mathbb Z$)
\begin{equation}
\label{6a}
\begin{aligned}
B_{ij}&:= E_{-i,j}-E_{-j,i},\\
 C_{i+\frac12,j-\frac12}&:= (-1)^j E_{-i,j}-(-1)^{i}E_{1-j,1+i}, \\
D_{i+\frac12, j-\frac12} &:=E_{-i,j}-E_{1-j,1+i},  
\end{aligned}
\end{equation}
respectively, span (if we allow infinite sums)  the corresponding Lie algebra $\overline x_\infty$. The Lie algebra $\overline a_\infty $ has the standard {\it  triangular decomposition}
\[
\overline a_\infty =\overline a_\infty^-\oplus \overline a_\infty^0\oplus \overline a_\infty^+,
\]
in a sum of subalgebras, consisting of strictly lower, diagonal and strictly upper triangular matrices. It induces one on $  a_\infty $ by adding $\mathbb CK$ to $\overline a_\infty^0$. This triangular decomposition induces one on all $x_\infty$.

\subsection{Affine Lie algebras  as subalgebras of classical  infinite matrix Lie algebras}

Let $\mathfrak g=gl_n$, then the loop algebra $\tilde{gl}_n$ naturally acts on the loop space $\mathbb C[t,t^{-1}]^n=\mathbb C^n\otimes C[t,t^{-1}]$.
Using  $v_1,v_2,\ldots v_n$,   the standard  basis of $\mathbb C^n$,   we can identify this loop space with $\mathbb C^\infty$ as follows:
\begin{equation} 
\label{standardemb}
e_{kn+j}=t^{-k}v_j, \quad k\in\mathbb Z,\  1\le j\le n,
\end{equation}
so that an element $t^k e_{ij}\in \tilde{gl}_n$ can be identified with an element of $\overline a_\infty$ 
as
\begin{equation}
\label{tkeij}
t^k e_{ij}= \sum_{\ell\in \mathbb Z}E_{(\ell-k)n+i, \ell n+j},
\end{equation}
producing an embedding 
 $\tilde{gl}_n\subset \overline a_\infty$. This  induces an embedding $\hat{gl}_n\subset   a_\infty$.  (It is easy to check that the cocycles match.) 

Let $\mathfrak g=so_{n}$. Then the loop algebra $\tilde{so}_{n}$ naturally acts on the loop space $\mathbb C[t,t^{-1}]^{n}$.
  We can use the 
identification \eqref{tkeij} to obtain 
\begin{equation}
\label{tkeij0}
\begin{aligned}
t^k e_{ij}-t^ke_{n+1-j, n+1-i}&= \sum_{\ell\in \mathbb Z}E_{(\ell-k)n+i, \ell n+j}- \sum_{\ell\in \mathbb Z}E_{(\ell-k+1)n+1-j, (\ell +1)n+1-i}\\
&=\sum_{\ell\in \mathbb Z}(E_{(\ell-k)n+i, \ell n+j}-E_{-\ell n+1-j,- (\ell-k) n+1-i})\\
& =
\sum_{\ell\in \mathbb Z}D_{-(\ell-k)n-i+\frac12, \ell n+j-\frac12}
 .
\end{aligned}
\end{equation}
Due to \eqref{6a}, this makes  $\tilde {so}_{n}$ a subalgebra of $\overline d_\infty$. This embedding induces an embedding of the affine Lie algebra $\hat{so}_n$ in $d_\infty$,  with the cocycle $\gamma_d\omega $, with  $\gamma_d=\frac12$, where $\omega$  is the restriction of the cocycle \eqref{omega}.

The triangular decompositions of $x_\infty$ induce the standard triangular decompositions \cite{Kacbook} on the corresponding affine subalgebras.

We can also choose a  a different identification from \eqref{standardemb} if $n=2m+1$ is odd, namely,  we  identify $\mathbb C[t,t^{-1}]^{2m+1 }$ with $\mathbb C^\infty$ via   
\begin{equation}
\label{2.14a}
 e_{k(2m+1)+j-m-1}=t^{-k}v_j,   \quad k\in\mathbb Z,\  1\le j\le 2m+1.
\end{equation}
Then we obtain the identification  
\begin{equation}
\label{tkeij000}
\begin{aligned}
t^k &e_{ij}-t^ke_{2m+2-j, 2m+2-i}= \\
&=\sum_{\ell\in \mathbb Z}
( 
E_{(\ell-k)(2m+1)+i-m-1, \ell (2m+1)+j-m-1} -  
E_{(\ell-k)(2m+1) +m+1-j, \ell (2m+1) +m+1-i})\\
&=\sum_{\ell\in \mathbb Z}
(
E_{(\ell-k)(2m+1)+i-m-1, \ell (2m+1)+j-m-1}-
E_{-\ell(2m+1) +m+1-j, (k-\ell )(2m+1) +m+1-i})\\
&=\sum_{\ell\in \mathbb Z}
B_{-(\ell-k)(2m+1)-i+m+1, \ell (2m+1)+j-m-1}
 .
\end{aligned}
\end{equation}
Due to \eqref{6a}, this makes   $\tilde {so}_{2m+1}$ a subalgebra of $\overline b_\infty$.  
Recall, that $\gamma_b=\frac12$.  This embedding induces an embedding of $\hat{so}_{2m+1}$ in $b_\infty$ with  "non-standard" cocycle 
\begin{equation}
\label{b-cocycle}
\frac12 \omega (t^k a, t^\ell b)= \frac{k}2\delta_{k,-\ell}\left( {\rm Tr}(ab)+
 \sum_{i=1}^m {\rm Tr}((e_{ii}-e_{2m+2-i,2m+2-i})[a,b])\right).
\end{equation}
Since the second term is a trivial cocycle, this non-standard cocycle is equivalent to $\frac12\omega$.

Finally, let $\mathfrak g=sp_{2n}$. Then the
identification \eqref{tkeij}, for $n$ replaced by $2n$, gives the following identification,
\begin{equation}
\label{cij}
\begin{aligned}
(-1)^j t^k e_{ij}-(-1)^it^ke_{2n+1-j, 2n+1-i}&= \sum_{\ell\in \mathbb Z}(-1)^jE_{2(\ell-k)n+i, 2\ell n+j}-\sum_{\ell\in \mathbb Z}(-1)^iE_{2(\ell-k+1)n+1-j, 2(\ell +1)n+1-i}\\
&=\sum_{\ell\in \mathbb Z}((-1)^jE_{2(\ell-k)n+i, 2\ell n+j}-(-1)^iE_{-2\ell n+1-j,- 2(\ell-k) n+1-i})\\
&=\sum_{\ell\in \mathbb Z}C_{-2(\ell-k)n-i+\frac12, 2\ell n+j-\frac12}
 .
\end{aligned}
\end{equation}
Due to \eqref{6a}, this  makes  $\tilde {sp}_{2n}$ a subalgebra of $\overline c_\infty$.  This embedding induces  an embedding of $\hat{sp}_{2n}$ in $c_\infty$ with the cocycle $\omega$, so that the coefficient of $\omega$ is $\gamma_c=1$.
The triangular decompositions of $x_\infty$ induce triangular decompositions \cite{Kacbook} of the classical affine Lie algebras  $\hat{\mathfrak g}$.

Using \eqref{bil1} and identifications  \eqref{standardemb} and \eqref{2.14a}
we obtain the following
\begin{lemma}
\label{Lemma1}
We have the following bilinear forms on the loop space $\mathbb C[t,t^{-1}]^N$, restricted from $\mathbb C^\infty$:
\[
\begin{aligned}
	(t^kv,t^\ell w)_b=&\delta_{k+\ell,1}(v,w)_{so_{2n+1}},\quad&\ \mbox{embedding \eqref{tkeij},}&\ N=2n+1,\\
(t^kv,t^\ell w)_c=&\delta_{k+\ell,1}(v,w)_{sp_{2n}},\quad&\ \mbox{embedding \eqref{tkeij},}&\ N=2n,\\
(t^kv,t^\ell w)_d=&\delta_{k+\ell,1}(v,w)_{so_{2n}},\quad&\ \mbox{embedding \eqref{tkeij},}&\ N=2n,\\
(t^kv,t^\ell w)_b=&\delta_{k+\ell,0}(v,w)_{so_{2n+1}},\quad  &\ \mbox{embedding \eqref{2.14a},} &\ N=2n+1,
\end{aligned}
\]
such that  for $x=b,\ c,$ or $d$   we have
\[
\hat {\mathfrak g}=\{ g\in\hat{gl}_N\,  |\, (gv,w)_{x}=-(v,gw)_{x}\ \mbox{for all }v,w\in\mathbb C[t,t^{-1}]^N \}.
\]
\end{lemma}
\subsection{Twisted affine Lie algebras  as subalgebras of classical  infinite matrix Lie algebras}
\label{twistaf}
In this subsection,   we  embed the twisted classical affine Lie algebras $\hat{\mathfrak g}^{(2)}$, for $\mathfrak g=gl_n$ and $so_{2n}$,  into either 
$d_\infty$  (the former case with $n$ even) or  $b_\infty$ (the other cases).  We proceed as in the non-twisted  case,
and relate the basis $e_j\in\mathbb C^\infty$,  $j\in\mathbb Z$,  with the basis of a subspace of the loop space $(\mathbb C[t, t^{-1}])^n$ (resp.  $(\mathbb C[t, t^{-1}])^{2n}$) for $\mathfrak g=gl_n$ (resp.  $\mathfrak g=so_{2n}$),  which is invariant under  $ \hat{\mathfrak g}^{(2)}$.  We choose a basis of $\mathbb C^n$  (resp.  $\mathbb C ^{2n}$)  consisting of eigenvectors of 
$J_n$ (resp.  $O$),  this gives 
 the following identification of  $\mathbb C^\infty$ and the subspace of the loop space:
\[
\begin{aligned}
gl_{2m}:\quad &e_{k(2m)+j-m}=c_{k}\frac{ t^{-k}}{\sqrt 2}(v_j+(-1)^kv_{2m+1-j}),\\
gl_{2m+1}:\quad &e_{k(2m+1)+j-m-1}=c_k\frac{t^{-k}}{\sqrt 2}(v_{j}+(-1)^kv_{2m+2-j}).
\\
so_{2n}:\quad &e_{2nk+i-n}= t^{-2k}v_i, \quad e_{2nk+n-i}= t^{-2k}v_{2n+1-i}, \qquad i=1,\ldots , n- 1  \\
 &e_{2nk}=\frac{t^{-2k}}{\sqrt 2}(v_n+v_{n+1}),\quad\\
&e_{2nk+n}=(-1)^k\frac{t^{-2k-1}}{\sqrt 2}(v_n-v_{n+1}),
\end{aligned}
\]
where
\begin{equation}
\label{bk}
c_k=\begin{cases}
1&\mbox{if }k\ \mbox{even},\\
\sqrt{-1}&\mbox{if } k\ \mbox{odd}.
\end{cases}
\end{equation}
Then we obtain the following identification:
\\
\
\\
{ $\hat{gl}_{2m}^{(2)}\subset d_\infty$:}
\begin{equation}
\label{twist2m}
\begin{aligned}
  t^{k}(e_{ij}-(-1)^k e_{2m+1-j,2m+1-i}) =
\sum_{\ell\in\mathbb Z}     (c_{\ell-k})^{-1}c_\ell D_{-2m(\ell-k)-i+m+\frac12,2m\ell+j-m-\frac12};
\end{aligned}
\end{equation}
{ $\hat{gl}_{2m+1}^{(2)}\subset b_\infty$:}
\begin{equation}
\label{twist2m+1}
\begin{aligned}
  t^{k}(e_{ij}-(-1)^ke_{2m+2-j,2m+2-i})=
\sum_{\ell\in\mathbb Z}(c_{\ell-k})^{-1}c_\ell B_{-(2m+1)(\ell-k)-i+m+1,(2m+1)\ell+j-m-1} ;
\end{aligned}
\end{equation}
{ $\hat{so}_{2n}^{(2)}\subset b_\infty$: ($1
\le i,j<n$)}
\begin{equation}
\label{twistD}
\begin{aligned}
t^{2k}(e_{ij}-e_{2n+1-j,2n+1-i})&=\sum_{\ell\in\mathbb Z}B_{-2n(\ell-k)+n-i,2n\ell+j-n},\\
t^{2k}(e_{i,2n+1-j}-e_{ j,2n+1-i})&=\sum_{\ell\in\mathbb Z}B_{-2n(\ell-k)+n-i,2n\ell+n-j},\\
t^{2k}(e_{2n+1-i,j}-e_{2n+1-j,i})&=\sum_{\ell\in\mathbb Z}B_{-2n(\ell-k)+i-n,2n\ell+j-n},\\
\frac{t^{2k}}{\sqrt 2}(e_{in}+ e_{i,n+1}-e_{n,2n+1-i}-e_{n+1,2n+1-i})&
=\sum_{\ell\in\mathbb Z}B_{-2n(\ell-k)+n-i,2n\ell},\\
\frac{t^{2k}}{\sqrt 2}(e_{2n+1-i,n}+ e_{2n+1-i,n+1}-e_{n,i}-e_{n+1,i})&
=\sum_{\ell\in\mathbb Z}B_{-2n(\ell-k)+i-n,2n\ell},\\
\frac{t^{2k+1}}{\sqrt 2}(e_{in}-e_{i,n+1}+e_{n,2n+1-i}-e_{n+1,2n+1-i})&
=(-1)^k\sum_{\ell\in\mathbb Z}B_{-2n(\ell-k)+n-i,2n\ell+n},\\
\frac{t^{2k+1}}{\sqrt 2}(e_{2n+1-i,n}- e_{2n+1-i,n+1}+e_{n,i}-e_{n+1,i})&
=(-)^k\sum_{\ell\in\mathbb Z}B_{-2n(\ell-k)+i-n,2n\ell +n},\\
t^{2k+1}(e_{nn}-e_{n+1,n+1})&=(-1)^{k}\sum_{\ell\in\mathbb Z}B_{-2n(\ell-k),2n\ell+n}.
\end{aligned}
\end{equation}
Note that in all cases the matrices are periodic of period $4m, \ 4m+2,\  2n$ for  $\hat{gl}_{2m}^{(2)}$,  $\hat{gl}_{2m+1}^{(2)}$, $\hat{so}_{2n}^{(2)}$, respectively. 
\begin{remark}
\label{Jremark1}
 The automorphism $\sigma$ induces an automorphism on $d_\infty$ (resp. $b_\infty$),
which is given by conjugation by $J$ for  $\hat{gl}_{n}^{(2)}$,   and $O$  for 
$\hat{so}_{2n}^{(2)}$,
where 
\begin{equation}
\label{Jmatrix}\begin{aligned}
J=& \sum_{k\in \mathbb Z} (-1)^k \sum_{j=1}^{2m}  E_{k(2m)+j-m, k(2m)+j-m}, 
\quad n=2m,
\\  
\big(\mbox{resp.  }J=&\sum_{k\in \mathbb Z} (-1)^k \sum_{j=1}^{2m+1} E_{k(2m+1)+j-m+1, k(2m+1)+j-m+1}, \quad n=2m+1\big),\\
O=&\sum_{k\in \mathbb Z}  
\sum_{ j=1 }^{2n} (-1)^{\delta_{jn}}E_{2kn+j,2kn+j}
.
\end{aligned}
\end{equation}
Note that the matrices $(a_{ij})_{i,j\in\mathbb Z}$ that correspond to an element of $\hat{gl}_n^{(2)}$    satisfy 
\begin{equation}
\label{Jperiodic}
a_{i+pn,j+pn}=J^pa_{i,j}J^p= c_i^{2p}c_j^{2p}a_{ij} ,\quad p\in \mathbb Z .
\end{equation}
\end{remark}

As before, we can restrict the cocycle of $x_\infty$ to $\hat{\mathfrak g}^{(2)}$.
One can check that this gives the  cocycle \eqref{b-cocycle}    for $\hat{gl}_{2m+1}^{(2)}$,  and the following cocycle in the remaining cases:
\[
\begin{aligned}
\hat{gl}_{2m}^{(2)}:&\qquad\frac12 \omega (t^k a, t^\ell b)= \frac{k}2\delta_{k,-\ell}\left( {\rm Tr}(ab)+
 \sum_{i=1}^m {\rm Tr}((e_{ii}-e_{2m+1-i,2m+1-i})[a,b])\right),\\
\hat{so}_{2n}^{(2)}:&\qquad\frac12 \omega (t^k a, t^\ell b)= \frac{k}2\delta_{k,-\ell}\left( {\rm Tr}(ab)+
 \sum_{i=1}^{n-1} {\rm Tr}((e_{ii}-e_{2n+1-i,2n+1-i})[a,b])\right).\
\end{aligned}
\]
\begin{lemma}
\label{Lemma1twist}
We have the following bilinear forms on the subspace of the loop space $\mathbb C[t,t^{-1}]^N$,    restricted from $\mathbb C^\infty$:
\[
\begin{aligned}
(t^kv,t^\ell w)_{ \hat{gl}_{2n}^{(2)}}=&(-1)^k\delta_{k,\ell}(v,w)_{so_{2n}},\quad &N={2n},\\
(t^kv,t^\ell w)_{\hat{gl}_{2n+1}^{(2)}}=&(-1)^k\delta_{k,\ell}(v,w)_{so_{2n+1}},\quad &N=2n+1,\\
(t^kv,t^\ell w)_{\hat{so}_{2n}^{(2)}}=&(-1)^k\delta_{k,\ell}(Ov,w)_{so_{2n}},\quad &N=2n.
\end{aligned}
\]
\end{lemma}

\section{Representations of classical infinite matrix Lie algebras}
In this section we will recall the spin and Weyl modules for the classical infinite matrix algebras. These modules will be fundamental for the description of representations of  the classical  affine Lie algebras, used in this paper.
\subsection{The spin module for $a_\infty$}
Consider the Clifford algebra $C\ell$  as a unital associative algebra  with generators $\psi_i^\pm$,  $i\in\frac12 +\mathbb Z$,   and relations
\begin{equation}
\label{rel1}
\psi^{\lambda}_{i} \psi^{\mu}_{j} + \psi^{\mu}_{j}
\psi^{\lambda}_{i} = \delta_{\lambda ,-\mu} \delta_{i,-j},\quad
i,j\in\frac12+\mathbb Z,\ \lambda, \mu=+\ \mbox{or }-.
\end{equation}
Its corresponding spin module $F=F_a$  is defined as the irreducible  $C\ell$-module, which admits  the vacuum vector $|0\rangle$, subject to the relation
$$\psi^{\pm}_{j} |0\rangle = 0 \ \text{for}\ j > 0 .$$ 
 
%

The well-known  representation   $r$ of $a_\infty$ on $F$ is defined by
\begin{equation}
\label{rEijk}
r(K)=I,\quad r(E_{ij}) = :\psi^{+}_{-i+\frac12} \psi^{-}_{j-\frac12 }:,\quad i,j\in\mathbb Z, 
\end{equation}
where  as usual the  normally  ordered product,  $:\psi_i\psi_j:$ is defined as equal to $\psi_i\psi_j$ if $i\le j$ and $-\psi_j\psi_i$ otherwise.
Define the charge  on $F$ by setting 
\[
 \text{charge}(|0\rangle ) = 0\ \text{and charge} (\psi^{\pm}_{j}) =
\pm 1. 
\]
Then $F$ decomposes into charge sectors 
\begin{equation}
\label{3.3}F = \bigoplus_{m \in {\mathbb Z}} F^{(m)}, \quad \text{
where }
F^{(m)}=\{ v\in F\, |\, \text{charge} (v)=m\}.
\end{equation}
Each subspace $F^{(m)}$ is an irreducible highest weight
$a_{\infty}$-module \cite{KP}, \cite{Kacbook}, for which 
\[
|\pm m\rangle=
\psi^\pm _{ \frac12-m}\psi^\pm _{ \frac32-m}\cdots \psi^\pm _{- \frac12}|0\rangle,\qquad m\in\mathbb Z_{\ge 0},
\]
 is its highest weight vector, i.e. 
$$
r(E_{ij})|\pm m\rangle = 0 \ \text{for}\ i <  j. 
$$
The highest weight is  given by  $r(E_{ii})|m\rangle=|m\rangle$ for $1\le i\le m$,      $r(E_{ii})|-m\rangle=-|-m\rangle$ for $1-m \le i\le 0$, if $m\ge 1$, and $r(E_{ii}|m\rangle=0$ in all other cases.

The elements 
\[
\psi_{i_1}^+\psi_{i_2}^+\cdots\psi_{i_p}^+\psi_{j_1}^-\psi_{j_2}^-\cdots \psi_{j_p}^-|m\rangle,
\]
with  $i_1<i_2<\cdots <i_p< -m $ and  $j_1<j_2<\cdots <j_p< m$ form a basis of $F^{(m)}$.

Finally, using the embedding \eqref{tkeij},  the spin module $F$  restricted to $a_\infty$ is $\hat{gl}_n$-module, such that  each $F^{(m)}$ is an irreducible  level  $r(K)=1$   highest weight $\hat{gl}_n$-module \cite{KP}. 
\subsection{The spin module for $b_\infty$ and $d_\infty$}
\label{spinBD}
Recall the bilinear forms \eqref{bil1} and the description of $x_\infty$ for $x=b$ and  $ d$ in Subsection \ref{clasinf}. Then 
the elements $B_{jk}=E_{-j,k}-E_{-k,j}=- B_{kj}$   with $j>k$, 
$C_{j+\frac12, k-\frac12}=(-1)^kE_{-j,k}-(-)^jE_{-k+1,j+1}=C_{k-\frac12, j+\frac12} $,
$D_{j+\frac12, k-\frac12}=E_{-j,k}-E_{-k+1,j+1}=- D_{k-\frac12, j+\frac12} $ with $j\ge k$,  given by \eqref{6a} together with $ K$,
 form a basis of $\oi$,  $c_\infty$, $\di$, respectively.
Note that the representation $r$ restricted to $x_\infty$, is given by
\[
\begin{aligned}
r(B_{jk})& =:\psi^+_{j+\frac12}\psi^-_{k-\frac12}-  \psi^+_{k+\frac12}\psi^-_{j-\frac12}:,\  &r(K )=I;\\
r (C_{j+\frac12, k-\frac12})&=(-1)^k:\psi^+_{j+\frac12}\psi^-_{k-\frac12}- (-1)^j\psi^+_{k-\frac12}\psi^-_{j+\frac12}:,\ &r(K )=I;\\
r (D_{j+\frac12, k-\frac12})&=:\psi^+_{j+\frac12}\psi^-_{k-\frac12}- \psi^+_{k-\frac12}\psi^-_{j+\frac12}:,\ &r(K )=I.
\end{aligned}
\]
This suggests to define   involutions $\iota_b$ and $\iota_d$ and an automorphism $\iota_c$ of order 4 on the Clifford algebra $C\ell$   (which respects the relations \eqref{rel1}):
\begin{equation}
\label{anti}
\begin{aligned}
\iota_b(\psi^+_{j+\frac12})&=  \psi^-_{j-\frac12},\quad&\iota_b( \psi^-_{k-\frac12})=  \psi^+_{k+\frac12},\\
\iota_c(\psi^+_{j+\frac12})&= (-1)^j\psi^-_{j+\frac12},\quad&\iota_c( \psi^-_{k-\frac12})=(-1)^k \psi^+_{k-\frac12},\\
\iota_d(\psi^+_{j+\frac12})&= \psi^-_{j+\frac12},\quad&\iota_d( \psi^-_{k-\frac12})= \psi^+_{k-\frac12}.
\end{aligned}
\end{equation}
This induces via $r$ the following   involution on $\overline a_\infty$
$$\iota_b(E_{jk})=-E_{-k,-j},
\quad 
\iota_c(E_{jk})=-(-1)^{j+k} E_{-k+1,-j+1},
\quad \mbox{and }
\iota_d(E_{jk})=-E_{-k+1,-j+1}, $$
thus
$$ x_\infty =\{g\in a_\infty |\iota_x(g)=g\}\quad \mbox{for } x=b, \ c\ \mbox{or } d.$$

Introduce  the following elements of $C\ell$:
\begin{equation}
\label{phi-hat-phi}
\begin{aligned}
\tilde\phi_i&= \frac{\psi_{i+\frac12}^++ \psi_{i-\frac12}^-}{\sqrt 2} \quad \mbox{for } i\in \mathbb{Z},\\
\phi_i&= \frac{\psi_{i}^++\psi_{i}^-}{\sqrt 2}    \quad \mbox{for } i\in\frac12+\mathbb{Z}.
\end{aligned}
\end{equation}
They generate the Clifford algebra  $C\ell_x$, where $x=b$ or $d$, subject to 
relations
\begin{equation}
\begin{split}
\tilde\phi_i\tilde\phi_j+\tilde\phi_j\tilde\phi_i= \delta_{i,-j}, \ i,j\in\mathbb{Z},\quad \mbox{if }x=b,\\
\phi_i\phi_j+\phi_j\phi_i= \delta_{i,-j},  \ i,j\in\frac12+\mathbb{Z},\quad  \mbox{if }x=d.
\end{split}
\end{equation} 
We define  the  spin module $F_x$ over $C\ell_x$ for $x=b$ and $d$,   and its dual  respectively, generated by the vacuum vector $|0\rangle$  (resp. $\langle 0|$), subject to the conditions:
\begin{equation}
\label{1_B}
\begin{split}
\tilde\phi_j|0\rangle &=\phi_j|0\rangle=0 \ \mbox{for }j>0\  (\mbox{resp. } \langle 0|\tilde\phi_j =\langle 0|\phi_j =0  \ \mbox{for }j<0),\\
\tilde\phi_0|0\rangle &=\frac{1}{\sqrt 2}|0\rangle\   (\mbox{resp. }\langle 0| \tilde\phi_0 =\frac1{\sqrt 2}\langle 0|).
\end{split}
\end{equation}
The elements $\tilde\phi_{j_1}\tilde\phi_{j_2}\cdots\tilde\phi_{j_p}|0\rangle$ (resp.
$\phi_{j_1}\phi_{j_2}\cdots\ \phi_{j_p}|0\rangle$)
with $j_1<j_2<\cdots <j_p< 0$ form a basis of $F_b$ (resp. $F_d$).
We define the spin  representation $r_b$ of $b_\infty$ in $F_b$,  and $r_d$ of $d_\infty$ in $F_d$, by   
\begin{equation}
\label{rphi}
\begin{aligned}
r_b(B_{jk})& =  :\tilde\phi_j\tilde\phi_k :
\ \mbox{for } i,j\in\mathbb{Z},\quad r_b(K)=\frac12 I\ (=\gamma_b I),\\  
r_d (D_{j k})&=:\phi_j\phi_k:
  \ \mbox{for } i,j\in\frac12+\mathbb{Z},\quad r_d(K)=\frac12 I\ (=\gamma_dI).
\end{aligned}
\end{equation}
Then $F_b$ is an irreducible highest weight  $b_\infty$-module with highest weight vector $|0\rangle$, but $F_d$ splits into two irreducible highest weight  $d_\infty$-modules $F_d^{\overline 0}\oplus F_d^{\overline 1}$, defined by the $\mathbb Z_2$-gradation given by ${\rm deg}(|0\rangle)=\overline 0$ and ${\rm deg}(\tilde\phi_i)=\overline 1$. The highest weight vector of  $F_d^{\overline 0}$ is  $|\overline 0\rangle=|0\rangle$ and of $ F_d^{\overline 1}$ is  $|\overline 1\rangle=\sqrt 2 \tilde\phi_{-\frac12}|0\rangle $.

The embeddings \eqref{tkeij0} of $\hat{so}_n$ in $d_\infty$ (resp.  \eqref{tkeij000}
of $\hat{so}_{2m+1}$ in $b_\infty$)  give, via restriction of the $x_\infty$ spin modules, 
 irreducible highest weight representations of $\hat{so}_{n}$ with highest weights $\Lambda_0$ and $\Lambda_1$ of $F_d^{\overline 0}$, respectively $ F_d^{\overline 1}$ (resp. an irreducible highest weight representation of $\hat{so}_{2m+1}$ with highest weight $\Lambda_m$)  \cite{tKLD}, \cite{tKLB}. Hereafter  the $\Lambda_j$ denote the fundamental weights of $\hat{\mathfrak g}$.
\begin{remark}
If we choose another embedding for $\tilde{so}_{2m}$, via the identification
\[
 e_{2km+jmn}=t^{-k}v_j,   \quad k\in\mathbb Z,\  1\le j\le 2m,
\]
we obtain an irreducible representations of $\hat{so}_{2m}$ with other highest weights, viz.   $\Lambda_m$ and $\Lambda_{m-1}$ of $F_d^{\overline 0}$, respectively $ F_d^{\overline 1}$. Since, the description in this case is similar, one can use an outer automorphism of the Lie algebra  to interchange these representations with the ones with highest weights  $\Lambda_0$ and $ \Lambda_1$. Hence, we will not describe this case here.
In the twisted cases, the embeddings  \eqref{twist2m+1} and \eqref{twistD} in $b_\infty$, give an irreducible representation  of $\hat{gl}_{2m+1}^{(2)}$,  resp.  $\hat{so}_{2n}^{(2)}$,
with highest weight $\Lambda_0$,  resp.  $\Lambda_{n-1}$.  The embedding \eqref{twist2m} of $\hat{gl}_{2m}^{(2)}$ into $d_\infty$ gives two irreducible highest weight representations of $\hat{gl}_{2m}^{(2)}$ with highest weights $\Lambda_0$ and $\Lambda_1$ in  $F_d^{\overline 0}$, respectively $ F_d^{\overline 1}$.  Again,  as for  $\hat{so}_{2m}$, we can define also for $\hat{so}_{2n}^{(2)}$ a different embedding, which produces the irreducible representation with highest weight $\Lambda_0$. However,  also here we can apply an outer automorphism wich interchanges this representation with the representation with highest weight $\Lambda_{n-1}$. 
\end{remark}
\begin{remark}
\label{Jremark2}
The automorphisms $\sigma$,  which define the twisted affine lie algebras,  induce an automorphism on  $x_\infty$ for $x=b$  or $d$.
Viewing $x_\infty$ as subalgebra of $a_\infty$,   we find that $\sigma (E_{ij}) =JE_{ij }J$ for $\hat{\mathfrak g}^{(2)} =\hat{gl}_n^{(2)}$,  where $J$ is given by \eqref{Jmatrix} and $\sigma (E_{ij}) =OE_{ij }O$ for $\hat{\mathfrak g}^{(2)} =\hat{so}_{2n}^{(2)}$,  where $O$  is given by \eqref{Jmatrix}.
 We can extend this automorphism to an  automorphism on the Clifford algebra $C\ell_d$ (resp $C\ell_b$),  which we denote by $\sigma $, viz.  $\sigma (\phi_j)=J\phi_j=\sigma_j \phi_j$,  for $j\in\frac12+\mathbb Z$ (resp.  $\sigma (\tilde \phi_i)=J  \phi_i=\sigma_i \tilde \phi_i$ for $i\in\mathbb Z$).  Note, that $J$ acts as matrix multiplication from the left on a column vector $\phi$. More explicitly,
\[
\begin{aligned}
\hat{gl}_{2m}^{(2)}:&\quad \sigma(\phi_{k(2m)+j-m-\frac12})=c_k^2\phi_{k(2m)+j-m-\frac12}, \quad k\in\mathbb Z, \  j=1,\ldots, 2n,\\
\hat{gl}_{2m+1}^{(2)}:&\quad\sigma (\tilde\phi_{k(2m+1)+j-m+1})=c_k^2\tilde\phi_{k(2m+1)+j-m+1},\quad k\in\mathbb Z, \  j=1,\ldots, 2n+1,\\
\hat{so}_{2n}^{(2)}:&\quad  \sigma (\tilde\phi_{k(2n)+j})=\begin{cases}
\tilde\phi_{k(2n)+j}& j=1,\ldots,n-1,n+1,\ldots 2n,\\
(-1)^k\tilde\phi_{k(2n)+n}&j=n,
\end{cases}\  k\in \mathbb Z.
\end{aligned}
\]
where the $c_k$ are given by \eqref{bk}.  Observe,  for later calculations that $\sigma_i=\sigma_{-i}$ This automorphism can be extended to the spin module by $\sigma(|0\rangle)=|0\rangle$,   since it leaves the annihilation space of $|0\rangle$ (which is the subspace of $\mathbb C^{\infty}$consisting of all the elements $\phi$ such that $\phi |0\rangle=0$) invariant.
\end{remark}

\subsection{The Weyl module for $c_\infty$ (cf. \cite{DJKMVI})}  
\label{spinC}
Introduce
free symplectic bosons    $b_i$,   $i\in\frac12+\mathbb Z$, that form the Weyl algebra $C\ell_c$ with
commutation relations
\[
b_i b_j- b_j b_i=(-1)^{i-\frac12}\delta_{i,-j},\quad i,j\in \frac12+\mathbb Z.
\]
Define the space $F_c$ as an irreducible $C\ell_c$-module,  generated by the  vacuum vector $|0\rangle$ (resp. its dual by  $\langle  0|$) subject to conditions
\[
b_j|0\rangle=0\qquad(\mbox{resp. }\langle 0|b_{-j}=0)\quad\mbox{for }j>0.
\]
The elements 
\[
b_{j_1} b_{j_2} \cdots b_{j_p} |0\rangle,\quad \mbox{with } j_1\le j_2\le \cdots j_p  <0 ,
\]
form a basis of $F_c$. Define a representation of  $c_\infty$ in $F_c$, by  
\begin{equation}
\label{cijc}
r_c(C_{ij})=
 :b_ib_j  : \ \mbox{for } i,j\in\frac12+ \mathbb{Z},\quad r_c (K)=- \frac12 I.
\end{equation}
As in the case of $F_d$, the $c_\infty$-module $F_c$ splits in two irreducible  modules with highest weights $-\frac12 \Lambda_0$ and $-\frac32 \Lambda_0+\Lambda_1$in $F_c^{\overline 0}$, respectively $ F_c^{\overline 1}$.
The embedding \eqref{cij} gives a  highest weight representations of $\hat{sp}_{2n}$ with the same highest weight vectors as for $c_\infty$.
\subsection{Generating fermionic and bosonic fields}
\label{sec3.2}
In view of Sections 3.1-3.3,
introduce the following generating fields of fermions and symplectic bosons
\begin{equation}
\label{FIELDS}
\begin{aligned}
\psi^{\pm}(z)&=\sum_{j\in\frac12+\mathbb Z}\psi_j^{\pm} z^{-j-\frac12}&(\mbox{charged free fermions}),\\
\tilde\phi(z)&=\sum_{j\in\mathbb Z}\tilde\phi_j z^{-j}&(\mbox{twisted neutral free fermions}),\\
b(z)&=\sum_{j\in\frac12+\mathbb Z}b_j z^{-j-\frac12}&(\mbox{neutral  free symplectic bosons}),\\
\phi(z)&=\sum_{j\in\frac12+\mathbb Z}\phi_jz^{-j-\frac12}&(\mbox{neutral free fermions}).
\end{aligned}
\end{equation}
In terms of these fields we can define generating series for the representation of $x_\infty$ in $F_x$  for $x=a,b,c,d$, viz.
\begin{equation}
\label{xFIELDS}
\begin{aligned}
a_\infty:&\qquad A(y,z)=\sum_{i,j\in\mathbb Z}  r_a(E_{ij}) y^{i-1}z^{-j}=:\psi^{+}(y)\psi^-(z):,\\
b_\infty:& \qquad B(y,z)= \sum_{i,j\in\mathbb Z} r_b(B_{ij})y^{-i}z^{-j}=:\tilde\phi(y) \tilde\phi(z):,\\
c_\infty:&\qquad C( y,z)= \sum_{i,j\in\frac12+\mathbb Z} r_c(C_{ij})y^{-i-\frac12}z^{-j-\frac12}=:b(y) b(z):,\\
d_\infty:&\qquad D(y,z)= \sum_{i,j\in\frac12+\mathbb Z} r_d(D_{ij})y^{-i-\frac12}z^{-j-\frac12}=:\phi(y) \phi(z):.
\end{aligned}
\end{equation}
Unfortunately, the fermionic fields that define $b_\infty$ and $d_\infty$ are not so convenient if one wants to express them in terms of vertex operators. 
For that reason we   define new fields.   In the case of $b_\infty$ and $d_\infty$  let
\[
\tilde\phi^1(z)=\sum_{j\in\mathbb Z}\tilde\phi^1_j z^{-j},
\]
where 
\begin{equation}
\label{redeffermion}
\begin{aligned}
\mbox{for }b_\infty:&\quad \tilde\phi^1_{2j}= \tilde\phi_{2j},\quad  \tilde\phi_{2j+1}^1= \sqrt{-1} \tilde\phi_{2j+1},\quad j\in\mathbb Z;\\
\mbox{for }d_\infty:&\quad   
\tilde\phi^1_0= \frac1{\sqrt 2}(\phi_{\frac12}+\phi_{-\frac12}),\quad 
\tilde\phi_{-2j}^1= \phi_{-2j-\frac12},\quad   \tilde\phi_{-2j+1}^1= \phi_{-2j+\frac12},\quad j\in\mathbb Z_{>0}.\\
&\quad  \tilde\phi_{2j}^1=\phi_{2j+\frac12} ,\quad  \tilde\phi^1_{2j-1}=\sqrt{-1} \phi_{2j-\frac12 } , \quad (j\in\mathbb Z_{>0}),
\quad \sigma_0=\frac{\sqrt{-1}}{\sqrt 2}(\phi_{\frac12}-\phi_{-\frac12}).
\end{aligned}
\end{equation}
Then these elements satisfy the anti-commutation relations \eqref{comb} of the twisted neutral fermionic field  of the Introduction:
\[
\tilde\phi^1_j\tilde\phi^1_k+\tilde\phi^1_k\tilde\phi^1_j=(-1)^j\delta_{j,-k}, \quad  \sigma_0^2=\frac12\quad\mbox{and }\sigma_0\tilde\phi^1_j+\tilde\phi^1_j\sigma_0=0
\]
and 
\begin{equation}
\label{phi-sigma}
(\tilde\phi_0^1-\sqrt{-1}\sigma_0)|0\rangle   =0,  \qquad  \tilde\phi^1_j|0\rangle= 0,\quad j>0.
\end{equation}
Letting $\tilde \phi^1(z)=\sum_{j\in z}\tilde\phi_j z^{-j}$, we find the following generating series for the basis of
\[
\begin{aligned}
b_\infty:&\quad E^{11}(y,z)=:\tilde\phi^1(y)\tilde\phi^1(z):,\\
d_\infty:&\quad E^{11}(y,z)=:\tilde\phi^1(y)\tilde\phi^1(z):,\quad\mbox{and }E^1(z)=:\tilde\phi^1(z)\sigma_0:.
\end{aligned}
\]
Note that in the latter case  the modes of $E^{11}(y,z)$ form a $b_\infty$  subalgebra inside $d_\infty$.

\subsection{Multicomponent realizations of $F_x$ and $x_\infty$}
\label{section 3.5}
It is possible to relabel the generators of the Clifford and Weyl algebra to obtain multicomponent realizations of the module $F_x$  for $x=a,b,c$ and $d$.
\\
\
\\
\indent In the case of $a_\infty$,  we can relabel in such a way that one obtains instead of one pair of  fields $\psi^{\pm}(z)$,  $s$ pairs   of such fields, where $s$ is a positive integer.
The simplest such a  relabeling is \eqref{relabela}  (cf.  \cite{KLmult}, p. 3253):
\begin{equation}
\label{3.15a}
\psi^{\pm a}_j=\psi^{\pm}_{sj\pm \frac12(s-2a+1)},\quad a=1,\ldots ,s,
\end{equation}
for which the subscripts on the right-hand side are $\equiv \pm \frac12 (1-2a)$.
Then we 
have
\begin{equation}
\label{aCOM}
\psi^{+a}_i\psi^{+b}_j+\psi^{+b}_j\psi^{+a}_i=\psi^{-a}_i\psi^{-b}_j+\psi^{-b}_j\psi^{-a}_i=0,\quad
\psi^{+a}_i\psi^{-b}_j+\psi^{-b}_j\psi^{-a}_i=\delta_{ab}\delta_{i,-j},
\end{equation}
and
\begin{equation}
\label{aVAC}
\psi^{\pm a}_j|0\rangle =0\qquad \mbox{for }j>0,\quad a=1,\ldots,s.
\end{equation}
We define the corresponding fields by
\begin{equation}
\label{aFERMIONS}
\psi^{\pm a}(z)=\sum_{j\in\frac12+\mathbb Z}\psi_j^{\pm a} z^{-j-\frac12},\qquad a=1,\ldots,  s.
\end{equation}
As a result, we obtain  
  $s^2$ generating fields for $a_\infty$, namely
\[
A^{ab}(y,z)=:\psi^{+a}(y)\psi^{-b}(z):,\quad \  1\le a,b\le s.
\]
\ 
\\
\indent In the cases  $x=b$ and $d $ the multicomponent realization of $F_x$ and $x_\infty$
can be given in terms of $s\ge 0$ pairs of charged fermionic fields and $r\ge 0$ neutral twisted fermionic fields, where $s+r>0$.  In addition one needs one fermion $\sigma_0$ in case $x=b$ (resp. $d$) if $r$ is even (resp.  odd).  
We relabel as follows.  In principle if we relabel the $\phi_i$ with $i>0$ and the $\tilde\phi_i$ with i$\ge 0$, the ones with a negative index are then  fixed.  Let's describe the case that $x=b$.  We  do the relabeling in blocks first we relabel $2s$ fermions to obtain the charged ones as follows
$\phi_{(2s+r)+j+1+q}=\psi^{-1}_{j+\frac12}, \phi_{(2s+r)+j+2+q}=\psi^{+1}_{j+\frac12},\phi_{(2s+r)+j+3+q}=\psi^{-2}_{j+\frac12},\phi_{(2s+r)+j+4+q}=\psi^{+2}_{j+\frac12},
\ldots, \phi_{(2s+r)+j+2s+q}=\psi^{+s}_{j+\frac12}$.  Here $q$  is some positive integer which depends on $r$. Then we relabel $r$ fermions to obtain $\tilde\phi^{s+c}_j$ with $j>0$.  We do this as follows
  $\tilde\phi_j^{s+1}={\rm constant}\tilde\phi_{(r+2s)j+2s+1+q}, \tilde\phi_j^{s+2}={\rm constant}\tilde\phi_{(r+2s)j+2s+2+q},
\ldots, \tilde\phi_j^{s+r}={\rm constant}\tilde\phi_{(r+2s)j+2s+r+q}$. We need the constant to get the right commutation relations.
Finally, we want to obtain the $\tilde\phi^{r+c}_0$'s.  Note that we want the following commutation relations for them
$(\tilde\phi^{r+c}_0)^2=\frac12$. There is however only one fermion which has this relation, viz.  $\tilde\phi_0$. So we have to create $r-1$ of such fermions. This can be done as follows, we can combine a creation and annihilation fermion $\tilde \phi_{-j}$ and $\tilde \phi_{j}$.  Note that we haven't relabeled the $\tilde\phi_j$ with $1\le j\le q$  yet .
Let 
\[
\tilde\phi^{s+1}_0=\frac{1}{\sqrt 2}(\tilde\phi_1+\tilde\phi_{-1}),\quad \tilde\phi^{s+2}_0=\frac{\sqrt{-1}}{\sqrt 2}(\tilde\phi_1-\tilde\phi_{-1}),
\]
Then these elements satisfy $(\tilde\phi^{s+1}_0)^2=(\tilde\phi^{s+2}_0)^2=\frac12$.  We can do a similar thing for the other $r-3$ ones.  If $r$ is odd, we obtain all the 
elements $\tilde\phi_0^{s+a}$ in this way,  where in particularly $\tilde\phi^{s+r}_0=\tilde\phi_0$.  So in this case $q=\frac{r-1}2$.  If $r$ is even,   we haven't relabeled $\tilde\phi_0$,
we thus set $\sigma_0=\tilde\phi_0$.  Note that $q=\frac{r}2$ in this case.  In the $x=d$, we do a similar thing,  the role of $r$ even and odd is interchanged in this case.

Explicitly,  we use the following 
relabeling of the generators $\tilde\phi_j$, $j\in\mathbb Z$ (resp. $\phi_j$, $j\in\frac12+\mathbb Z$) of $C\ell_b$ (resp.   $C\ell_d$).  For $x=b$ we let $r=2q+1$ if $r\ge 1$  is odd,  and $r=2q$ if $r\ge 0$ is even. and introduce the following generators of $C\ell_b$, where $j\in\mathbb Z_{\ge 0}$,  $1\le a\le s$,  $1\le c\le r$:
\begin{equation}
\label{b-relabel}
\begin{aligned}
&\tilde\phi^{s+b}_{\pm (2j+1)}=\sqrt{-1}\tilde\phi_{\pm(2j (2s+r)+q+2s+b)},\quad \tilde\phi^{s+b}_{\pm 2j+2}=\tilde\phi_{\pm((2j+1)(2s+r)+q+2s+b)},\\
&\psi^{\mp a}_{\pm( j+\frac12)}=\tilde\phi_{\pm (j(2s+r)+q+2a-1)},\quad \psi^{\pm a}_{\pm( j+\frac12)}=\tilde\phi_{\pm (j(2s+r)+q+2a)},\\
&\tilde\phi^{s+2c-1}_0=\frac{1}{\sqrt 2}(\tilde\phi_c+\tilde\phi_{-c}),\quad \tilde\phi^{s+2c}_0=\frac{\sqrt{-1}}{\sqrt 2}(\tilde\phi_c-\tilde\phi_{-c}),\\
&\tilde\phi^{s+r}_0=\tilde\phi_0,\quad\mbox{if } r=2q+1, \sigma_0=\tilde\phi_0,\quad\mbox{if } r=2q.
\end{aligned}
\end{equation}
For $x=d$, we let $r=2q-1$ if $r\ge 1$ is odd, and $r=2q$ if $r\ge 0$ is even, and introduce the following generators of $C\ell_d$,  where $j\in\mathbb Z_{\ge 0}$,  $1\le a\le s$,  $1\le b\le r$,  $1\le c\le q$:
\begin{equation}
\label{d-relabel}
\begin{aligned}
&\tilde\phi^{s+b}_{\pm (2j+1)}=\sqrt{-1}\phi_{\pm(2j (2s+r)+q+2s+b-\frac12)},\quad \tilde\phi^{s+b}_{\pm 2j+2}=\phi_{\pm((2j+1)(2s+r)+q+2s+b-\frac12)},\\
&\psi^{\mp a}_{\pm( j+\frac12)}=\phi_{\pm (j(2s+r)+q+2a-\frac32)},\quad \psi^{\pm a}_{\pm( j+\frac12)}=\phi_{\pm (j(2s+r)+q+2a-\frac12)},\\
&\tilde\phi^{s+2c-1}_0=\frac{1}{\sqrt 2}(\phi_{c-\frac12}+\phi_{\frac12-c}),\quad \tilde\phi^{s+2c}_0=\frac{\sqrt{-1}}{\sqrt 2}(\phi_{c-\frac12}-\phi_{\frac12-c}),\\
&\tilde\phi^{s+r}_0= \frac{1}{\sqrt 2}(\phi_{q-\frac12}+\phi_{\frac12-q}),\quad \sigma_0= \frac{\sqrt{-1}}{\sqrt 2}(\phi_{q-\frac12}-\phi_{\frac12-q}),
\quad\mbox{if } r=2q-1.
\end{aligned}
\end{equation}
As a result we obtain in both cases $s$ pairs of charged fermionic fields 
\[
\psi^{\pm a}(z)=\sum_{j\in\frac12+\mathbb Z}\psi^{\pm a}_j z^{-j-\frac12}, \quad 1\le a\le s,
\]
and $r$ twisted neutral fields 
\[
\tilde\phi^{\pm b}(z)=\sum_{j\in\mathbb Z}\tilde\phi^{b}_j z^{-j }, \quad s< b\le r+s.
\]
Besides we have an extra fermion $\sigma_0$ if $x=b$ and $r$ is even, and if $x=d$ and $r$ is odd. 

The commutation relations between the charged fermions, as well as their action on the vacuum are \eqref{aCOM} and \eqref{aVAC}. The remaining commutation relations are 
\begin{equation}
\label{bCOM}
\begin{aligned}
&\sigma_0\psi^{\pm a}_i+\psi^{\pm a}_i\sigma_0=\sigma_0\tilde\phi^{b}_i+\tilde\phi^{b}_i\sigma_0=0, \  \sigma_0^2=\frac12,\ 
\psi^{\pm a}_i\tilde\phi^{b}_j+\tilde\phi^{b}_j\psi^{\pm a}_i=0,\\
&\tilde\phi^{b}_j\tilde\phi^{c}_k+\tilde\phi^{c}_k\tilde\phi^{b}_=(-1)^j\delta_{bc}\delta_{j,-k},\quad i\in \frac12+\mathbb Z,\  j,k\in\mathbb Z,\ 1\le a\le s< b,c\le r+s,
\end{aligned}
\end{equation}
and we have  the following action on the vacuum 
\begin{equation}
\label{BVAC}
(\tilde\phi_0^{s+2k-1}-\sqrt{-1}\tilde\phi_0^{s+2k}) |0\rangle= \tilde\phi^{b}_j|0\rangle =0,\qquad \mbox{for }j>0,\  1\le k \le  \frac{r}2,
\end{equation}
together with 
\begin{equation}
\label{BVA2}
\sigma_0|0\rangle =\frac1{\sqrt 2}|0\rangle, \quad\mbox{if } r\ \mbox{even},  
\quad \tilde\phi_0^{s+r}|0\rangle =\frac1{\sqrt 2}|0\rangle,\quad \mbox{if } r\ \mbox{odd},
 \end{equation}
if $x=b$ 
and   (cf.    \eqref{phi-sigma}):
\begin{equation}
\label{BVAC3}
(\tilde\phi_0^{s+r}-\sqrt{-1} \sigma_0)|0\rangle=0,\quad \mbox{if } r\ \mbox{ is   odd},
 \end{equation}
if $x=d$.

The  generating fields for the representation  $r_x$ for  $x=b$ or $d$ then are respectively ($\delta,\epsilon=+$ or $-$, $1\le a,b\le s<c,d\le s+r$):
\begin{equation}
\label{XFIELDS}
\begin{aligned}
&E^{\delta a ,\epsilon b}(y, z)=:\psi^{\delta a}(y)\psi^{\epsilon b}(z):,\quad\mbox{if }r>0, \\
&E^{\delta a,c}(y,z)=:\psi^{\delta a}(y)\tilde\phi^{c}(z):, \quad\mbox{if }r,s>0, \\
&E^{c,d }(y,z)=:\tilde\phi^{ c}(y)\tilde\phi^{d}(z):, \quad\mbox{if }s>0,\\
&E^{\delta a}(z)=:\psi^{\delta a}(z)\sigma_0:,\quad\mbox{if }s>0,\ (x=b\ \mbox{and }r\ \mbox{even})\ \mbox{or } (x=d\ \mbox{and }r\ \mbox{odd}), 
 \\
&E^{c }(z)=:\tilde\phi^{ c}(z)\sigma_0:, \quad\mbox{if }r>0,\ (x=b\ \mbox{and }r\ \mbox{even})\ \mbox{or } (x=d\ \mbox{and }r\ \mbox{odd}).
\end{aligned}
\end{equation}
\\
\ 
\\
\indent For $c_\infty$,   we  can choose a different basis of the Weyl algebra such that we get $r\ge 0$ neutral symplectic bosonic fields as in the previous section. In analogy with the $x=b,d$ case, one can also obtain $s\ge 0$ so-called charged symplectic bosonic fields $b^{+a}(z)$ and $b^{-a}(z)$, for $1\le a\le s$.
One can obtain this for instance by relabeling as follows.  If $r$ is even ($j>0$, $1\le a\le s$ ,$1\le c\le \frac{r}2$),
\begin{equation}
\label{c-relabel-even}
\begin{aligned}
&b_{\pm(2j+\frac12)}^{s+2c-1}=b_{\pm (2j(r+2s)+2c-\frac32)},\quad
b_{\pm(2j+ \frac12)}^{s+2c}=\sqrt{-1}b_{\pm (2j(r+2s)+2c-\frac12)},\\
&b_{\pm(2j+\frac32)}^{s+2c-1}=\sqrt{-1}b_{\pm ((2j+1)(r+2s)+2c-\frac32)},\quad
b_{\pm(2j+ \frac32)}^{s+2c}=b_{\pm ((2j+1)(r+2s)+2c-\frac12)},\\
&b_{\pm (j+\frac12)}^{\pm a}=b_{\pm(j(r+2s)+r+2a-\frac32)},\quad
b_{\mp (j+\frac12)}^{\pm a}= b_{\mp(j(r+2s)+r+2a-\frac12)}.
\end{aligned}
\end{equation}
 If $r$ is odd ($j>0$,  $1\le a\le s$ ,$1\le c<\frac{r}2$),
\begin{equation}
\label{c-relabel-odd}
\begin{aligned}
&b_{\pm(j+\frac12)}^{s+r}=b_{\pm (j(r+2s)+r-\frac12)},\\
&b_{\pm(j+\frac12)}^{s+2c-1}=b_{\pm (j(r+2s)+2c-\frac32)},\quad
b_{\pm(j+ \frac12)}^{s+2c}=\sqrt{-1}b_{\pm (j(r+2s)+2c-\frac12)},\\
&b_{\pm (2j+\frac12)}^{\pm a}=\sqrt{-1}b_{\pm(2j(r+2s)+r+2a-\frac32)},\quad
b_{\mp (2j+\frac32)}^{\pm a}= \sqrt{-1}b_{\mp((2j+1)(r+2s)+r+2a-\frac32)},\\
&b_{\pm (2j+\frac12)}^{\mp a}=b_{\pm(2j(r+2s)+r+2a+\frac12)},\quad
b_{\mp (2j+\frac32)}^{\mp a}=\sqrt{-1} b_{\mp((2j+1)(r+2s)+r+2a+\frac12)}.
\end{aligned}
\end{equation}

For $1\le a\le s$ we have 
\[
b^{\pm a}(z)=\sum_{i\in \frac12+\mathbb Z} b^{\pm a}_i z^{-i-\frac12},\quad b^{c}(z)=\sum_{i\in\frac12+\mathbb Z} b^{c}_i z^{-i-\frac12},\quad 1\le a\le s<c \le r+s,
\]
which satisfy
\begin{equation}
\label{cCOM}
\begin{aligned}
&b^{+a}_ib^{+c}_j-b^{+c}_jb^{+a}_i=b^{-a}_ib^{-c}_j-b^{-c}_jb^{-a}_i=0,\
b^{+a}_ib^{-c}_j-b^{-c}_jb^{-a}_i=\delta_{ab}\delta_{i,-j},\\
&
b_i^e b_j^d- b_j^db_i^e=(-1)^{i-\frac12}\delta_{de}\delta_{i,-j},\ b^{\pm a}_ib^{d}_j-b^{d}_jb^{\pm a}_i=0,\\
\end{aligned}
\end{equation}
 for $ 1\le a,c\le s< d$, $e\le s+r,\ i,j\in \frac12+\mathbb Z$.
\\
The action on the vacuum vector is as follows
\begin{equation}
\label{cVAC}
b_j^{\pm a}|0\rangle =b_j^{c}|0\rangle=0 \quad\mbox{for }j>0.
\end{equation}
The  generating fields for the representation  $r_c$   then are ($\delta,\epsilon=+$ or $-$, $1\le a,c\le s<d,e\le s+r$)
\begin{equation}
\label{cFIELDS}
\begin{aligned}
&C^{\delta a ,\epsilon c}(y, z)=:b^{\delta a}(y)b^{\epsilon c}(y):,\  C^{\delta a,d}(y,z)=:b^{\delta a}(y)b^{d}(z):,\ 
C^{de }(y,z)=:b^{ d}(y)b^{e}(z):.
\end{aligned}
\end{equation}
\begin{remark}
Note that the relabeling that we have used here is different from the relabeling in Section \ref{sec4}. There the relabeling is induced by the embedding of  $\hat{\mathfrak g}$ and $\hat{\mathfrak g}^{(2)}$ into $x_\infty$ and the element in the Weyl group of $\mathfrak g$ on which the construction is based.  If we would use the relabeling as above,  then the construction of Section \ref{sec4} would not give matrices in $\overline x_\infty$.
\end{remark}

\subsection{Vertex operator realizations of $F_x$}
\subsubsection{Bosonization of type $a$,    $b$ and $d$}
\label{FermVO}
In Section \ref{section 3.5} we have obtained  fermionic fields of two types, viz. the charged fermionic fields $\psi^{\pm a}(z)$,  $1\le a\le s$, and  the twisted neutral fermionic fields $\tilde\phi^b(z)$,  $s< b\le r+s$,      and in some cases an additional operator $\sigma_0$.    To the charged and twisted neutral fermionic fields, we can associate  free bosonic fields
\begin{equation}
\label{11-2}
\begin{aligned}
\alpha^a(z)&= \sum_{j\in\mathbb Z} \alpha_j^b z^{-j-1}= :\psi^{+b}(z)\psi^{-b}(z): ,\ 1\le b\le s,\\
\alpha^{s+b}(z)&= \sum_{j\in 1+2\mathbb Z}\alpha_j^{s+b} z^{-j}=\frac12 :\tilde\phi^{s+b}(z)\tilde\phi^{s+b}(-z):,\ 1\le b\le r.
\end{aligned}
\end{equation}
Their modes satisfy the  relations of a Heisenberg Lie algebra:  
\begin{equation}
\label{11-3}
\begin{aligned}
[\alpha^{a}_j,\alpha^b_k]&=j\delta_{ab}\delta_{jk}\ 1\le a\le s,\ 1\le b\le s+r,\  j,k \in\mathbb  Z, \ k\  \mbox{odd }\mbox{if }b>s,\\
[\alpha^a_j,\alpha^b_k]&=\frac{j}2 \delta_{ab}\delta_{jk},\  s+1\le a\le r+s,\ 1\le b\le s+r,
\  j,k \in\mathbb  Z, \ j\ \mbox{odd},\ k\  \mbox{odd }\mbox{if }b>s.\\
\end{aligned}
\end{equation}
The action on the vacuum vector is
\[
 \alpha_j^a|0\rangle
 =0,\quad\mbox{for }j\ge 0. 
\]

Note that 
\begin{equation}
\label{alpha3}
[\alpha_i^b,\psi^{\pm a}(z)]=\pm \delta_{ab}   z^i\psi^{\pm a}(z)\quad\mbox{for }i\in\mathbb Z, \  i\  \mbox{odd }\mbox{if }b>s .
\end{equation}
This makes it possible (see \cite{tKLA} or \cite{KLmult}) to express these fermionic fields  almost completely in terms of the Heisenberg algebra, viz.
for $1\le a\le s$ we have
\begin{equation}
\label{FVO}
\psi^{\pm a}(z)=Q_a^{\pm 1}
z^{\pm \alpha^a_0}
\exp\left(
\mp \sum_{k<0} \frac{\alpha^a_k}k z^{-k}
\right)
\exp\left(
\mp \sum_{k>0} \frac{\alpha^a_k}k z^{-k}
\right),
\end{equation}
where the operators $Q_a$ are uniquely defined by
\begin{equation}
\label{QQ}
Q_a|0\rangle=\psi^{+a}_{-\frac12}|0\rangle,\quad Q_a\psi^{\pm b}_i=(-1)^{\delta_{ab}-1}\psi^{\pm b}_{i\mp \delta_{ab}}Q_a.
\end{equation}
It is straightforward to check that for $a< b$, $ 1\le a,b,c \le s$ we have 
\begin{equation}
\label{QQQ}
Q_aQ_b=-Q_bQ_a \quad [\alpha^a_i ,Q_c]=\delta_{ac}\delta_{i0}Q_c.
\end{equation}

If $x=a$, we  have $s$ pairs of  charged free fermions.
When $x=b$ or $d$ we have in addition also $r$ twisded neutral free fermionic fields.

For the neutral twisted fields,  the commutation relations of the Heisenberg elements $\alpha^{b}_k$,  $k$ odd, with $s< b\le s+r$:
\begin{equation}
\label{alpha-phi}
[ \alpha^b_k, \tilde\phi^c(z)]= \delta_{b,c}z^k \tilde\phi^c (z)\quad\mbox{for }k\in \mathbb Z_{odd}.
\end{equation}
Thus for $s<b\le r+s$ we have
\begin{equation}
\label{NFVO}
\tilde\phi^{  b}(z)=R_{b-s}
\exp\left(
- 2\sum_{k<0,\, {\rm odd}} \frac{\alpha^{b}_k}{k} z^{-k}
\right)
\exp\left(
- 2\sum_{k>0,\, {\rm odd}} \frac{\alpha^{b}_k}{k} z^{-k}
\right),
\end{equation}
\begin{equation}
\label{QQQQ}
R_b|0\rangle=\tilde\phi^{s+b}_0|0\rangle,\quad\mbox{and } R_bR_c+R_cR_b=\delta_{bc},\ R_bQ_a+Q_aR_b=0.
\end{equation}

If $x=b$ and $r$ is even, or $x=d$ and $r$ is odd we have in addition one extra operator $\sigma_0$ such that $\sigma_0^2=\frac12$.
This operator commutes with all the operators $\alpha^a_k$, for $1\le a\le r+s$ of the Heisenberg algebra and it anti-commutes with the operators $Q_a$ ($1\le b\le s$) and $R_b$ ($s<b\le r+s$).

If $x=b$ and $r$ is odd,  then $\sigma_0|0\rangle=\frac1{\sqrt 2}|0\rangle$.  If $x=d$ and $r$ is even,  then $\sigma_0 |0\rangle=-\sqrt{-1}\tilde \phi_0^{r+s}|0\rangle$.

The $q$ dimension formula for the $a_\infty$ module $F^{(m)}$ is as follows,
\begin{equation}\label{QDIMA}
\dim_q F^{(m)}=\sum_{
\genfrac{}{} {0pt}{2}{j_1, \ldots j_s\in\mathbb Z}
{j_1+\cdots +j_s=m}
}
q^{\frac{j_1^2+j_2^2+\cdots +j_s^2}2}
\prod_{i=1}^\infty \frac1{(1-q^i)^s}.
\end{equation}
In the $b$ case,  the spin module $F$ is irreducible for $b_\infty$ and in that case,  we find the following $q$-dimension formula.
\[
\dim_q F =2^{[\frac{r}2]}\sum_{{j_1, \ldots j_s\in\mathbb Z} }
q^{\frac{j_1^2+j_2^2+\cdots +j_s^2}2}
\prod_{i=1}^\infty \frac1{(1-q^i)^s (1-q^{2i-1})^r}
.
\]
$F$ splits in to two irreducible  $d_\infty$ modules $F^{\overline 0}$ and $F^{\overline 1}$, which both have the same $q$-dimension formula,  except when $r=0$.   We have 
\begin{equation}
\label{QDIMD}
\dim_q F^{\overline \epsilon} =
\begin{cases}
2^{[\frac{r-1}2]}\sum_{{j_1, \ldots j_s\in\mathbb Z} }
q^{\frac{j_1^2+j_2^2+\cdots +j_s^2}2}
\prod_{i=1}^\infty \frac1{(1-q^i)^s (1-q^{2i-1})^r},&\mbox{for }r\ne 0,\\
&\
\\
  \sum_{\genfrac{}{} {0pt}{2}{j_1, \ldots j_s\in\mathbb Z}{j_1+\cdots +j_s\in \epsilon+\mathbb Z}}
q^{\frac{j_1^2+j_2^2+\cdots +j_s^2}2}
\prod_{i=1}^\infty \frac1{(1-q^i)^s },& \mbox{when }r=0.
\end{cases}
\end{equation}
\subsubsection{Bosonization of type $c$}
\label{3.6.2}
Recall that in the multicomponent construction of $C\ell_c$,  we have two types of symplectic bosonic fields,  viz.    charged (resp. neutral)   fields $b^{\pm a}(z)$,  $1\le a\le s$ (resp. $b^c (z)$,  $s <c\le  r+s$), satisfying \eqref{cCOM}
and \eqref{cVAC}.
\begin{equation}
\label{symplectic-alpha}
\begin{aligned}
\alpha^a(z)&= \sum_{j\in\mathbb Z} \alpha_j^a z^{-j-1}= :b^{+a}(z)b^{-a}(z): \quad 1\le a\le s,\\
\alpha^{c}(z)&= \sum_{j\in 1+2\mathbb Z}\alpha_j^{c} z^{-j}=\frac12 :b^{c}(z)b^{c}(-z):\quad s< c\le s+r,
\end{aligned}
\end{equation}
 and their modes satisfy the commutation relations of a Heisenberg Lie algebra: 
\begin{equation}
\label{symplectic-alpha-rel}
\begin{aligned}
[\alpha^{a}_j,\alpha^b_k]&=-j\delta_{ab}\delta_{jk}\ 1\le a\le s,\ 1\le b\le s+r,\  j,k \in\mathbb  Z, \ k\  \mbox{odd }\mbox{if }b>s,\\
[\alpha^a_j,\alpha^b_k]&=-\frac{j}2 \delta_{ab}\delta_{jk},\  s+1\le a\le r+s,\ 1\le b\le s+r,
\  j,k \in\mathbb  Z, \ j\ \mbox{odd},\ k\  \mbox{odd }\mbox{if }b>s,\\
 \alpha_j^a|0\rangle&
 =0,\quad\mbox{for }j\ge 0. 
\end{aligned}
\end{equation}
We start with the charged fields.
The commutation relations can be rewritten as
\[
b^{\pm a}(y)b^{\pm c}(z)-b^{\pm c}(z)b^{\pm a}(y)=0,\quad
b^{+ a}(y)b^{- c}(z)-b^{- c}(z)b^{+ a}(y)=\delta_{ac}\delta(y-z).
\]
Then
\begin{equation}
\label{I1}
[ \alpha^i_k,b^{\pm a}(z)]=\mp z^k\delta_{i,a}b^{\pm a}(z),\quad \mbox{for } k\in\mathbb Z,
\end{equation}
hence, 
\begin{equation}
b^{\pm a}(z) = E_-^a(z)^{\pm 1} Q^{\pm a}(z) E_+^a(z)^{\pm 1} 
\end{equation}
where
\begin{equation}
\label{eee}
E_-^{ a}(z)=\exp\left(
- \sum_{k<0} \frac{\alpha^a_k}k z^{-k}
\right),\qquad E_+^{  a}(z)=\exp\left(
- \sum_{k>0} \frac{\alpha^a_k}k {z^{-k}}
\right),
\end{equation}
and 
\[
[\alpha_k^i, Q^{\pm a}(z)]=\mp \delta_{i,a}\delta_{k,0}Q^{\pm a}(z)\quad \mbox{for } k\in\mathbb Z.
\]
Following  \cite{A}, we set 
\begin{equation}
\label{THETA}
\Theta^{\pm a}(z)=\sum_{k\in\mathbb Z}\Theta^{\pm a}_kz^{-k-1}:=E_-^a(z)^{\mp 1}b^{\pm a}(z) z^{\mp \alpha_0^a}E_+^a(z)^{\mp1},
\end{equation}
Then 
\[
\Theta^{\pm a}(z)|0\rangle=E_-^a(z)^{\mp 1}\sum_{i>0}b^{\pm a}_{-i}
|0\rangle z^{i-\frac12} \in F_c [[z]],
\]
hence,  
\[
\Theta^{\pm a}_k|0\rangle=0\quad\mbox{for }    k\in \mathbb Z_{> 0}.
\]
\begin{lemma}
\label{LL}
Let $i_{z,y} R(y,z)$ denote  the power series expansion of a rational function $R(y,z)$ in the domain $|z|>|y|$ and 
let $\lambda, \mu=\pm 1$. Then  
\[
\begin{aligned}
E_-^a(y)^{\lambda}E_+^b(z)^{\mu}&=i_{z,y}(1-\frac{y}{z})^{\lambda\mu\delta_{ac}}E_+^b(z)^{\mu}E_-^a(y)^{\lambda},\\
E_+^a(z)b^{\pm c}(y)&=i_{z,y}(1-\frac{y}{z})^{\mp\delta_{ac}}b^{\pm c}(y)E_+^a(z),\\
E_-^a(y)b^{\pm c}(z)&=i_{z,y}(1-\frac{y}{z})^{\pm\delta_{ac}}b^{\pm c}(z)
E_-^a(y),\\
[ \alpha^i(y),\theta^{\pm a}(z)]&=    \mp  \frac{\delta_{i,a}}{y}\theta^{\pm a}(z)
.
\end{aligned}
\]  
\end{lemma}
 {\bf Proof } Let $|z|>|y|$, then
\[
\begin{aligned}
E_-^a(y)^{\lambda}E_+^b(z)^{\mu}&=\exp\left(-\sum_{k=1}^\infty \frac{\lambda\mu}{k^2}\left(\frac{y}{z}\right)^k [\alpha^a_{-k},\alpha^b_k] \right) E_+^b(z)^{\mu}E_-^a(y)^{\lambda}\\
&=\exp\left(-\delta_{ab}\sum_{k=1}^\infty \frac{\lambda\mu}{k}\left(\frac{y}{z}\right)^k  \right) E_+^b(z)^{\mu}E_-^a(y)^{\lambda}\\
&=i_{z,y}(1-\frac{y}{z})^{\lambda\mu\delta_{ab}}E_+^b(z)^{\mu}E_-^a(y)^{\lambda}.
\end{aligned}
\]
Next, since 
\[
\alpha^a_k b^{\pm c}(z)=b^{\pm c}(z)(\alpha^a_k\mp \delta_{ac}z^k),\quad \mbox{for } k\in \mathbb Z,
\]
also
\[
\begin{aligned}
E_+^a(z)b^{\pm c}(y)&=b^{\pm c}(y)\exp \left(- \sum_{k>0} \frac{1}k {z^{-k}}(\alpha^a_k \mp \delta_{ac}y^k)
\right)\\
&=\exp \left(\pm \delta_{ac}\sum_{k>0}  \frac1 k   \left(\frac {y}{z}\right)^k 
\right)b^{\pm c}(y)E_+^a(z)\\
&=i_{z,y}(1-\frac{y}{z})^{\mp\delta_{ac}}b^{\pm c}(y)E_+^a(z).
\end{aligned}
\] 
The last identity follows from \eqref{I1}.\hfill$\square$
\\
 
Using this lemma and the expression \eqref{THETA}, we find that for $\lambda=\pm$ we have
\[
\begin{aligned}
\Theta^{+ a}(y)\Theta^{\lambda b}(z)&=
i_{y,z}(y-z)^{\lambda\delta_{ab}}
 E_-^a(y)^{-1}E_-^b(z)^{-\lambda 1}b^{+ a}(y)b^{\lambda b}(z)y^{-\alpha_0^a}z^{-\lambda \alpha_0^b} E_+^a(y)^{-1}E_+^b(z)^{-\lambda 1}
\\
&= i_{y,z}(y-z)^{\lambda\delta_{ab}}
E_-^a(y)^{-1}E_-^b(z)^{-\lambda 1}(:b^{+ a}(y)b^{\lambda b}(z):+\\
&\quad +\delta_{ab}\delta_{\lambda,-}i_{y,z}(y-z)^{\lambda1})y^{-\alpha_0^a}z^{-\lambda \alpha_0^b}E_+^a(y)^{-1}E_+^b(z)^{-\lambda 1}
 ,\\
\Theta^{\lambda b}(z)\Theta^{+ a}(y)&= i_{z,y}(z-y)^{\lambda\delta_{ab}}
E_-^a(y)^{-1}E_-^b(z)^{-\lambda 1}(:b^{\lambda b}(z)b^{+a}(y):\\
&\quad -\delta_{ab}\delta_{\lambda,-}i_{z,y}(z-y)^{\lambda1})y^{-\alpha_0^a}z^{-\lambda \alpha_0^b}E_+^a(y)^{-1}E_+^b(z)^{-\lambda 1}
.
\end{aligned}
\]
From \eqref{I1}, we deduce that 
\[
\begin{aligned}
\Theta^{+ a}(y)\Theta^{+b}(z)-(-1)^{\delta_{ab}}\theta^{ +b}(z)\Theta^{+ a}(y)&=0,\\
\Theta^{+ a}(y)\Theta^{- b}(z)-\Theta^{- b}(z)\Theta^{+ a}(y)&=0, \quad\mbox{if } a\ne  b,
\end{aligned}
\]
and 
\begin{equation}
\label{thet1}
\begin{aligned}
&\Theta^{+ a}(y)\Theta^{- a}(z)+\Theta^{- a}(z)\Theta^{+ a}(y)=\\
&E_-^a(y)^{-1}E_-^a(z)
(\delta(y-z)  :b^{+ a}(y)b^{-a}(z):+\partial_z\delta(y-z)
)E_+^a(y)^{-1}E_+^a(z)
y^{-\alpha_0^a}z^{ \alpha_0^a}=\\
&\delta(y-z)\alpha^a(z)+\partial_z\delta(y-z)- \delta(y-z)\alpha^a(z)\\
&=\partial_z\delta(y-z).
\end{aligned}
\end{equation}
Here we have used Taylor's formula.    In a    similar way one can prove that
\begin{equation}
\label{thet2}
\Theta^{- a}(y)\Theta^{-b}(z)-(-1)^{\delta_{ab}}\Theta^{ -b}(z)\Theta^{- a}(y)=0.
\end{equation}
Note that still
\begin{equation}
\label{alphatheta}
[\alpha_k^a,\Theta^{\pm b}(z)]=\mp \delta_{ab}\delta_{k0}\Theta^{\pm b}(z).
\end{equation}
For convenience of  counting the energy degree,  we introduce an operator $Q_a$ by setting $ \Theta^{\pm a}=Q_a^{\mp 1}
\theta^{\pm a}(z)$,   where $[\alpha_0, Q_a]=Q_a$,  then $\theta^{\pm a}(z)$ satisfies the same commutation relations as 
$\Theta^{\pm a}(z)$, the only difference is that $[\alpha_k^a,\theta^{\pm b}(z)]=0$.
Thus,
\begin{equation}
\label{thet3}
b^{\pm a}(z)=Q_a^{\mp 1}z^{\pm \alpha_0^a}E_-^a(z)^{\pm 1 }\theta^{\pm a}(z)E_+^a(z)^{\mp1}.
\end{equation}
This last formula was  already present  in \cite{FMS} in the situation when there is only one such pair of symplectic boson  fields.
Note that $Q_a^{\pm 1}$ is an operator that always  appears together with each $\theta^{\pm a}_j$.  In principle we do not need to introduce it,  but  for the moment this operator will be  convenient when calculating the energy decomposition of $F$.  We will show this in the next paragraph.  Note also that  $Q_a b^{\pm c}(z)= z^{\pm \delta_{ac}}b^{\pm c}(z)Q_a$.

Assume for the moment that there is only one such pair $b^{\pm}(z)$.
Then 
$(b^{\pm}_{-\frac12})^n|0\rangle$, which has energy  $\frac{n}2$, is equal to the constant coefficient of $b^{\pm}(z_n)\cdots b^{\pm}(z_1)|0\rangle$.  Using 
\eqref{thet3}, this is equal to 
\begin{equation}
\label{TET3}
\theta^\pm_{-n}\theta^\pm_{1-n}\cdots\theta^\pm_{-2}\theta^\pm_{-1}Q^{\mp n}|0\rangle.
\end{equation}
More generally  the energy of 
\begin{equation}
\label{thetabasis}
\theta^-_{-j_m}\cdots \theta^-_{-j_1}\theta^+_{-k_n}\cdots \theta^+_{-k_1}Q^{m-n}|0\rangle,\quad\mbox{where $j_m>\cdots j_1>0$ and $k_n>\cdots >k_1>0$, }
\end{equation}
  is equal to 
\begin{equation}
\label{energ}
j_1+\cdots+ j_m+k_1+\cdots+k_n-\frac{(m-n)^2}2.
 \end{equation}
This can be seen as follows.  Consider, the element   
\[ 
b^-_{m-n-j_m+\frac12}\cdots b^-_{-n-j_1+\frac12}b^+_{n-k_n-\frac12}\cdots b^+_{-k_1+\frac12}|0\rangle,
\]
which, clearly has  energy   \eqref{energ}.  It  is the coefficient of 
 $$   y_m^{j_m+n-m}\cdots y_2^{j_2+n-2} y_1^{j_1+n-1}  z_n^{k_n-n}\cdots z_2^{k_2-2} z_1^{k_1-1}$$ in
\[
\begin{aligned}
&b^-(y_m)\cdots b^-(y_1)b^+(z_n)\cdots b^+(z_1)|0\rangle=\\
&z_2^{-1}z_3^{-2}z_n^{-n-1}y_1^n y_2^{n-1}\cdots y_m^{n-m+1}\theta^-(y_m)\cdots \theta^-(y_1)\theta^+(z_n)\theta^+(z_1)Q^{m-n}|0\rangle+\cdots=\\&
y_m^{j_m+n-m}\cdots  y_1^{j_1+n-1}z_n^{k_n-n}\cdots  z_1^{k_1-1} 
\theta^+_{-j_m}\cdots \theta^+_{-j_1}
\theta^+_{-k_n}\cdots \theta^+_{-k_1}Q^{m-n}|0\rangle
+\cdots\ 
\end{aligned}
\]
Hence,
\[
{\rm energy}(\theta^\pm_{-j})=j, \quad {\rm energy}(Q^m|0\rangle)=-\frac{m^2}2,\qquad j,m\in\mathbb Z, \ j>0.
\]
An element of the form \eqref{thetabasis} can be obtained by action of several elements of the form  
$\theta_i^\pm \theta_k^\mp$
%
on an element \eqref{TET3}.
\begin{lemma}Let 
\[
F_{ann}=\{ w\in F|\, \alpha_k w=0, \ \mbox{ for all }k>0\}.
\]
Then the elements \eqref{thetabasis} form a basis for $F_{ann}$.
\end{lemma}
{\bf Proof}
It suffices  to consider a linear combination of elements of the form \eqref{TET3},  where the energy,  say $N$, is fixed and where the charge $m-n$,    the eigenvalue of $\alpha_0$,   is fixed.   Since the energy is $N$,   there can only be a finite number of elements of the form \eqref{thetabasis} which have this energy.  Now let 
\begin{equation}
\label{TET4}
\sum_{i=1}^p \lambda_i \theta^-_{-j_{m_i}^i}\cdots \theta^-_{-j_1^i} \theta^+_{-k_{n_i}^i}\cdots \theta^+_{-k_1^i}Q^{m-n}|0\rangle=0,
\end{equation}
where we take  a linear combination of all elements wich have energy $N$ and charge $m-n$.   If we compare two elements in this sum, then there always exists a pair  $\theta^\pm_a$ and  $ \theta^\pm_b$,    such that one appears in one of the two elements and is missing in the other, and vice versa.  Hence,  we can show that the coefficient $\lambda_i =0$,  by letting 
\[
 \theta^-_{-j_{m_i}^i}    \theta^+_{j_{m_i}^i} \cdots \theta^-_{-j_1^i}   \theta^+_{j_{1}^i}  
\theta^+_{-k_{n_i}^i}   \theta^-_{k_{n_i}^i}  \cdots \theta^+_{-k_1^i}  \theta^-_{k_1^i}  
\]
act on this sum, which kills of all terms except for $\lambda_i \theta^-_{-j_{m_i}^i}\cdots \theta^-_{-j_1^i} \theta^+_{-k_{n_i}^i}\cdots \theta^+_{-k_1^i}$.  Hence, elements of the form \eqref{thetabasis} form a basis of $F_{ann}$.\hfill$\square$

So using \eqref{energ},  we have the following:
\begin{corollary}
The 
$q,t$-dimension of   $F_{ann}$ is equal to
\begin{equation}
\label{thetadim}
\begin{aligned}
\sum_{m,n\in\mathbb Z_{\ge 0}} t~^{m-n}q^{-\frac{(m-n)^2}2}s_m(q)s_{n}(q),\\
\end{aligned}
\end{equation}
where $s_m(q)$ is the power series  in $q$, which is defined as  the coefficient of $s^m$ in the expansion of $\prod_{j=1}^\infty (1+sq^j)=\sum_{m=0}^\infty s_m(q) s^m$.  
\end{corollary}

Note that  the coefficients of $q^k$ in $s_m(q)$ are the possible ways that the number $k$ can be written as the sum of $m$ distinct positive integers. 
It is well known that
\[
 s_m(q)= \frac{q^\frac{m^2+m}2} {(1-q)\cdots(1-q^m)}.
\]
Thus we find that
\begin{equation}
\label{TETdim1}
\dim_{q,t}F=\prod_{i\in\mathbb Z_{>0}}
\frac1{1-q^i}
\sum_{j,k=0}^\infty \frac{q^{\frac{k}2}t^k} {(1-q)\cdots(1-q^k)}
q^{jk} \frac{q^{\frac{j}2}t^{-j} }{(1-q)\cdots(1-q^j)}.
\end{equation}
 On the other hand, by looking at the energy and charge of $F$ in terms of the symplectic bosons $b^\pm _j$, we find that
\begin{equation}
\label{TETdim2}
\dim_{q,t}F=\prod_{i\in\mathbb Z_{>0}}
\frac1{(1-tq^{i-\frac12})(1-t^{-1}q^{i-\frac12})}
 .
\end{equation}
 This means that the right-hand side of  \eqref{TETdim2} is equal to the right-hand side of  \eqref{TETdim1}, which gives a remarkable identity, which we could not find in the literature.  In \cite{A}, Section 3,  and \cite{GR},  Section 7.2 and the Appendix,  there were some other formula's for the q-dimensions of symplectic bosons.  

Next, we want to describe the neutral fields, see \cite{A} or \cite{LOS} for more details.
The commutation relations can be rewritten to
\[
b^a(y)b^{c}(z)-b^{c}(z)b^{a}(y)=\delta_{ac}\sum_{k\in\mathbb Z}\frac1{y}\left(-\frac {y}{z}\right)^k=\delta_{ac}\delta(y+z).
\]
Since  for $a>s$,
\[
[\alpha_k^a ,b^c(z)]=-\delta_{ac}z^kb^c(z),  \quad [\alpha_k^{a},  
 \alpha_\ell^{c}]=-\frac{\delta_{ac}}{2} k, \quad\mbox{for } k\in1+2\mathbb{Z}, \ \ell\in\mathbb Z, \ \mbox{odd if }c>s.
\]
We define
\[
\Theta^{a}(z)=\sum_{k\in\frac12 +\mathbb Z}\theta^{a}_kz^{-k-\frac12}:=V_-^{a}(z)^{-1}b^{a}(z)V_+^{a}(z)^{-1},
\]
where (cf.  \eqref{introV})
\begin{equation}
\label{vVv}
V_-^{ a}(z)=\exp\left(
-2 \sum_{k<0,\,\rm odd} \frac{\alpha^{a}_k}k z^{-k}
\right),\qquad V_+^{  a}(z)=\exp\left(
-2 \sum_{k>0,\,\rm odd} \frac{\alpha^{a}_k}k {z^{-k}}
\right),
\end{equation}
Then 
\[
\Theta^{a}(z)|0\rangle =V_-^{a}(z)^{-1}\sum_{i>0}b^{a}_{-i}z^{i-\frac12}|0\rangle\in F_c[[z]],
\]
hence 
\[
\theta^{a}_k|0\rangle=0\quad \mbox{for }k\in \frac12+ \mathbb Z_{\ge 0}.
\]
In a similar way as before,  see \cite{LOS} for  details (note $\alpha_k=-J_k$), 
we obtain that 
\[
\Theta^{a}(y)\Theta^b(-z)-(-1)^{\delta_{ab}}\Theta^b(-z)\Theta^{a}(y)=2
\delta_{ab}z^{\frac12}\partial_z(z^{\frac12}\delta(y-z)).
\]
Note that we also have 
\[
\Theta^{\pm a}(y)\Theta^b(z)- \Theta^b(z)\Theta^{\pm a}(y)=0 \quad\mbox{for } a\le s,\ b>s.
\]
We thus have
\begin{proposition}
For $a\le s$ and $j>s$ we have 
\begin{equation}
\label{form3.4}
\begin{aligned}
b^{\pm a}(z)=&Q_a^{\pm 1}z^{\pm \alpha_0^a}E_-^a(z)^{\pm 1 }\theta^{\pm a}(z)E_+^a(z)^{\mp1}
=E_-^a(z)^{\pm 1 }\Theta^{\pm a}(z)E_+^a(z)^{\mp1}z^{\pm \alpha_0^a},\\
b^{j}(z)=&V_-^{j}(z) \Theta^{j}(z)V_+^{j}(z),
\end{aligned}
\end{equation}
with $E_\pm ^a(z)$ ($V_\pm^{j}(z)$) given by \eqref{eee} (resp. \eqref{vvv}), such that the last equation of Lemma \ref{LL} holds. The modes of the fields 
$\theta^{\pm a}(z)=\sum_{k\in\mathbb Z}\theta^{\pm a}_kz^{-k-1}$ and 
$\Theta^{j}(z)=\sum_{k\in\frac12 +\mathbb Z}\theta^{j}_kz^{-k-\frac12}$ satisfy the following (anti-)commutation relations
\[
\begin{aligned}
\theta^{\pm a}_k\theta^{\pm b}_\ell-\theta^{\pm b}_\ell\theta^{\pm a}_k=
\theta^{\pm a}_k\theta^{j}_\ell-\theta^{j}_\ell\theta^{\pm a}_k=0,\\
\theta^{+ a}_k\theta^{- b}_\ell-(-1)^{\delta_{ab}}\theta^{- b}_\ell\theta^{+ a}_k=\delta_{ab}\delta_{k,-\ell}k,\\
\theta^{i}_k\theta^{j}_\ell-(-1)^{\delta_{ij}}\theta^{j}_\ell\theta^{i}_k=
(-1)^{k-\frac12}\delta_{ij}\delta_{k,-\ell}2k.
\end{aligned}
\] 
\end{proposition}
\begin{remark}
In  \cite{A} I. Anguelova  bosonizes the fields   $\Theta^{\pm a}(z)$  further, following an idea of \cite{FMS}.  We will not do that here, but continue to work with the fields $\Theta^{\pm a}(z)$ and find a realization for them, more or less as in \cite{Luminy}.
\end{remark}

The Weyl module $F$ splits in  two irreducible $c_\infty$ modules $F^{\overline \epsilon}$,  ${\overline\epsilon}\in \mathbb Z_2$
The $q$-dimension formula can be most conveniently  described by using the variable $x$ such that $x^2=1$.  We have $s$ (resp.  $r$) charged (resp.  neutral) symplectic bosonic fields. Therefore we combine $s$ copies of \eqref{TETdim1} and $r$ copies copies of \eqref{Cqidentity}, which gives
\begin{equation}
\label{QDIMC}
\begin{aligned}
\dim_q F^{\overline 0}+x F^{\overline 1}&=
\prod_{j=1}^\infty \frac{1}{(1-xq^{j-\frac12})^{r+2s}}\\
&=\left( 
\sum_{j,k=0}^\infty \frac{q^{\frac{k}2}x^k} {(1-q)\cdots(1-q^k)}
q^{jk} \frac{q^{\frac{j}2}x^j}{(1-q)\cdots(1-q^j)}\right)^s\times\\
&\qquad\quad\prod_{j=1}^\infty \frac{(1+xq^{j-\frac12})^r} {(1-q^j)^s(1-q^{2j-1})^r}
.
\end{aligned}
\end{equation}

\subsection{The bosonic realization of $F_x$}
\label{BRF}
\subsubsection{The bosonic realization of $F_x$,  the Clifford algebra case}
\label{BRF1}
In Subsection   \ref{FermVO}   we have found bosonizations of the multicomponent fermionic fields that generate the Clifford algebra $C\ell_x$ for $x=a,b,d$ and  described the spin module $F_x$.   The description was based on the Heisenberg algebra  and some additional operators $Q_a$ and $R_b$.
In this subsection we will give the corresponding  bosonic description of $F_x$,  using the  Fock space representation of the Heisenberg    Lie algebra:
\begin{equation}
\label{alphaAA}
\alpha_k^j=\frac{\partial}{\partial t^{(j)}_k},\quad \alpha_{-k}^j= k t_k^{(j)},\quad k\in\mathbb Z_{>0} ,\ 1\le j\le s,
\end{equation}
 for the charged fields and 
\begin{equation}
\label{alphaBB}
\alpha_k^j =\frac{\partial}{\partial t^{(j)}_k},\quad  \alpha_{-k}^j = \frac{k}2  t_k^{(j)},\quad\ k\in\mathbb Z_{odd>0} ,\quad   s< j\le r+s,\\
\end{equation}
for the neutral fields.

For $x=a$,  we  let  
\[
 B_s=\mathbb C[q_j,q_j^{-1}, t_1^{(j)}, t_2^{(j)},\ldots |\, 1\le j\le s].
\]
Recalling the operators $Q_j$ on $F_a$, defined by \eqref{QQ},
we define an isomorphism $\sigma:F_a\to B_s$ by the following properties
\begin{equation}
\label{3.48a}
\sigma(Q_1^{k_1}Q_2^{k_2}\cdots Q_s^{k_s}|0\rangle)=q_1^{k_1}q_2^{k_2}\cdots q_s^{k_s},
\end{equation}
and 
\begin{equation}
\label{alphaA}
\begin{aligned}
\sigma r(\alpha_0^j)\sigma^{-1}= q_j\frac{\partial}{\partial q_j},\quad
\sigma r(\alpha_k^j)\sigma^{-1}=\frac{\partial}{\partial t^{(j)}_k},\quad \sigma r(\alpha_{-k}^j)\sigma^{-1}= k t_k^{(j)},\quad k>0,
\end{aligned}
\end{equation}
see \cite{tKLA} or \cite{KLmult}, for more details.  One also has 
\begin{equation}
\label{3.49a}
\sigma Q_j \sigma^{-1}= (-1)^{\sum_{k=1}^{j-1}q_k\frac{\partial}{\partial q_k}} q_j.
\end{equation}
Thus substituting \eqref{alphaA} in \eqref{FVO},  we obtain \eqref{intro1.18}.
It  is straightforward to check (see \cite{tKLA} or \cite{KLmult}) that  for $1\le a,b \le s$,  with $a\neq  b$:
\begin{equation}
\label{VOA}
\begin{aligned}
\sigma  E^{+a,-b}(w,z)\sigma^{-1}=&
\sigma :\psi^{+a}(w)\psi^{-b}(z):\sigma^{-1} \\
=q_a (-1)^{\sum_{\ell=1}^{a-1}  q_\ell\frac{\partial}{\partial q_\ell}}&
 q_b^{-1} (-1)^{\sum_{\ell=1}^{b-1}q_\ell\frac{\partial}{\partial q_\ell}}
w^{ q_a \frac{\partial}{\partial q_a}}z^{- q_b \frac{\partial}{\partial q_b}}
e^{z\cdot t^{(a)}-w\cdot t^{(b)}}e^{w^{-1}\cdot \tilde\partial_{t^{(b)}}-z^{-1}\tilde\partial_{t^{(a)}}},
\\
\sigma  E^{+a,-a}(w,z)\sigma^{-1}=&
\sigma :\psi^{+a}(w)\psi^{-a}(z):\sigma^{-1} \\
=& i_{w,z}\frac1{w-z}
\big[  \left(\frac{w}{z}\right)^{ q_a \frac{\partial}{\partial q_a}}
e^{z\cdot t^{(a)}-w\cdot t^{(a)}}e^{w^{-1}\cdot \tilde\partial_{t^{(a)}}-z^{-1}\cdot \tilde\partial_{t^{(a)}}}
-1\big].
\end{aligned}
\end{equation}

For $x=b$ or $d$,  for which there is in general a combination of $r$ neutral and $s$ charged fields, we have to extend the   Fock space $B_s$ by $p$  variables  $\theta_i$,  $1\le i\le p$,  for  $r=2p$ or $r=2p+1$, in the $b$-case, and $r=2p$ or $r=2p-1$ in the $d$-case,
which anticommute: 
  $\theta_i\theta_j=-\theta_j\theta_i$.  
So   for $x=b$ or $d$ we let 
\begin{equation}
\label{FockD}
 B_{s,r}=\mathbb C[\theta_i, q_j,q_j^{-1} , t^{(\ell)} |\, 1\le i\le p,\ 1\le j\le s ,\  1\le \ell\le s+r],
\end{equation}
where $t^{(\ell)}=(t^{(\ell)}_1,t^{(\ell)}_2,t^{(\ell)}_3,\ldots)$ if $1\le \ell\le s$ and $t^{(\ell)}=(t^{(\ell)}_1,t^{(\ell)}_3,t^{(\ell)}_5,\ldots)$ if $\ell>s$.
We define a vector space   isomorphism $\sigma:F_x\to B_{s,r}$ by 
$$
\begin{aligned}
\sigma(Q_1^{a_1} \cdots Q_s^{a_s}R_1^{b_1} \cdots R_{\tilde r}^{b_{\tilde r}}|0\rangle)= 
q_1^{a_1}q_2^{a_2}\cdots q_s^{a_s}
\left(R_1(\theta)\right)^{b_1}\left(R_2(\theta)\right)^{b_2}
\cdots  \left(R_{\tilde r}(\theta)\right)^{b_{\tilde r}}1,
\end{aligned}
$$
where 
$\tilde r=r$ in the $b$-case and $\tilde r=r$ if $r$ is even and $\tilde r=r+1$ if $r$ is odd in the $d$-case, and 
\begin{equation}
\label{Rtheta}
R_{2j-1}(\theta):=\frac{\frac{\partial}{\partial \theta_j}+\theta_j}{\sqrt 2}, \quad
R_{2j}(\theta):=\sqrt{-1}\frac{\frac{\partial}{\partial \theta_j}-\theta_j}{\sqrt 2},\quad 1\le j\le p .
\end{equation}
Then $\sigma r_x(\alpha_k^j)\sigma^{-1}$,  for $j\le s$,  can be realized as in
\eqref{alphaAA}, and for $j>s$ we set:
\begin{equation}
\label{neutralalpha}
\sigma r_x(\alpha_k^j)\sigma^{-1}=\frac{\partial}{\partial t^{(j)}_k},\quad \sigma r(\alpha_{-k}^j)\sigma^{-1}= \frac{k}2  t_k^{(j)},\quad j>s,\ k>0.\\
\end{equation}
The  action of $Q_j$ is is given by \eqref{3.49a}.
The action of $R_j$ for $1\le j\le 2p$ is given by
\begin{equation}
\label{R}
\begin{aligned}
\sigma R_{j} \sigma^{-1}
= (-1)^{\sum_{k=1}^{s}q_k\frac{\partial}{\partial q_k}} R_j(\theta), 
\end{aligned}
\end{equation}
where $R_j(\theta)$ is given by \eqref{Rtheta}.  For $b_\infty$ and $r$ odd, we also have 
\[
R_{r}(\theta)=(-1)^{\sum_{j=1}^{p} \theta_j\frac{\partial}{\partial \theta_j}}
\]
Thus, as before, $ \sigma\psi^{\pm j}(z)\sigma^{-1}$, for $1\le j\le r$, is given by 
\eqref{intro1.18}.
For $1\le j\le r$  (see e.g.  \cite{KvdLB} or \cite{tKLB}) we have  that $\sigma\tilde\phi^{s+j}(z)\sigma^{-1}$ is given by 
\begin{equation}
\label{intro1.18b}
\sigma \tilde \phi^{s+b} (z)\sigma_{s,r}^{-1}= (-1)^{\sum_{j=1}^{s}    q_j\frac{\partial}{\partial q_j} }R_b(\theta)
e^{z\circ t^{(s+b)}}e^{-2(z^{-1}\circ\tilde \partial_{ t^{(s+b)}})}.
\end{equation}
Finally,  if $r=2p$ is even for $b_\infty$ or $r=2p-1$ is odd for $d_\infty$,
there is also an operator $\sigma_0$.  One has
\begin{equation}
\label{S(theta)}
\sigma\sigma_0\sigma^{-1}=S(\theta)=\begin{cases}
\frac1{\sqrt 2}(-1)^{\sum_{k=1}^{s}q_k\frac{\partial}{\partial q_k}+\sum_{j=1}^{p} \theta_j\frac{\partial}{\partial \theta_j}},& x=b, \\
\sqrt{-1}(-1)^{\sum_{k=1}^{s}q_k\frac{\partial}{\partial q_k}} \frac{\frac{\partial}{\partial \theta_p}-\theta_p}{\sqrt 2},& x=d.
\end{cases}
\end{equation}
Using the results of Subsection \ref{FermVO}  we obtain a representation of $b_\infty$ or $d_\infty$ on the Fock space $B_{r,s}$.  It is straightforward to check that  for $1\le a,b \le s$,   the equations \eqref{VOA} still hold.   
Let  $a, d\le s$ and $1\le b,c\le r$, then 
\begin{equation}
\label{VOD}
\begin{aligned}
&\sigma  E^{\pm a,\pm d}(w,z)\sigma^{-1}=
\sigma :\psi^{\pm a}(w)\psi^{\pm d}(z):\sigma^{-1} =\\
&\qquad =(w-z)^{\delta_{ad}}
q_a^{\pm 1} (-1)^{\sum_{\ell=1}^{a-1}q_\ell\frac{\partial}{\partial q_\ell}}
 q_d^{\pm 1} (-1)^{\sum_{\ell=1}^{d-1}q_\ell\frac{\partial}{\partial q_\ell}}
w^{ \pm q_a \frac{\partial}{\partial q_a}}z^{-\pm q_d \frac{\partial}{\partial q_d}}\times\\
&\qquad\qquad   e^{\pm (z\cdot t^{(a)}+w\cdot t^{(b)})}e^{\mp(z^{-1}\cdot\tilde\partial_{t^{(a)}}+w^{-1}\cdot \tilde\partial_{t^{(b)}})},
\\
&\sigma  E^{s+b,s+c}(w,z)\sigma^{-1}=\sigma :\tilde\phi^{s+b}(w)\tilde\phi^{s+c}(z):\sigma ^{-1}\\
&\qquad = i_{w,z}\left(\frac{w-z}{w+z}\right)^{\delta_{bc}}\big[
 R_b(\theta) R_c(\theta)
e^{ (z\circ t^{(s+b)}+w\circ t^{(s+c)})}e^{-2(z^{-1}\circ\tilde\partial_{t^{(s+b)}}+w^{-1}\circ \tilde\partial_{t^{(s+c)}})}
-\delta_{bc}\big],
\\
&\sigma  E^{\pm a,s+b}
(w,z)\sigma^{-1}
=\sigma :\psi^{\pm a}(w)\tilde\phi^{s+b}(z):\sigma ^{-1}=\\
&\qquad=q_a(-1)^{\sum_{i=a}^s q_i\frac{\partial}{\partial q_i}}  w^{\pm q_a\frac{\partial}{\partial q_a}}  R_b(\theta) 
e^{\pm z\cdot t^{(a)}+w\circ t^{(s+b)}}e^{\mp z^{-1}\cdot\tilde\partial_{t^{(a)}}-2(w^{-1}\circ \tilde\partial_{t^{(s+b)}})}.
%
%
\end{aligned}
\end{equation}
If $\sigma_0$ is present, we also have 
\begin{equation}
\label{VODsigma}
\begin{aligned}
&\sigma  E^{\pm a}(z)\sigma^{-1}=
\sigma :\psi^{\pm a}(z)\sigma_0:\sigma^{-1} 
= \ q_a^{\pm 1} (-1)^{\sum_{\ell=1}^{a-1}q_\ell\frac{\partial}{\partial q_\ell}}
S(\theta)
z^{ \pm q_a \frac{\partial}{\partial q_a}} 
e^{\pm z\cdot t^{(a)} }e^{\mp z^{-1}\cdot\tilde\partial_{t^{(a)}}}
%
,
\\
&\sigma  E^{s+b }(z)\sigma^{-1}=
\sigma :\tilde\phi^{s+b}(z)\sigma_0:\sigma ^{-1} 
=R_b(\theta) S(\theta)
e^{z\circ t^{(s+b)}}e^{ -2(z^{-1}\circ \tilde\partial_{t^{(s+b)}})},
%
\end{aligned}
\end{equation}
where $S(\theta)$ is given by \eqref{S(theta)}.

\subsubsection{The bosonic realization of $F_c$,  the Weyl algebra case}
\label{3.7.2}
We use the results of Subsection
\ref{3.6.2}  to obtain a bosonic realization of $F_c$.
Let
\begin{equation}
\label{FockC}
\begin{aligned}
C_{s,r}=&\mathbb C[    t^{(a)}_i, t^{(c)}_j, \xi^a_k, \xi^c_\ell |\, 1\le a\le s<c\le s+r;, i\in\mathbb Z_{>0}, \,
	j\in\mathbb Z_{odd>0},  \,
 k\in\mathbb Z_{\ne 0},\, 
\ell\in\frac12+\mathbb Z_{\ge 0}],
\end{aligned}
\end{equation}
where $\xi^a_k$ are anticommuting variables,
and
  define  the vector space isomorphism $\sigma:F_c\to C_{s,r}$ as follows.
Again we use the standard realization of the Heisenberg algebra ($1\le a\le s<c \le s+r$,  $i\in\mathbb Z_{>0}$ and $j\in \mathbb Z_{odd>0}$)
\[
\begin{aligned}\sigma \alpha^a_{-i}\sigma^{-1} &=it^{(a)}_i, 
\quad
\sigma \alpha^a_{i}\sigma^{-1} =-\frac{\partial}{\partial t_i^{(a)}},
\\
\sigma \alpha^c_{-j}\sigma^{-1}& =\frac12 jt^{(c)}_j, 
\quad
\sigma \alpha^c_{j}\sigma^{-1} =-\frac{\partial}{\partial t_j^{(c)}}.
\end{aligned}
\]
We realize the modes of the fields $\Theta^{\pm a}(z)=\sum_{ j\in \mathbb Z} \Theta^{\pm a} _j z^{-j-1}=  Q_a^{\mp a} \theta^{\pm a}(z) $  for $1\le a\le s$ and
the modes of $\Theta^c(z)$ for $s< c\le r$,   as multiplication and differentiation of Grassmann variables,  viz.   for $a=1,\ldots,s$,
\[
\sigma \Theta^{\pm a}(z)
\sigma^{-1}=
\Xi^{\pm a}(z)S_a,\qquad \sigma \Theta^{c}(z)\sigma^{-1}=\Xi^{c}(z)S_c
,
\]
where ($a\le s< c$)
\begin{equation}
\label{Xipma}
\begin{aligned}
\Xi^{+ a}(z)&=\frac{\partial}{\partial \xi_{0}^a}z^{-1}+\sum_{k=1}^\infty -k\xi_{k}^az^{k-1}+\sum_{k=1}^\infty \frac{\partial}{\partial \xi_{-k}^a}z^{-k-1},\\
\Xi^{- a}(z)&=-\frac{\partial}{\partial \xi_{0}^a}z^{-1}+\sum_{k=1}^\infty k\xi_{-k}^az^{k-1}+\sum_{k=1}^\infty \frac{\partial}{\partial \xi_{k}^a}z^{-k-1},\\
\Xi^{c}(z)&
= \sum_{i\in  \frac12+\mathbb Z_{\ge  0}}\left((-1)^{i+\frac12} 2i\xi^c_i z^{i-\frac12}
+\frac{\partial}{\partial \xi^c_i}z^{- i-\frac12}\right).
\end{aligned}
\end{equation}
Here the operator $S_a$ acts as 
\[
S_a (1)=1, \qquad S_a\xi_k^bS_a^{-1}=\begin{cases}
-\xi_k^b&\mbox{if } a> b,\\
\xi_k^b&\mbox{if } a\le b,
\end{cases}
,\qquad S_a\frac{\partial }{\partial \xi_k^b}S_a^{-1}=\begin{cases}
-\frac{\partial }{\partial \xi_k^b}&\mbox{if } a> b,\\
\frac{\partial }{\partial \xi_k^b}&\mbox{if } a\le b.
\end{cases}
\]
Hence,  for the   modes  we find that
\[
\begin{aligned}
\sigma \Theta^{\pm a}_{0}\sigma^{-1} &=\pm \frac{\partial}{\partial \xi_{\mp j}^a}S_a,\quad
\sigma \Theta^{\pm a}_{-j}\sigma^{-1} =\mp j\xi^a_{\pm j} S_a, 
\quad
\sigma \Theta^{\pm a}_{j}\sigma^{-1} =-\frac{\partial}{\partial \xi_{\mp j}^a}S_a,\quad j\in\mathbb Z_{>0},
\\
\sigma \theta^c_{-k}\sigma^{-1} &=(-1)^{ k+\frac12}2 k \xi^c_kS_c,
\quad
\sigma \theta^c_{k}\sigma^{-1} =\frac{\partial}{\partial \xi_k^c}S_c,  \quad k\in\frac12+ \mathbb Z_{>0}
\end{aligned}
 \]
We still need to define $\sigma \alpha_0^a\sigma^{-1}$ for $1\le a\le s$.  Formula \eqref{alphatheta} suggest the following:
\begin{equation}
\label{Ta}
\sigma \alpha_0^a\sigma^{-1}=T_a:=-\sum_{k=1}^\infty \xi_k^a\frac{\partial}{\partial \xi_{k}^a}+
\sum_{k=0}^\infty \xi_{-k}^a\frac{\partial}{\partial \xi_{-k}^a}.
\end{equation}
\begin{remark}
Note that here we use the fields $\Theta^{\pm a}(z)$ instead of $\theta^{\pm a}(z)$,  which  means that we  do not consider  the operators $Q_a$.  When using $\theta^{\pm a}(z)$ instead of $\Theta^{\pm a}(z)$, the operator $Q_a$ appears in the  field 
$\theta^{\pm a}(z)$ and we can assume that $\sigma Q_a \sigma^{-1}=q_a$,  but then $\sigma \alpha_0^a\sigma^{-1}= q_a\frac{\partial}{\partial q_a}$.
\end{remark}

Using all this and formulas \eqref{form3.4},  we obtain for $1\le a\le s<c\le r+s$:
\begin{equation}
\label{Bos-sympfield}
\begin{aligned}
\sigma b^{\pm a}(z)\sigma^{-1}&= e^{\pm z\cdot t^{(a)}}
\Xi^{+a}(z)z^{ \pm T_a}
 e^{\pm z\cdot \tilde\partial_{t^{(a)}}}
S_a,\\
\sigma b^c(z)\sigma^{-1}&=e^{z\circ t^{(c)}}
\Xi^c(z) 
e^{2 (z\circ \tilde\partial_{t^{(c)}})}
S_c.
\end{aligned}
\end{equation}
\begin{remark}
In this way we always want to present the Grassmann variables in a prefered order, namely the lexicographic ordering, i.e.
$\xi_i^k <_{\rm lex}\xi_j^\ell$, when $k<\ell$, or when $k=\ell$  and $i<j$.
Hence we always  write a monomial in the following ordered way.
\[
\xi_{i_1}^{k_1} \xi_{i_2}^{k_2}\cdots \xi_{i_m}^{k_m},\quad\mbox{with }
(i_1,k_1) <_{\rm lex}(i_2,k_2) <_{\rm lex}\cdots<_{\rm lex}(i_m,k_m) .
\]
\end{remark}

We use the vertex operators \eqref{Bos-sympfield}   to derive for $1\le a\le d\le s$
\begin{equation}
\label{VOAC1}
\begin{aligned}
&\sigma C^{\pm a,  \pm d}(z,w)\sigma^{-1}= 
\sigma :b^{\pm a}(z)b^{\pm d}(w):\sigma^{-1}\\
&= 
i_{z,w}\left(\frac{1}{z-w} \right)^{\delta_{ad}}
e^{\pm (z\cdot  {t^{(a)}}+ w\cdot {t^{(d)}}}
\Xi^{\pm a}(z)\Xi^{\pm d}(w)z^{\pm T_a}w^{\pm T_d}
e^{\pm (z^{-1}\cdot \tilde\partial_{t^{(a)}}+w^{-1}\cdot \tilde\partial_{t^{(d)}})}
S_aS_d,\\
&\sigma C^{+a,- d}(z,w)\sigma^{-1}=
\sigma: b^{+a}(z)b^{- d}(w):\sigma^{-1}\\
& =\left(z-w\right)^{\delta_{ad}}
e^{ z\cdot  {t^{(a)}}-w\cdot {t^{(d)}}}
\Xi^{+a}(z)\Xi^{-d}(w)z^{T_a}w^{-T_d}
e^{ (z^{-1}\cdot \tilde\partial_{t^{(a)}}- w^{-1}\cdot \tilde\partial_{t^{(d)}}}
S_aS_d -i_{z,w}\left(\frac{\delta_{ad}}{z-w} \right) ,
\end{aligned}
\end{equation}
and for $s< a\le c\le r+s$, we obtain
\begin{equation}
\label{VOAC2}
\begin{aligned}
&\sigma C^{a, c}(z,w):\sigma^{-1}=
\sigma :b^a(z)b^c(w):\sigma^{-1}\\
&=
i_{z,w}\left(\frac{z+w}{w-z}\right)^{\delta_{ac}}
\Xi^a(z)\Xi^b(w)
e^{z\circ t^{(a)}+w\circ t^{(c)}}
e^{2 (z^{-1}\circ \tilde\partial_{t^{(a)}}+w^{-1}\circ \tilde\partial_{t^{(c)}})}
S_aS_c
  -i_{z,w}\frac{\delta_{ac}}{z+w}
 .
\end{aligned}
\end{equation}
Finally,   the mixed terms, when $1\le a\le r<c\le r+s$,  are
\begin{equation}
\label{VOAC3}
\begin{aligned}
\sigma C^{\pm a,c}(z,w):\sigma^{-1}&=
\sigma:b^{\pm a}(z)b^c(w):\sigma^{-1}\\
&= 
e^{\pm z\cdot  t^{(a)}+w\circ t^{(c)}}
\Xi^{+a}(z)\Xi^c(w)z^{\pm T_a}
e^{\pm z^{-1}\cdot \tilde\partial_{t^{(a)}}+w^{-1}\circ \tilde\partial_{t^{(c)}})}
%
S_aS_c.
\end{aligned}
\end{equation}

\section{Hierarchies of KP type}
\label{sec4} 
\subsection{The fermionic description  of the KP, BKP  and DKP hierarchies}
\label{subsec4.1}
{\bf The KP hierarchy.} Recall the charge decomposition \eqref{3.3} of the module $F$  .
By definition, a non-zero element  $\tau\in F^{(0))}$,  satisfies the  KP  hierarchy if (cf. \eqref{intro1.5})
\begin{equation}
\label{KP}
S(\tau\otimes \tau)=
\sum_{i\in\mathbb{Z}+
\frac{1}{2}} \psi_i^+ \tau\otimes \psi_{-i}^-\tau =0.
\end{equation}
Note that \eqref{KP} can be rewritten   in terms of $s$ pairs of  charged free fermionic fields $\psi^{\pm j}(z)$,  defined by  \eqref{aFERMIONS}, as follows.:
\begin{equation}
\label{S}
S(\tau\otimes \tau)= \sum_{j=1}^s\sum_{i\in\mathbb{Z}+
\frac{1}{2}} \psi_i^{+j}\tau \otimes \psi_{-i}^{-j}\tau = {\rm Res}_{z=0}\sum_{j=1}^s \psi^{+j}(z)\tau\otimes \psi^{-j}(z) \tau dz.
\end{equation}
Equation (\ref{KP}) or \eqref{S}  is the KP hierarchy in the fermionic picture. It becomes a hierarchy of differential equations if we apply  the isomorphism $\sigma$,  constructed in  section \ref{BRF1}. Taking $\tau\in F^{(m)}$ produces the same hierarchy.
\\
\ 
\\
\noindent {\bf The BKP hierarchy.}
Recall from Section \ref{spinBD} that the   spin module  $F_b$ is an irreducible $C\ell_b$-module $F_b$.
The following equation on a  non-zero $\tau\in F_b$ is called the BKP hierarchy in the fermionic picture  (cf.  \eqref{intro1.6b}):
\begin{equation}
\label{BKP}
S_B(\tau\otimes \tau)=
\sum_{i\in\mathbb{Z}
} \tilde\phi_i \tau\otimes \tilde\phi_{-i}\tau =\frac 12 \tau\otimes \tau.
\end{equation}
This equation describes the $B_\infty$-orbit of the vacuum vector $|0\rangle$,  see \cite{KvdLB} for more details.  
Equation (\ref{BKP}) can be rewritten in terms of $s$ pairs of charged free fermionic  fields $\psi^{\pm b}(z)$,  $b=1,\ldots , s$, and $r$  neutral twisted free fermionic fields $\tilde\phi^{c}(z)$, $c=s+1,\ldots, s+r$,   as follows
If $r=2p+1$ is odd,  then \eqref{BKP} becomes
\begin{equation}
\label{SB2oddfield}
\begin{aligned}
{\rm Res}_{z=0}\,& 
\left(\sum_{b=1}^s  \left( \psi^{+b}(z) \otimes \psi^{-b}(z)
+   \psi^{-b}(z) \otimes \psi^{+b}(z)\right)+\right.
\\
&\qquad+\left.
\sum_{c=s+1}^{s+2p+1}z^{-1} \tilde\phi^c(z)\otimes \tilde\phi^c(-z)\right)(\tau\otimes \tau)dz=\frac12\tau\otimes \tau  .
\end{aligned}
\end{equation}
However, if $r=2p$ is even,  we  have the extra generator  $\sigma_0$  so that   (\ref{BKP})  turns into:
\begin{equation}
\label{SB2evenfield}
\begin{aligned}
\big\{ \sigma_0 \otimes \sigma_0+
{\rm Res}_{z=0}\,& 
\left(\sum_{b=1}^s  \left( \psi^{+b}(z) \otimes \psi^{-b}(z)
+   \psi^{-b}(z) \otimes \psi^{+b}(z)\right)+\right.
\\
&\qquad+\left.
\sum_{c=s+1}^{s+2p}z^{-1} \tilde\phi^c(z)\otimes \tilde\phi^c(-z)\right)dz
\big\}
(\tau\otimes \tau)=\frac12\tau\otimes \tau .
\end{aligned}
\end{equation}
\ 
\\
{\bf The DKP hierarchy} Recall from Section \ref{spinBD} that the   spin module  $F_d$ splits into two irreducible $d_\infty$-modules $F_d^{\overline 0}\oplus F_d^{\overline 1}$, . The highest weight vector of  $F_d^{\overline 0}$ is  $|\overline 0\rangle=|0\rangle$ and of $ F_d^{\overline 1}$ is  $|\overline 1\rangle=\sqrt 2 \phi_{-\frac12}|0\rangle $.
The following equation on the non-zero  $\tau_a\in F_d^a$ is called DKP hierarchy in the fermionic picture:
\begin{equation}
\label{DKP}
S_D(\tau_a\otimes \tau_a)=
\sum_{i\in\frac12 +\mathbb{Z}
} \phi_i \tau_a\otimes \phi_{-i}\tau_a =0,  \quad a=\overline 0 \ \mbox{or }\overline 1.
\end{equation}
This equation describes the $D_\infty$-orbit of the highest weight vector $|\overline a\rangle$, see \cite{KvdLB} for more details. This group consists of (even) invertible elements $g$ such that $g\phi_jg^{-1}=\sum_{i} a_{ij}\phi_i$ where  $\sum_{k} a_{kj}a_{-k,j}=\delta_{i,-j}$, i.e. that $g$   leaves the bilinear form $(\cdot,\cdot)_d$ of \eqref{bil1}  invariant.  

Equation (\ref{DKP}) can be  expressed in terms of $s$ pairs of charged free fermionic fields $\psi^{\pm b}(z)$ and $r$  neutral twisted fields $\tilde\phi^{c}(z)$,  introduced in Section \ref{section 3.5}. If $r=2p$ is even,  then \eqref{DKP} becomes
\begin{equation}
\label{SD2evenfield}
\begin{aligned}
{\rm Res}_{z=0}\,& 
\left(\sum_{b=1}^s  \left( \psi^{+b}(z) \otimes \psi^{-b}(z)
+   \psi^{-b}(z) \otimes \psi^{+b}(z)\right)+\right.
\\
&\qquad+\left.
\sum_{c=s+1}^{s+2p}z^{-1} \tilde\phi^c(z)\otimes \tilde\phi^c(-z)\right)(\tau_a\otimes \tau_a)dz=0 .
\end{aligned}
\end{equation}
However, if $r=2p-1$ is odd,  there is an  extra generator   $\sigma_0$.
Then (\ref{DKP})  turns into:
\begin{equation}
\label{SD2oddfield}
\begin{aligned}
\big\{ \sigma_0 \otimes \sigma_0+
{\rm Res}_{z=0}& 
\left(\sum_{b=1}^s  \left( \psi^{+b}(z) \otimes \psi^{-b}(z)
+   \psi^{-b}(z) \otimes \psi^{+b}(z)\right)+\right.
\\
&\qquad+\left.
\sum_{c=s+1}^{s+2p-1}z^{-1} \tilde\phi^c(z)\otimes \tilde\phi^c(-z)\right)dz
\big\}
(\tau_a\otimes \tau_a)=0.
\end{aligned}
\end{equation}
\subsection{The symplectic bosons description of the  CKP hierarchy}
\label{subsec4.2}
Recall from Section \ref{spinC} that the   Weyl module  $F_c$ splits into two irreducible $c_\infty$-modules $F_c^{\overline 0}$ and   $F_c^{\overline  1}$ with highest weight vectors $|\overline 0\rangle =|0\rangle $, respectively $|\overline 1\rangle = b_{-\frac12}|0\rangle $.
The  CKP hierarchy (see \cite{LOS}) in the fermionic picture is the following equation on $\tau\in F_c^{\overline 0}$:
\begin{equation}
\label{CKP}
S_C(\tau\otimes \tau)=
\sum_{i\in \frac12+\mathbb{Z}
} (-1)^{i+\frac12}b_i \tau\otimes b_{-i}\tau=0.
\end{equation}
Equation (\ref{CKP}) can be rewritten  in terms of $s$ charged fields $b^{\pm a}(z)$ and $r$  neutral fields $b^{c}(z)$.:
\begin{equation}
\label{CKP2}
\begin{aligned}
\sum_{a=1}^s\sum_{i\in\frac{1}{2}+\mathbb{Z}}&\left( b^{-a}_i\tau \otimes b_{-i}^{+a}\tau- b^{+a}_i\tau \otimes b_{-i}^{-a}\tau\right)+
\sum_{c=s+1}^{r+s}\sum_{i\in\frac12+\mathbb{Z}}(-1)^{i+\frac12}b^c_i\tau\otimes b^c_{-i}\tau=0.
\end{aligned}
\end{equation}
In terms of the  symplectic bosonic fields,  it becomes
\begin{equation}
\label{CKP3}
\begin{aligned}
{\rm Res}_{z=0}\big\{
\sum_{a=1}^s & \left(b^{-a}(z)\tau \otimes b^{+a}(z)\tau- b^{+a}(z)\tau \otimes b ^{-a}(z)\tau\right)
+
\sum_{c=s+1}^{r+s} b^c(z)\tau\otimes b^c (-z)\tau\big\} dz =0.
\end{aligned}
\end{equation}
Unfortunately, taking $\tau\in F_c^{\overline  1}$, does not give the same hierarchy.

\subsection{Bosonic description of the multicomponent KP,  BKP,  CKP  and DKP hierarchies}
\label{subsec4.3}
{\bf The KP hierarchy}
Equation (\ref{KP}) or \eqref{S}  is the KP hierarchy in the fermionic picture. It becomes a hierarchy of differential equations if we apply  the isomorphism $\sigma:F \xrightarrow{\sim}  B_s$,  constructed in \eqref{BRF1} so that  for $\tau \in F^{(0)}$
\[
\sigma(\tau)=\tau(q, t)=\sum_{|\underline k|=0}\tau_{\underline k}(t)q^{\underline k}
\]
becomes a function in $t=(t_1^{(i)},t_2^{(i)},\ldots)$,  $i=1,\dots ,s$, and in the  $sl_s$-root lattice, 
with lattice points $\underline k=(k_1,k_2,\ldots k_s)$ with $|\underline k|=k_1+k_2+\cdots+k_s=0$,  where  $q^{\underline k}=q_1^{k_1}q_2^{k_2}\cdots q_s^{k_s}$. If $s=1$ one  obtains the classical KP-hierarchy. For $s>1$ we obtain the $s$-component KP hierarchy.
Equation \eqref{KP} describes the $GL_\infty$-group orbit of $|0\rangle$ \cite{KP}.  Note that in the fermionic picture this equation is the same.  It is only after we apply the isomorphism $\sigma$ and  $\tau\in B_s^{(0)}$ that we obtain different descriptions for different choices of $s$. 
Using the fermionic vertex operators \eqref{FVO} and \eqref{S} we rewrite \eqref{KP} and obtain the $s$-component KP hierarchy. Let $\underline e_i$ be the basis vector in $\mathbb Z^s $, wich has a 1 on the $i$-th place and zeros elsewhere and let $|\underline k|_j=k_1+k_2+\cdots+k_{j}$.  Then, for each pair  $\underline k,\underline\ell$ with $|\underline k|=|\underline \ell|=0$ we have the  $s$-component KP hierarchy of Hirota bilinear equations \eqref{BosKP} on $\tau_{\underline k}(t) $,  $|\underline k|=0$ (cf. \eqref{S}),
describing the orbit $GL_\infty|0\rangle$ in the bosonic picture. Taking $\tau\in F^{(m)}$, we obtain the same equations on $\tau_{\underline k}(t) $, $|\underline k|=m$ for the $GL_\infty$   orbit of $|m\rangle$.
\\
\
\\
{\bf The BKP hierarchy}
In this case $F_b$ is irreducible and $\sigma:F_b\xrightarrow{\sim} B_{s,r}$, where $B_{s,r}$, $r=2p$ or $r=2p+1$, is given by \eqref{FockD} and $\sigma$ constructed in \eqref{BRF1}.
We write 
\[
\sigma(\tau)=\tau(q,t,\theta) =\sum_{\underline k\in \mathbb Z^s,\underline \ell\in \mathbb Z_2^{p}}\tau_{\underline k,\underline \ell}(t)q^{\underline k}\theta^{\underline\ell},
\]
where $q^{\underline k}=q_1^{k_1}q_2^{k_2}\cdots q_s^{k_s}$ and  $\theta^{\underline\ell}=\theta_1^{\ell_1}\theta_2^{\ell_2}\cdots\theta_{p}^{\ell_p}$.  
Now using $\sigma$ and substituting the vertex operators \eqref{intro1.18},  \eqref{intro1.18b} 
and if $r=2p$, then use formula \eqref{S(theta)} for
$\sigma \sigma_0\sigma^{-1}$,
we obtain:
\begin{equation}
\label{BosBKP}
\begin{aligned}
&{\rm Res}_{z=0}\big(\sum_{b=1}^s \big(
(-1)^{|\underline k+\overline{\underline k}|_{b-1}} z^{ k_b-\overline k_b-2+2\delta_{bs}}
e^{z\cdot( t^{(b)}-\overline t^{(b)})}e^{z^{-1}\cdot( \tilde\partial_{\overline  t^{(b)}}- \tilde\partial_{ t^{(b)}} )}
\tau_{\underline k+\underline e_s-\underline e_b,\underline\ell}(t)\tau_{\overline{\underline k}+\underline e_b-\underline e_s,\overline{\underline \ell}}(\overline t)+
\\
&+(-1)^{|\underline k+\overline{\underline k}|_{b-1}} z^{ -k_b+\overline k_b-2-2\delta_{bs}}
e^{z\cdot(\overline  t^{(b)}- t^{(b)})}e^{z^{-1}\cdot( \tilde\partial_{ t^{(b)}}-\tilde \partial_{\overline t^{(b)}} )}
\tau_{\underline k+\underline e_s+\underline e_b,\underline\ell}(t)\tau_{\overline{\underline k}-\underline e_b-\underline e_s,\overline{\underline\ell}}(\overline t)\big)+\\
&+\frac12
\sum_{c=1}^{2p} (-1)^{{|\underline k+\overline{\underline k}|}+|\underline\ell+\overline{\underline\ell}|_{\left[\frac{c-1}2\right]}+(c-1)(\ell_{\left[\frac{c+1}2\right]}+\overline\ell_{\left[\frac{c+1}2\right]}+1)}z^{-1}
e^{z\circ( t^{(s+c)}-\overline t^{(s+c)})}\times \\
&\quad e^{2 (z^{-1}\circ( \tilde\partial_{ \overline t^{(s+c)}}- \tilde\partial_{  t^{(s+c)}}) )}
%
\tau_{\underline k+\underline e_s ,\underline\ell+ \underline e_{\left[\frac{c+1}2\right]}}(t)
\tau_{\overline{\underline k} -\underline e_s,\overline{\underline\ell}+ \underline e_{\left[\frac{c+1}2\right]}}(\overline t)\\
&+\frac{\delta_{r,2p+1}}2
(-1)^{{|\underline k+\overline{\underline k}|}+|\underline\ell+\overline{\underline\ell}|}
z^{-1}
e^{z\circ( t^{(s+r)}-\overline t^{(s+r)})} 
e^{2 (z^{-1}\circ( \tilde\partial_{ \overline t^{(s+r)}}- \tilde\partial_{ t^{(s+r)}}) )}
%
\tau_{\underline k+\underline e_s ,\underline\ell}(t)
\tau_{\overline{\underline k} -\underline e_s,\overline{\underline\ell}}(\overline t)
\big) dz\\
=&\left( \frac12-\frac{\delta_{r,2p}}2 (-1)^{{{|\underline k+\overline{\underline k}|}+|\underline\ell+\overline{\underline\ell}|}}\right)
\tau_{\underline k+\underline e_s ,\underline\ell}(t)
	\tau_{\overline{\underline k} -\underline e_s,\overline{\underline \ell}}(\overline t)
.
\end{aligned}
\end{equation}
Here $[x]$ stands for the   largest integer smaller or equal to $x$.\\
\
\\
{\bf The DKP hierachy}
This case is similar to the BKP case, however in this case $F_d$ splits into two irreducible $d_\infty$ modules.  Again we have two cases,
viz $r=2p$ and $r=2p-1$,  and we 
   write
\[
\sigma(\tau_a)=\tau^a(q,t,\theta) =\sum_{\underline k\in \mathbb Z^s,\underline \ell\in \mathbb Z_2^{p},\, |\underline k|+|\underline\ell|=a\, {\rm mod}\,  2}\tau_{\underline k,\underline \ell}(t)q^{\underline k}\theta^{\underline\ell},
\]
Then the fermionic DKP hierarchy turns into the following set of equations for all $\underline k,\overline{\underline k}\in \mathbb Z^s$, $\underline \ell,\overline{\underline\ell}\in \mathbb Z_2^{p}$, such that $|\underline k|+|\underline\ell|+1=a\mod 2$:
\begin{equation}
\label{BosDKP}
\begin{aligned}
&{\rm Res}_{z=0}\big(\sum_{b=1}^s \big(
(-1)^{|\underline k+\overline{\underline k}|_{b-1}} z^{ k_b-\overline k_b-2+2\delta_{bs}}
e^{z\cdot( t^{(b)}-\overline t^{(b)})}e^{z^{-1}\cdot( \tilde\partial_{ \overline t^{(b)}}- \tilde\partial_{ t^{(b)}} )}
\tau_{\underline k+\underline e_s-\underline e_b,\underline\ell}(t)\tau_{\overline{\underline k}+\underline e_b-\underline e_s,\overline{\underline \ell}}(\overline t)+
\\
&\qquad+(-1)^{|\underline k+\overline{\underline k}|_{b-1}} z^{ -k_b+\overline k_b-2-2\delta_{bs}}
e^{z\cdot(\overline t^{(b)}-  t^{(b)})}e^{z^{-1}\cdot( \tilde\partial_{ t^{(b)}}-\tilde \partial_{\overline t^{(b)}} )}
\tau_{\underline k+\underline e_s+\underline e_b,\underline\ell}(t)\tau_{\overline{\underline k}-\underline e_b-\underline e_s,\overline{\underline\ell}}(\overline t)\big)+\\
&\qquad+\frac12
\sum_{c=1}^{r} (-1)^{{|\underline k+\overline{\underline k}|}+|\underline\ell+\overline{\underline\ell}|_{\left[\frac{c-1}2\right]}+(c-1)(\ell_{\left[\frac{c+1}2\right]}+\overline\ell_{\left[\frac{c+1}2\right]}+1)}z^{-1}
e^{z\circ( t^{(s+c)}-\overline t^{(s+c)})}\times \\
&\qquad\qquad\qquad\qquad\qquad e^{2 (z^{-1}\circ( \tilde\partial_{ \overline t^{(s+c)}}- \tilde\partial_{  t^{(s+c)}}) )}
%
\tau_{\underline k+\underline e_s ,\underline\ell+ \underline e_{\left[\frac{c+1}2\right]}}(t)
\tau_{\overline{\underline k} -\underline e_s,\overline{\underline\ell}+ \underline e_{\left[\frac{c+1}2\right]}}(\overline t)\big)dz\\
&\qquad=\frac{\delta_{r,2p-1}}2 
\tau_{\underline k+\underline e_s ,\underline\ell+\underline e_{p}}(t)
\tau_{\overline{\underline k} -\underline e_s,\overline{\underline\ell}+\underline e_{ p}}(\overline t)
.
\end{aligned}
\end{equation}
In the above, we have used that $(-1)^{{{|\underline k+\overline{\underline k}|}+|\underline\ell+\overline{\underline\ell}|}}=1$.
\begin{remark}
\label{BDcorresp}
Note that the equations of  DKP hierarchy for $r=2p$ appear  as a subset of the equations of the BKP hierarchy for the same $r=2p$.  
Namely,  one only takes those $\underline k$,  $\underline\ell$, and $\overline{\underline k}$,  $\overline{\underline\ell}$, for which
$|\underline k|+|\underline\ell|+1=|\underline{\overline k}|+|\underline{\overline\ell|}+1=a\mod 2$. 
Since both $a=\overline 0$ and $a=\overline 1$, appear in the BKP hierarchy we also find, besides these two DKP hierarchies,  the so-called modified DKP equations.  See (\cite{KvdLB},  Section 2).
\end{remark}
\ 
\\
{\bf The CKP hierarchy}
Now using the isomorphism $\sigma$, see
Subsection  \ref{3.7.2}, 
we assume that $\tau(t,y,\xi)\in \sigma(F_0)$, thus 
\begin{equation}
\begin{aligned}
&\tau(t,\xi)
\in\\
 & \mathbb C[    t^{(a)}_i, t^{(c)}_j, \xi^a_k, \xi^c_\ell |\, 1\le a\le s<c\le s+r,\, i\in\mathbb Z_{>0}, \,
j\in\mathbb Z_{>0,odd}, \,
0\ne k\in\mathbb Z,\, 
\ell\in\frac12+\mathbb Z_{\ge 0}].
\end{aligned}
\end{equation}
with 
\[
\tau(t,\xi)=\sum_{w}
\tau_{w}(t) w,
\]
where 
\begin{equation}
\label{w}
w=\xi^1_{k_1^1}\cdot\cdot \xi^1_{k_{i_1}^1}
\xi^1_{\ell_1^1}\cdot\cdot \xi^1_{\ell_{j_1}^1}
\xi^2_{k_2^1}\cdot\cdot \xi^2_{k_{i_2}^2}
\xi^2_{\ell_2^1}\cdot\cdot \xi^2_{k_{j_2}^2}\cdot\cdot \xi^s_{k_1^s}\cdot\cdot  \xi^s_{k_{i_s}^s}\xi^s_{\ell_1^s}\cdot\cdot  \xi^s_{\ell_{j_s}^s}
\xi^{s+1}_{m_1^{s+1}}\cdots \xi^{s+1}_{m_{j_{s+1}}}
\cdot\cdot \xi^{s+r}_{m_1^{s+r}}\cdot\cdot \xi^{s+r}_{m_{j_{s+r}}},
\end{equation}
such that $i_a,j_b\in\mathbb Z_{\ge 0}$, 
$k_p^a\in\mathbb Z_{<0}$,  $\ell_p^a\in\mathbb Z_{>0}$, and  $m^c_q\in \frac12+\mathbb Z_{\ge 0}$,
with  $k_p^a< k_{p+1}^a$, $\ell_p^a< \ell_{p+1}^a$  and $m^c_q<m^c_{q+1}$   and 
 $i_1+\cdots i_s+j_1 +\cdots +j_{s+r}\in 2\mathbb Z$.

To describe the CKP hierarchy of differential equations, we will use the following notations.  
Let $\xi^b_j$ (resp.  $\xi_k^c$) be one of the Grassmann variables appearing  (resp.  not appearing)  in the expression \eqref{w}.  We write  $w\backslash \xi^b_j$ (resp $w\cup \xi^c_k$ for the monomial that we get by removing $\xi^b_j$ from $w$ (resp.  by adding $\xi^c_k$ to $w$ and writing it in the preferred lexicographical ordering).  We substitute \eqref{Bos-sympfield} in \eqref{CKP3} and  and thus obtain the 
multicomponent CKP hierarchy in the bosonic picture.
Let $w$ be given by \eqref{w} and $\overline{w}$ by the same expression but with all $\xi^b_j$ replaced by ${\overline\xi}^b_{\overline j}$, such that the number of Grassmann variables appearing in $w$ and $\overline w$ is odd,  then  we obtain 
\begin{equation}
\label{BosCKP}
\begin{aligned}
{\rm Res}_{z=0}&\bigg\{
\sum_{a=1}^s \bigg [
-z^{i_a-\overline i_a-j_a+\overline j_a}
e^{z\cdot (t^{(a)}-\overline t^{(a)})}
e^{z^{-1}\cdot (\tilde\partial_{t^{(a)}}-\tilde\partial_{\overline t^{(a)}})}
\times
\\
&\left(
\sum_{k\in\mathbb Z_{<0},\, k\ne  k_1^a,\ldots, k^a_{  i_a}}
\!\!\!\!\!\!\!\!\!\!\!
N_a( w, \xi_k^a)z^{k}\tau_{  w\cup  \xi_k^a}( t)-
\sum_{\ell=  \ell_1^a,\ldots, \ell^a_{  j_a}}
\!\!\!\!\!\!
 \ell N_a( w,  \xi^a_\ell)z^{\ell}\tau_{ w\backslash  \xi^a_\ell}(  t)
\right)
\times
\\
&\left(
\sum_{\ell\in\mathbb Z_{>0},\, \ell\ne \overline\ell_1^a,\ldots,\overline\ell^a_{\overline j_a}}
\!\!\!\!\!\!\!\!\!\!\!\!
N_a(\overline w,\overline \xi_\ell^a)z^{-\ell}\tau_{\overline w\cup \overline \xi_\ell^a}(\overline t)-
\sum_{k=\overline k_1^a,\ldots,\overline k^a_{\overline i_a}}
\!\!\!\!\!\!
kN_a(\overline w, \overline \xi^a_k)z^{-k}\tau_{\overline w\backslash \overline \xi^a_k}(\overline t)
\right)\\
&\qquad
-(-z)^{j_a-\overline j_a-i_a+\overline i_a}
e^{-z\cdot (t^{(a)}-\overline t^{(a)})}e^{-z^{-1}\cdot (\tilde\partial_{t^{(a)}}-\tilde\partial_{\overline t^{(a)}})}
\times
\\
&\left(
\sum_{\ell\in\mathbb Z_{>0},\, \ell\ne \ell_1^a,\ldots,\ell^a_{j_a}}
\!\!\!\!\!\!\!\!\!\!\!\!
N_a(w, \xi_\ell^a)(-z)^{-\ell}\tau_{w\cup \xi_\ell^a}(t)-
\sum_{k=k_1^a,\ldots,k^a_{i_a}}
\!\!\!\!\!\!
kN_a(w, \xi^a_k)(-z)^{-k}\tau_{w\backslash \xi^a_k}(t)
\right)\times
\\
&\left(
\sum_{k\in\mathbb Z_{k<0},\, k\ne \overline k_1^a,\ldots,\overline k^a_{\overline i_a}}
\!\!\!\!\!\!\!\!\!\!\!
N_a(\overline w, \overline \xi_k^a)(-z)^{k}\tau_{\overline w\cup \overline \xi_k^a}(\overline t)-
\sum_{\ell=\overline \ell_1^a,\ldots,\overline\ell^a_{\overline j_a}}
\!\!\!\!\!\!
 \ell N_a(\overline w, \overline\xi^a_\ell)(-z)^{\ell}\tau_{\overline w\backslash \overline \xi^a_\ell}(\overline t)
\right)
\bigg]+
\\
+&\sum_{c=s+1}^{r+s} \bigg [
e^{z\circ (t^{(c)}-\overline t^{(c)})} 
 e^{2(z^{-1}\circ (\tilde\partial_{t^{(c)}}-\tilde\partial_{\overline t^{(c)}}))}
\times\\
&\left(
\sum_{
\genfrac{}{} {0pt}{2}{m\in\frac12+\mathbb Z_{\ge 0},}{m\ne  m_1^c,\ldots, m^c_{  j_c}}
}
\!\!\!\!\!\!\!\!\!\!
N_c(  w,  \xi_m^c)z^{-m-\frac12}\tau_{  w\cup  \xi_m^c}( t)-
\!\!\!\!\!\!\!\!
\sum_{m=  m_1^c,\ldots, m^c_{  j_c}}
\!\!\!\!\!\!\!\!\!
(-1)^{m+\frac12}2m N_c( w,  \xi^c_m)z^{m-\frac12}\tau_{ w\backslash  \xi^c_m}(  t)
\right)
\times\\
& \!\!\!\!\!\!\!
\left(
\sum_{\genfrac{}{} {0pt}{2}{m\in\frac12+\mathbb Z_{\ge 0},}{ m\ne  \overline m_1^c,\ldots, \overline m^c_{ \overline j_c}}}
\!\!\!\!\!\!\!\!\!\!
N_c(\overline w, \overline\xi_m^c)(-z)^{-m-\frac12}\tau_{  \overline w\cup  \overline \xi_m^c}( \overline t)-
\!\!\!\!\!\!\!\!
\sum_{m=  \overline m_1^c,\ldots, \overline m^c_{  \overline j_c}}
\!\!\!\!\!\!\!\!\!
(-1)^{m+\frac12}2m N_c( \overline w,  \overline \xi^c_m)(-z)^{m-\frac12}\tau_{ \overline w\backslash  \overline \xi^c_m}(  t)
\right)\bigg] \bigg\}dz\\
&=0,
\end{aligned}
\end{equation}
where 
 $N_b(w ,\xi_n^b)=(-1)^{\{\mbox{the  number of $\xi^b_p$ appearing in $w$ with index $p<n$}\}}.$

\subsection{The wave function,  Sato-Wilson and Lax equations for the KP,  BKP,  CKP and DKP hierarchies}
\label{section4}
{\bf The KP hierarchy}\\ We follow the exposition \cite{KLmult},  Section IV.
The first step is that we replace  in \eqref{BosKP} all $t_1^{(j)}$ and $\overline t_1^{(j)}$ by 
$t_1^{(j)}+x$,  respectively $\overline t_1^{(j)}+\overline x$, so that the tau-function also depends on the variable $x$.  Introduce the $s\times s$-matrices \[
V^\pm (\underline \alpha; x,t,z)= P^\pm (\underline \alpha;x,t,\pm z)R^\pm (\underline\alpha;\pm z)S^\pm( t,\pm z)e^{\pm xz},
\]
where
\begin{equation}
\begin{aligned}
\label{VV}
P^\pm (\underline \alpha;x,t,\pm z)_{ij}&=
\frac{(-1)^{|\underline e_i|_{j-1}}z^{\delta_{ij}-1}}{\tau_{\underline\alpha}(x,t)}
e^{\mp z^{-1}\cdot \tilde\partial_{t^{(j)}}}
\left(
\tau_{\underline \alpha\pm(\underline e_i-\underline e_j)}(x,t)\right),\\
R^\pm (\underline\alpha;\pm z)&=\sum_{j=1}^s (-1)^{|\underline \alpha|_{j-1}}z^{\pm \alpha_j}E_{jj},\\
S^\pm( t,\pm z)&=\sum_{j=1}^s 
e^{\pm z\cdot  t^{(j)}}
E_{jj}.
\end{aligned}
\end{equation}
Then \eqref{BosKP} turns into 
\begin{equation}
\label{BosKP2}
{\rm Res}_{z=0}
\sum_{j=1}^s
V^+(\underline \alpha; x,t,z)E_{jj}V^-(\underline \beta;\overline x, \overline t,z)^t dz=0.
\end{equation}
Let $\partial=\partial_x$ and using the fundamental lemma (see \cite{KLmult},  Lemma 4.1),  we deduce the following expression for the matrix pseudo-differential operators
\begin{equation}
\label{VV2}
\sum_{j=1}^s
(P^+(\underline \alpha;x,t,\partial) R^+(\underline \alpha;\partial)E_{jj}R^-(\underline\beta;x,t,\partial)^*
P^-(\underline \beta; x,t,\partial)^*)_-=0.
\end{equation}
(Note  that one cannot simply substitute $\partial$ for $\pm z$ in the formulas \eqref{VV}.  One first has to write $P^\pm (\pm z)$ as $P^\pm (\pm z)= I+\sum_{k=1}^\infty P^\pm_k (\pm z)^{-k}$, then $P^\pm (\partial)= 1+\sum_{k=1}^\infty P^\pm_k\partial^{-k}$.  This means that $P^\pm ( \pm z)$ is the symbol of pseudo-differential operator $P^\pm (\partial)$.)

Taking $\underline\alpha=\underline \beta$,  we deduce the Sato-Wilson equations
\begin{equation}
\label{sato}
\begin{aligned}
P^-(\underline \beta; x,t,\partial)^*&=P^+(\underline \beta; x,t,\partial)^{-1},\\
\frac{\partial P^+(\underline \beta; x,t,\partial)}{\partial t_i^j}&=
-(P^+(\underline \beta; x,t,\partial)\partial^i E_{jj}P^+(\underline \beta; x,t,\partial)^{-1})_-
P^+(\underline \beta; x,t,\partial).
\end{aligned}
\end{equation}
Now introducing the Lax operators 
\begin{equation}
\label{lax}
\begin{aligned}
L&=L(\underline\beta;t,x,\partial)=P^+(\underline \beta; x,t,\partial)\partial P^+(\underline \beta; x,t,\partial)^{-1},\\
C^j&=C^j(\underline\beta; x,t,\partial)=P^+(\underline \beta; x,t,\partial) E_{jj}P^+(\underline \beta; x,t,\partial)^{-1},
\end{aligned}
\end{equation}
we deduce from \eqref{sato}   that
\begin{equation}
\label{lax2}
\frac{\partial L}{\partial t^{(j)}_i}=[(L^iC^j)_+,L],\quad \frac{\partial C^k}{\partial t_i^{(j)}}=[(L^iC^j)_+,C^k].
\end{equation}
\
\\
{\bf The BKP hierarchy (first approach)}\\
{
In general one can define an $(2s+r)\times (2s+r)$-matrix valued wave function as in the previous case.  However, in general the corresponding wave operators  $P^{
\pm}(\partial)$ are not invertible. This is a problem.  Therefore we take a different approach.  If we  assume that $s\ne 0$, and $r=0$ or $s$ arbitrary and $r=1$,  there is no problem.  In those cases the wave operators $P^{
\pm}(\partial)$ are  invertible.  The case that $r=0$  suggest to take the following approach (the case $r=1$, suggests the second approach, which we will present later in this section).  So assume from now on that $s\ne 0$.
If  $r\ge 1$,  we set in  equation \eqref{BosBKP}  all the variables $t_{j}^{(c)}=\overline t_{j}^{(c)}=0$ for $c>s$ and $j\in\mathbb Z_{\ge 1}$.  Next replace all $t_1^{(a)}$ for $1\le a\le s$ by $t_1^{(a)}+x $ and similarly $
\overline t_1^{(a)}$ by by $\overline t_1^{(a)}+\overline x $.  
Define the $2s\times 2s$-matrix valued wave function $V^\pm (\alpha; x, t,z)$ 
by 
\begin{equation}
\label{VPMB}
\tilde V^\pm (\underline \alpha; x,t,z)= P^\pm (\underline \alpha;x,t,\pm z)R^\pm (\underline\alpha;\pm z)S^\pm( t,\pm z)e^{\pm xz},
\end{equation}
where for $1\le i,j\le s$ and all $a=1, \ldots ,r$
\begin{equation}
\label{VVV}
\begin{aligned}
P^\pm (\underline \alpha;x,t,\pm z)_{ij}&=
\frac{(-1)^{|\underline e_i|_{j-1}}z^{\delta_{ij}-1}}{\tau_{\underline\alpha}(x,t)}
e^{\mp z^{-1}\cdot \tilde\partial_{t^{(j)}}}
\left(
\tau_{\underline \alpha\pm(\underline e_i-\underline e_j)}(x,t)\right)\big|_{t^{(s+a)}=0},\\
P^\pm (\underline \alpha;x,t,\pm z)_{s+i,j}&=
\frac{(-1)^{|\underline e_i|_{j-1}}z^{-\delta_{ij}-1}}{\tau_{\underline\alpha}(x,t)}
e^{\mp z^{-1}\cdot \tilde\partial_{t^{(j)}}}
\left(
\tau_{\underline \alpha\mp(\underline e_i+\underline e_j)}(x,t)\right)\big|_{t^{(s+a)}=0},\\
P^\pm (\underline \alpha;x,t,\pm z)_{i,s+j}&=
\frac{(-1)^{|\underline e_i|_{j-1}}(-z)^{-\delta_{ij}-1}}{\tau_{\underline\alpha}(x,t)}
e^{\pm 
((- z^{-1})\cdot \tilde\partial_{t^{(j)}})}
\left(
\tau_{\underline \alpha\pm(\underline e_i+\underline e_j)}(x,t)\right)\big|_{t^{(s+a)}=0},\\
P^\pm (\underline \alpha;x,t,\pm z)_{s+i,s+j}&=
\frac{(-1)^{|\underline e_i|_{j-1}}(-z)^{\delta_{ij}-1}}{\tau_{\underline\alpha}(x,t)}
e^{\pm ((- z^{-1})\cdot \tilde\partial_{t^{(j)}})}
\left(
\tau_{\underline \alpha\mp(\underline e_i-\underline e_j)}(x,t)\right)\big|_{t^{(s+a)}=0},
\end{aligned}
\end{equation}
and
\begin{equation}
\begin{aligned}
\label{VVVV}
R^\pm (\underline\alpha;\pm z)&= \sum_{j=1}^s (-1)^{|\underline \alpha|_{j-1}}\left(z^{\pm \alpha_j}E_{jj}
+(-z)^{\mp \alpha_j}E_{s+j,s+j}
\right)
,\\
S^\pm( t,\pm z)&= 
\sum_{j=1}^s \big(
e^{\pm z\cdot t^{(j)}}
E_{jj}+
e^{\mp (- z)\cdot t^{(j)}}
E_{s+j,s+j}
 \big).
\end{aligned}
\end{equation}
For $r\ge 1$ define  the $2s\times r$-matrix valued   functions 
\[
W^{\pm}(\underline\alpha;x,t,z)= (
W^{\pm}(\underline\alpha;x,t)_{ij},z)_{1\le i\le 2s,1\le j\le r},=\sum_{n=0}^\infty  W^{\pm}(\underline\alpha;x,t)^n z^{-n},
\]
with 
\begin{equation}
\label{WW1}
\begin{aligned}
W^{\pm}(\underline\alpha;x,t,z)_{i,2c-1}=\sqrt{-1}(-1)^{|\underline k|+|\underline \ell|_{c-1}} 
\frac{
e^{\pm 2(( z^{-1})\circ\tilde\partial_{t^{(s+2c-1)}})}
\tau_{\underline k\pm\underline e_i,\underline \ell+\underline e_c}(x,t)}{\tau_{\underline k ,\underline \ell }(x,t)}\big|_{t^{(s+a)}=0},\\
W^{\pm}(\underline\alpha;x,t,z)_{s+i,2c-1}=\sqrt{-1}(-1)^{|\underline k|+|\underline \ell|_{c-1}} 
\frac{
e^{\pm 2(( z^{-1})\circ\tilde\partial_{t^{(s+2c-1)}})}
\tau_{\underline k\mp\underline e_i,\underline \ell+\underline e_c}(x,t)}{\tau_{\underline k ,\underline \ell }(x,t)}\big|_{t^{(s+a)}=0},\\
W^{\pm}(\underline\alpha;x,t,z)_{i,2c}=
(-1)^{|\underline k|+|\underline \ell|_{c}} 
\frac{
e^{\pm 2(( z^{-1})\circ\tilde\partial_{t^{(s+2c)}})}
\tau_{\underline k\pm\underline e_i,\underline \ell+\underline e_c}(x,t)}{\tau_{\underline k ,\underline \ell }(x,t)}\big|_{t^{(s+a)}=0}\big|_{t^{s+a}=0},\\
W^{\pm}(\underline\alpha;x,t,z)_{s+i,2c}=
(-1)^{|\underline k|+|\underline \ell|_{c}} 
\frac{
e^{\pm 2(( z^{-1})\circ\tilde\partial_{t^{(s+2c)}})}
\tau_{\underline k\mp\underline e_i,\underline \ell+\underline e_c}(x,t)}{\tau_{\underline k ,\underline \ell }(x,t)}\big|_{t^{(s+a)}=0},\\
\end{aligned}
\end{equation}
for $1\le c\le p$.  If $r=2p$ that is all,  but if $r=2p+1$    we define 
\begin{equation}
\begin{aligned}
\label{WW2}
W^{\pm}(\underline\alpha;x,t,z)_{i, r}=\sqrt{-1} (-1)^{|\underline k|+|\underline \ell|} 
\frac{
e^{\pm 2(( z^{-1})\circ\tilde\partial_{t^{(s+r)}})}
\tau_{\underline k\pm\underline e_i,\underline \ell+\underline e_{p+1}}(x,t)}{\tau_{\underline k ,\underline \ell }(x,t)}\big|_{t^{(s+a)}=0},\\
W^{\pm}(
\underline\alpha;x,t,z)_{s+i,r}=\sqrt{-1}   (-1)^{|\underline k|+|\underline \ell|} 
\frac{e^{\pm 2(z^{-1}\circ\tilde\partial_{t^{(s+r)}})}
\tau_{\underline k\mp\underline e_i,\underline \ell+\underline e_{p+1}}(x,t)} {\tau_{\underline k ,\underline \ell }(x,t)}
\big |_{t^{(s+a)}=0},\\
\end{aligned}
\end{equation}
and the $2s\times 1$ matrix $T^{\pm}(\alpha; x,t)$
\begin{equation}
\label{Tmatrix}
\begin{aligned}
T^{\pm}(\alpha; x,t)_{i1}&=\frac{\tau_{\underline k\pm\underline e_i,\underline \ell }(x,t)}{\tau_{\underline k ,\underline \ell }(x,t)}\big|_{t^{s+a}=0},\\
T^{\pm}(\alpha; x,t)_{s+i,1}&=
\frac{\tau_{\underline k\mp\underline e_i,\underline \ell }(x,t)}{\tau_{\underline k ,\underline \ell }(x,t)}\big|_{t^{s+a}=0}.
\end{aligned}
\end{equation}
We thus obtain the following bilinear identity for the wave function:
\begin{equation}
\label{BosDKPr ne 0}
\begin{aligned}
&{\rm Res}_{z=0}
\sum_{j=1}^s
V^+(\underline \alpha; x,t,z)
\left( E_{jj}
- E_{s+j,s+j}
\right)
V^-(\underline \beta;\overline x, \overline t,z)^t dz=\\ 
 (\frac12-\frac{\delta_{r0}}2)&W^+(\underline \alpha; x,t)^0
W^-(\underline \beta;\overline x, \overline t)^{0t}
+ 
\left( \frac12-\frac{\delta_{r,2p}}2 (-1)^{{{|\underline \alpha- \underline \beta}|}}\right)
T^{+}(\alpha; x,t) 
T^{-}(\beta; \overline x,\overline t)^t
.
\end{aligned}
\end{equation}

Using the fundamental lemma,  we obtain that 
\begin{equation}
\label{BosDKPr ne 0 2}
\begin{aligned}
\sum_{j=1}^s&
\left( P^+(\underline \alpha; x,t,\partial)R^+(\underline \alpha-\underline\beta,\partial)
\left( E_{jj}
- E_{s+j,s+j}
\right)
P^-(\underline \beta; x,  t,\partial)^*\right)_-=\\
 & (\frac12-\frac{\delta_{r0}}2)W^+(\underline \alpha; x,t)^0\partial^{-1}
W^-(\underline \beta; x,  t)^{0t}+\\ 
&+
\left( \frac12-\frac{\delta_{r,2p}}2 (-1)^{{{|\underline \alpha- \underline \beta}|}}\right)
T^{+}(\alpha; x,t) \partial^{-1}
T^{-}(\beta; x, t)^t
.
\end{aligned}
\end{equation}
Taking $\underline \alpha=\underline\beta$,  it is straightforward to check that the right-hand side of \eqref{BosDKPr ne 0 2} is equal to $0$.  

Let $\partial=\partial_x$ and using the fundamental lemma (see \cite{KLmult},  Lemma 4.1),  we deduce the following expression for the matrix pseudo-differential operators
\begin{equation}
\label{DD2}
\sum_{j=1}^s
(P^+(\underline \alpha;x,t,\partial) R^+(\underline \alpha;\partial)\left( E_{jj}- E_{s+j,s+j}\right)R^-(\underline\beta;x,t,\partial)^*
P^-(\underline \beta; x,t,\partial)^*)_-=0.
\end{equation}
Letting
\begin{equation}
\label{J}
J= \sum_{j=1}^s E_{jj}-E_{s+j,s+j}=J^{-1},
\end{equation}
 we  thus get
\begin{equation}
\label{sato3}
\begin{aligned}
JP^-(\underline \beta; x,t,\partial)^*&=P^+(\underline \beta; x,t,\partial)^{-1}J,\\
\frac{\partial P^+(\underline \beta; x,t,\partial)}{\partial t_i^j}&=
-(P^+(\underline \beta; x,t,\partial)\left(\partial^i E_{jj}-(-\partial)^i E_{s+j,s+j}\right)
P^+(\underline \beta; x,t,\partial)^{-1})_-
P^+(\underline \beta; x,t,\partial).
\end{aligned}
\end{equation}
Now introducing the Lax operators 
\begin{equation}
\label{laxd}
\begin{aligned}
L&=L(\underline\beta;t,x,\partial)=P^+(\underline \beta; x,t,\partial)J\partial   P^+(\underline \beta; x,t,\partial)^{-1},\\
D^j&=D^j(\underline\beta; x,t,\partial)=P^+(\underline \beta; x,t,\partial)\left( E_{jj}-E_{s+j,s+j}\right)P^+(\underline \beta; x,t,\partial)^{-1},\\
E^j&=(D^j)^2=E^j(\underline\beta; x,t,\partial)=P^+(\underline \beta; x,t,\partial)\left( E_{jj}+E_{s+j,s+j}\right)P^+(\underline \beta; x,t,\partial)^{-1},
\end{aligned}
\end{equation}
we deduce from \eqref{sato3}   that
\begin{equation}
\label{lax22}
\frac{\partial L}{\partial t_i^{(j)}}=[(L^iD^j)_+,L],\quad \frac{\partial D^k}{\partial t_i^{(j)}}=[(L^iD^j)_+,D^k] ,\quad \frac{\partial E^k}{\partial t_i^{(j)}}=[(L^iD^j)_+,E^k].
\end{equation}
Let 
\begin{equation}
\label{N}
N=\sum_{j=1}^s \left(E_{j,s+j}+E_{s+,j}\right),
\end{equation}
then it is straightforward to check, using \eqref{VVV}, that
\[
P^-(\underline \beta; x,t,\partial)=N P^+(\underline \beta; x,t,\partial)N
\]
Now let 
\begin{equation}
\label{K}
K=JN=\sum_{j=1}^s\left( E_{j,s+j}-E_{s+j,j}\right),
\end{equation}
then 
\[
L(\partial)^*=KL(\partial)K^{-1},\quad D^{j}(\partial)^*=-KD^j (\partial)K^{-1}, \quad E^{j}(\partial)^*=KE^j (\partial)K^{-1}.
\]

\begin{example}
Let's consider the special case $s=1$.
In this case 
\[
\begin{aligned}
L=&\begin{pmatrix}1&0\\ 0&-1
\end{pmatrix}\partial +
\begin{pmatrix}u_{11}^{(1)}&u_{12}^{(1)}\\
u_{21}^{(1)}&-u_{11}^{(1)}
\end{pmatrix}\partial^{-1}+
\begin{pmatrix}
u_{11}^{(2)}&-\frac12 u_{12x}^{(1)}\\
-\frac12 u_{21x}^{(1)}&u_{11}^{(2)}+u_{11x}^{(1)}
\end{pmatrix}\partial^{-2}+\\
&\qquad+
\begin{pmatrix}
u_{11}^{(3)}&u_{12}^{(3)}\\
u_{21}^{(3)}
&-u_{11}^{(3)}-2u_{11x}^{(2)}-u_{11xx}^{(1)}
\end{pmatrix}\partial^{-3}+\cdots,
\end{aligned}
\]
\[
D=\begin{pmatrix}1&0\\ 0&-1
\end{pmatrix}+
\begin{pmatrix}0&u_{12}^{(1)}\\
u_{21}^{(1)}&0
\end{pmatrix}\partial^{-2}+
\begin{pmatrix}
0&-u_{12x}^{(1)}\\
-u_{21x}^{(1)}&0
\end{pmatrix}\partial^{-3}+\cdots
\]
We obtain the following set of equations from the Lax equations for $L$:
\[
\begin{aligned}
\frac{\partial u_{11}^{(1)}}
{\partial t_2}&=\left(2u_{11}^{(2)}+ u_{11x}^{(1)}\right)_x\, ,
\\
\frac{\partial u_{12}^{(1)}}
{\partial t_2}&=2(u_{11}^{(1)}u_{12}^{(1)}+ u_{12}^{(3)})\, ,
\\
\frac{\partial u_{21}^{(1)}}
{\partial t_2}&=-2(u_{21}^{(1)}u_{11}^{(1)}+ u_{21}^{(3)})\, ,
\\
\frac{\partial u_{11}^{(1)}}
{\partial t_3}&=\frac12\left(6(u_{11}^{(1)})^2+6u_{11}^{(3)}+3 u_{12}^{(1)}u_{21}^{(1)}+
6u_{11x}^{(2)}+2u_{11xx}^{(1)}
\right)_x\, ,
\\
\frac{\partial u_{21}^{(1)}}
{\partial t_3}&=\frac12\left(-12u_{11}^{(2)} u_{21}^{(1)} +6u_{21x}^{(3)}-
u_{21xxx}^{(1)} 
\right)\, ,
\\
\frac{\partial u_{12}^{(1)}}
{\partial t_3}&=\frac12\left(12  (u_{11}^{(2)} +u_{11x}^{(1)}) u_{12}^{(1)}+6u_{12x}^{(3)}-
u_{12xxx}^{(1)} 
\right)\, .
\end{aligned}
\]
\end{example}
{\bf The  BKP hierarchy (second approach)}\\
In this case we assume that $r\ge 1$, in this case $s$ can be 0.
We follow the approach of the previous case,  setting $t^{(s+c)}_j=0$,  but only for $1\le c<r$,  we do not put $t^{(s+r)}_j=0$,  but leave them in the equation.   Next we replace  $t^{(a)}_1$ by $t_1^{(a)}+x$, for $1\le a\le s$ and $a=s+r$ and  define the $(2s+1)\times(2s+1)$-matrix valued wave functions
\begin{equation}
\label{VPMB2}
\tilde V^\pm (\underline \alpha; x,t,z)= P^\pm (\underline \alpha;x,t,\pm z)R^\pm (\underline\alpha;\pm z)S^\pm( t,\pm z)e^{\pm xz},
\end{equation}
where for $1\le i,j\le 2s$, the coefficients $P^\pm (\underline \alpha;x,t,\pm z)_{ij}$ are given by \eqref{VVV}.
The other coefficients are as follows ($a=1,  \ldots ,r-1$)
\begin{equation}
\label{vvv}
\begin{aligned}
P^\pm (\underline \alpha;x,t,\pm z)_{2s+1,j}&=
 \frac{z^{-1}}{\tau_{\underline\alpha}(x,t)}
e^{\mp z^{-1}\cdot \tilde\partial_{t^{(j)}}}
\left(
\tau_{\underline \alpha\mp \underline e_j}(x,t)\right)\big|_{t^{(s+a)}=0},\\
P^\pm (\underline \alpha;x,t,\pm z)_{2s+1,s+j}&=
 \frac{(-z)^{-1}}{\tau_{\underline\alpha}(x,t)}
e^{\pm ((-z^{-1})\cdot \tilde\partial_{t^{(j)}})}
\left(
\tau_{\underline \alpha\pm \underline e_j}(x,t)\right)\big|_{t^{(s+a)}=0},\\
P^\pm (\underline \alpha;x,t,\pm z)_{i,2s+1}&=\frac{1}{ \tau_{\underline\alpha}(x,t)}
 e^{\pm 2(z^{-1}\circ\tilde\partial_{t^{(s+r)}})}
%
\left(\tau_{\underline \alpha\pm \underline e_i  }(t)\right)\big|_{t^{(s+a)}=0},
\\
P^\pm (\underline \alpha;x,t,\pm z)_{s+i,2s+1}&=
\frac{1}{ \tau_{\underline\alpha}(x,t)}
e^{\pm 2(( z^{-1})\circ\tilde\partial_{t^{(s+r)}})}
%
\left(\tau_{\underline \alpha\mp \underline e_i  }(t)\right)\big|_{t^{(s+a)}=0},
\\
P^\pm (\underline \alpha;x,t,\pm z)_{2s+1,2s+1}&=
\frac{1}{ \tau_{\underline\alpha}(x,t)}
e^{\pm 2(( z^{-1})\circ\tilde\partial_{t^{(s+r)}})}
%
\left(\tau_{\underline \alpha }(t)\right)\big|_{t^{(s+a)}=0},
\end{aligned}
\end{equation}
where $\underline \alpha=(\underline k,\underline \ell)$ and
$|\underline \alpha|=|\underline k|+|\underline \ell|$.
Here,
\begin{equation}
\begin{aligned}
\label{VVVVV}
R^\pm (\underline\alpha;\pm z)&=(-1)^{|\underline \alpha|}E_{2s+1,2s+1}+\sum_{j=1}^s (-1)^{|\underline \alpha|_{j-1}}\left(z^{\pm \alpha_j}E_{jj}
+(-z)^{\mp \alpha_j}E_{s+j,s+j}
\right)
,\\
S^\pm( t,\pm z)&= 
e^{2(z\circ t^{(s+r)})}
E_{2s+1,2s+1}+
\sum_{j=1}^s \big(
e^{\pm(z\cdot t^{(j)})}E_{jj}+e^{\mp((-z)\cdot t^{(j)})}E_{s+j,s+j}
 \big).
\end{aligned}
\end{equation}

We define the $(2s+1)\times (r-1)$-matrix $W^{\pm}(\underline\alpha;x,t,z)=\left(W^{\pm}(\underline\alpha;x,t,z)\right)_{i,j}$, where
$W^{\pm}(\underline\alpha;x,t,z)_{i,j}$ for $i=1, \dots 2s$ is given as in 
\eqref{WW1}
and 
define 
\begin{equation}
\label{WW3}
\begin{aligned}
W^{\pm}(\underline\alpha;x,t,z)_{2s+1,2c-1}=&\sqrt{-1}  (-1)^{|\underline k|+|\underline \ell|_{c-1}}
\frac{
e^{\pm 2(( z^{-1})\circ\tilde\partial_{t^{(s+2c-1)}})}
\tau_{\underline k,\underline \ell+\underline e_c}(x,t)}{\tau_{\underline k ,\underline \ell }(x,t)}\big|_{t^{(s+a)}=0},\\
W^{\pm}(\underline\alpha;x,t,z)_{2s+1,2c}=&
(-1)^{|\underline k|+|\underline \ell|_{c}} 
\frac{
e^{\pm 2(( z^{-1})\circ\tilde\partial_{t^{(s+2c)}})}
\tau_{\underline k,\underline \ell+\underline e_c}(x,t)}{\tau_{\underline k ,\underline \ell }(x,t)}\big|_{t^{(s+a)}=0} .\\
\end{aligned}
\end{equation}
Define also the $(2s+1)\times 1 $-matrix $T^{\pm}(\alpha; x,t)$,  where $T^{\pm}(\alpha; x,t)_{2s+1,1}=1$ and all the other coefficients are given by 
\eqref{Tmatrix}.
Then we obtain 
\begin{equation}
\label{BosBKPHIR}
\begin{aligned}
{\rm Res}_{z=0}&
\tilde V^+(\underline \alpha; x,t,z)
  J(z) 
\tilde V^-(\underline \beta;\overline x, \overline t,z)^t dz=\\ 
 &\frac12 W^+(\underline \alpha; x,t)^{0}W^-(\underline \beta;\overline x, \overline t)^{0t}     
+ 
\left( \frac12-\frac{\delta_{r,2p}}2 (-1)^{{{|\underline \alpha- \underline \beta}|}}\right)
T^{+}(\alpha; x,t) 
T^{-}(\beta; \overline x,\overline t)^t
,
\end{aligned}
\end{equation}
where \begin{equation}
\label{J(z)}
J(z)=\frac12 E_{2s+1,2s+1}z^{-1}  + \sum_{i=1}^s(E_{jj}
- E_{s+j,s+j}).
\end{equation} 
We thus obtain 
\begin{equation}
\label{BosDKPr ne 0 2!!}
\begin{aligned}
\left( P^+(\underline \alpha; x,t,\partial)R^+(\underline \alpha-\underline\beta,\partial)J(\partial)
P^-(\underline \beta;x,  t,\partial)^*\right)_-=\\
 \frac12 W^+(\underline \alpha; x,t)^0\partial^{-1}
W^-(\underline \beta; x, t)^{0t}+ 
\left( \frac12-\frac{\delta_{r,2p}}2 (-1)^{{{|\underline \alpha- \underline \beta}|}}\right)
T^{+}(\alpha; x,t) 
\partial^{-1}
T^{-}(\beta; x, t)^t
.
\end{aligned}
\end{equation}
Now taking $\underline \alpha=\underline\beta$,  it is straightforward to check that  if $r=2p+1$ the part with the $W^\pm$ is equal to zero, 
and that the $T^\pm$ are equal to the constant coefficients $P^\pm-I+E_{2s+1,2s+1}$.   If  $r=2p$, the $T^\pm$ part drops out and many things cancel in the $W^\pm$ expression, except for the last column,  which is the $2p-1$-th one,  of both $W^\pm$.   A careful  inspection gives for both cases that   
\begin{equation}
\label{BosBKPHir}
\begin{aligned}
{\rm Res}_{z=0}&
\tilde V^+(\underline \alpha; x,t,z)
R^+(\underline \alpha;z) J(z) 
  R^-(\underline \beta;-z) 
\tilde V^-(\underline \beta;\overline x, \overline t,z)^t dz= \\
&={\rm Res}_{z=0}P^+(\underline \alpha; x,t)^0 J(z) P^-(\underline \beta;\overline x, \overline t)^{0t}dz ,
\end{aligned}
\end{equation}
where  $P^\pm(\underline \alpha; x,t)^0$ is the  constant coefficient of $P^\pm(\underline \alpha; x,t,z)$.
Now choosing  $\underline \alpha=\underline \beta$ and using the fundamental lemma,  we deduce that
\[
\left(
P^{+}(\underline\alpha;x,t,\partial)J(\partial)P^{-}(\underline\alpha;x,t,\partial)^*\right)_-=\frac12 P^+(\underline \alpha; x,t)^0
E_{2s+1,2s+1}\partial^{-1}
P^-(\underline \alpha;x, t)^{0t},
\]
which gives that
\begin{equation}
\label{sato1B}
 P^{+}(\underline\alpha;x,t,\partial)J(\partial)P^{-}(\underline\alpha;x,t,\partial)^*=P^+(\underline \alpha; x,t)^0J(\partial)P^-(\underline \alpha;x, t)^{0t}.
\end{equation}
Note that $P^\pm(\underline \alpha; x,t)^{0}$ is invertible  and 
hence it makes sense to redefine the wave function  and  $P^\pm (\underline \alpha;x,t,\partial)$
\begin{equation}
\begin{aligned}
\label{hatV}
\hat V^\pm (\underline \alpha; x,t,z)&=
(P^\pm (\underline \alpha; x,t)^0)^{-1}
\tilde V^\pm (\underline \alpha; x,t,z), \\
\hat P^\pm (\underline\alpha;x,t,\partial)&=(P^\pm (\underline \alpha; x,t)^0)^{-1} P^\pm (\underline\alpha;x,t,\partial).
\end{aligned}
\end{equation}
Then \eqref{BosBKPHir} turns into
\begin{equation}
\label{BosBKPHIR2}
\begin{aligned}
{\rm Res}_{z=0}&
 \hat V^+(\underline \alpha; x,t,z)
R^+(\underline \alpha;z)J(z) R^-(\underline \alpha;-z) 
 \hat V^-(\underline \alpha;\overline x, \overline t,z)^t dz=  
{\rm Res}_{z=0} \, 
 J(z)dz,
\end{aligned}
\end{equation}
which means   that
\begin{equation}
\label{SATO1B}
\hat P^{-}(\underline\alpha;x,t,\partial)^*=  J(\partial)^{-1}\hat P^{+}(\underline\alpha;x,t,\partial)^{-1}J(\partial).
\end{equation}
Next differentiate \eqref{BosBKPHIR2} by $t_j^{a}$, where $j$ is odd if $a=s+r$,  then we obtain for $\underline \alpha=\underline \beta$
\begin{equation}
\label{SATOB}
\begin{aligned}
\big[\left(
\frac{\partial \hat P^+(\underline \alpha)}{\partial t_j^a}+\hat P^+(\underline \alpha)
\left( (1-\delta_{a,s+r})(E_{aa}\partial^j-E_{s+a,s+a}(-\partial)^j)+2\delta_{a,s+r}E_{2s+1,2s+1}\partial^j\right)\right)\times\\
J(\partial)\hat P^-(\underline \alpha)^* 
\big]_-=0.
\end{aligned}
\end{equation}
Using \eqref{sato1B}, this turms into
\begin{equation}
\label{SATOB2}
\begin{aligned}
\frac{\partial \hat P^+(\underline \alpha)}{\partial t_j^{(a)}}\hat P^+(\underline \alpha)^{-1}
=
&-\bigg(\hat P^+(\underline \alpha)
\bigg( (1-\delta_{a,s+r})(E_{aa}\partial^j-E_{s+a,s+a}(-\partial)^j)+\\
&\qquad +2\delta_{a,s+r}E_{2s+1,2s+1}\partial^j\bigg)\hat P^+(\underline \alpha)^{-1} J(\partial) 
\bigg)_-J(\partial)^{-1}.
\end{aligned}
\end{equation}
Finally,  following \cite{KvdLB},  we define a new splitting of the algebra of matrix pseudo-differential operators, viz.
\begin{equation}
\label{Pge}
P(\partial)_{\ge}=(P(\partial) J(\partial))_+ J(\partial)^{-1}, \quad P(\partial)_<   =P(\partial)-P(\partial)_{\ge},
\end{equation}
Then \eqref{SATOB2}, turns into 
\begin{equation}
\label{SATOB3}
\begin{aligned}
\frac{\partial \hat P^+(\underline \alpha)}{\partial t_j^{(a)}}
&=-\bigg(\hat P^+(\underline \alpha)
\bigg( (1-\delta_{a,s+r})(E_{aa}\partial^j-E_{s+a,s+a}(-\partial)^j)+\\
&\qquad\qquad+2\delta_{a,s+r}E_{2s+1,2s+1}\partial^j\bigg)\hat P^+(\underline \alpha)^{-1} 
\bigg)_<  \hat P^+(\underline \alpha).
\end{aligned}
\end{equation}
Define the following Lax operators  ($1\le a\le s$):
\begin{equation}
\label{BLax}
\begin{aligned}
L(\underline \alpha)&=\hat P^+(\underline \alpha)\left(E_{2s+1,2s+1}+\sum_{i=1}^s \left(E_{ii}-E_{s+i,s+i}\right)\right) \partial \hat P^+(\underline \alpha)^{-1}, \\\ 
D^{s+r}(\underline \alpha)&= 2  \hat P^+(\underline \alpha)E_{2s+1,2s+1}\hat P^+(\underline \alpha)^{-1},\\
 D^a(\underline \alpha)&= \hat P^+(\underline \alpha)(E_{aa}-E_{s+a,s+a})\hat P^+(\underline \alpha)^{-1}\\
E^a(\underline \alpha)&=D^a(\underline \alpha)^2= \hat P^+(\underline \alpha)(E_{aa}+E_{s+a,s+a})\hat P^+(\underline \alpha)^{-1}.
\end{aligned}
\end{equation}
Using \eqref{SATOB3}, it is straightforward to verify that
\begin{equation}
\label{BLax2}
\begin{aligned}
\frac{\partial L(\underline \alpha)}{\partial t_j^{(a)}}=&[\left( L(\underline \alpha)^j D^a(\underline \alpha)\right)_{\ge},L(\underline \alpha)],\\
\frac{\partial D^b(\underline \alpha)}{\partial t_j^{(a)}}=&[\left( L(\underline \alpha)^j D^a(\underline \alpha)\right)_{\ge},D^b(\underline \alpha)],\\
\frac{\partial E^b(\underline \alpha)}{\partial t_j^{(a)}}=&[\left( L(\underline \alpha)^j D^a(\underline \alpha)\right)_{\ge},E^b(\underline \alpha)]
.
\end{aligned}
\end{equation}
\begin{remark}
Note that one gets the wave-functions  and the Lax equations  of the first approach of  the BKP here as the $2s\times2s$-principal minor, which one gets by deleting the last row and column.
\end{remark}
Since (cf. \eqref{N})
\[
P^-(\underline \alpha; x,t,\partial)=(N +E_{2s+1,2s+1})P^+(\underline \alpha; x,t,\partial)(N +E_{2s+1,2s+1}),
\]
this also holds for $\hat P^\pm(\underline \alpha; x,t,\partial)$, and thus 
\[
\hat P^+(\underline \alpha; x,t,\partial)^{*-1}= K(\partial)\hat P^+(\underline \alpha; x,t,\partial)K(\partial)^{-1},
\]
where
\[
K(\partial)=-J(\partial)^{*-1}(N +E_{2s+1,2s+1})=2\partial E_{s+1,s+1}+\sum_{a=1}^s E_{s+a,a}-E_{a,s+a}.
\]
Then 
\[
 D^{s+r * }=K(\partial)D^{s+r}K(\partial)^{-1},\quad D^{a*}=-K(\partial)D^aK(\partial)^{-1}, \quad E^{a*}=K(\partial)E^aK(\partial)^{-1}, 
\]
$a=1,\ldots, s$, 
and
\[
L^{*}=(L(D^{s+r}+\sum_{a=1}^s E^a))^*=K(\partial)(-D^{s+r}+\sum_{a=1}^s E^a) LK(\partial)^{-1}.
\]
In particular if $s=0$,  then $L^*=-\partial L\partial^{-1}$. Note that the classical 1-component BKP hierarchy is the case that $s=0$ and $r=1$.
\\
\ 
\\
{\bf The  DKP hierarchy (first approach)}
\\
We assume that if $r\ne 0$  $r=2p$ or $r=2p-1$ and take equation \eqref{BosDKP}  and set all the variables $t_{2j+1}^{(c)}=\overline t_{2j+1}^{(c)}=0$ for $c>s$ and $j\in\mathbb Z_{\ge 0}$.  Next replace all $t_1^{(a)}$ for $1\le a\le s$ by $t_1^{(a)}+x $ and similarly $
\overline t_1^{(a)}$ by by $\overline t_1^{(a)}+\overline x $.
Define the wave function $V^\pm (\alpha; x, t,z)$ as in  the first approach of the BKP case case by \eqref{VPMB},  but now for $\underline\alpha=(\underline k, \underline \ell)  $, where we also substitute   $t_{2j+1}^{(s+c)}=0$.  Define the $s\times r$-matrix, which we assume to be zero if $r=0$,
\[
W^{\pm}(\underline\alpha;x,t,z)= (
W^{\pm}(\underline\alpha;x,t)_{ij},z)_{1\le i\le 2s,1\le j\le r},=\sum_{n=0}^\infty  W^{\pm}(\underline\alpha;x,t)^n z^{-n},
\]
with coefficients as in \eqref{WW1} and the matrix $T^\pm$ by 
\begin{equation}
\label{Tmatrix2}
\begin{aligned}
T^{\pm}(\alpha; x,t)_{i1}&=\frac{\tau_{\underline k\pm\underline e_i,\underline \ell +\underline e_p}(x,t)}{\tau_{\underline k ,\underline \ell }(x,t)}\big|_{t^{(s+a)}=0},\\
T^{\pm}(\alpha; x,t)_{s+i,1}&=
\frac{\tau_{\underline k\mp\underline e_i,\underline \ell +\underline e_p}(x,t)}{\tau_{\underline k ,\underline \ell }(x,t)}\big|_{t^{(s+a)}=0}.
\end{aligned}
\end{equation}
We thus obtain the following bilinear identity for the wave function:
\begin{equation}
\label{BosDKPr ne 0!}
\begin{aligned}
{\rm Res}_{z=0}
\sum_{j=1}^s
V^+(\underline \alpha; x,t,z)
\left( E_{jj}
- E_{s+j,s+j}
\right)
V^-(\underline \beta;\overline x, \overline t,z)^t dz=\\
 \frac12 W^+(\underline \alpha; x,t)^0
W^-(\underline \beta;\overline x, \overline t)^{0t}
+\frac{\delta_{r,2p-1}}2  T^{+}(\alpha; x,t)  
T^{-}(\beta; \overline x,\overline t)^t
.
\end{aligned}
\end{equation}
Now using the fundamental lemma, we obtain that 
\begin{equation}
\label{BosDKPr ne 0 2!}
\begin{aligned}
\sum_{j=1}^s
\left( P^+(\underline \alpha; x,t,\partial)R^+(\underline \alpha-\underline\beta,\partial)
\left( E_{jj}
- E_{s+j,s+j}
\right)
P^-(\underline \beta; x,  t,\partial)^*\right)_-=\\
 \frac12 W^+(\underline \alpha; x,t)^0\partial^{-1}
W^-(\underline \beta;x,  t)^{0t}+\frac{\delta_{r,2p-1}}2  T^{+}(\alpha; x,t) 
  \partial^{-1}
T^{-}(\beta; x, t)^t
.
\end{aligned}
\end{equation}
Choosing $\underline \alpha=\underline \beta$, gives that the right-hand side of \eqref{BosDKPr ne 0 2!} is equal to 0,   hence we obtain the same equations as in the first approach of the BKP case.  
\\
\ 
\\
{\bf The DKP hierarchy (second approach)}
Now we assume that $r\ge 1$, in this case $s$ can be 0.  $r$ is either $r=2p$ or $r=2p-1$. 
We follow the approach of the previous case,s, now setting $t^{(s+c)}_j=0$,  but only for $1\le c<r$,  we do not put $t^{(s+r)}_j=0$.  Next we replace  $t^{(a)}_1$ by $t_1^{(a)}+x$, for $1\le a\le s$ and $a=s+r$ and  define the $(2s+1)\times(2s+1)$-matrix valued wave functions
\begin{equation}
\label{VPMB22}
\tilde V^\pm (\underline \alpha; x,t,z)= P^\pm (\underline \alpha;x,t,\pm z)R^\pm (\underline\alpha;\pm z)S^\pm( t,\pm z)e^{\pm xz},
\end{equation}
Where the coefficients  $P^\pm (\underline \alpha;x,t,\pm z)$ are given by \eqref{VVV} and \eqref{vvv} and the other matrices by \eqref{VVVV}. The $(2s+1)\times (r-1)$ is 
$W^{\pm}(\underline\alpha;x,t,z) $ is again defined as  in \eqref{WW2} and \eqref{WW3} and $T^{\pm}\underline\alpha;x,t)$ as in the BKP second approach case.
Then we obtain the following bilinear identity:
\begin{equation}
\label{BosBKPHIR2D}
\begin{aligned}
{\rm Res}_{z=0}&
\tilde V^+(\underline \alpha; x,t,z)
  J(z) 
\tilde V^-(\underline \beta;\overline x, \overline t,z)^t dz=\\ 
 &\frac12 W^+(\underline \alpha; x,t)^{0}W^-(\underline \beta;\overline x, \overline t)^{0t}     
+ \frac{\delta_{r,2p-1}}2 T^{+}(\alpha; x,t) 
T^{-}(\beta; \overline x,\overline t)^t
,
\end{aligned}
\end{equation}
where $J(z)$ is as in \eqref{J(z)}.
One can now continue as in the second approach of the BKP case and nothing  changes. This is not surprising considering the first approach in the DKP case and considering Remark \ref{BDcorresp}.
\begin{remark}
Note that the first approach gives the same wave-functions and Lax operators in both the BKP as DKP hierarchy as in  the case  $r=0$.  The second approach with $r>1$ does not differ from the case  $r=1$.  Hence both approaches seem to be  superfluous. This however will not be the case if we consider reductions to affine Lie algebras. The case $r=0$ will produce a differential operator like in the Gelfand-Dickey case,  but the first approach  for $r>0$, will produce a Lax operator which is of constrained KP-type.
\end{remark}

\noindent
{\bf The CKP hierarchy} 
As in the previous cases,  we replace $t^b_1$ by $t_1^b+x$, for $1\le b\le r+s$, and introduce  $(2s+r)\times(2s+r)$-matrix valued wave functions
\begin{equation}
\label{VPMC}
 V^\pm (x,t,z)= P^\pm ( x,t,\pm z) S^\pm( t,\pm z)e^{\pm xz},
\end{equation}
where 
\[
\begin{aligned}
S^\pm( t,\pm z)=&\sum_{a=1}^s\big(
e^{\pm (z\cdot t^{(a)})}
E_{aa}+e^{\mp ((-z)\cdot t^{(a)})}E_{s+a,s+a}\big)
+\sum_{c=s+1}^{r+s}
e^{\pm (z\circ t^{(c)})}
E_{a+c,a+c},
\\
P^\pm ( x,t,\pm z)=&\left(P^\pm ( x,t,\pm z)_{i,j}\right)_{1\le i,j\le r+2s},
\end{aligned}
\]
and ($1\le a\ne  b\le s\le c,d \le r+s$)   
\begin{equation}
\label{P-C}
\begin{aligned}
\tau_0(x,t)P^\pm ( x,t,\pm z)_{a,a}&=
e^{\pm z^{-1}\cdot \tilde\partial_{t^{(a)}}}
 \left(\tau_0(x,t)-\sum_{k=1}^\infty \tau_{\xi^a_{\mp k} \cup \xi^a_{\pm 1}}(x,t)z^{-k-1}\right),\\
\tau_0(x,t)P^\pm ( x,t,\pm z)_{a,b}&=
e^{\pm z^{-1}\cdot \tilde\partial_{t^{(b)}}}
 \left(\mp\sum_{k=1}^\infty \tau_{\xi^b_{\mp k} \cup \xi^a_{\pm 1}}(x,t)z^{-k}\right),\\
\tau_0(x,t)P^\pm ( x,t,\pm z)_{a,s+a}&=
e^{\mp ((-z^{-1})\cdot \tilde\partial_{t^{(a)}})}
\left(-\sum_{k=2}^\infty \tau_{\xi^a_{\pm  k} \cup \xi^a_{\pm 1}}(x,t)(-z)^{1-k}\right),
\\
\tau_0(x,t)P^\pm ( x,t,\pm z)_{a,s+b}&=
e^{\mp ((-z^{-1})\cdot \tilde\partial_{t_k^{(b)}})}
\left(\pm\sum_{k=1}^\infty \tau_{\xi^b_{\pm  k} \cup \xi^a_{\pm 1}}(x,t)(-z)^{-k}\right),
\\
\tau_0(x,t)P^\pm ( x,t,\pm z)_{a,s+c}&=
e^{\pm 2(z^{-1}\circ\tilde\partial_{t^{(s+c)}})}
\left(-
\sum_{m\in\frac12+\mathbb Z_{\ge 0} }
\!\!\!\!
 \tau_{  \xi_{\pm 1}^a\cup  \xi_m^c}( t)(\pm z)^{-m-\frac12}
\right),
\\
\tau_0(x,t)P^\pm ( x,t,\pm z)_{s+a,a}&=
e^{\pm z^{-1}\cdot \tilde\partial_{t^{(a)}}}
 \left(-\sum_{k=2}^\infty \tau_{\xi^a_{\mp k} \cup \xi^a_{\mp 1}}(x,t)z^{1-k}\right),\\
%
%
%
%
\tau_0(x,t)P^\pm ( x,t,\pm z)_{s+c,a}&=
e^{\pm z^{-1}\cdot \tilde\partial_{t^{(a)}}}
 \left(\mp\sum_{k=1}^\infty \tau_{\xi^a_{\mp k} \cup \xi^c_{\frac12}}(x,t)z^{-k}\right),\\
%
%
\tau_0(x,t)P^\pm ( x,t,\pm z)_{s+c,s+c}&=
e^{\pm 2(z^{-1}\circ\tilde\partial_{t^{(s+c)}})}
\left(\tau_0(x,t)-
\sum_{m\in\frac12+\mathbb Z_{> 0} }
\!\!\!\!
 \tau_{  \xi_{\frac12}^c\cup  \xi_m^c}( t)(\pm z)^{-m-\frac12}
\right),
\\
\tau_0(x,t)P^\pm ( x,t,\pm z)_{s+c,s+d}&=
e^{\pm 2(z^{-1}\circ\tilde\partial_{t^{(s+r)}})}
\left(
-\sum_{m\in\frac12+\mathbb Z_{\ge 0} }
\!\!\!\!
 \tau_{  \xi_{\frac12}^c\cup  \xi_m^d}( t)(\pm z)^{-m-\frac12}
\right)
.
\end{aligned}
\end{equation}
The other coefficients are given by:
\begin{equation}
\begin{aligned}
\label{P-C2}
P^\pm ( x,t,\partial)_{s+a,s+a}&=P^\mp ( x,t,\partial)_{a,a},\\
 P^\pm ( x,t,,\partial)_{s+a,s+b}&=P^\mp ( x,t,\partial)_{a,b},\\
P^\pm ( x,t,\partial)_{s+a,b}&=P^\mp ( x,t,\partial)_{a,s+b},\\
P^\pm ( x,t,\partial)_{s+a,s+c}&=P^\mp ( x,t,\partial)_{a,s+c},
\\
 P^\pm ( x,t,\partial)_{s+c,s+a}&=P^\mp ( x,t,\partial)_{s+c,a}.
\end{aligned}
\end{equation}
Then \eqref{BosCKP} gives 
\begin{equation}
\label{waveCKP}
{\rm Res}_{z=0} V^+(x,t,z)V^-(\overline x,\overline t,z)^Tdz=0.
\end{equation}
Note that $P^\pm ( x,t,\partial)=I_{r+2s}+ \sum_{k=1}^\infty P^\pm_k ( x,t)\partial^{-k}$ and that 
\begin{equation}
\label{P-C22}
P^\pm ( x,t,\partial)=(N+I)P^\mp ( x,t,\partial)(N+I)
,  
\end{equation}
where (cf.  \eqref{N})
\[
N=\sum_{a=1}^s \left(E_{a,s+a}+E_{s+a,a}\right)\quad\mbox{and } I=\sum_{c=s+1}^{s+r} E_{s+c,s+c}.
\]
Hence,  applying the fundamental lemma gives
that
\begin{equation}
\label{NI}
\begin{aligned}
P^\mp( x,t,\partial)^{-1}&=P^\pm( x,t,\partial)^{*}\\
&=\left((N+I)P^\mp ( x,t,\partial)(N+I) \right)^*\\
&=(N+I)P^\mp ( x,t,\partial)^*(N+I)
.
\end{aligned}
\end{equation}
Next differntiate \eqref{VPMC} by $t^a_k$,  $t^c_\ell$ for $1\le a\le s$, $s+1\le c\le r+s$, $k\in\mathbb Z_{>0}$ and $\ell\in\mathbb Z_{odd>0}$ and apply the fundamental lemma, this gives the following Sato-Wilson equation:
\begin{equation}
\label{SatoC}
\begin{aligned}
\frac{\partial P^+( x,t,\partial)}{\partial t^{(a)}_k}&=-\left(
P^+( x,t,\partial)\left(\partial^k E_{aa}-(-\partial)^kE_{s+a,s+a}\right)P^+( x,t,\partial)^{-1}
\right)_- P^+( x,t,\partial),\\
\frac{\partial P^+( x,t,\partial)}{\partial t^{(c)}_\ell}&=-\left(
P^+( x,t,\partial) \partial^\ell E_{s+c,s+c}P^+( x,t,\partial)^{-1}
\right)_- P^+( x,t,\partial).
\end{aligned}
\end{equation} 
Define the following Lax operators
\begin{equation}
\label{Lax-op-C}
\begin{aligned}
L&=L(x,t,\partial)=P^+( x,t,\partial) \left(\sum_{a=1}^s \left(\partial E_{aa}-\partial E_{s+a,s+a}\right)+\sum_{c=s+1}^{s+c}\partial E_{s+c,s+c}\right)
P^+( x,t,\partial)^{-1},\\
C^b&=C^b( x,t,\partial) =P^+( x,t,\partial) E_{bb} P^+( x,t,\partial)^{-1},\qquad 1\le b\le r+2s,\\
D^a&=D^a( x,t,\partial) =P^+( x,t,\partial) (E_{aa} -E_{s+a,s+a})P^+( x,t,\partial)^{-1},\\
E^a&=(D^a)^2=E^a( x,t,\partial) =P^+( x,t,\partial) (E_{aa} +E_{s+a,s+a})P^+( x,t,\partial)^{-1}
.
\end{aligned}
\end{equation}
Now let $a=1,\ldots s$,   $c=s+1,\ldots r+s$, $b=1,\ldots r+2s$, then we have the following Lax Equations 
\begin{equation}
\label{Lax}
\begin{aligned}
\frac{\partial  L}{\partial t^{(a)}_j}&=[(L^j D^a_+,L],\quad \frac{\partial  C^b}{\partial t^{(a)}_j}=[(L^j D^a)_+,C^b],\\
\frac{\partial  L}{\partial t^{(c)}_j}&=[(L^j C^{s+c})_+,L],\qquad \frac{\partial  C^b}{\partial t^{(c)}_j}=[(L^j C^{s+c}_+,C^b].
\end{aligned}
\end{equation}
From \eqref{NI} we deduce that
\[
C^{s+b*}=(N+I) C^{s+b}(N+I),\quad E^{a*}=(N+I) E^a(N+I),\quad D^{a*}=-(N+I) D^a(N+I),
\]
1$\le a\le s< b\le r+s$,
and
\[
\begin{aligned}
L^*&=((\sum_{a=1}^s E^a+\sum_{b=s+1}^{r+s} C^{s+b})L)^*
=(\sum_{a=1}^s E^a+\sum_{b=s+1}^{r+s} C^{s+b})L
\\
&=(N+I)((\sum_{a=1}^s E^a-\sum_{b=s+1}^{r+s} C^{s+b})L)(N+I).
\end{aligned}
\]
There are two special cases, viz.  if $s=0$ or $r=0$, which give that 
\[
\begin{aligned}
s=0:\quad&L(x,t,\partial)^*=-L(x,t,\partial),\qquad C^c( x,t,\partial)^*=C^c( x,t,\partial),\\
r=0:\quad&L(x,t,\partial)^*=NL(x,t,\partial)N, \qquad D^a( x,t,\partial)^*=-ND^{a}( x,t,\partial)N ,
\end{aligned}
\]
Note that the classical 1-component CKP case (cf.  \cite{LOS}, \cite{KZ}) corresponds to $s=0 $ and $r=1$,  hence the case that $L^*=-L$.
%
%
%
%

\section{Conjugacy classes of the Weyl group for classical Lie algebras and generating fields for $\hat{\mathfrak g}$ and $\hat{\mathfrak g}^{(2)}$}
\label{sec5}
\subsection{Weyl group elements and the twisted realizations of $\hat{\mathfrak g}$}
\label{twistweyl}
Let $\mathfrak g$ be one of the classical Lie algebras over $\mathbb C$  and let  $G$ be the corresponding classical complex Lie group. Choose a Cartan subalgebra  $\mathfrak h$  in $\mathfrak g$ and let $W\in GL(\mathfrak h) $ be the Weyl group. Fix  $w\in W$ and let $\mathfrak h_0$ be the fixed point set of $w$ in $\mathfrak h$. We can lift $w$ (non-uniquely)   to an  element $\tilde  w\in G $ of order $N$, such that 
\begin{equation}
\label{4.1}
\tilde w=S e^{2\pi \sqrt{-1} h_w} S^{-1},
\end{equation}
where $S\in G$ and $h_w\in\mathfrak h$ is such that 
\begin{equation}
\label{4.2}
(h_w|\mathfrak  h_0)=0, 
\end{equation}
in all cases, except for $so_{2n+1}$. For the level one $so_{2n+1}$-module with the highest weight $\Lambda_n=\frac12 \sum_{i=1}^n \epsilon_i$,  condition \eqref{4.2} is replaced by
\begin{equation}
\label{4.2b}
(h_w-\Lambda_n|\mathfrak  h_0)=0.
\end{equation}

Consider the $\mathbb Z_N=\mathbb Z/ N\mathbb Z$-gradation of $\mathfrak g$ defined by $\sigma=e^{2\pi \sqrt{-1} {\rm ad}\, h_w}$:
\begin{equation}
\label{4.3}
\mathfrak g=\bigoplus_{\overline j\in\mathbb Z_N}\mathfrak g_{\overline j} ,\quad\mbox{where }
\mathfrak g_{\overline j}=
\{ g\in\mathfrak g\, |\,
[h_w, g]=\frac{j }N g\ \mbox{ for some } j\in \mathbb Z, \ \overline j \equiv j\,{\rm mod } \,N\}.
\end{equation}

Define the twisted affine Lie algebra  $ \hat L(\mathfrak g, \sigma)$ as the following  subalgebra of $\hat{\mathfrak g}$:
\begin{equation}
\label{4.4}
\hat L(\mathfrak g, \sigma)=\mathbb C K\oplus \bigoplus_{j\in\mathbb Z} t^j\mathfrak g_{j\,{\rm mod}\, N} .
\end{equation}
The Lie algebra $\hat{\mathfrak g}$ and $ \hat L(\mathfrak g,\sigma)$  are isomorphic for any inner automorphism $\sigma$ of $\mathfrak g$ (\cite{Kacbook}, Chapter 8).  Explicitly the isomorphism $\tilde\phi_{h_w}:\hat L(\mathfrak g, \sigma)\to \hat{\mathfrak g}$ is given  by \cite{KP112}, \cite{tKLA} 
\begin{equation}
\label{4.5}
\tilde\phi_{h_w}(t^j g)=\hat g(j):=g(j)- \delta_{j,0}(h_w| g)K\quad
\mbox{for }g\in\mathfrak g_{j\,{\rm mod }N},
\end{equation}
where 
\begin{equation}
\label{4.6}
g(j)=t^{-{\rm ad}\, h_w} (t^{\frac jN}g)\in\hat{\mathfrak g},\quad g\in \mathfrak g_{j\,{\rm mod }N}.
\end{equation}

Each element $g\in \mathfrak g$ decomposes with respect to the $\mathbb Z_N$-gradation \eqref{4.3}
as $g=\sum_{  \overline  j\in \mathbb Z_N} g_{\overline j}$, and we consider the corresponding generating    field
\begin{equation}
\label{below5.7}
g(z)= \sum_{j\in \mathbb Z} \hat g_{j\,{\rm mod}\, N}(j) z^{-j-1}.
\end{equation}

Let  $d= \dim \mathfrak g$ and let  $g^1,g^2,\ldots, g^d$  be the standard basis of $\mathfrak g$, i.e. the root vectors together with a basis of $\mathfrak h$. Then we obtain a new basis $a^i=S^{-1}g^iS$, which has the same commutation relations as the $g^i$. 
One of the main constructions of this paper will be the construction of  the vertex operators for $a^i(z)$. Note that the standard Cartan subalgebra $\mathfrak h$ gets conjugated to a non-standard one 
$\mathfrak h_\sigma$
by $S^{-1}$.  Hence if $h^1, h^2,\ldots h^n$ is  a basis of $\mathfrak h $,  the modes of the vertex operators for $
(S^{-1} h^i S)(z)$,  for $1\le i\le n$ (see formula \eqref{below5.7}),  together with $K$ span a non-standard Heisenberg subalgebra of $\hat{\mathfrak g}$. Of course for $w=1$ we get the standard homogeneous 
Heisenberg subalgebra $(\mathfrak h\otimes\mathbb C[t,t^{-1}])\oplus\mathbb C K$.

\begin{example}{\bf (The Lie algebra $\mathfrak g=gl_2$)}
There are two elements in the Weyl group, namely the identity
$id$  and the element $w$ that interchanges $\epsilon_1$ and $\epsilon_2$. The first case is related to the homogeneous realization and the second  to the principal realization of the subalgebra $\hat{sl}_2$. 
It is easy to find a lift in both cases satisfying \eqref{4.2}:
\[
I=\begin{pmatrix} 1&0\\0&1\end{pmatrix},\quad 
\tilde w=\begin{pmatrix} 0&-i\\-i&0\end{pmatrix},\ \mbox{respectively}.
\] 
Then 
$S=I$   in the first case and $S=\frac{1}{\sqrt 2}\begin{pmatrix} -1&1\\1&1\end{pmatrix}$ for the second one, which leads respectively to the following choices of $h_w$:
$$
I=\exp (2\pi \sqrt{-1} h_1), \ \mbox{where $h_1=0$,  and }
\tilde w=\exp (2\pi \sqrt{-1} h_w),  \ \mbox{ where }h_w= \begin{pmatrix} \frac14&0\\0&-\frac14\end{pmatrix}.
$$
In the first case we have 4 fields,  for $\hat{gl}_2$,  that generate the algebra, namely $e_{ij}(z)
=\sum_{k\in\mathbb Z} t^ke_{ij}z^{-k-1}$, $1\le i,j\le 2$.  For $\hat{sl}_2$ we have 3 fields,  
namely   $e_{12}(z)$, $e_{21}(z)$ and $(e_{11}-e_{22})(z)$.
In the second case we have only two fields,  for $\hat{gl}_2$, namely 
\[
\begin{aligned}
\left(S^{-1}e_{22}S\right)(z)=-\left(S^{-1}e_{11}S\right)(-z)&=\sum_{k\in\mathbb Z} \frac{t^k }2\begin{pmatrix} 
1& z^{-1}\\t z^{-1}&1\end{pmatrix}z^{-2k-1},\\
\left(S^{-1}e_{12}S\right)(z)=-\left(S^{-1}e_{21}S\right)(-z)&=\sum_{k\in\mathbb Z} \frac{t^k}2 \begin{pmatrix} 
-1&-z^{-1}\\t z^{-1}&1\end{pmatrix}z^{-2k-1}.
\end{aligned}
\]
For $\hat{sl}_2$ we have two fields: $(S^{-1}e_{12}S)(z)$ and 
\[
\left(S^{-1}(e_{22}-e_{11})S\right)(z)=\sum_{k\in\mathbb Z} {t^k }\begin{pmatrix} 
0& 1\\
t  &0\end{pmatrix}z^{-2k-2}
\]
The modes of the second  field, together with $K$, form a basis of the so called principal Heisenberg subalgebra of $\hat{sl}_2$.
\end {example}

\subsection{$gl_n$}

Recall that the conjugacy classes of the Weyl group $W=S_n$  for $gl_n$ are in one-to-one correspondence with the partitions 
\begin{equation}
\label{lambdaa}
\lambda=(\lambda_1\ge \lambda_2\ge \cdots\ge \lambda_s>0)
\end{equation}
 of $n$,
i.e. 
\begin{equation}
\label{lambdal}
|\lambda|=\lambda_1+\lambda_2+\cdots+\lambda_s=n. 
\end{equation} 
We can decompose each element $w\in W$ in a product of    disjoint cycles of length $\lambda_i$ . Then two elements with the same cycle type are conjugate. 
Let $|\lambda|_0=0$ and denote  for $j= 1, 2\ldots s$,
\begin{equation}
\label{lambdaj}
|\lambda|_j=\lambda_1+\lambda_2+\cdots+\lambda_j,
\end{equation}
thus $|\lambda|=|\lambda |_s$.
We can associate to a partition $\lambda$ of $n$ the standard element $w_\lambda\in W$, namely in the basis $\{ \epsilon_i\}$ of $\mathfrak h$:
\begin{equation}
\label{wlambda}
w_\lambda: \epsilon_{|\lambda|_{j}} \mapsto \epsilon_{|\lambda|_{j-1}+1},\ \mbox{and }
\epsilon_{|\lambda|_{j-1}+i}\mapsto \epsilon_{|\lambda|_{j-1}+i+1},\
i=1,2,\ldots,\lambda_j-1, 
\quad 1\le j\le s.
\end{equation}
We obtain this $w_\lambda$ if we restrict  
${\rm Ad}\, \tilde w_\lambda$ to the Cartan subalgebra $\mathfrak h$, where
\begin{equation}
\label{Wlambda}
\tilde w_\lambda=\sum_{a=1}^s  \left(e_{ 1,\lambda_{a} }^{aa}+
\sum_{i=1}^{\lambda_a-1} e_{ i+1,  i}^{aa}\right)\in GL_n,
\end{equation}
where we redefine the basis vectors as follows:
\begin{equation}
\label{redefbasis}
e_{ij}^{ab}=e_{|\lambda|_{a-1}+i,|\lambda|_{b-1}+j},\quad 1\le i\le \lambda_a,\ 1\le j\le \lambda_b,\ 1\le a,b\le s.
\end{equation}

So for the diagonal block related to the part of $\lambda_j$ this is the Coxeter element in the Weyl group of $gl_{\lambda_j}$. Hence, it will be useful to study this case first.
\begin{example}{ \bf  (One cycle, the principal Heisenberg algebra)}
Assume that $s=1$, i.e. $\lambda=(n)$ and let 
\[
\tilde w_{(n)}'=e_{ 1n}+
\sum_{i=1}^{n-1} e_{i+1,  i}.
\]
Let $\omega =e^{\frac {2\pi \sqrt{-1}}n}$. The vectors $(\omega^j, \omega^{2j}, \cdots,\omega^{nj})^T$ are eigenvectors of $\tilde w$ with  the eigenvalue $\omega^{-j}$. Thus let 
\[
S=\frac1{\sqrt n}\sum_{i,j=1}^n \omega^{ij}e_{ij}, \quad \mbox{  then }S^{-1}=
\frac1{\sqrt n}\sum_{i,j=1}^n \omega^{-ij}e_{ij}
\]
and 
\[
S^{-1}  \tilde w_{(n)}'S= \sum_{j=1}^n \omega^{-j}e_{jj}=\exp\left( 2\pi \sqrt{-1} \sum_{j=1}^n -\frac{j }ne_{jj}\right).
\]
Hence we could  choose $h_w'=\sum_{j=1}^n -\frac{j}n e_{jj}$, then ${\rm Ad}\, \tilde w_{(n)}=S
e^{ 2\pi \sqrt{-1} {\rm ad}\, h_w}S^{-1}$. However, $(h_w'|I_n)\ne 0$, but we can always add a multiple of $I_n$ to $h_w'$, hence we choose
\[
h_w=\sum_{j=1}^n \frac{n-2j+1}{2n} e_{jj},\quad \mbox{so that } (h_w|\mathfrak h_0)=0.
\]
Note that this means that we let  $$\tilde w_{(n)}=\omega^{\frac{n+1}2}\tilde w_{(n)}'.$$
Now, let
\[
a_{k\ell}=S^{-1} e_{k\ell}S= \frac1n\sum_{i,j=1}^n \omega^{j\ell-ik}e_{ij}\in gl_n,
\]
then 
\[
a_{k\ell}(z)=\frac1n\sum_{m\in\mathbb Z}\sum_{i,j=1}^n \omega^{j\ell-ik}
(t^{m}e_{ij})z^{-mn+i-j-1}.
\]
Using \eqref{tkeij} and \eqref{rEijk},
  we find that 
\[
r(a_{k\ell})(z)=\frac{\omega^{-k}}n\sum_{m,\ell \in\mathbb Z}\sum_{i,j=1}^n \omega^{j\ell-ik}
   :\psi^+_{-\ell n-i+\frac12}\psi^-_{ (\ell +m)n+j-\frac12}:z^{-mn+i-j-1}.
\]
In terms of the charged  fermionic fields
\begin{equation}
\label{FFprinc}
\psi^\pm(z)=\sum_{i\in\frac12+\mathbb Z}\psi_i^\pm z^{-i-\frac12},
\end{equation}
this becomes 
\[
r(a_{k\ell})(z)=\frac1n :\psi^+(\omega^{-k}z)\psi^-(\omega^{-\ell} z):,
\]
and in particular
\[
\alpha(z)=\sum_{j\in\mathbb Z}\alpha_j z^{-j-1}=nr( a_{nn})(z)=\sum_{j\in\mathbb Z} r((te_{n1}+\sum_{i=1}^{n-1} e_{i, i+1})^j)
z^{-j-1}=:\psi^+(z)\psi^-(z):
\]
will be the generating field for the principal Heisenberg algebra, i.e. 
\[
[\alpha_j,\alpha_k]=j \delta_{jk},\quad  \alpha_j|0\rangle
=0,\quad\mbox{for }j\ge 0.
\]
\end{example}

The general construction follows from this example.  
Consider the partition\eqref{lambdaa} of $n$. Introduce the associated relabeling,  $j\equiv |\lambda|_{a-1}+1,\ldots ,|\lambda|_a \mod n$, induced by \eqref{redefbasis},  \eqref {tkeij} and \eqref{rEijk}:
\begin{equation}
\begin{aligned}
\label{psiaa}
\psi^{+a}_{j\lambda_a -i+\frac12}=\psi^{+}_{j n-|\lambda|_{a-1}-i+\frac12},\quad 
\psi^{-a}_{j\lambda_a +i-\frac12}=\psi^{-}_{j n+|\lambda|_{a-1}+i-\frac12},
\end{aligned}
\end{equation}
for $1\le i\le \lambda_a$,  $j\in\mathbb Z$, $a=1,\ldots, s$. 
 Let $\omega_a=e^{\frac{2\pi\sqrt{-1}}{\lambda_a}}$. It is straightforward to deduce the following proposition from the above example.
\begin{proposition}
\label{P5.3}
(a) The element 
\[
\tilde w_\lambda=\left(\sum_{a=1}^s \omega_a^{\frac{\lambda_a+1}{2}}\sum_{j=1}^{\lambda_a}e^{a a}_{jj}\right)\tilde w_\lambda'\in GL_n
\]
 for  $\tilde w_\lambda'$ as in \eqref{Wlambda}, is equal to $S^{-1}\exp ( 2\pi \sqrt{-1}
h_w)S$,   where 
\[
h_w=\sum_{a=1}^s \sum_{i=1}^{\lambda_a}  \frac{\lambda_a-2j+1}{2\lambda_a} e_{jj}^{aa}\in\mathfrak h
\]
satisfies
\eqref{4.2},
and
\[
S=\sum_{a=1}^s \sum_{i=1}^{\lambda_a}\frac1{\sqrt \lambda_a}\sum_{i,j=1}^{\lambda_a} \omega_a^{ij}e_{ij}^{aa}, \quad S^{-1}=\sum_{a=1}^s \sum_{i=1}^{\lambda_a}\frac1{\sqrt \lambda_a}\sum_{i,j=1}^{\lambda_a} \omega_a^{-ij}e_{ij}^{aa}.
\]
(b) The order  $N$   of $\exp (2\pi \sqrt{-1} {\rm ad}\, h_w)$  is equal to the least common multiple $N'$ of $\lambda_1, \lambda_2, \ldots, \lambda_s$ if $N'(\frac1{\lambda_a}+\frac1{\lambda_b})\in2\mathbb Z$ for all $1\le a,b\le s$ and $2N'$ otherwise.\\
(c) The elements 
\[ 
a_{k,\ell}^{ab}:=S^{-1}e_{k\ell}^{ab} S=\frac1{\sqrt{ \lambda_a\lambda_b}}\sum_{i=1}^{\lambda_a}\sum_{j=1}^{\lambda_b} \omega_a^{-ik}\omega_b^{j\ell}e_{ij}^{ab}
\]
form a new basis of $gl_n$ and the representation $r$ in $F$ of their generating fields
\[
a_{k,\ell}^{ab}(z)=\frac1{\sqrt{ \lambda_a\lambda_b}}\sum_{m\in\mathbb Z}\sum_{i=1}^{\lambda_a}\sum_{j=1}^{\lambda_b} \left(\omega_a^{-ik}\omega_b^{j\ell}t^{m}e_{ij}^{ab}\right)z^{-mN+ {i}{\frac{N}{\lambda_a}}- {j}{\frac{N}{\lambda_b}}-1},
\]
are equal to
\begin{equation}
\label{gl-a}
r(a_{k,\ell}^{ab})(z)=\frac {\omega_a^{-k}}{\sqrt{ \lambda_a\lambda_b}}:\psi^{+a}(\omega_a^{-k}z^{\frac{N}{\lambda_a}})\psi^{-b}(\omega_b^{-\ell}z^{\frac{N}{\lambda_b}}):z^{\frac{N}{\lambda_a}-1},
\end{equation}
where 
\[
\psi^{\pm a}(z)=\sum_{i\in\frac12+\mathbb Z}\psi_i^{\pm a}z^{-i-\frac12}
\]
are the generating fields 
for the relabeld generators $\psi_j^{\pm a} $of the Clifford algebra $C\ell$,  $j\in\frac12+\mathbb Z$ and $1 \le a\le s$.\\
(d) The representation of $\hat{gl}_n$ is given in terms of vertex operators by formula \eqref{VOA},  by substitution of roots of unity $\omega_a$ and powers of $z$ as given 
in formula \eqref{gl-a}.  In particular,  the free bosonic fields are 
\[
\alpha^b(z)= \sum_{j\in\mathbb Z} \alpha_j^b z^{-j-1}= :\psi^{+b}(z)\psi^{-b}(z): \quad 1\le b\le s,
\]
so that $\alpha^b(z)= \lambda_b z^{\frac{\lambda_a}N-1}r(a_{\lambda_b\lambda_b}^{bb})(z^{\frac{\lambda_a}N})$ and their modes satisfy the commutation relations of a Heisenberg Lie algebra of "rank" $s$:
\[
[\alpha^{a}_j,\alpha^b_k]=j\delta_{ab}\delta_{jk},\quad j,k\in\mathbb Z, \ a,b=1,\ldots,s,\quad  \alpha_j^a|0\rangle
=0,\quad\mbox{for }j\ge 0.\qquad\qquad\qquad\hfill\square
\]
\end{proposition}

Let 
$\hat{\mathfrak a}_\lambda$ be the span of $K$ and the elements appearing as the modes of the fields $a^{bb}_{\lambda_b\lambda_b}$ for $1\le b\le s$.
It is a Heisenberg subalgebra of $\hat{gl}_n$.
Then the $\hat{gl}_n$-module 
$F^{(m)}$, when restricted to the subalgebra $\hat{\mathfrak a}_\lambda$ 
remains irreducible when $\lambda=(n)$ and  is not irreducible, otherwise.  
It was shown  \cite{KP112}, that    the representation remains irreducible under $\hat{\mathfrak a}_\lambda$ and the centralizer of $\hat{\mathfrak a}_\lambda$   in $SL_n(\mathbb C[t,t^{-1}])$. This means in this case that we have to  add  the operators that describe the group action of the elements  
\[
\sum_{b=1}^s \frac1{\lambda_b}a^{bb}_{\lambda_b\lambda_b}(k_b), \quad k_b\in\mathbb Z, \quad k_1+\cdots+ k_s=0.
\]
These are the operators $Q_1^{k_1}\cdots Q_s^{k_s}$,  wich are products of $Q_iQ_{i+1}^{-1}$ for $i=1,\ldots s-1$ and their inverses.
This gives  the (twisted) group algebra of the root lattice of $sl_s$.  

The $q$-dimension formula for the $\hat{gl}_n$ module $F$ now differs from the formula of the $a_\infty$-module \eqref{QDIMA}, since the gradation is different, see e.g. \cite{tKLA}.
Let $N$ be the least common multiple of $\lambda_1,\ldots,\lambda_s$, then \cite{tKLA}
\[
\dim_q F^{(m)}=\sum_{\genfrac{}{} {0pt}{2}{j_1, \ldots, j_s\in\mathbb Z}{j_1+\cdots +j_s=m}}
q^{\frac{N}2(\frac{j_1^2}{\lambda_1^2}+\frac{j_2^2}{\lambda_2^2}+\cdots +\frac{j_s^2}{\lambda_s^2})}
\prod_{a=1}^s \prod_{i=1}^\infty \frac1{1-q^{\frac{jN}{\lambda_a}}} .
\]
The module $F^{(m)}$ when restricted to $\hat{sl}_n$ is no longer irreducible.  However,  it is straightforward to calculate the $q$-dimension  formula of the irreducible component $F_{sl_n}^{(m)}$, that contains $|m\rangle$.  For that  one only has to remove all 
\[
\alpha_{j\lambda_1}^1+\alpha_{j\lambda_2}^2+\cdots+\alpha_{j\lambda_s}^1s \quad \mbox{for all }j\in\mathbb Z.
\]
Since such an element has energy $jN$, this gives,
\[
\dim_q F_{sl_n}^{(m)}=\sum_{\genfrac{}{} {0pt}{2}{j_1, \ldots, j_s\in\mathbb Z}{j_1+\cdots +j_s=m}}
q^{\frac{N}2(\frac{j_1^2}{\lambda_1^2}+\frac{j_2^2}{\lambda_2^2}+\cdots +\frac{j_s^2}{\lambda_s^2})}\prod_{j=1}^\infty
(1-q^{jN})
\prod_{a=1}^s \prod_{i=1}^\infty \frac1{1-q^{\frac{jN}{\lambda_a}}} .
\]
\subsection{$sp_{2n}$}
\label{S5.3}
Recall that the conjugacy classes of the Weyl group $W$  of $sp_{2n}$ are in one-to-one correspondence with the pair of partitions 
\begin{equation}
\label{lambda,mi}
\lambda=(\lambda_1\ge \lambda_2\ge \cdots\ge \lambda_s>0)\quad\mbox{and }
\mu=(\mu_1\ge \mu_2, \ge \cdots\ge \mu_r>0),
\end{equation}
 of $n$,
i.e. 
\begin{equation}
\label{lambdamul}
|\lambda|+|\mu|=\lambda_1 +\cdots+\lambda_s+\mu_1+\cdots+\mu_r=n. 
\end{equation} 
Namely, each element $w\in W$    decomposes in a product of $s$ disjoint positive cycles of length $\lambda_i$,  and $r$ disjoint negative cycles of length $\mu_j$, such that \eqref{lambdamul} holds. Two elements with the same cycle type are conjugate. 

We can associate to a pair of  partition $(\lambda, \mu)$ of $n$ a standard element $w_{(\lambda,\mu)}\in W$, namely:
\begin{equation}
\begin{aligned}
\label{wlambdamu}
w_{(\lambda,\mu)}:& \epsilon_{|\lambda|_{i}} \mapsto \epsilon_{|\lambda|_{i-1}+1},\ 
  \epsilon_{|\lambda|+|\mu|_{j}} \mapsto -\epsilon_{|\lambda|+|\mu|_{j-1}+1} 
,\\
&\epsilon_{|\lambda|_{i-1}+k}\mapsto \epsilon_{|\lambda|_{i-1}+k+1},\quad
k=1,2,\ldots,\lambda_i-1, 
\quad 1\le i\le s,\\
&\epsilon_{|\lambda|+|\mu|_{j-1}+k}\mapsto \epsilon_{|\lambda|+|\mu|_{j-1}+k+1},\quad
k=1,2,\ldots,\mu_j-1, 
\quad 1\le j\le r.
\end{aligned}
\end{equation}
This suggests to redefine the elements  that span $sp_{2n}$ as follows.
Let $m_a=\lambda_a$, $M_a=|\lambda|_{a-1}$ for $1\le a\le s$ and $m_a=\mu_{a-s}$, $M_a=|\lambda|+|\mu|_{a-s-1}$ for $s<a\le r+s$  and $M_{r+s+1}=|\lambda|+ |\mu|$,. Let ($1\le a,b\le s+r$, $ 1\le i\le m_a$, $ 1\le j\le m_b$)
\begin{equation}
\label{redefbasisC}
\begin{aligned}
e_{ij}^{+a,-b}=&
  e_{M_a+i,M_b+j}- (-1)^{M_a +M_b+i+j}e_{2n+1-M_b-j, 2n+1-M_a-i}\\
 e_{ij}^{+a,+b}=&(-1)^{M_a+M_b+i+j}e_{ji}^{+b,+a}= e_{M_a+i,2n+1-M_b-j}- (-1)^{M_a+M_b+i+j+1 }e_{ M_b+j, 2n+1-M_a-i}\\
 e_{ij}^{-a,-b}&=(-1)^{M_a+M_b+i+j}e_{ji}^{-b,-a}= e_{2n+1-M_a-i, M_b+j}- (-1)^{M_a+M_b +i +j+1}e_{ 2n+1-M_b-j,  M_a+i}.
\end{aligned}
\end{equation}
Before finding the lift of the element \eqref{wlambdamu}, we first investigate two special cases, which we present as examples.
\begin{example}
{\bf  (One positive cycle)}\\
In this case $s=1$ and $r=0$, hence 
\begin{equation}
\label{onepossp}
w_{((n),\emptyset )}: \epsilon_{n} \mapsto \epsilon_{ 1},\quad
\epsilon_{ k}\mapsto \epsilon_{ k+1},\quad
k=1,2,\ldots ,  n. 
\end{equation}
We let 
\[
  \tilde w=\tilde w_{((n),\emptyset )}=  e_{1,n}- (-1)^n e_{2n,n+1}
+\sum_{i=1}^{n-1} (e_{i+1,i}-e_{2n-i,2n+1-i}).
\]
Then it is straightforward to check that $
(  \tilde w_{((n),\emptyset )}u,  \tilde w_{((n),\emptyset )}v)_{sp_{2n}}=(u,  v)_{sp_{2n}}
$ for $u,v\in\mathbb C^{2n}$, i.e. $\tilde w\in Sp_{2n}$, and that  ${\rm Ad}\,  \tilde w_{((n),\emptyset )}|_{\mathfrak h}=w_{((n),\emptyset )}$.
 
The matrix $\tilde w$ has eigenvectors $(\omega^j, \omega^{2j},\ldots\omega^{nj},0,\cdots,0)^T$ and \break $(0,\ldots, 0,(-\omega^j)^n,(-\omega^j)^{n-1},\ldots, -\omega^j)^T$ with eigenvalue $\omega^{-j}$, where as before $\omega =e^{\frac {2\pi \sqrt{-1}}n}$.
Set 
\[
S=\frac{1}{\sqrt n}\sum_{i,j=1}^n( \omega^{ij}e_{ij}+(-1)^{i+j}\omega^{-ij}  e_{2n+1-i,2n+1-j}),
\]
then 
\[
S^{-1}=\frac{1}{\sqrt n}\sum_{i,j=1}^n( \omega^{-ij}e_{ij}+(-1)^{i+j}\omega^{ij}  e_{2n+1-i,2n+1-j}),
\]
and
\[
\begin{aligned}
(S v_k,Sv_{2n+1-\ell})_{sp_{2n}}&=\frac1{n}\sum_{i,j=1}^n \omega^{ki}(-1)^{i+\ell}\omega^{-j\ell} 
(v_i, v_{2n+1-j})_{sp_{2n}}\\
&= \frac1{n}\sum_{i=1}^n (-1)^{\ell}\omega^{(k-\ell)i}\\
&=(-1)^{\ell}\delta_{k\ell}=(v_k,v_{2n+1-\ell})_{sp_{2n}}.
\end{aligned}
\]
Thus also $S\in Sp_{2n}$ and
\[
\tilde w_{((n),\emptyset )}=S \sum_{j=1}^n \omega^{-j} (e_{jj}-e_{2n+1-j,2n+1-j})S^{-1}.
\]
So if we define, analogously to the principal case for $gl_n$,
\[
h_w=\sum_{j=1}^n \frac{n-2j+1}{2n} (e_{jj}-e_{2n+1-j,2n+1-j}),
\]
then $S^{-1} {\rm Ad}\, \tilde w S=\exp (2\pi \sqrt{-1}{\rm ad}\, h_w )$.
Note that condition \eqref{4.2} is satisfied since $\mathfrak h_0=\mathbb C\sum_{i=1}^n (e_{ii}-e_{2n+1-i,2n+1-i})$.

Let 
\[
\begin{aligned}
a_{k\ell}^{+-}&=S^{-1}e_{k\ell}^{+-}S=\frac1n\sum_{i,j=1}^n (\omega^{-k})^{i}(\omega^\ell)^j e_{ij}^{+-},\\
a_{k\ell}^{++}&=S^{-1}e_{k\ell}^{++}S=(-1)^{\ell}\frac1n\sum_{i,j=1}^n (\omega^{-k})^{i}(-\omega^{-\ell})^{j}e_{ij}^{++},\\
a_{k\ell}^{--}&=S^{-1}e_{k\ell}^{--}S=(-1)^{k}\frac1n\sum_{i,j=1}^n (-\omega^k)^i(\omega^\ell)^j e_{ij}^{--}.\\
\end{aligned}
\] 
Then 
\[
\begin{aligned}
a_{k\ell}^{+-}(z)&= \frac1n\sum_{m\in\mathbb Z}\sum_{i,j=1}^n\left( (\omega^{-k})^{i}(\omega^{-\ell})^{-j }t^me_{ij}^{+-}\right)z^{-mn+i-j-1},\\
a_{k\ell}^{++}(z)& =(-1)^{\ell}\frac1n\sum_{m\in\mathbb Z}\sum_{i,j=1}^n\left( (\omega^{-k})^{i}(-\omega^{-\ell})^{j}t^me_{ij}^{++}\right)z^{-(m+1)n+i+j-2},\\
a_{k\ell}^{--}(z)&=(-1)^{k}\frac1n\sum_{m\in\mathbb Z}\sum_{i,j=1}^n\left( (-\omega^{-k})^{-i}(\omega^{-\ell})^{-j} t^me_{ij}^{--}\right)z^{-(m-1)n-i-j}.\\
\end{aligned}
\] 
Using \eqref{cij} and \eqref{cijc}, we can calculate 
\[
\begin{aligned}
r(a_{k\ell}^{+-})(z)&= \frac1n\sum_{\ell,m\in\mathbb Z}\sum_{i,j=1}^n (\omega^{-k})^{i}(-\omega^{-\ell})^{-j } 
:b_{-2\ell n-i+\frac12}b_{2(\ell+m)n+j-\frac12}:
 z^{-mn+i-j-1},\\
r(a_{k\ell}^{++})(z)& =(-1)^{\ell+1}\frac1n\sum_{\ell,m\in\mathbb Z}\sum_{i,j=1}^n (\omega^{-k})^{i}(\omega^{-\ell})^{j}
:b_{-2\ell n-i+\frac12}b_{2(\ell+m+1)n-j+\frac12}:
z^{-(m+1)n+i+j-2},\\
r(a_{k\ell}^{--})(z)&=(-1)^{k}\frac1n\sum_{m\in\mathbb Z}\sum_{i,j=1}^n (-\omega^{-k})^{-i}(-\omega^{\ell})^{-j}
:b_{-2(\ell-1) n+i-\frac12}b_{2(\ell+m)n+j-\frac12}:
z^{-(m-1)n-i-j}.\\
\end{aligned}
\] 
Thus, relabeling the $b_j$ as follows:
\begin{equation}\
\label{20a}
b^+_{\ell n-i+\frac12}=b_{2\ell n-i+\frac12}, \quad
b^-_{\ell n+i-\frac12}=(-1)^ib_{2\ell n+i-\frac12},
\end{equation}
and their generating series as in \eqref{FFprinc}, we find that 
\[
\begin{aligned}
r(a_{k\ell}^{+-})(z)&= \frac{\omega^{-k}}n
:b^+(\omega^{-k}z)b^-(\omega^{-\ell}z):,
\\
r(a_{k\ell}^{++})(z)& =(-1)^{\ell+1}\frac{\omega^{-k-\ell}}n 
:b^+(\omega^{-k}z)b^+(\omega^{-\ell}z):,
,\\
r(a_{k\ell}^{--})(z)&=(-1)^{k}\frac1n 
:b^-(\omega^{-k}z)b^-(\omega^{-\ell}z):
.
\end{aligned}
\] 
Note that the $b^\pm_j$ defined by \eqref{20a} satisfy
\[
b^\pm_ib^\pm_j-b^\pm_jb^\pm_i=0,\quad
b^+_ib^-_j-b^-_jb^+=\delta_{i,-j}\quad i,j\in \frac12+\mathbb Z,
\]
and 
\[
b^\pm_i|0\rangle=0\quad\mbox{for }i>0.
\]
Again letting $\alpha(z)=\sum_{j\in \mathbb Z}\alpha_j z^{-j-1}=n r(a_{nn}^{+-})(z)$, this defines the Heisenberg subalgebra of $ r(\hat{sp}_{2n})$, with commutation relations 
\[
[\alpha_j,\alpha_k]=-j \delta_{jk},\quad  \alpha_j|0\rangle=0
 ,\quad\mbox{for }j\ge 0.
\]
Finally, 
\[
[\alpha_j, b^{\pm}(z)]= \mp z^jb^{\pm}(z).
\]
 \end{example}
 \begin{example}{\bf (One negative cycle, the principal Heisenberg subalgebra)}
 In this case $r=1$ and $s=0$, hence 
\[
w_{(\emptyset ,(n))}: \epsilon_{n} \mapsto -\epsilon_{ 1},\quad
\epsilon_{ k}\mapsto \epsilon_{ k+1},\quad
k=1,2,\ldots ,  n-1. 
\]
We choose 
\[
 \tilde w=\tilde w_{(\emptyset,(n) )}=  e_{2n,n}+ (-1)^n e_{1,n+1}
+\sum_{i=1}^{n-1} (e_{i+1,i}-e_{2n-i,2n+1-i}).
\]
Then it is straightforward to check that ${\rm Ad}\, \tilde w|_{\mathfrak h}=w_{(\emptyset,(n) )}$ and  that this element satisfies 
\[
( \tilde w u, \tilde w v)_{sp_{2n}}=(u,v)_{sp_{2n}}\quad 
u,v\in\mathbb C^{2n}.
\]
Thus  $ \tilde w\in  {Sp}_{2n}$.
Let $\omega=e^{\frac{2\pi \sqrt{-1}}{4n}}$.  The  matrix  $\tilde w$ has eigenvectors 
 \[
 (\omega^j,\omega^{2j},\ldots,\omega^{nj},(-1)^{n} ,( -1)^{n}(- \omega^{-j})^1,(-1)^{n}(-\omega^{ - j})^2,  \ldots,
 (-1)^n (-\omega^{ -j})^{n-2},(-1)^n(-\omega^{-j})^{n-1})^T,
 \]  
with eigenvalues $\omega^{-j}$, for $j=1,3,5,\ldots 4n-1$.
Hence, we can diagonalize $  \tilde w$ by $S=(S_{ij})_{1\le i,j\le 2n}$, where
\[
 S_{i,j}=\frac{ \omega^{\frac{n}2} (-\omega^{2j-1})^i}{\sqrt {2n}},\quad  S_{2n+1-i,j}=\frac{ \omega^{\frac{n}2}(\omega^{1-2j})^{n-i}}
 {\sqrt {2n}},\quad\mbox{for }1\le i\le n,\ 1\le j\le  2n,
\]
so that ${\rm Ad}\,  \tilde w= S(\exp(2\pi \sqrt{-1} {\rm ad}\, h_w)S^{-1}$, for
\[
h_w=\sum_{i=1}^n \frac{2n-2i+1}{4n}(e_{ii}-e_{2n+1-i,2n+1-i})
\]
Note that $S^{-1}=(T_{ij})_{1\le i,j\le 2n}$, where
\[
 T_{i,j}=\frac{ \omega^{-\frac{n}2} (-\omega^{2i-1})^{-j}}{\sqrt {2n}},\quad  T_{i,2n+1-j}=\frac{ \omega^{-\frac{n}2}(\omega^{1-2i})^{j-n}}
 {\sqrt {2n}},\quad\mbox{for }1\le i\le 2n,\ 1\le j\le  n.
\]
Moreover, it is straightforward to check that $(Su,Sv)_{sp_{2n}}=(u,v)_{sp_{2n}}$ for $u,v\in\mathbb C^{2n}$, hence  $S\in Sp_{2n}$.
 Thus
 \[
\begin{aligned}
a_{k\ell}^{+-}&=S^{-1}e_{k\ell}^{+-}S=\frac{(-\omega)^{k-\ell}}{2n}\sum_{i,j=1}^{2n} (\omega^{-2k})^{i}(\omega^{2\ell})^j (e_{ij}-(-1)^{i+j} e_{2n+1-j,2n+1-i}),\\
a_{k\ell}^{++}&=S^{-1}e_{k\ell}^{++}S=\frac{(-\omega)^{k}\omega^{n-\ell}}{2n}\sum_{i,j=1}^{2n} (\omega^{-2k})^{i}(-\omega^{2\ell})^j (e_{ij}-(-1)^{i+j} e_{2n+1-j,2n+1-i}) \\
a_{k\ell}^{--}&=S^{-1}e_{k\ell}^{--}S=\frac{(-\omega)^{-\ell}\omega^{k-n}}{2n}\sum_{i,j=1}^{2n} (-\omega^{-2k})^{i}(\omega^{2\ell})^j (e_{ij}-(-1)^{i+j} e_{2n+1-j,2n+1-i}) ,
\end{aligned}
\] 
and the corresponding generating fields are
 \[
\begin{aligned}
a_{k\ell}^{+-}(z)&= \frac{(-\omega)^{k-\ell}}{2n}\sum_{m\in\mathbb Z}\sum_{i,j=1}^{2n}\left( (\omega^{-2k})^{i}(\omega^{2\ell})^j t^m(e_{ij}-(-1)^{i+j} e_{2n+1-j,2n+1-i})\right)z^{-2mn+i-j-1},\\
a_{k\ell}^{++}(z)& =\frac{(-\omega)^{k}\omega^{n-\ell}}{2n}\sum_{m\in\mathbb Z}
\sum_{i,j=1}^{2n}\left( (\omega^{-2k})^{i}(-\omega^{2\ell})^j t^m(e_{ij}-(-1)^{i+j} e_{2n+1-j,2n+1-i}) \right)
 z^{-2mn+i-j-1 },\\
a_{k\ell}^{--}(z)&=\frac{(-\omega)^{-\ell}\omega^{k-n}}{2n}\sum_{m\in\mathbb Z}
\left((-\omega^{-2k})^{i}(\omega^{2\ell})^j t^m(e_{ij}-(-1)^{i+j} e_{2n+1-j,2n+1-i})\right)
z^{ -2mn+i-j-1}.\\
\end{aligned}
\] 
Using \eqref{cij} and \eqref{cijc}, we can express them in terms of  the $b_j$:
 \[
\begin{aligned}
r(a_{k\ell}^{+-})(z)&= \frac{(-\omega)^{k-\ell}}{2n}\sum_{\ell,m\in\mathbb Z}\sum_{i,j=1}^{2n} (\omega^{-2k})^{i}(-\omega^{2\ell})^j 
 :b_{-2\ell n-i+\frac12}b_{2(\ell+m)n+j-\frac12}:z^{-2mn+i-j-1}
 ,\\
r(a_{k\ell}^{++})(z)& =\frac{(-\omega)^{k}\omega^{n-\ell}}{2n}\sum_{\ell, m\in\mathbb Z}
\sum_{i,j=1}^{2n} (\omega^{-2k})^{i}(\omega^{2\ell})^j 
:b_{-2\ell n-i+\frac12}b_{2(\ell+m)n+j-\frac12}:
 z^{-2mn+i-j-1 },\\
r(a_{k\ell}^{--})(z)&=\frac{(-\omega)^{-\ell}\omega^{k-n}}{2n}\sum_{\ell,m\in\mathbb Z}
(-\omega^{-2k})^{i}(-\omega^{2\ell})^j 
:b_{-2\ell n-i+\frac12}b_{2(\ell+m)n+j-\frac12}:
z^{ -2mn+i-j-1 }.\\
\end{aligned}
\] 
Defining the generating field of the $\tilde\phi_j$ as $b(z)=\sum_{j\in\frac12+\mathbb Z}b_j z^{-j-\frac12}$, we find that for $\delta,\epsilon=+,-$:
\[
r(a^{\delta,\epsilon}_{k,\ell})(z)= \frac{c^{\delta,\epsilon}_{k,\ell}}{2n}:b(\delta \omega^{-2k} z)b(\epsilon\omega^{-2\ell} z):,
\]
where
\[
c^{+-}_{k,\ell}=(-\omega)^{-k-\ell}, \quad
c^{++}_{k,\ell}=\sqrt{-1}(-\omega)^{-k}\omega^{-\ell},\quad
c^{--}_{k,\ell}=\sqrt{-1}(-\omega)^{-\ell}\omega^{-k}.
\]
Furthermore,
\[
\alpha(z):=nr(a^{+-}_{nn})(z)= \sum_{j\in 1+2\mathbb Z}\alpha_j z^{-j}=\frac12 :b(z)b(-z):
\]
is the generating field for the Heisenberg subalgebra. The modes 
of this field satisfy
 \[
[\alpha_j,\alpha_k]=-\frac{j}2 \delta_{jk},\quad  \alpha_j|0\rangle=0
 \quad\mbox{for }j> 0.
\]
Finally,
\[
[\alpha_j,b(z)]=z^jb(z).
\]
 \end{example}

The general construction follows from these two examples.  Namely,    it is straightforward to deduce the following proposition.
\begin{proposition}
\label{prop-sp}
let 
$w_{(\lambda,\mu)}$ be given by \eqref{wlambdamu}.
Set  $\lambda_{s+a}=:\mu_a$ and 
  $\omega_a=e^{\frac {2\pi \sqrt{-1}}{\lambda_a}}$ if $1\le a\le s$ and   
$\omega_a=e^{\frac {2\pi \sqrt{-1}}{2\lambda_a}}$  if $s+1\le a\le r+s$
. Then \\
(a) The element 
\[
\begin{aligned}
\tilde w_{(\lambda, \mu)}&= \sum_{a=1}^s  \omega_a^{\frac{\lambda_a+1}2}
\big(e_{M_a+1,M_{a+1}}-(-1)^{\lambda_a} e_{2n-M_a, 2n-M_{a+1}+1}+\\
&\qquad\qquad+\sum_{i=1}^{\lambda_a-1}(e_{M_a+i+1, M_a+i}-E_{2n-M_a-i,2n+1-Ma-i})\big)+\\
&\quad +\sum_{a=s+1}^{r+s} \big(
e_{2n-M_a,M_{a+1}} +(-1)^{\lambda_a}e_{M_a+1,N-M_{a+1}+1}+\\
&\qquad\qquad +\sum_{i=1}^{\lambda_a-1} (e_{M_a+i+1, M_a+i}-e_{2n-M_a-i,2n+1-Ma-i})
\big)
\in Sp_{2n}
\end{aligned}
\] 
is a lift of $ w_{(\lambda, \mu)}$ and  
 $\tilde w_{(\lambda, \mu)}=S^{-1}\exp ( 2\pi \sqrt{-1}
h_w)S$  , where 
\[
h_w=\sum_{a=1}^s \sum_{j=1}^{\lambda_a}  \frac{\lambda_a-2j+1}{2\lambda_a} \epsilon_{M_a+j} +\sum_{b=s+1}^{r+s}
\sum_{j=1}^{\lambda_b}  \frac{\lambda_b-j+\frac12}{2\lambda_b} \epsilon_{M_b+j}
\in\mathfrak h
\]
satisfies 
\eqref{4.2},
and
\[
\begin{aligned}
S=&\sum_{a=1}^s \sum_{i=1}^{\lambda_a}
\frac{1}{\sqrt \lambda_a}\sum_{i,j=1}^{\lambda_a}( \omega_a^{ij}e_{M_a+i,M_a+j}+(-1)^{i+j}\omega^{-ij}  e_{2n+1-M_a-i,2n+1-M_a-j})+\\
&+\sum_{a=s+1}^{r+s}
\frac{1}{\sqrt {2\lambda_a}}
\sum_{i,j=1}^{\lambda_a} 
\big((-1)^i
   \omega_a^{\frac{\lambda_a}4} (\omega_a^{-i})^{\frac12-j}
 e_{M_a+i,M_a+j}
 +
 (-1)^i\omega_a^{\frac{\lambda_a}4} (\omega_a^{-i})^{\frac12-\lambda_a-j}
 e_{M_a+i,2n-M_{a+1}+j}+\\
 &+
 \omega_a^{\frac{\lambda_a}4}(\omega^{\lambda_a-i})^{\frac12-j} e_{2n+1-M_a-i,M_a+j}+
 \omega_a^{\frac{\lambda_a}4}(\omega^{\lambda_a-i})^{\frac12-\lambda_a-j}e_{2n+1-M_a-i,2n-M_{a+1}+j}
 \big)
.
\end{aligned}
\]
(b) The order  $N$   of $\exp (2\pi \sqrt{-1} {\rm ad}\, h_w)$  is equal to the least common multiple $N'$ of $\lambda_1, \lambda_2, \ldots, \lambda_s, 2\lambda_{s+1}, 2\lambda_{s+2},\ldots 2\lambda_{r+s}$ if $N'(\frac1{\lambda_a}+\frac1{\lambda_b})\in2\mathbb Z$ for all $1\le a,b\le s$, $N'(\frac1{2\lambda_a}+\frac1{2\lambda_b})\in2\mathbb Z$ for all $s+1\le a,b\le r+s$,  $N'(\frac1{\lambda_a}+\frac1{2\lambda_b})\in2\mathbb Z$ for all $ 1\le a\le s$, $s+1\le b\le r+s$ and $2N'$ otherwise.
\\
(c) The elements 
$
a_{k,\ell}^{\delta a,\epsilon b}:=S^{-1}e_{k\ell}^{\delta a, \epsilon b} S=
$
form a new basis of $sp_{2n}$ and the representation $r$ in $F_c$ of their generating fields
$
a_{ k \ell}^{\delta a,\epsilon b}(z)
$
are equal to
\[
r(a_{k,\ell}^{\delta_a ,\epsilon b})(z)= L_k^{\delta a}R_\ell^{\epsilon b}:B^{\delta a}_k(z)B^{\epsilon b}_\ell (z):,
\]
where 
\[
\begin{aligned}
L_k^{+a}=&
\begin{cases}
\frac{(-1)^{k+M_a}}{\sqrt{\lambda_a}}&1\le a\le s,\\
\frac{\gamma_a^{-1}}{\sqrt{2\lambda_a}}\omega_a^{-\frac{k}2+\frac{\lambda_a}4}&s<a\le r+s,\\
\end{cases}\qquad
L_k^{-a}=
\begin{cases}
\frac{\omega_a^{-k}}{\sqrt{\lambda_a}}&1\le a\le s,\\
\frac{(-1)^k\gamma_a^{-1}}{\sqrt{2\lambda_a}}\omega_a^{-\frac{k}2-\frac{\lambda_a}4}&s<a\le r+s\\
\end{cases}\\
R_k^{+a}=&
\begin{cases}
\frac{(-1)^{k+M_a+1}}{\sqrt{\lambda_a}}\omega_a^{-k}&1\le a\le s,\\
\frac{\gamma_a}{\sqrt{2\lambda_a}}\omega_a^{-\frac{k}2+\frac{3\lambda_a}4}&s<a\le r+s,\\
\end{cases}\qquad
R_k^{-a}=
\begin{cases}
\frac{1}{\sqrt{\lambda_a}}&1\le a\le s,\\
\frac{(-1)^k\gamma_a}{\sqrt{2\lambda_a}}\omega_a^{-\frac{k}2+\frac{\lambda_a}4}&s<a\le r+s,\\
\end{cases}
\end{aligned}
\]
where $\gamma_{a}=1$ if $M_{a}$ is even and $\gamma_{a}=i$ if $M_{a}$ is odd,and 
\[
B_k^{\pm a}(z)=
\begin{cases}
b^{\pm a}(\omega_a^{-k} z^{\frac{N}{\lambda_a}})z^{\frac{N}{2\lambda_a}-\frac12}&1\le a\le s,\\
b^{ a}(\pm \omega_a^{-k} z^{\frac{N}{2\lambda_a}})z^{\frac{N}{4\lambda_a}-\frac12}&s< a\le s+r,\\
\end{cases}
\]
where the 
\[
b^{\pm a}(z)=\sum_{i\in\frac12+\mathbb Z}b_i^{\pm a}z^{-i-\frac12},\quad
b^{s+b}(z)=\sum_{i\in\frac12+\mathbb Z}b^{s+b}z^{-i-\frac12},\quad
1\le a\le s,\ 1\le b\le r,
\]
are the generating fields 
for the relabeld generators of the Weyl algebra $C\ell_c$ ($1\le a\le s$, $1\le b\le r$, $1\le i\le \lambda_a$, $1\le j\le \lambda_{s+b}$, $\ell\in\mathbb Z$):
\begin{equation}
\begin{aligned}
b^{+a}_{\ell \lambda_a-i+\frac12}=b_{2\ell n-M_a-i+\frac12}, \quad
b^{-a}_{\ell \lambda_a+i-\frac12}=(-1)^{M_a+i}b_{2\ell n+M_a+i-\frac12},\\
b^{s+b}_{2\ell \lambda_{s+b}-j+\frac12}=\gamma_{s+b}b_{2\ell n-M_{s+b}-j+\frac12},
\quad 
b^{s+b}_{2\ell \lambda_{s+b}+j-\frac12}=\gamma_{s+b}b_{2\ell n+M_{s+b}+j-\frac12}.
\end{aligned}
\end{equation}
They satisfy ($1\le a,c\le s,\ 1\le b,d\le r,\quad i,j\in \frac12+\mathbb Z$)
\begin{equation}
\label{symplectic-rel}
\begin{aligned}
b^{\pm a}_ib^{\pm c}_j-b^{\pm c}_jb^{\pm a}_i&=0,\\
b^{+ a}_ib^{-c}_j-b^{-c}_jb^{+a}&=\delta_{ac}\delta_{i,-j}\\
b_i^{s+b}b_j^{s+d}-b_j^{s+d}b_i^{s+b}&=(-1)^{i-\frac12}\delta_{bd}\delta_{i,-j},\\
b^{\pm a}_ib^{s+b}_j-b^{s+b}_jb^{\pm a}_i&=0,\\
b^{\pm a}_i|0\rangle =b_i^{s+b}|0\rangle &=0\quad i>0.
\end{aligned}
\end{equation}
(d) The representation of $\hat{sp}_n$ is given in terms of vertex operators by formulas \eqref{VOAC1},  \eqref{VOAC2} and \eqref{VOAC3} by substitution of roots of unity  $\omega_a$ and powers of $z$ as given 
in the above formulas.  In particular, the Heisenberg algebra, which is described 
by the fields $\alpha^a(z)$ are defined as in \eqref{symplectic-alpha}
and they satisfy \eqref{symplectic-alpha-rel}.
\end{proposition}
The $q$ dimension formula differs from the one given in \eqref{QDIMC}, since the gradation is different.  Let $x^2=1$ and $N$ be the least common multiple of $\lambda_1,\ldots,\lambda_s$,$2\mu_1,\ldots, 2\mu_r$, then 
\[
\begin{aligned}
\dim_q F^{\overline 0}&+x F^{\overline 1}=
\prod_{a=1}^s \prod_{j=1}^\infty \frac{1}{(1-xq^{\frac{(j-\frac12)N}{\lambda_a})^2}}
\prod_{b=1}^r\prod_{j=1}^\infty \frac{1}{1-xq^{\frac{(j-\frac12)N}{\mu_b}}}
\\
&=\prod_{a=1}^s   \prod_{j=1}^\infty  \frac{1}{1-q^{\frac{jN}{\lambda_a}}} 
\sum_{j,k=0}^\infty \frac{q^{\frac{kN}{2\lambda_a}}x^k}{(1-q^{\frac{N}{\lambda_a}})\cdots(1-q^{\frac{kN}{\lambda_a}})}
q^{jk \frac{N}{\lambda_a}} \frac{q^{\frac{jN}{2\lambda_a}}x^j}{(1-q^{\frac{N}{\lambda_a}})\cdots(1-q^{\frac{jN}{\lambda_a}})}\times\\
&\qquad\quad\prod_{b=1}^r\prod_{j=1}^\infty \frac{1+xq^{\frac{N(j-\frac12)}{\mu_b} }}  {1-q^{\frac{(2j-1)N}{\mu_b}}}
.
\end{aligned}
 \]

\subsection{$so_{2n}$}
Recall that the conjugacy classes of the Weyl group for $so_{2n}$ are in one-to-one correspondence with the pair of partitions 
\begin{equation}
\label{lambda,mi2}
\lambda=(\lambda_1\ge \lambda_2\ge \cdots\ge \lambda_s>0),
\mu=(\mu_1\ge \mu_2, \ge \cdots\ge \mu_{r}>0),
\end{equation}
 of $n$, with $r\in 2\mathbb Z$,
i.e. 
\begin{equation}
\label{lambdamul2}
|\lambda|+|\mu|=\lambda_1 +\cdots+\lambda_s+\mu_1+\cdots+\mu_{r}=n. 
\end{equation} 
We can decompose each element $w\in W$ in $s$   disjoint positive cycles of length $\lambda_i$,  and $r$ disjoint negative cycles of length $\mu_j$ such that \eqref{lambdamul2} holds. It is straightforward to show that two elements with the same cycle type are conjugate under $O_{2n}$. 

We relabel the basis as follows.
Let $m_a=\lambda_a$, $M_a=|\lambda|_{a-1}$ for $1\le a\le s$ and $m_{a}=\mu_{a-s}$,  $M_a=|\lambda|+|\mu|_{a-s-1}$ for $s<a\le r+s$ and $M_{r+s+1}=|\lambda|+ |\mu|$, then set ($1\le a,b\le s+r$, $ 1\le i\le m_a$, $ 1\le j\le m_b$)
\begin{equation}
\label{redefbasisD}
\begin{aligned}
e_{ij}^{+a,-b}=&
  e_{M_a+i,M_b+j}-  e_{2n+1-M_b-j, 2n+1-M_a-i}\\
 e_{ij}^{+a,+b}=&-= e_{M_a+i,2n+1-M_b-j}-  e_{ M_b+j, 2n+1-M_a-i}\\
 e_{ij}^{-a,-b}&=-e_{ji}^{-b,-a}= e_{2n+1-M_a-i, M_b+j}- e_{ 2n+1-M_b-j,  M_a+i}.
\end{aligned}
\end{equation}

We can associate, as for $sp_{2n}$ to a pair of  partition $(\lambda, \mu)$ of $n$ a standard element $w_{(\lambda,\mu)}\in W$, namely the one given by \eqref{wlambdamu}, but now we want to lift this to $SO_{2n}$.
As in the case of $sp_{2n}$ we will first investigate two examples.
\begin{example}{\bf (One positive cycle)}
In this case $s=1$ and $r=0$ and we have the same element as in 
\eqref{onepossp}. However now we have to lift this to $SO_{2n}$.
Let $\omega= e^{\frac{2\pi \sqrt{-1}}{n}}$ choose 
\[
  \tilde w_{((n),\emptyset )}= \omega^{\frac{n+1}2} e_{1,n}+ \omega^{-\frac{n+1}2} e_{2n,n+1}
+\sum_{i=1}^{n-1} (\omega^{\frac{n+1}2} e_{i+1,i}+\omega^{-\frac{n+1}2} e_{2n-i,2n+1-i}).
\]
Clearly, $(\tilde w_{((n),\emptyset )}v,\tilde w_{((n),\emptyset )}w)_{so_{2n}}=(v,w)_{so_{2n}}$ and we can diagonalize $\tilde w_{((n),\emptyset )}$ by choosing
\[
S=\sum_{i,j=1}^n  (\omega^{ij}e_{ij}+\omega^{-ij}e_{2n+1-i,2n+1-j}).
\]
Then 
\[
S^{-1}=\sum_{i,j=1}^n  (\omega^{-ij}e_{ij}+\omega^{ij}e_{2n+1-i,2n+1-j})
\]
and 
$\tilde w_{((n),\emptyset )}=S e^{2\pi \sqrt{-1}  h_w}S^{-1}$ with 
\[
h_w=\sum_{j=1}^n \frac{n-2j+1}{2n}(e_{jj}-e_{2n+1-j,2n+1-j}).
\]
Let $a_{k\ell}^{\delta\epsilon}=S^{-1}e_{k\ell}^{\delta\epsilon}S$, then 
\begin{equation}
\label{so2naa}
\begin{aligned}
a_{k\ell}^{+-}(z)&= \frac1n\sum_{m\in\mathbb Z}\sum_{i,j=1}^n (\omega^{-k})^{i}(\omega^{-\ell})^{-j }t^me_{ij}^{+-}z^{-mn+i-j-1},\\
a_{k\ell}^{++}(z)& := \frac1n\sum_{m\in\mathbb Z}\sum_{i,j=1}^n (\omega^{-k})^{i}(\omega^{-\ell})^{j}t^me_{ij}^{++}z^{-(m+1)n+i+j-2},\\
a_{k\ell}^{--}(z)&:= \frac1n\sum_{m\in\mathbb Z}\sum_{i,j=1}^n (\omega^{-k})^{-i}(\omega^{-\ell})^{-j} t^me_{ij}^{--}z^{-(m-1)n-i-j}.\\
\end{aligned}
\end{equation}
Using \eqref{cij} and \eqref{tkeij0}, we can calculate 
\[
\begin{aligned}
r(a_{k\ell}^{+-})(z)&= \frac1n\sum_{\ell,m\in\mathbb Z}\sum_{i,j=1}^n (\omega^{-k})^{i}(\omega^{-\ell})^{-j } 
:\phi_{-2\ell n-i+\frac12}\phi_{2(\ell+m)+j-\frac12}:
 z^{-mn+i-j-1},\\
r(a_{k\ell}^{++})(z)& = \frac1n\sum_{\ell,m\in\mathbb Z}\sum_{i,j=1}^n (\omega^{-k})^{i}(\omega^{-\ell})^{j}
:\phi_{-2\ell n-i+\frac12}\phi_{2(\ell+m+1)-j+\frac12}:
z^{-(m+1)n+i+j-2},\\
r(a_{k\ell}^{--})(z)&= \frac1n\sum_{m\in\mathbb Z}\sum_{i,j=1}^n (\omega^{-k})^{-i}(\omega^{-\ell})^{-j}
:\phi_{-2(\ell-1) n+i-\frac12}\phi_{2(\ell+m)+j-\frac12}:
z^{-(m-1)n-i-j}.\\
\end{aligned}
\] 
Thus, relabeling the $\tilde\phi_j$ as follows:
\begin{equation}
\label{24a}
\psi^+_{\ell n-i+\frac12}=\phi_{2\ell n-i+\frac12}, \quad
\psi^-_{\ell n+i-\frac12}= \phi_{2\ell n+i-\frac12},
\end{equation}
and their generating series as in \eqref{FFprinc}, we find that 
\begin{equation}
\label{so2na}
\begin{aligned}
r(a_{k\ell}^{+-})(z)&= \frac{\omega^{-k}}n
:\psi^+(\omega^{-k}z)\psi^-(\omega^{-\ell}z):,
\\
r(a_{k\ell}^{++})(z)& = \frac{\omega^{-k-\ell}}n 
:\psi^+(\omega^{-k}z)\psi^+(\omega^{-\ell}z):,
,\\
r(a_{k\ell}^{--})(z)&= \frac1n 
:\psi^-(\omega^{-k}z)\psi^-(\omega^{-\ell}z):
.
\end{aligned}
\end{equation}
Note that the $\psi^\pm_j$ defined by \eqref{24a}  satisfy
\[
\psi^\pm_i\psi^\pm_j+\psi^\pm_j\psi^\pm_i=0,\quad
\psi^+_i\psi^-_j+\psi^-_j\psi^+=\delta_{i,-j}\quad i,j\in \frac12+\mathbb Z,
\]
and 
\[
\psi^\pm_i|0\rangle=0\quad\mbox{for }i>0.
\]
Again defining $\alpha(z)=\sum_{j\in \mathbb Z}\alpha_j z^{-j-1}=n r(a_{nn}^{+-})(z)$, this defines the Heisenberg algebra, with commutation relations 
\[
[\alpha_j,\alpha_k]=j \delta_{jk},\quad  \alpha_j|0\rangle=0
 ,\quad\mbox{for }j\ge 0.
\]
Finally, 
\[
[\alpha_j, \psi^{\pm}(z)]= \pm z^j\psi^{\pm}(z).
\]
\end{example}
\begin{example}{\bf (Two negative cycles)}
\label{tnc}
In this case $s=0$ and $r=2$, hence $\mu=(\mu_1,\mu_2)$ with $\mu_1+\mu_2=n$.
The special case that $\mu=(n-1,1)$ corresponds to the   principal realization.
Take the following lift of $w_{(\emptyset,(\mu_1,\mu_2))}$ in $O_{2n}$:
\[
\begin{aligned}
\tilde w_{(\emptyset,(\mu_1,\mu_2))}= &e_{1,2n+1-\mu_1}+e_{2n,\mu_1}+
e_{\mu_1+1, 2n+1-\mu_1-\mu_2}+e_{2n-\mu_1, \mu_1+\mu_2}+\\
&+ \sum_{i=1}^{\mu_1-1} e_{i+1,i}+e_{2n-i,2n+1-i}+
 \sum_{i=1}^{\mu_2-1} e_{\mu_1+i+1,\mu_1+i}+e_{2n-i-\mu_1,2n+1-i-\mu_1}.
\end{aligned}
\]
Let $\omega_j=e^{\frac{\pi i}{\mu_j}}$, and $\gamma_i=1$ if $\mu_i$ is even and $\gamma_i=\sqrt{-1}$ if $\mu_i$ is odd. 
Let 
\[
\begin{aligned}
S=&
\frac1{\sqrt{4\mu_1}}\sum_{k=1}^{\mu_1} [(-1)^k (\gamma_1 e_{k1}+\gamma_1^{-1} e_{2n+1-k,1}+\gamma_1^{-1}
e_{2n+1-k,2n}+\gamma_1 e_{k,2n})
+
\\
&\quad\qquad+
(e_{k,\mu_1+\mu_2}+  e_{2n+1-k,\mu_1+\mu_2}+
e_{2n+1-k,2n+1-\mu_1-\mu_2}+  e_{k,2n+1-\mu_1-\mu_2})]
+
\\
&+\frac1{\sqrt{4\mu_2}}\sum_{k=1}^{\mu_2} 
\sqrt{-1}[(-1)^k (\gamma_2 e_{\mu_1+k,1}+\gamma_2^{-1} e_{2n+1-\mu_1-k,1}+
-\gamma_2^{-1}e_{2n+1-\mu_1-k,2n}-\gamma_2 e_{\mu_1+k,2n})
+
\\
&\quad\qquad+
(e_{\mu_1+k,\mu_1+\mu_2}+  e_{2n+1-\mu_1-k,\mu_1+\mu_2}-
e_{2n+1-\mu_1-k,2n+1-\mu_1-\mu_2}- e_{\mu_1+k,2n+1-\mu_1-\mu_2})]
+
\\
&+
\frac1{\sqrt {2\mu_1}}\sum_{i=1}^{\mu_1}\sum_{j=2}^{\mu_1}
(\omega_1^{(j-\mu_1-1)i}e_{ij}+
\omega_1^{(j-\mu_1-1)(\mu_1+i)}e_{2n+1-i,j}+
\omega_1^{-(j-\mu_1-1)(i-\mu_1)}e_{i,2n+1-j}+
\\
&\qquad \qquad+ \omega_1^{-(j-\mu_1-1)i} e_{2n+1-i,2n+1-j})
+ \\
&+\frac1{\sqrt {2\mu_2}}\sum_{i=1}^{\mu_2}\sum_{j=1}^{\mu_2-1}
 (\omega_2^{(j-\mu_2)i}e_{\mu_1+i,\mu_1+j}+ \omega_2^{(j-\mu_2)(\mu_2+i)}e_{2n+1-\mu_1-i,\mu_1+j}+ 
\\
&\qquad \qquad+
\omega_2^{-(j-\mu_2)(i-\mu_2)} e_{\mu_1+i,2n+1-\mu_1-j}
+\omega_2^{-(j-\mu_2)i}e_{2n+1-\mu_1-i,2n+1-\mu_1-j}).
\end{aligned}
\]
It is straightforward to check that $S$ is indeed in $O_{2n}$ and hence,  the inverse of this matrix is equal to
\[
\begin{aligned}
S^{-1}=&
\frac1{\sqrt{4\mu_1}}\sum_{k=1}^{\mu_1} [(-1)^k (\gamma_1 e_{1,k }+  \gamma_1 e_{2n,k}+\gamma_1^{-1}e_{1,2n+1-k}+\gamma_1^{-1}e_{2n,2n+1-k}
)
+
\\
&\quad\qquad+
e_{\mu_1+\mu_2,k }+  e_{2n+1-\mu_1-\mu_2,k}+e_{ \mu_1+\mu_2, 2n+1-k}+e_{2n+1-\mu_1-\mu_2,2n+1-k})]
+
\\
&-\frac 1{\sqrt{4\mu_2}}\sum_{k=1}^{\mu_2} 
\sqrt{-1}[(-1)^k (-\gamma_2 e_{1,   \mu_1+k }
+\gamma_2 e_{2n, \mu_1+k}-\gamma_2^{-1} e_{1,2n+1-\mu_1-k}+
\gamma_2^{-1}e_{2n,2n+1-\mu_1-k})
+
\\
&\quad\qquad+
(-
e_{\mu_1+\mu_2, \mu_1+k}
+e_{2n+1-\mu_1-\mu_2,\mu_1+k}- e_{\mu_1+\mu_2, 2n+1-\mu_1-k }
+e_{2n+1-\mu_1-\mu_2,2n+1-\mu_1-k })]
+
\\
&+
\frac1{\sqrt {2\mu_1}}\sum_{i=1}^{\mu_1}\sum_{j=2}^{\mu_1}
(\omega_1^{(j-\mu_1-1)i} e_{ji }
+\omega_1^{-(j-\mu_1-1)i}e_{2n+1-j,2n+1-i}+
\omega_1^{-(j-\mu_1-1)(\mu_1+i)}e_{2n+1-j,i}
&\qquad \qquad+ 
+ \\
&\qquad \qquad+
\omega_1^{(j-\mu_1-1)(i-\mu_1)}e_{j,2n+1-i}+
\\
&+\frac1{\sqrt {2\mu_2}}\sum_{i=1}^{\mu_2}\sum_{j=1}^{\mu_2-1}
( \omega_2^{(j-\mu_2)i}e_{ \mu_1+j, \mu_1+i}
+
\omega_2^{-(j-\mu_2)i}e_{2n+1-\mu_1-j,2n+1-\mu_1-i}
\\
&\qquad \qquad
+ \omega_2^{-(j-\mu_2)(\mu_2+i)}e_{2n+1-\mu_1-j,\mu_1+i}+ 
\omega_2^{(j-\mu_2)(i-\mu_2)} e_{\mu_1+j,2n+1-\mu_1-i}
).
\end{aligned}
\]
Then
$
S^{-1}\tilde w_{(\emptyset,(\mu_1,\mu_2))}S= \exp\left( 2\pi \sqrt{-1} h_w\right)
$, where 
\[
h_w=
\sum_{i=1}^{\mu_1} \frac{\mu_1-i+1}{2\mu_1} (e_{ii}-e_{2n+1-i,2n+1-i})
+
\sum_{i=1}^{\mu_2} \frac{\mu_2-i}{2\mu_2} (e_{\mu_1+i,\mu_1+i}-e_{2n+1-\mu_1-i,2n+1-\mu_1-i})
.
\]
$S$ and $S^{-1}$ suggest  the following relabeling of the $\tilde\phi_j$, namely  ($1\le j\le \mu_1-1$, $1\le k\le \mu_2-1$):
\begin{equation}
\label{26a}
\begin{aligned}
\tilde\phi^1_{ 2\ell \mu_1-j}&=\gamma_1^{-1} \phi_{2\ell n-j-\frac12},
\qquad
\tilde\phi^1_{ 2\ell \mu_1+j}=\gamma_1  (-1)^j\phi_{2\ell n+j+\frac12},
\\
\tilde\phi^1_{2\ell\mu_1 +\mu_1}&=\frac{\gamma_1^{-1}}{\sqrt 2}(\phi_{2\ell n+\mu_1+\mu_2-\frac12}+
\phi_{2(\ell +1)-\mu_1-\mu_2+\frac12}),\\
\tilde\phi^1_{2\ell\mu_1}&=\frac{1}{\sqrt 2}
(\phi_{2\ell n+\frac12}+\phi_{2\ell n-\frac12}),\\
\tilde\phi^2_{2\ell \mu_2-k}&=\gamma_2^{-1}\phi_{ 2\ell n-\mu_1-k+\frac12},
\qquad
\tilde\phi^2_{2\ell \mu_2+k}=\gamma_2 (-1)^k\phi_{ 2\ell n+\mu_1+k-\frac12},
\\
\tilde\phi^2_{2\ell\mu_2}&=  \sqrt{\frac{-1}{  2}}(\phi_{2\ell  -\frac12}
- \phi_{2\ell n+\frac12})
 ,\\
\tilde\phi^2_{2\ell\mu_2+\mu_2}&=\frac{\gamma_2^{-1} \sqrt{-1}}{\sqrt 2}(\phi_{2(\ell+1) n-\mu_1-\mu_2+\frac12}-\phi_{2\ell n+\mu_1+\mu_2-\frac12}).
\end{aligned}
\end{equation}
We find that the new $\tilde\phi^a_j$ defined by \eqref{26a} satisfy
\[
\tilde\phi^a_j\tilde\phi^b_k+\tilde\phi_k^b\tilde\phi_j^a=(-1)^{j}\delta_{ab}\delta_{j,-k}\quad
a,b=1,2,\ j,k\in \mathbb Z,
\]
and 
\[
\tilde\phi_j^a|0\rangle=0,\qquad j>0.
\]
Defining the fields
\[ 
\tilde\phi^b(z)=\sum_{j\in\mathbb Z} \tilde\phi_j^bz ^{-j},
\]
then for $a^{\delta b,\epsilon c}_{k,\ell}:=S^{-1}e^{\delta b,\epsilon c}_{k,\ell}S$, with
$b,c=1,2$, $ \delta,\epsilon=+,-$, $1\le k\le \mu_b$ and $1\le \ell\le \mu_c $, we have 
\[
\begin{aligned}
r(a^{\delta b,\epsilon c}_{k,\ell})(z)
&=\sum_{ j\in\mathbb Z}r(a^{\delta b,\epsilon c}_{k,\ell}(j)_{j\, {\rm mod}\, N})z^{-j-1}\\
&=\frac {\gamma_b^{\delta 1}\gamma_c^{\epsilon 1}(-1)^{k+\ell}z^{-1}}{2\sqrt {\mu_b\mu_c}}:\tilde\phi^b(\delta \omega_b^{-k}z^{\frac{N}{2\mu_b}})
\tilde\phi^b(\epsilon  \omega_c^{-\ell}z^{\frac{N}{2\mu_c}}): \qquad b,c=1,2,\ \delta,\epsilon=+,-.
\end{aligned}
\]
where   $N=2\, {\rm lcm}(\mu_1,\mu_2)$, is the order of the automorphism $\exp (2\pi \sqrt{-1} {\rm ad}\, h_w)$ (see \cite{tKLB} for more details).
Let 
\[
\alpha^b(z)=\sum_{k\in 1+2\mathbb Z}
\alpha_k^b z^{-k}=
 -\lambda_b z^{\frac{2\lambda_b}N}r( a^{+b,-b}_{\lambda_b,\lambda_b})(z^{\frac{2\lambda_b}N})=:\frac12 :\tilde\phi^b(z)\tilde\phi^b(-z):.
\]
be the generating field for the Heisenberg algebra. The modes 
of the field satisfy
 \[
[\alpha_j,\alpha_k]=\frac{j}2 \delta_{jk},\quad  \alpha_j|0\rangle=0
 ,\quad\mbox{for }j\ge 0.
\]
Finally,
\[
[\alpha_j,\tilde\phi(z)]=z^j\tilde\phi(z).
\]
\end{example}

The general description follows from the above two examples.
It  is straightforward to deduce the following proposition.
\begin{proposition}
\label{prop-so-2n}
Let 
$w_{(\lambda,\mu)}$ be given by \eqref{wlambdamu}, where now $r$ is even.
Set  $\lambda_{s+a}=:\mu_a$ and 
  $\omega_a=e^{\frac {2\pi \sqrt{-1}}{\lambda_a}}$ if $1\le a\le s$ and 
$\omega_a=e^{\frac {2\pi \sqrt{-1}}{2\lambda_a}}$  is $s+1\le a\le r+s$. 
Then \\
(a) The element  
\[
\begin{aligned}
\tilde w_{(\lambda, \mu)}&= \sum_{a=1}^s  \big(
\omega_a^{\frac{\lambda_a+1}2} e_{M_a+1,M_{a+1}}+ \omega_a^{-\frac{\lambda_a+1}2} e_{2n-M_a,2n-M_{a+1}+1}
+\sum_{i=1}^{\lambda_a-1} (\omega_a^{\frac{\lambda_a+1}2} e_{M_a+i+1,M_a+i}+\\
&\qquad+\omega^{-\frac{\lambda_a+1}2} e_{2n-M_a-i,2n+1-M_a-i})\big) 
+\sum_{a=1}^{\frac{r}2} \big(
e_{M_{s+2a-1}+1,2n+1-M_{s+2a}}+ \\
&\qquad+e_{2n-M_{s+2a},M_{s+2a}}+
e_{M_{s+2a}+1, 2n+1-M_{s+2a+1}}
+e_{2n-M_{s+2a}, M_{s+2a+1}}\\
&\qquad
+ \sum_{i=1}^{\lambda_{s+2a-1}-1} e_{M_{s+2a-1}+i+1,M_{s+2a-1}+i}+
e_{2n-M_{s+2a-1}-i,2n+1-M_{s+2a-1}-i}+\\
&\qquad
+
 \sum_{i=1}^{\lambda_{s+2a}-1} e_{M_s+{2a}+i+1,M_{s+2a}+i}+e_{2n-M_{s+2a}-i,2n+1-M_{s+2a}-i}.
\big)
\in O_{2n}
\end{aligned}
\] 
is a lift of $ w_{(\lambda, \mu)}$ and  
 $\tilde w_{(\lambda, \mu)}=S^{-1}\exp ( 2\pi \sqrt{-1}
h_w)S$,  where 
\[
h_w=\sum_{a=1}^s \sum_{i=1}^{\lambda_a}  \frac{\lambda_a-2j+1}{2\lambda_a} \epsilon_{M_a+j} +\sum_{a=1}^{\frac{r}2}
\sum_{i=1}^{\mu_{2a-1}} \frac{\mu_{2a-1}-i+1}{2\mu_{2a-1} }\epsilon_{M_{s+2a-1}+i}
+
\sum_{i=1}^{\mu_{2a}} \frac{\mu_{2a}-i}{2\mu_{2a}} \epsilon_{M_{s+2a}+i}
\in\mathfrak h
\]
satisfies 
\eqref{4.2},
and
\[
\begin{aligned}
S=&\sum_{a=1}^s \sum_{i=1}^{\lambda_a}
\sum_{i,j=1}^{\lambda_a}  (\omega_a^{ij}e_{M_a+i,M_a+j}+\omega_a^{-ij}e_{2n+1-M_a-i,2n+1-M_a-j})\\
&+\sum_{a=1}^{\frac{r}2}
\frac1{\sqrt{4\mu_{2a-1}}}\sum_{k=1}^{\mu_{2a-1}} [(-1)^k (\gamma_{s+2a-1} e_{M_{s+2a-1}+k, M_{s+2a-1}+1}+\gamma_{s+2a-1}^{-1} e_{2n+1-M_{s+2a-1}-k,M_{s+2a-1}+1}+
\\
&\qquad\qquad\qquad+
\gamma_{s+2a-1}^{-1}
e_{2n+1-M_{s+2a-1}-k,2n-M_{s+2a-1}}+\gamma_{s+2a-1}e_{M_{s+2a-1}+k,2n-M_{s+2a-1}})
+
\\
&\qquad\qquad\qquad+
(e_{M_{s+2a-1}+k,M_{s+2a+1}}+  e_{2n+1-M_{s-2a-1}-k,M_{s+2a+1}}+\\
&\qquad\qquad\qquad+
e_{2n+1-M_{s+2a-1}k,2n+1-M_{s+2a+1}}+  e_{M_{s+2a-1}+k,2n+1-M_{s+2a+1}})]
+
\\
&+\frac1{\sqrt{4\mu_{2a}}}\sum_{k=1}^{\mu_{2a}} 
\sqrt{-1}[(-1)^k (-\gamma_{s+2a} e_{M_{s+2a}+k,M_{s+2a-1}+1}+\gamma_{s+2a}^{-1} e_{2n+1-M_{s+2a}-k,M_{s+2a-1}+1}\\
&\qquad\qquad\qquad
-\gamma_{s+2a}^{-1}e_{2n+1-M_{s+2a}-k,2n-M_{s+2a-1}}
+
\gamma_{s+2a} e_{M_{s+2a}+k,2n-M_{s-2a-1}})
+
\\
&\qquad\qquad\qquad+
(e_{M_{s+2a}+k,M_{s+2a+1}}+  e_{2n+1-M_{s+2a}-k,M_{s+2a+1}}
\\
&\qquad\qquad\qquad
-
e_{2n+1-M_{s+2a}-k,2n+1-M_{s+2a+1}}- e_{M_{s+2a}+k,2n+1-M_{s+2a+1}})]
+
\\
&+
\frac1{\sqrt {2\mu_{2a-1}}}\sum_{i=1}^{\mu_{2a-1}}\sum_{j=2}^{\mu_{2a-1}}
(\omega_{s+2a-1}^{(j-\mu_{2a-1}-1)i}e_{M_{s+2a-1}+i,M_{s+2a-1}+j}+
\\
&\qquad\qquad\qquad+
\omega_{s+2a-1}^{(j-\mu_{2a-1}-1)(\mu_{2a-1}+i)}e_{2n+1-M_{s+2a-1}-i,M_{s+2a-1}+j}+
\\
&\qquad\qquad\qquad+
\omega_{s+2a-1}^{-(j-\mu_{2a-1}-1)(i-\mu_{2a-1})}e_{M_{s+2a-1}+i,2n+1-M_{s+2a-1}-j}+
\\
&\qquad \qquad\qquad+ \omega_{s+2a-1}^{-(j-\mu_{2a-1}-1)i} e_{2n+1-M_{s+2a-1}+i,2n+1-M_{s+2a-1}-j})
+ \\
&+\frac1{\sqrt {2\mu_{2a}}}\sum_{i=1}^{\mu_{2a}}\sum_{j=1}^{\mu_{2a}-1}
 (\omega_{s+2a}^{(j-\mu_{2a})i}e_{M_{s+2a}+i,M_{s+2a}+j}+ \omega_{s+2a}^{(j-\mu_{2a})(\mu_{2a}+i)}e_{2n+1-M_{s+2a}-i,M_{s+2a}+j}+ 
\\
&\qquad \qquad+
\omega_{s+2a}^{-(j-\mu_{2a})(i-\mu_{2a})} e_{M_{s+2a}+i,2n+1-M_{s+2a}-j}
+\omega_{s+2a}^{-(j-\mu_{2a})i}e_{2n+1-M_{s+2a}-i,2n+1-M_{s+2a}-j})
,
\end{aligned}
\]
where $\gamma_{a}=1$ if $M_{a}$ is even and $\gamma_{a}=i$ if $M_{a}$ is odd.
\\
(b)
The order  $N$   of $\exp (2\pi \sqrt{-1} {\rm ad}\, h_w)$  is equal to the least common multiple $N'$ of $\lambda_1, \lambda_2, \ldots, \lambda_s, 2\lambda_{s+1}, 2\lambda_{s+2},\ldots 2\lambda_{r+s}$ if $N'(\frac1{\lambda_a}+\frac1{\lambda_b})\in2\mathbb Z$ for all $1\le a,b\le s$, $N'(\frac1{2\lambda_a}+\frac1{2\lambda_b})\in2\mathbb Z$ for all $s+1\le a,b\le r+s$,  $N'(\frac1{\lambda_a}+\frac1{2\lambda_b})\in2\mathbb Z$ for all $ 1\le a\le s$, $s+1\le b\le r+s$ and $2N'$ otherwise.
\\
(c) The elements 
$
a_{k,\ell}^{\delta a,\epsilon b}:=S^{-1}e_{k\ell}^{\delta a, \epsilon b} S=
$
form a new basis of $so_{2n}$ and the representation $r$ in $F_d$ of their generating fields
$
a_{ k \ell}^{\delta a,\epsilon b}(z)
$
are equal to
\[
r(a_{k,\ell}^{\delta_a ,\epsilon b})(z)= C_k^{\delta a}C_\ell^{\epsilon b}:\Psi^{\delta a}_k(z)\Psi^{\epsilon b}_\ell (z):,
\]
where 
\[
\begin{aligned}
C_k^{+a}=&
\begin{cases}
\frac{\omega_a^{-k}}{\sqrt{\lambda_a}}&1\le a\le s,\\
\frac{\gamma_a(-1)^k}{\sqrt{2\lambda_a}} &s<a\le r+s,\\
\end{cases}\qquad
C_k^{-a}=
\begin{cases}
\frac{1}{\sqrt{\lambda_a}}&1\le a\le s,\\
\frac{\gamma_a^{-1}(-1)^k}{\sqrt{2\lambda_a}} &s<a\le r+s,\\
\end{cases}
\end{aligned}
\]
and 
\[
\Psi_k^{\pm a}(z)=
\begin{cases}
\psi^{\pm a}(\omega_a^{-k} z^{\frac{N}{\lambda_a}})z^{\frac{N}{2\lambda_a}-\frac12}&1\le a\le s,\\
\tilde\phi^{ a}(\pm \omega_a^{-k} z^{\frac{N}{2\lambda_a}})z^{-\frac12}&s< a\le s+r,\\
\end{cases}
\]
where the 
\[
\psi^{\pm a}(z)=\sum_{i\in\frac12+\mathbb Z}\psi_i^{\pm a}z^{-i-\frac12},\quad
\tilde\phi^{s+b}(z)=\sum_{i\in\mathbb Z}\tilde\phi^{s+b}z^{-i},\quad
1\le a\le s,\ 1\le b\le r,
\]
are the generating fields 
for the relabeld generators of the Clifford algebra $C\ell_d$ ($1\le a\le s$,  $1\le b\le \frac{r}2$, $1\le i\le \lambda_a$, $1\le j< \lambda_{s+2b-1}$ or $\lambda_{s+2b}$):
\begin{equation}
\begin{aligned}
\psi^{+a}_{\ell \lambda_a-i+\frac12}&=\phi_{2\ell n-M_a-i+\frac12}, \qquad\qquad\qquad
\psi^{-a}_{\ell \lambda_a+i-\frac12}=\phi_{2\ell n+M_a+i-\frac12},\\
\tilde\phi^{s+2b-1}_{ 2\ell\mu_{2b-1} -j}&=\gamma_{s+2b-1}^{-1} \phi_{2\ell n-M_{s+2b-1}-j-\frac12},
\quad
\tilde\phi^{s+2b-1}_{ 2\ell\mu_{2b-1} +j}=\gamma_{s+2b-1}  (-1)^j\phi_{2\ell n+M_{s+2b-1}+j+\frac12},
\\
\tilde\phi^{s+2b-1}_{2\ell\mu_{2b-1}+\mu_{2b-1}}&=\frac{\gamma_{s+2b-1}^{-1}}{\sqrt 2}(\phi_{2\ell n+M_{s+2b+1}-\frac12}+
\phi_{2(\ell +1)n-M_{s+2b+1}+\frac12}),\\
\tilde\phi^{s+2b-1}_{2\ell \mu_{2b}}&=\frac{1}{\sqrt 2}
(\phi_{2\ell n+M_{s+2b-1}+\frac12}+\phi_{2\ell n-M_{s+2b-1}-\frac12}),\\
\tilde\phi^{s+2b}_{2\ell \mu_{2b}-j}&=\gamma_{s+2b}^{-1}\phi_{ 2\ell n-M_{s+2b}-j+\frac12},
\quad
\tilde\phi^{s+2b}_{2\ell \mu_{2b}+j}=\gamma_{s+2b}(-1)^j\phi_{ 2\ell n+M_{s+2b}+j-\frac12},
\\
\tilde\phi^{s+2b}_{2\ell\mu_{2b}}&=  \sqrt{\frac{-1}{  2}}(\phi_{2\ell n-M_{s+2b-1} -\frac12}
- \phi_{2\ell n+M_{s+2b-1}+\frac12})
 ,\\
\tilde\phi^{s+2b}_{2\ell\mu_{s+2b}+\mu_{s+2b}}&=\frac{\gamma_{s+2b}^{-1} \sqrt{-1}}{\sqrt 2}(\phi_{2(\ell+1) n-M_{s+2b+1}+\frac12}-\phi_{2\ell n+M_{s+2b+1}-\frac12}).
\end{aligned}
\end{equation}
They satisfy ($1\le a,c\le s,\ 1\le b,d\le r,\quad i,j\in \frac12+\mathbb Z$)
\begin{equation}
\label{11-1}
\begin{aligned}
\psi^{\pm a}_i\psi^{\pm c}_j+\psi^{\pm c}_j\psi^{\pm a}_i&=0,\\
\psi^{+ a}_i\psi^{-c}_j+\psi^{-c}_j\psi^{+a}&=\delta_{ac}\delta_{i,-j}\\
\tilde\phi_i^{s+b}\tilde\phi_j^{s+d}+\tilde\phi_j^{s+d}\tilde\phi_i^{s+b}&=(-1)^{i}\delta_{bd}\delta_{i,-j},\\
\psi^{\pm a}_i\tilde\phi^{s+b}_j+\tilde\phi^{s+b}_j\psi^{\pm a}_i&=0,\\
\psi^{\pm a}_i|0\rangle =\tilde\phi_i^{s+b}|0\rangle &=0\quad i>0.
\end{aligned}
\end{equation}
(d) The representation of $\hat{so}_{2n}$ is given in terms of vertex operators by formulas \eqref{VOA} and \eqref{VOD}   by substitution of roots of unity  $\omega_a$ and powers of $z$ as given 
in the above formulas.   
In particular,  we have the Heisenberg algebra \eqref{11-2}, which 
are equal to $\lambda_a z^{\frac{\lambda_a}N-1}r(a_{\lambda_a\lambda_a}^{+a,-a})(z^{\frac{\lambda_a}N})$,
respectively 
$ \lambda_{s+b} z^{\frac{2\lambda_{s+b}}N}r(a_{\lambda_{s+b}\lambda_{s+b}}^{+(s+b),-(s+b)})(z^{\frac{2\lambda_{s+b}}N})$
 and their modes satisfy \eqref{11-3}.
\end{proposition}

$F$ splits in to two irreducible  $\hat{so}_{2n}$ modules $F^{\overline 0}$ and $F^{\overline 1}$,  which differ from the formulas \eqref{QDIMD},  since the gradation is different in this case.   
Let $N$ be the least common multiple of $\lambda_1,\ldots,\lambda_s$,$2\mu_1,\ldots, 2\mu_r$,  then the $q$-dimension formulas in this case combines $s$ copies of the Jacobi triple product identity,  which is the equality between  
  \eqref{JTI1}, and \eqref{JTI2}, where we replace in each copy $q$ by $q^{\frac{N}{\lambda_a}}$,  and $r$ copies of \eqref{qb},
where we replace in each copy $q$ by $q^{\frac{N}{\mu_b}}$,   and this multiplied by  some factor (see \cite{tKLD}):
\begin{equation}
\label{QDIMDred}
\dim_q F^{\overline \epsilon} =
\begin{cases}
2^{[\frac{r-1}2]}
\sum_{\genfrac{}{} {0pt}{2}{j_1, \ldots, j_s\in\mathbb Z}{j_1+\cdots +j_s\in \mathbb Z}}
q^{\frac{N}2(\frac{j_1^2}{\lambda_1^2}+\frac{j_2^2}{\lambda_2^2}+\cdots +\frac{j_s^2}{\lambda_s^2})}
\prod_{a=1}^s \prod_{i=1}^\infty \frac1{1-q^{\frac{jN}{\lambda_a}}}
\prod_{b=1}^r \prod_{j=1}^\infty \frac1{1-q^{\frac{(2j-1)N}{\mu_b}}}\\
\hfill \mbox{for }r\ne 0,\\
  \sum_{\genfrac{}{} {0pt}{2}{j_1, \ldots j_s\in\mathbb Z}{j_1+\cdots +j_s\in \epsilon+\mathbb Z}}
q^{\frac{N}2(\frac{j_1^2}{\lambda_1^2}+\frac{j_2^2}{\lambda_2^2}+\cdots +\frac{j_s^2}{\lambda_s^2})}
\prod_{a=1}^s \prod_{i=1}^\infty \frac1{1-q^{\frac{jN}{\lambda_a}}},\qquad  \mbox{when }r=0.
\end{cases}
\end{equation}
\subsection{$so_{2n+1}$}
\label{S5.5}
Recall that the conjugacy classes of the Weyl group for $so_{2n+1}$ are in one-to-one correspondence with the pair of partitions 
\begin{equation}
\label{lambda,mi3}
\lambda=(\lambda_1\ge \lambda_2\ge \cdots\ge \lambda_s>0),
\mu=(\mu_1\ge \mu_2\ge \cdots\ge \mu_{r}>0),
\end{equation}
 of $n$, i.  e.  \eqref{lambdamul2} holds.
We can decompose each element $w\in W$ in $s$   disjoint positive cycles of length $\lambda_i$,  and $r$ disjoint negative cycles of length $\mu_j$ such that \eqref{lambdamul} holds. It is straightforward to show that two elements with the same cycle type are conjugate under $O_{2n}$. 

We relabel the basis as follows.
Let $m_a=\lambda_a$, $M_a=|\lambda|_{a-1}$ for $1\le a\le s$,  $m_a=\mu_{a-s}$, $M_a=|\lambda|+|\mu|_{a-s-1}$ for $s<a\le r+s$,  and $M_{r+s+1}= |\lambda|+|\mu|$,then set ($1\le a,b\le s+r$, $ 1\le i\le m_a$, $ 1\le j\le m_b$)
\begin{equation}
\label{redefbasisB}
\begin{aligned}
e_{ij}^{+a,-b}=&
  e_{M_a+i,M_b+j}-  e_{2n+2-M_b-j, 2n+2-M_a-i},\\
 e_{ij}^{+a,+b}=&- e_{M_a+i,2n+2-M_b-j}-  e_{ M_b+j, 2n+2-M_a-i},\\
 e_{ij}^{-a,-b}=&-e_{ji}^{-b,-a}= e_{2n+2-M_a-i, M_b+j}- e_{ 2n+2-M_b-j,  M_a+i},\\
 e_{i}^{+a}=&e_{M_a+i,n+1}-e_{n+1,2n+2-Ma-i},\\
 e_{i}^{-a}=&e_{2n+2-M_a-i,n+1}-e_{n+1,M_a+i}.
\end{aligned}
\end{equation}
 
 We will embed this affine algebra into $d_\infty$ and into $b_\infty$. In the first case this will give the representation of the affine Lie algebra of $so_{2n+1}$ with highest weights 
 $\Lambda_0$ and $\Lambda_1$, here we assume that $h_w$ satisfies \eqref{4.2}. For the embedding in $b_\infty$, we obtain the highest weight representation with highest weight $\Lambda_n$. In this case we assume that $h_w$ satisfies \eqref{4.2b}. To distinguish these two cases  we set $x$ either to $d$ or to $b$.

Again we will first investigate two examples. 
\begin{example}{\bf (One positive cycle)}
In this case $s=1$ and $r=0$ and we have the same element as in 
\eqref{onepossp}. However now we have to lift this to $O_{2n+1}$.
Let $\omega= e^{\frac{2\pi \sqrt{-1}}{n}}$ choose 
\[
  \tilde w_{((n),\emptyset )}= e_{n+1,n+1}+(-1)^{\delta_{xb}}(\omega^{\frac{n+1}2} e_{1,n}+ \omega^{-\frac{n+1}2} e_{2n+1,n+2}
+\sum_{i=1}^{n-1} (\omega^{\frac{n+1}2} e_{i+1,i}+\omega^{-\frac{n+1}2} e_{2n+1-i,2n+2-i})).
\]
Clearly, we are in the same situation as in the case of one positive cycle of $so_{2n}$. 
And the same construction holds. In this case,
\[
S=e_{n+1,n+1}+\sum_{i,j=1}^n  (\omega^{ij}e_{ij}+\omega^{-ij}e_{2n+2-i,2n+2-j}),
\]
and 
$\tilde w_{((n),\emptyset )}=S e^{2\pi \sqrt{-1}   h_w}S^{-1}$ with 
\[
h_w=\sum_{j=1}^n \frac{(1+\delta_{xb})n-2j+1+\delta_{xb}n}{2n}(e_{jj}-e_{2n+2-j,2n+2-j}).
\]
{\bf Embedding in $d_\infty$:} This situation is similar to the case of one positive cycle in the $so_{2n}$-case.
We now relabel the elements $\tilde\phi_j$ as follows
\[
\psi^+_{\ell n-i+\frac12}=\phi_{\ell(2n+1)-i+\frac12}, \quad
\psi^-_{\ell n+i-\frac12}= \phi_{\ell(2n+1)+i-\frac12},\quad \sigma_{\ell+\frac12}=
\phi_{\ell(2n+1)+n+\frac12},
\]
and define their fields as in \eqref{FFprinc} and as
\[ 
\sigma(z)=\sum_{j\in\frac12+\mathbb Z} \sigma_j z^{-j-\frac12}.
\]
Here ($i,j\in\frac12+\mathbb Z$)
\begin{equation}
\label{sigma}
\sigma_i\sigma_j+\sigma_j\sigma_i=\delta_{i,-j}.
\end{equation}
Let $a_{k\ell}^{\delta\epsilon}=S^{-1}e_{k\ell}^{\delta\epsilon}S$, 
and $a_k^\delta=S^{-1}e_{k}^{\delta}S$.
Then we  have \eqref{so2na}, and 
\begin{equation}
\label{SOO}
a_{k}^{\pm}(z)=\frac1{\sqrt n}\sum_{m\in \mathbb Z}\sum_{i=1}^n
\omega^{\mp ki} r(t^m e_i^\pm)z^{-mn\mp\frac12 (n-2i+1)-1}
\end{equation}
and 
\[
r(a_{k}^{+})(z) 
= \frac{\omega^kz^{-\frac12}}{\sqrt n}
:\psi^+(\omega^{-k}z)\sigma(z^n):,\quad
r(a_{k }^{-})(z) = \frac{z^{-\frac12}} {\sqrt n}
:\psi^-(\omega^{-k}z)\sigma(z^n):
.
\]
Note that we get a half integer power of $z$, this is because the  order of the automorphism is $2n$ instead of $n$.\\
{\bf Embedding in $b_\infty$:}
Again we have \eqref{so2naa} and \eqref{SOO}, but now the embedding \eqref{tkeij000}, which suggest the relabeling
 \[ 
\psi^\pm_{\ell n\pm (n+1-i-\frac12)}=\tilde\phi_{\ell(2n+1)\pm(n+1- i)},
\quad
 \sigma_{\ell }=
\tilde\phi_{\ell(2n+1) }.
\]
Define their generating fields as in  \eqref{FFprinc} and
\[ 
\sigma(z)=\sum_{i\in\mathbb Z}\sigma_i z^{-i}.
\]
Since $h_w$ is slightly different from the usual case, we have 
\begin{equation}
\label{so2naab}
\begin{aligned}
a_{k\ell}^{+-}(z)&= \frac1n\sum_{m\in\mathbb Z}\sum_{i,j=1}^n (\omega^{-k})^{i}(\omega^{-\ell})^{-j }t^me_{ij}^{+-}z^{-mn+i-j-1},\\
a_{k\ell}^{++}(z)& := \frac1n\sum_{m\in\mathbb Z}\sum_{i,j=1}^n (\omega^{-k})^{i}(\omega^{-\ell})^{j}t^me_{ij}^{++}z^{-(m+2)n+i+j-2},\\
a_{k\ell}^{--}(z)&:= \frac1n\sum_{m\in\mathbb Z}\sum_{i,j=1}^n (\omega^{-k})^{-i}(\omega^{-\ell})^{-j} t^me_{ij}^{--}z^{-(m-2)n-i-j},\\
a_{k}^{\pm}(z)&=\frac1{\sqrt n}\sum_{m\in \mathbb Z}\sum_{i=1}^n
\omega^{\mp ki} r(t^m e_i^\pm)z^{-mn\mp\frac12 (2n-2i+1)-1}
\end{aligned}
\end{equation}
It is then straightforward to check that
\begin{equation}
\label{so2nab}
\begin{aligned}
r(a_{k\ell}^{+-})(z)&= \frac{(-\omega)^k(-1)^\ell}n
:\psi^+(\omega^{-k}z)\psi^-(\omega^{-\ell}z):,
\\
r(a_{k\ell}^{++})(z)& = \frac{(-\omega)^{k+\ell}}n 
:\psi^+(\omega^{-k}z)\psi^+(\omega^{-\ell}z):
,\\
r(a_{k\ell}^{--})(z)&= \frac{(-1)^{k+\ell}}n 
:\psi^-(\omega^{-k}z)\psi^-(\omega^{-\ell}z):,\\
r(a_{k}^{+})(z) 
&= \frac{(-\omega)^kz^{-\frac12}}{\sqrt n}
:\psi^+(\omega^{-k}z)\sigma(z^n):,\\
r(a_{k }^{-})(z) &= \frac{(-1)^k z^{-\frac12}} {\sqrt n}
:\psi^-(\omega^{-k}z)\sigma(z^n):
.
\end{aligned}
\end{equation}
\end{example}
\begin{example}{\bf (One negative cycle, the principal Heisenberg algebra)}
In this case $r=1$ and $s=0$, hence 
\[
w_{(\emptyset ,(n))}: \epsilon_{n} \mapsto -\epsilon_{ 1},\quad
\epsilon_{ k}\mapsto \epsilon_{ k+1},\quad
k=1,2,\ldots ,  n. 
\]
This case is analogous to the principal case of $sp_{2n}$.
We choose 
\[
 \tilde w=\tilde w_{(\emptyset,(n) )}= -e_{n+1,n+1}+ e_{2n+1,n}+ e_{1,n+2}
+\sum_{i=1}^{n-1} (e_{i+1,i}+e_{2n+1-i,2n+2-i})\in SO_{2n+1}.
\]
Let $\omega=e^{\frac{2\pi \sqrt{-1}}{2n}}$ and  $\gamma=1$ if $n$ is even and 
$\gamma=\sqrt{-1}$ if $n$ is odd. Let
\[
\begin{aligned}
S&=\sum_{i=1}^n\big(
\frac{(-1)^i}{\sqrt{4n}}(e_{i1}+(-1)^ne_{i,2n+1}+(-1)^ne_{2n+2-i,1}+e_{2n+2-i,2n+1})
+\\ 
 &\qquad
 +
  \frac1{\sqrt{2n}}(e_{i,n+1}+ e_{ 2n+1-i, n+1})+\frac1{\sqrt {2n}}\sum_{j=2}^n (
\omega^{(j-n-1)i}e_{ij}+\omega^{(n-j+1)(n+i)}e_{i,2n+2-j}+
\\
&\qquad +
\omega^{(j-n-1)(n+i)}e_{2n+2-i,j}+
\omega^{(n-j+1)i}e_{2n+2-i,2n+2-j})
\big)+ 
\sqrt{\frac{-1}{2}}(\gamma^{-1}e_{n+1,1}- \gamma e_{n+1,2n+1}),
\end{aligned}
\]
such that $   \tilde w_{(\emptyset,(n) )}= S\exp(2\pi \sqrt{-1}  h_w)S^{-1}$, for
\[
h_w=\sum_{i=1}^n \frac{n-i+1}{2n}(e_{ii}-e_{2n+2-i,2n+2-i}).
\]
It is straightforward to check that $(Sv,Sw)_{so_{2n+1}}=(v,w)_{so_{2n+1}}$, thus $S\in O_{2n}$ and 

\[
\begin{aligned}
S^{-1}&=
\sum_{i=1}^n\big(
\frac{(-1)^i}{\sqrt{4n}}(e_{1i}+(-1)^ne_{2n+1,i}+(-1)^ne_{1,2n+2-i}+e_{2n+1,2n+2-i} )+\\
&\qquad + \frac1{\sqrt{2n}}(e_{n+1,i}+ e_{n+1,2n+1-i})+\frac1{\sqrt {2n}}\sum_{j=2}^n (\omega^{(n-j+1)i}e_{ji}+
\omega^{(j-n-1)(i+n)}e_{2n+2-j,i}+
\\
&\qquad +
\omega^{(n-j+1)(n+i)}e_{j,2n+2-i}
+
\omega^{(j-n-1)i}e_{2n+2-j,2n+2-i}
\big)
+ 
\sqrt{\frac{-1}{2}}(\gamma^{-1}e_{2n+1,n+1}-\gamma e_{1,n+1}).
\end{aligned}
\]
$S$ and $S^{-1}$ suggest the following relabelling of the $\tilde\phi_i$.\\
{\bf Embedding in $d_\infty$:} ($\ell\in\mathbb Z$)
\begin{equation}
\label{28d}
\begin{aligned}
\sigma_\ell&=
\sqrt{\frac{-1}{2}}(\gamma^{-1} \tilde\phi_{(2n+1)\ell - \frac12}-\gamma\tilde\phi_{(2n+1)\ell +\frac12})
\\
\tilde\phi^1_{2n\ell }&=\frac1{\sqrt 2}(\gamma^{-1}\phi_{(2n+1)\ell -\frac12}+\gamma\phi_{(2n+1)\ell +\frac12}),\quad 
\tilde\phi_{2n \ell +n}^1= \gamma^{-1}\phi_{(2n+1)\ell +n+\frac12}
\\
\tilde\phi^1_{2n\ell-k}&=\gamma^{-1}\phi_{(2n+1)\ell-k-\frac12},\quad \phi^1_{2n\ell+k}= \gamma(-1)^{k}\phi_{(2n+1)\ell+k+\frac12}\quad k=1,\ldots, n-1.
\end{aligned}
\end{equation}
Define the corresponding generating fields as
\[
\sigma(z)=\sum_{i\in\mathbb Z} \sigma_iz^{-i},\quad \tilde\phi^1(z) =\sum_{i\in\mathbb Z} \tilde\phi^1_iz^{-i}.
\]
For $\delta,\epsilon=+,-$ and $1\le k,\ell\le n$,
\[
\begin{aligned}
r(a^{\delta,\epsilon}_{k,\ell})(z)&= \frac{(-1)^{k+\ell }\gamma^{\delta 1+\epsilon 1}z^{-1} }{2n}:\tilde\phi(\delta \omega^{-k} z)\tilde\phi(-\epsilon\omega^{-\ell} z):,
\\
r(a^{\delta }_{k })(z)&=\frac{(-1)^{k}\gamma^{\delta 1} z^{-1} }{\sqrt{2n}}:\tilde\phi(\delta \omega^{-k} z)\sigma( z^{2n}):
.
\end{aligned}
\]
{\bf Embedding in $b_\infty$:}
($\ell\in\mathbb Z$)
\begin{equation}
\label{28b}
\begin{aligned}
\sigma_{\ell+\frac12}&=\sqrt{\frac{-1}{  2}}(\gamma^{-1}\tilde\phi_{(2n+1)\ell +n+1}-\gamma \tilde\phi_{(2n+1)\ell +n}),\\
\tilde\phi^1_{2n\ell +n}&= \frac1{\sqrt 2}(\tilde\phi_{(2n+1)\ell +n}+(-1)^n \tilde\phi_{(2n+1)\ell +n+1}),\quad 
\tilde\phi_{2n \ell  }^1=\tilde\phi_{(2n+1)\ell  }
\\
\tilde\phi^1_{2n\ell+ k}&= \tilde\phi_{(2n+1)\ell+k},\qquad 
\tilde\phi^1_{2n\ell- k}= (-1)^k\tilde\phi_{(2n+1)\ell-k}
\quad k=1,\ldots, n-1,
\end{aligned}
\end{equation}
 and define the corresponding generating fields as
\[
\sigma(z)=\sum_{i\in\frac12 +\mathbb Z} \sigma_iz^{-i-\frac12},\quad \tilde\phi^1(z) =\sum_{i\in\mathbb Z} \tilde\phi^1_iz^{-i},
\]
then  for $\delta,\epsilon=+,-$ and $1\le k,\ell\le n$
\[
\begin{aligned}
r(a^{\delta,\epsilon}_{k,\ell})(z)&=\frac{ z^{-1} }{2n}:\tilde\phi(\delta \omega^{-k} z)\tilde\phi(\epsilon\omega^{-\ell} z):,
\\
r(a^{\delta }_{k })(z)&=\frac{ z^{-1} }{\sqrt{2n}}:\tilde\phi(\delta \omega^{-k} z)\sigma( z^{2n}):
.
\end{aligned}
\]
In both cases the $\tilde\phi^1_j$ and $\sigma_j$, defined by \eqref{28d} and \eqref{28b}
satisfy
\[
\tilde\phi_i^1\tilde\phi_j^1+\tilde\phi_j^1\tilde\phi_i^1=(-1)^i\delta_{i,-j},\quad \sigma_i\sigma_j+\sigma_j\sigma_i=\delta_{i,-j},\quad \tilde\phi_i^1\sigma_j+\sigma_j\tilde\phi_i^1=0,
\]
and
\[
\sigma_i|0\rangle=\tilde\phi^1_i|0\rangle=0,\qquad i>0.
\]
\end{example}
The general case follows from these two examples and Example \ref{tnc}.  
 There are in general both neutral and charged fermionic fields. However, we also have an additional field $\sigma(z)$. 
Hence we have to extend  the Fock space $B_{(\lambda,\mu)}$ with extra variables $\xi$,  viz. let for $x=b$ or $d$:
\begin{equation}
\label{FockB}
\sigma(F_x)=B_{(\lambda,\mu)}(\xi)=\mathbb C[\theta_i,\xi_k, q_j,q_j^{-1} , t^\ell ;1\le j\le s ,\ 1\le i\le p, \ 1\le \ell\le s+r, k\in \delta+\mathbb Z_{> 0}],
\end{equation}
where the $\xi_i$ are again Grassmann variables that anti-commute with the $\theta_j$'s.\\
{\bf We have 4 cases}
\begin{enumerate}
\item $d_\infty$,  $r$ is even , then $p=\frac{r}2$ and $\delta=-\frac12$;
\item $d_\infty$, $r$ odd,  then $p=\frac{r+1}2$  and $\delta=0$;
\item $b_\infty$, $r$ odd,  then $p=\frac{r-1}2$ and $\delta=-\frac12$;
\item $b_\infty$, $r$ even, then $p=\frac{r}2$ and $\delta=0$.
\end{enumerate}
It  is straightforward to deduce the following proposition.
\begin{proposition}
\label{prop-so-2n+1}
Let 
$w_{(\lambda,\mu)}$ be given by \eqref{wlambdamu}.
Set  $\lambda_{s+a}=:\mu_a$ and 
  $\omega_a=e^{\frac {2\pi \sqrt{-1}}{\lambda_a}}$ if $1\le a\le s$ and 
$\omega_a=e^{\frac {2\pi \sqrt{-1}}{2\lambda_a}}$  is $s+1\le a\le r+s$. 
Let $[a]$ be the largest integer greater or equal to $a$ and let $\eta=1$ if $r$ is odd and $\eta=0$ if $r$ is even.     Then \\
(a) The element  
\[
\begin{aligned}
\tilde w_{(\lambda, \mu)}&= \sum_{a=1}^s 
(1-2\eta)e_{n+1,n+1}+(-1)^{\delta_{xb}}
 \big(
\omega_a^{\frac{\lambda_a+1}2} e_{M_a+1,M_{a+1}}+ \omega_a^{-\frac{\lambda_a+1}2} e_{2n+1-M_a,2n+1-M_{a+1}+1}+\\
&\qquad
+\sum_{i=1}^{\lambda_a-1} (\omega_a^{\frac{\lambda_a+1}2} e_{M_a+i+1,M_a+i}+
\omega^{-\frac{\lambda_a+1}2} e_{2n+1-M_a-i,2n+2-M_a-i})\big) +\\
&\qquad +\sum_{a=1}^{\left[\frac{r}2 \right]}\big(
e_{M_{s+2a-1}+1,2n+2-M_{s+2a}}+e_{2n+1-M_{s+2a},M_{s+2a}}+
e_{M_{s+2a}+1, 2n+2-M_{s+2a+1}}+\\
&\qquad
+e_{2n+1-M_{s+2a}, M_{s+2a+1}}
+ \sum_{i=1}^{\lambda_{s+2a-1}-1} e_{M_{s+2a-1}+i+1,M_{s+2a-1}+i}+
e_{2n+1-M_{s+2a-1}-i,2n+2-M_{s+2a-1}-i}+\\
&\qquad
+
 \sum_{i=1}^{\lambda_{s+2a}-1} e_{M_s+{2a}+i+1,M_{s+2a}+i}+e_{2n+1-M_{s+2a}-i,2n+2-M_{s+2a}-i}.
\big)\\
&\qquad
+
\eta\big(
e_{2n+1-M_{s+r},n}+ e_{M_{s+r}+1,n+2}
+\sum_{i=1}^{\mu_r-1} (e_{M_{s+r}+i+1,M_{s+r}+i}+e_{2n+1-M_{s+r}-i,2n+2-M_{s+r}-i})
\big)\\
&\qquad
\in O_{2n+1}
\end{aligned}
\]
is a lift of $ w_{(\lambda, \mu)}$ and  
 $\tilde w_{(\lambda, \mu)}=S^{-1}\exp ( 2\pi \sqrt{-1}
h_w)S$,  where 
\[
\begin{aligned}
h_w&=\sum_{a=1}^s \sum_{i=1}^{\lambda_a}  \frac{(1+\delta_{x,b})\lambda_a-2j+1}{2\lambda_a} \epsilon_{M_a+j} +\sum_{a=1}^{\left[ \frac{r}2\right]}
\sum_{i=1}^{\mu_{2a-1}} \frac{\mu_{2a-1}-i+1}{2\mu_{2a-1} }\epsilon_{M_{s+2a-1}+i}
+\\
&\qquad\quad+\sum_{i=1}^{\mu_{2a}} \frac{\mu_{2a}-i}{2\mu_{2a}} \epsilon_{M_{s+2a}+i}
+\eta \sum_{i=1}^{\mu_{r}} \frac{\mu_{r}-i+1}{2\mu_{r}} \epsilon_{M_{s+r}+i}
\in\mathfrak h
\end{aligned}
\]
satisfies 
\eqref{4.2},  if $x=d$ and \eqref{4.2b} if $x=b$
and
\[
\begin{aligned}
S=&(1-\eta)e_{n+1,n+1}+\sum_{a=1}^s \sum_{i=1}^{\lambda_a}
\sum_{i,j=1}^{\lambda_a}  (\omega_a^{ij}e_{M_a+i,M_a+j}+\omega_a^{-ij}e_{2n+2-M_a-i,2n+2-M_a-j})\\
&+\sum_{a=1}^{\left[ \frac{r}2\right]}
\frac1{\sqrt{4\mu_{2a-1}}}\sum_{k=1}^{\mu_{2a-1}} [(-1)^k (\gamma_{s+2a-1} e_{M_{s+2a-1}+k, M_{s+2a-1}+1}+\gamma_{s+2a-1}^{-1} e_{2n+2-M_{s+2a-1}-k,M_{s+2a-1}+1}+
\\
&\qquad\qquad\qquad+
\gamma_{s+2a-1}^{-1}
e_{2n+2-M_{s+2a-1}-k,2n+1-M_{s+2a-1}}+\gamma_{s+2a-1}e_{M_{s+2a-1}+k,2n+1-M_{s+2a-1}})
+
\\
&\qquad\qquad\qquad+
(e_{M_{s+2a-1}+k,M_{s+2a+1}}+  e_{2n+2-M_{s-2a-1}-k,M_{s+2a+1}}+\\
&\qquad\qquad\qquad+
e_{2n+2-M_{s+2a-1}k,2n+2-M_{s+2a+1}}+  e_{M_{s+2a-1}+k,2n+2-M_{s+2a+1}})]
+
\\
&+\frac1{\sqrt{4\mu_{2a}}}\sum_{k=1}^{\mu_{2a}} 
\sqrt{-1}[(-1)^k (-\gamma_{s+2a} e_{M_{s+2a}+k,M_{s+2a-1}+1}+\gamma_{s+2a}^{-1} e_{2n+2-M_{s+2a}-k,M_{s+2a-1}+1}\\
&\qquad\qquad\qquad
-\gamma_{s+2a}^{-1}e_{2n+2-M_{s+2a}-k,2n+1-M_{s+2a-1}}
+
\gamma_{s+2a} e_{M_{s+2a}+k,2n+1-M_{s-2a-1}})
+
\\
&\qquad\qquad\qquad+
(e_{M_{s+2a}+k,M_{s+2a+1}}+  e_{2n+2-M_{s+2a}-k,M_{s+2a+1}}
\\
&\qquad\qquad\qquad
-
e_{2n+2-M_{s+2a}-k,2n+2-M_{s+2a+1}}- e_{M_{s+2a}+k,2n+2-M_{s+2a+1}})]
+
\\
&+
\frac1{\sqrt {2\mu_{2a-1}}}\sum_{i=1}^{\mu_{2a-1}}\sum_{j=2}^{\mu_{2a-1}}
(\omega_{s+2a-1}^{(j-\mu_{2a-1}-1)i}e_{M_{s+2a-1}+i,M_{s+2a-1}+j}+
\\
&\qquad\qquad\qquad+
\omega_{s+2a-1}^{(j-\mu_{2a-1}-1)(\mu_{2a-1}+i)}e_{2n+2-M_{s+2a-1}-i,M_{s+2a-1}+j}+
\\
&\qquad\qquad\qquad+
\omega_{s+2a-1}^{-(j-\mu_{2a-1}-1)(i-\mu_{2a-1})}e_{M_{s+2a-1}+i,2n+2-M_{s+2a-1}-j}+
\\
&\qquad \qquad\qquad+ \omega_{s+2a-1}^{-(j-\mu_{2a-1}-1)i} e_{2n+2-M_{s+2a-1}+i,2n+2-M_{s+2a-1}-j})
+ \\
&+\frac1{\sqrt {2\mu_{2a}}}\sum_{i=1}^{\mu_{2a}}\sum_{j=1}^{\mu_{2a}-1}
 (\omega_{s+2a}^{(j-\mu_{2a})i}e_{M_{s+2a}+i,M_{s+2a}+j}+ \omega_{s+2a}^{(j-\mu_{2a})(\mu_{2a}+i)}e_{2n+2-M_{s+2a}-i,M_{s+2a}+j}+ 
\\
&\qquad \qquad+
\omega_{s+2a}^{-(j-\mu_{2a})(i-\mu_{2a})} e_{M_{s+2a}+i,2n+2-M_{s+2a}-j}
+\omega_{s+2a}^{-(j-\mu_{2a})i}e_{2n+2-M_{s+2a}-i,2n+2-M_{s+2a}-j})+\\
&\qquad+\eta\big\{
\sum_{i=1}^{\mu_r}\big(
\frac{(-1)^i}{\sqrt{4\mu_r}}(e_{M_{s+r}+i1}+(-1)^{\mu_r}e_{M_{s+r}+i,2n+1-M_{s+r}}+(-1)^{\mu_r}e_{2n+2-M_{s+r}-i,M_{s+r}+1}+\\
&\qquad\qquad+e_{2n+2-M_{s+r}-i,2n+1-M_{s+r}})
 +
  \frac1{\sqrt{2\mu_r}}(e_{i,n+1}+ e_{ 2n+1-M_{s+r}-i, n+1})+\\
&\qquad\qquad+\frac1{\sqrt {2\mu_r}}\sum_{j=2}^{\mu_r} (
\omega^{(j-\mu_r-1)i}e_{M_{s+r}+i,M_{s+r}+j}+\omega^{(\mu_r-j+1)(\mu_r+i)}e_{M_{s+r}+i,2n+2-M_{s+r}-j}+
\\
&\qquad\qquad +
\omega^{(j-\mu_r-1)(\mu_r+i)}e_{2n+2-M_{s+r}-i,M_{s+r}+j}+
\omega^{(\mu_r-j+1)i}e_{2n+2-M_{s+r}-i,2n+2-M_{s+r}-j})
\big)+ \\
&\qquad\qquad
\sqrt{\frac{-1}{2}}(\gamma_{s+r}^{-1}e_{n+1,M_{s+r}+1}- \gamma_{s+r} e_{n+1,2n+1-M_{s+r}})\big\}
,
\end{aligned}
\]
where $\gamma_{a}=1$ if $M_{a}$ is even and $\gamma_{a}=i$ if $M_{a}$ is odd.
\\
(b) 
 The order  $N$   of $\exp (2\pi \sqrt{-1} {\rm ad}\, h_w)$  is equal to the least common multiple $N'$ of $\lambda_1, \lambda_2, \ldots, \lambda_s, 2\lambda_{s+1}, 2\lambda_{s+2},\ldots 2\lambda_{r+s}$ if $s=0$ or  $N'\frac{2\lambda_a (1-\frac{\delta_{x,b}}2)+1}{\lambda_a}\in 2\mathbb Z$ for all $1\le a\le s$,and $2N'$ otherwise.
\\
(c) The elements 
$
a_{k,\ell}^{\delta a,\epsilon b}:=S^{-1}e_{k\ell}^{\delta a, \epsilon b} S 
$
and
$
a_{k}^{\delta a}:=S^{-1}e_{k\ell}^{\delta a, \epsilon b} S 
$
form a new basis of $so_{2n+1}$ and the representation $r$ in $F_d$ and $F_b$ of their generating fields
$
a_{ k \ell}^{\delta a,\epsilon b}(z)
$ and 
$
a_{ k}^{\delta a}(z)
$ 
are equal to
\[
r(a_{k,\ell}^{\delta_a ,\epsilon b})(z)= C_k^{\delta a}C_\ell^{\epsilon b}:\Psi^{\delta a}_k(z)\Psi^{\epsilon b}_\ell (z):,\qquad r(a_{k}^{\delta_a })(z)= C_k^{\delta a}z^{-\frac12}:\Psi^{\delta a}_k(z)\sigma(z^N):,
\]
where $ C_k^{\delta a}\in\mathbb C$ and  the fields $\Psi^{\delta a}_k(z)$ and $\sigma(z)$ are described below.
\\
(1) For the embedding in $d_\infty$,
\[
\begin{aligned}
C_k^{+a}=&
\begin{cases}
\frac{\omega_a^{-k}}{\sqrt{\lambda_a}}&1\le a\le s,\\
\frac{\gamma_a(-1)^k}{\sqrt{2\lambda_a}} &s<a\le r+s,\\
\end{cases}\qquad
C_k^{-a}=
\begin{cases}
\frac{1}{\sqrt{\lambda_a}}&1\le a\le s,\\
\frac{\gamma_a^{-1}(-1)^k}{\sqrt{2\lambda_a}} &s<a\le r+s,\\
\end{cases}
\end{aligned}
\]
and 
\[
\Psi_k^{\pm a}(z)=
\begin{cases}
\psi^{\pm a}(\omega_a^{-k} z^{\frac{N}{\lambda_a}})z^{\frac{N}{2\lambda_a}-\frac12}&1\le a\le s,\\
\tilde\phi^{ a}(\pm \omega_a^{-k} z^{\frac{N}{2\lambda_a}})z^{-\frac12}&s< a\le s+r,\\
\end{cases}
\]
where 
\begin{equation}
\label{14field}
\psi^{\pm a}(z)=\sum_{i\in\frac12+\mathbb Z}\psi_i^{\pm a}z^{-i-\frac12},\quad
\tilde\phi^{s+b}(z)=\sum_{i\in\mathbb Z}\tilde\phi^{s+b}_iz^{-i},\quad
1\le a\le s,\ 1\le b\le r,
\end{equation}
and 
\[
\sigma(z)=\begin{cases}
\sum_{i\in \mathbb Z} \sigma_i z^{-i}&r \ \mbox{odd},\\
\sum_{i\in\frac12+ \mathbb Z} \sigma_i z^{-i-\frac12}&r \ \mbox{even},
\end{cases}
\]
are the following generating fields 
for the relabeld generators of the Clifford algebra $C\ell_d$ ($1\le a\le s$,  $1\le b\le \left[\frac{r+1}2\right]$,  with $2b\ne r+1$ and $1\le i\le \lambda_a$, $1\le j< \lambda_{s+2b-1}$ or $\lambda_{s+2b}$):
\begin{equation}
\begin{aligned}
\sigma_{\ell+\frac{1-\eta}2}&=
\begin{cases}
\phi_{(\ell+\frac12)(2n+1)}
&r\ \mbox{even},\\
\sqrt{\frac{-1}{2}}(\gamma_r^{-1} \phi_{(2n+1)\ell - \frac12}-\gamma_r\phi_{(2n+1)\ell +\frac12})
&r\ \mbox{odd},
\end{cases}
\\
\psi^{+a}_{\ell \lambda_a-i+\frac12}&=\phi_{\ell(2n+1)-M_a-i+\frac12}, \qquad\qquad\qquad
\psi^{-a}_{\ell \lambda_a+i-\frac12}=\phi_{\ell(2n+1) +M_a+i-\frac12},
\\
\tilde\phi^{s+2b-1}_{ 2\ell\mu_{2b-1} -j}&=\gamma_{s+2b-1}^{-1} \phi_{\ell(2n+1)-M_{s+2b-1}-j-\frac12},
\quad
\tilde\phi^{s+2b-1}_{ 2\ell \mu{2b-1}+j}=\gamma_{s+2b-1}  (-1)^j\phi_{\ell(2n+1)+M_{s+2b-1}+j+\frac12},
\\
\tilde\phi^{s+2b-1}_{2\ell\mu_{2b-1}+\mu_{2b-1}}&=
\begin{cases}
\frac{\gamma_{s+2b-1}^{-1}}{\sqrt 2}(\phi_{\ell(2n+1)+M_{s+2b+1}-\frac12}+
\phi_{(\ell+1)(2n+1)-M_{s+2b+1}+\frac12})
&2b-1\ne r,
\\
 \gamma_r^{-1}\phi_{(2n+1)\ell  +n +\frac12}
&2b-1=r,
\end{cases}
\\
\tilde\phi^{s+2b-1}_{2\ell \mu_{2b}}&=
\begin{cases}
\frac{1}{\sqrt 2}
(\phi_{\ell(2n+1)+M_{s+2b-1}+\frac12}+\phi_{\ell(2n+1)-M_{s+2b-1}-\frac12})
&2b-1\ne r,\\
\frac1{\sqrt 2}(\gamma_r^{-1}\phi_{(2n+1)\ell -\frac12}+\gamma_r\phi_{(2n+1)\ell +\frac12})
&2b-1=r,
\end{cases}
\\
\tilde\phi^{s+2b}_{2\ell \mu_{2b}-k}&=\gamma_{s+2b}^{-1}\phi_{ \ell(2n+1)-M_{s+2b}-k+\frac12},
\quad
\tilde\phi^{s+2b}_{2\ell \mu_{2b}+k}=\gamma_{s+2b}(-1)^k\phi_{ \ell(2n+1)+M_{s+2b}+k-\frac12},
\\
\tilde\phi^{s+2b}_{2\ell\mu_{2b}}&=  \sqrt{\frac{-1}{  2}}(\phi_{\ell(2n+1)-M_{s+2b-1} -\frac12}
- \phi_{\ell(2n+1)+M_{s+2b-1}+\frac12})
 ,\\
\tilde\phi^{s+2b}_{2\ell\mu_{s+2b}+\mu_{s+2b}}&=\frac{\gamma_{s+2b}^{-1} \sqrt{-1}}{\sqrt 2}(\phi_{(\ell+1)(2n+1)-M_{s+2b+1}+\frac12}-\phi_{\ell(2n+1)+M_{s+2b+1}-\frac12}).
\end{aligned}
\end{equation}
(2) For the embedding in $b_\infty$,
\[
\begin{aligned}
C_k^{+a}=&
\begin{cases}
\frac{(-\omega_a)^{-k}}{\sqrt{\lambda_a}}&1\le a\le s,\\
\frac{1}{\sqrt{2\lambda_a}} &s<a\le r+s,\\
\end{cases}\qquad
C_k^{-a}=
\begin{cases}
\frac{(-1)^k}{\sqrt{\lambda_a}}&1\le a\le s,\\
\frac{1}{\sqrt{2\lambda_a}} &s<a\le r+s,\\
\end{cases}
\end{aligned}
\]
and 
\[
\Psi_k^{\pm a}(z)=
\begin{cases}
\psi^{\pm a}(\omega_a^{-k} z^{\frac{N}{\lambda_a}})z^{\frac{N}{2\lambda_a}-\frac12}&1\le a\le s,\\
\tilde\phi^{ a}(\pm \omega_a^{-k} z^{\frac{N}{2\lambda_a}})z^{-\frac12}&s< a\le s+r,\\
\end{cases}
\]
where 
\begin{equation}
\label{14field2}
\psi^{\pm a}(z)=\sum_{i\in\frac12+\mathbb Z}\psi_i^{\pm a}z^{-i-\frac12},\quad
\tilde\phi^{s+b}(z)=\sum_{i\in\mathbb Z}\tilde\phi^{s+b}z^{-i},\quad
1\le a\le s,\ 1\le b\le r,
\end{equation}
and 
\[
\sigma(z)=\begin{cases}
\sum_{i\in \mathbb Z} \sigma_i z^{-i}&r \ \mbox{even},\\
\sum_{i\in\frac12+ \mathbb Z} \sigma_i z^{-i-\frac12}&r \ \mbox{odd}
\end{cases}
\]
are the generating fields 
for the relabeld generators of the Clifford algebra $C\ell_b$ ($1\le a\le s$,  $1\le b\le \left[\frac{r+1}2\right]$,  with $2b\ne r+1$ and $1\le i\le \lambda_a$, $1\le j\le \lambda_{s+2b-1}$ or $\lambda_{s+2b}$):
\begin{equation}
\begin{aligned}
\sigma_{\ell+\frac{\eta}2}&=
\begin{cases}
\tilde\phi_{\ell(2n+1)}
&r\ \mbox{even},\\
\sqrt{\frac{-1}{  2}}(\gamma_r^{-1}\tilde\phi_{(2n+1)\ell +n+1}-\gamma_r \tilde\phi_{(2n+1)\ell +n})
&r\ \mbox{odd},
\end{cases}
\\
\psi^\pm_{\ell \lambda_a\pm (\lambda_a+1-i-\frac12)}&=\tilde\phi_{\ell(2n+1)\pm(n+1- M_a-i)},\\
\tilde\phi^{s+2b-1}_{ 2\ell\mu_{2b-1}+\mu_{2b-1} -j}& \tilde\phi_{\ell(2n+1)+n -M_{s+2b-1}-j},
\quad
\tilde\phi^{s+2b-1}_{ 2\ell\mu_{2b-1}+j-\mu_{2b-1}}=  (-1)^{\mu_{2b-1}-j}\tilde\phi_{\ell(2n+1)+M_{s+2b-1}+j-n},
\\
\tilde\phi^{s+2b-1}_{2\ell\mu_{2b-1}+\mu_{2b-1}}&=
\begin{cases}
\frac{\gamma_{s+2b-1}}{\sqrt 2}(\tilde\phi_{\ell(2n+1)+n-M_{s+2b-1}}+
\tilde\phi_{(\ell+1)(2n+1)+ n+1+M_{s+2b-1}})
&2b-1\ne r,
\\
\frac1{\sqrt 2}(\tilde\phi_{(2n+1)\ell +n}+(-1)^{\mu_r}\tilde\phi_{(2n+1)\ell +n+1})
&2b-1=r,
\end{cases}
\\
\tilde\phi^{s+2b-1}_{2\ell \mu_{2b}}&=
\begin{cases}
\frac{1}{\sqrt 2}
(\tilde\phi_{\ell(2n+1)+n+1-M_{s+2b+1}}+\tilde\phi_{\ell(2n+1)-n-1+M_{s+2b+1}})
&2b-1\ne r,\\
\tilde\phi_{\ell(2n+1)}
&2b-1=r,
\end{cases}
\\
\tilde\phi^{s+2b}_{2\ell \mu_{2b}+\mu_{2b}-j}&=\tilde\phi_{ \ell(2n+1)+n+1-M_{s+2b}-j},
\quad
\tilde\phi^{s+2b}_{2\ell \mu_{2b}-\mu_{2b}+k}=(-1)^{j-\mu_{2b}}\tilde\phi_{ \ell(2n+1)+M_{s+2b}+j-n-1},
\\
\tilde\phi^{s+2b}_{2\ell\mu_{2b}}&=  \sqrt{\frac{-1}{  2}}(\tilde\phi_{\ell(2n+1)-M_{s+2b+1} }
- \tilde\phi_{\ell(2n+1)-n+M_{s+2b+1}})
 ,\\
\tilde\phi^{s+2b}_{2\ell\mu_{s+2b}+\mu_{s+2b}}&=\frac{\gamma_{s+2b}\sqrt{-1}}{\sqrt 2}(\tilde\phi_{\ell(2n+1)+n-M_{s+2b-1}}-\tilde\phi_{\ell(2n+1)+n+1+M_{s+2b-1}}).
\end{aligned}
\end{equation}
In both cases,
the relabeled fermions satisfy \eqref{11-1} and 
\begin{equation}
\label{11-11}
\begin{aligned}
\sigma_i\sigma_j+\sigma_j\sigma_i&=\delta_{i,-j},\quad
\psi^{\pm a}_i \sigma_j+\sigma_j\psi^{\pm a}_i=0,\\
\tilde\phi^{s+b}_i\sigma_j+\sigma_j\tilde\phi^{s+b}_i &=0,\qquad
\sigma_i|0\rangle  =0\quad i>0.
\end{aligned}
\end{equation}
and 
 \eqref{11-1},   for $1\le a,c\le s,\ 1\le b,d\le r,\quad i,j\in \frac12+\mathbb Z$.
\\
(d) The representation of $\hat{so}_{2n+1}$ is given in terms of vertex operators by formulas \eqref{VOA}  and \eqref{VOD}, by substitution of roots of unity  $\omega_a$ and powers of $z$ as given 
in the above formulas,  and by
\begin{equation}
\label{psifieldsigma}
\begin{aligned}
\sigma
:&\psi^{\pm a}
(\omega_a^{-k}z^{\frac{N}{\lambda_a}} ):\sigma(z^{N}):
\sigma^{-1}=(-1)^{\sum_{k=a}^{s}q_k\frac{\partial}{\partial q_k}}q_a^{\pm 1}
(\omega_a^{-k}z^{\frac{N}{\lambda_a}})^{\pm q_a\frac{\partial}{\partial q_a}}\times \\
&\qquad \exp\left(
\pm \sum_{j>0}t^a_j(\omega_a^{-k}z^{\frac{N}{\lambda_a}})^j
\right)\exp\left(
\mp \sum_{j>0} \frac{\partial }{\partial t^a_j}\frac{ (\omega_a^{-k}z^{\frac{N}{\lambda_a}})^{-j}}j
\right)\Xi(z^{N}),
\end{aligned}
\end{equation}
\begin{equation}
\label{pPHI}
\begin{aligned}
\sigma:\tilde\phi^{s+b}(\omega_{s+b}^{-k}z^{\frac{N}{2\mu_b}})\sigma(z^{N}):\sigma^{-1}=& 
(-1)^{\sum_{k=1}^{s}q_k\frac{\partial}{\partial q_k}} R_b(\theta)
\exp\left(
\sum_{j>0,\ \rm odd}t^{s+b}_j (\omega_{s+b}^{-k}z^{\frac{N}{2\mu_b}})^j
\right)\times \\
&\qquad\exp\left(
-2 \sum_{j>0,\ \rm odd } \frac{\partial }{\partial t^{s+b}_j}\frac{ (\omega_{s+b}^{-k}z^{\frac{N}{2\mu_b}})^{-j}}j
\right)\Xi(z^{N}),
\end{aligned}
\end{equation}
where
\begin{equation}
\label{sigmaXi}
\Xi[z]=(1+2\delta) R_{r+1}(\theta)+\sum_{j\in \delta +\mathbb Z_>0}\left(
\xi_j z^{\delta +j} +\frac{\partial}{\partial \xi_j} z^{\delta-j}
\right).
\end{equation}
In particular,    the Heisenberg algebra \eqref{11-2} is given by
 $\alpha_a (z)=\lambda_a z^{\frac{\lambda_a}N-1}r(a_{\lambda_a\lambda_a}^{+a,-a})(z^{\frac{\lambda_a}N})$ for $1\le a\le s$,
respectively 
$ \alpha_{s+b}=\lambda_{s+b} z^{\frac{2\lambda_{s+b}}N}r(a_{\lambda_{s+b}\lambda_{s+b}}^{+(s+b),-(s+b)})(z^{\frac{2\lambda_{s+b}}N})$ for $1\le b\le r$,
 and their modes satisfy 
\eqref{11-3}.
\end{proposition}
The $q$-dimension formulas in this case  resemble the formula given in \eqref{QDIMDred}.  See \cite{tKLB}, for the explicit formulas.
%
 %
%
%

\subsection{Weyl group elements and the twisted realization of $\hat{\mathfrak g}^{(2)}$}
Let $\mathfrak g$ be one of the classical Lie algebras $gl_n$ or $so_{2n}$.  We copy the construction of Subsection \ref{twistweyl}, except that we replace the lift of the element of the Weyl group $\tilde w$ by $\tilde\sigma:= \tilde w\sigma $, where $\sigma$ is the outer automorphism that acts as \eqref{out},  i.e.  
$\tilde\sigma (g)=-{\rm Ad}\, \tilde w({\rm Ad}\, J_n( g^t))$ for $\mathfrak g=gl_n$ and $\tilde\sigma (g)= {\rm Ad}\, \tilde w({\rm Ad}\, O(g))$ for $\mathfrak g=so_{2n}$.
\\
\ 
\\
{\bf In the case of $gl_n$},  this $\tilde\sigma$,  produces not only a permutation of the elements  $\epsilon_i$,  but also multiplies it with $-1$.  Thus we get both positive and negative cycles, i.e. $\tilde \sigma $ acts as
\[
\epsilon_{i_1}\to -\epsilon_{i_2}\to\epsilon_{i_3}\to
 \cdots \to (-1)^m\epsilon_{i_{m-1}}\to(-1)^{m+1}\epsilon_{i_m}\to (-1)^m\epsilon_{i_1}.
\] 
We observe that if $m$ is even,  we obtain a positive cycle, but if $m$ is odd we get a negative cycle.
 Hence, as before  $\tilde\sigma$ is related to a pair of partitions $(\lambda,\mu)=((\lambda_1,\lambda_2,\ldots,\lambda_s),(\mu_1,\mu_2, \ldots\mu_r))$,  where all the parts of $\lambda$ are even  and all the parts of $\mu$ are odd, such that $|\lambda|+|\mu|=n$. This means that if $n$ is even (resp. odd), $r$ must also be even (resp. odd).  As before,  see
\cite{tKLtwist} for more details,  a part $\lambda_i$ is related to a pair of charged free fermionic fields $\psi^{\pm i}(z)$ and a part $\mu_j$ to a twisted neutral free fermionic field $\tilde\phi^j(z)$.  This means that we get a realization of $\hat{gl}_{2m+1}^{(2)}\subset b_\infty$  (resp.  $\hat{gl_{2m}}^{(2)}\subset d_\infty$) which has an odd (resp. even) number of twisted neutral free fermionic fields.
The $q$-dimension formulas in the  case are  given by \eqref{QDIMDred} for $\hat{gl_{2m}}^{(2)}\subset d_\infty$ and by 
by the following formula in the $\hat{gl}_{2m+1}^{(2)}\subset b_\infty$ case, see \cite{tKLtwist} for more information.
\[
2^{[\frac{r}2]}
\sum_{{j_1, \ldots, j_s\in\mathbb Z} }
q^{\frac{N}2(\frac{j_1^2}{\lambda_1^2}+\frac{j_2^2}{\lambda_2^2}+\cdots +\frac{j_s^2}{\lambda_s^2})}
\prod_{a=1}^s \prod_{i=1}^\infty \frac1{1-q^{\frac{jN}{\lambda_a}}}
\prod_{b=1}^r \prod_{j=1}^\infty \frac1{1-q^{\frac{(2j-1)N}{\mu_b}}}
\]

\ 
\\
{\bf In the case of $so_{2n}$},  this $\tilde\sigma$  gives a permutation of all the $\epsilon_i$,  $1\le i\le n$, together with sign changes.  But since ${\rm Ad}\, O(\epsilon_i)=
(-)^{\delta_{in}}\epsilon_i$, the cycle of ${\rm Ad}\,\tilde w$ that contains $\epsilon_n$ changes sign, i.e.,  if the cycle is  positive (resp. negative) for  ${\rm Ad}\,\tilde w$ it becomes
negative  (resp. positive).  For instance, if ${\rm Ad}\,\tilde w$ acs on a positive cycle as
\[
\epsilon_{n}\to \epsilon_{i_2}\to\epsilon_{i_3}\to
 \cdots \to  \epsilon_{i_m}\to \epsilon_{n},
\] 
Then $\tilde\sigma$ acts on these elements as
\[
\epsilon_{n}\to - \epsilon_{i_2}\to -\epsilon_{i_3}\to
 \cdots \to  -\epsilon_{i_m}\to -\epsilon_{n}\to \epsilon_{i_2}\to \epsilon_{i_3}\to
 \cdots \to  \epsilon_{i_m}\to \epsilon_{n}.
\] 
Hence, instead of an even number of negative cycles, as for the conjugacy classes of the Weyl group of $so_{2n}$,  we obtain an odd number of negative cycles.
Thus we obtain a realization of $\hat{so}_{2n}^{(2)}\subset b_\infty$ which is related again to two partitions $(\lambda,\mu)=((\lambda_1,\lambda_2,\ldots,\lambda_s),(\mu_1,\mu_2, \ldots\mu_r))$,  where a 
part 
 $\lambda_i$ is related to a pair of charged free fermionic fields $\psi^{\pm i}(z)$ and a part $\mu_j$ to a twisted neutral free fermionic field $\tilde\phi^j(z)$, but in this case the parts are arbitrary, such that $|\lambda|+|\mu|=n$,  and $r$ is odd.
The $q$-dimension formulas in this case are again given by \eqref{QDIMDred} (see \cite{tKLtwist}).
\begin{proposition}
The representation of these twisted algebras on the Fock space $B_{(\lambda,\mu)}$ is given in terms of the following (twisted) fermionic fields.
Let $\omega_a=e^{\frac{2\pi\sqrt{-1}}{\lambda_a}}$,  $\omega_b=e^{\frac{2\pi\sqrt{-1}}{\lambda_b}}$,  $\omega_c=e^{\frac{\pi\sqrt{-1}}{\mu_{c-s}}}$ and $\omega_d=e^{\frac{\pi\sqrt{-1}}{\mu_{d-s}}}$,  and $1\le a,b\le s< c,d\le r+s$, then 
\\
(a) For $\hat{gl}_n^{(2)}$: 
\[
\begin{aligned}
& \sigma: \psi^{(-)^j a}(\omega_a^j z^{\lambda_b}) \psi^{(-)^{k+1} b}(\omega_b^k z^{\lambda_a}):\sigma^{-1},&\quad  
1\le j\le \lambda_a,\ 1\le k\le \lambda_b,\\
&\sigma: \psi^{(-)^j a}(\omega_a^j z^{2\mu_{c-s}})  \tilde\phi^{c}((-1)^{k+1}\omega_c^k z^{\lambda_a}) :\sigma^{-1},&\quad
1\le j\le \lambda_a,\ 1\le k\le \mu_{c-s},\\
& \sigma: \tilde\phi^{c}((-1)^{k}\omega_c^k z^{\lambda_a}) \psi^{(-)^{j+1}a}(\omega_a^j z^{2\mu_{c-s}}) :\sigma^{-1},&
\quad 1\le j\le \lambda_a,\ 1\le k\le \mu_{c-s}, 
\quad\mbox{and}\\
& \sigma: \tilde \phi^{c} ((-1)^j\omega_c^j z^{2\mu_{d-s}})\tilde\phi^{d}((-1)^{k+1}\omega_d^k z^{2\mu_{c-s}}):\sigma^{-1},&
\quad
1\le j\le \mu_{c-s},\ 1\le k\le \mu_{d-s}.\\
\end{aligned}
\]
(b) For $\hat{so}_{2n}^{(2)}$: 
($\delta, \epsilon=+$ or $-$, $j,k\in\mathbb Z$) 
\[
\begin{aligned}
& \sigma: \psi^{\epsilon a}(\omega_a^j z^{\lambda_b}) \psi^{\delta b}(\omega_b^k z^{\lambda_a}):\sigma^{-1}, \quad  
\sigma: \psi^{\epsilon a}(\omega_a^j z^{2\mu_{c-s}})  \tilde\phi^{c}(\omega_c^k z^{\lambda_a}) :\sigma^{-1}\quad\mbox{and}\\
& \sigma: \tilde \phi^{c} (\omega_c^j z^{2\mu_{d-s}})\tilde\phi^{d}(\omega_d^k z^{2\mu_{c-s}}):\sigma^{-1},
\end{aligned}
\]
where one uses the formulas \eqref{VOA} and \eqref{VOD} to express them as vertex operators.  In particular,  we have the Heisenberg algebra \eqref{11-2}, which 
are equal to $\sigma\alpha^{a} (z)\sigma^{-1}=\sigma :\psi^{+a}(z)\psi^{-a}(z):\sigma^{-1}$,
respectively 
$ \sigma\alpha^c(z)\sigma^{-1}=\frac {1}2 \sigma :\tilde\phi^c(z)\tilde\phi^c(-z):\sigma^{-1}$
 and their modes satisfy \eqref{11-3}.
\end{proposition}
Here we use the notation $(-)^j a$ as upper index in the charged fermionic fields,  meaning that $(-)^j a=+a$ (resp.  $=-a$)  if $j$ is even (resp. odd).
For more detailed information on vertex operators for these  twisted affine Lie algebras
we refer to \cite{tKLtwist}.

%

\section{A map from the conjugacy classes of the Weyl group  
to nilpotent elements in $\mathfrak g$}
\label{S6}
\subsection{Construction based on Section 5}
\label{S6.1}
In the previous chapter we have constructed for every conjugacy class of the Weyl group of a classical Lie algebra $\mathfrak g$ (equivalently, for a  partion $\lambda$ for $gl_n$,   and a  pair of partitions $(\lambda,\mu)$ for other classical Lie algebras)  the  corresponding Heisenberg subalgebra in $\hat{\mathfrak g}$.  This Heisenberg subalgebra is the span of $K$ and the elements 
$a^{bb}_{\lambda_b\lambda_b}$ for $1\le b\le s$
appearing in the modes of the fields
 for $gl_n$ 
(see Proposition \ref{P5.3}),
and of $a^{+b,-b}_{\lambda_b\lambda_b}$ 
and $a^{+(s+c),-(s+c)}_{\mu_c\mu_c}$ for $1\le b\le s$,  $1\le c\le r$,  for the other classical algebras
(see Subsections \ref{S5.3}--\ref{S5.5}).
This Heisenberg algebra is the centralizer in $\tilde{\mathfrak g}$ of the  cyclic element
\[
\begin{aligned}
c_w(t)=c_{(\lambda,\mu)}(t):
&=r^{-1}\left(\sum_{b=1}^s \frac1{\lambda_a}\alpha^{b}_1 \pm 
\sum_{c=1}^r  \frac1{\mu_c}\alpha^{s+c}_1\right)\\
&=
\sum_{b=1}^s a^{+b,-b}_{\lambda_b\lambda_b}(\frac{N}{\lambda_b})+
\sum_{c=1}^r a^{+(s+c),-(s+c)}_{\mu_c\mu_c}(\frac{N}{2\mu_c})
\in \mathfrak g(\mathbb C[t]).
\end{aligned}
\]
Note that this cyclic element $c_w(t)$ is a linear combination of the smallest positive degree Heisenberg element $\alpha^a_1$,  in each 
component $a=1,2,\ldots, s+r$ of the decomposition $n=\lambda_1+\cdots+\lambda_s+\mu_1+\cdots+\mu_r$.

The element $c_w(0)=c_{(\lambda,\mu)}(0)$ is a nilpotent element in the classical Lie algebra $\mathfrak g$. Thus we have constructed a map from the conjugacy classes of the Weyl group $W$ to the nilpotent elements in $\mathfrak g$, viz.  $w\mapsto c_w(0)$.  Below we describe this map explicitly for all the cases.

Let $w_\lambda\in W$ of $gl_n$ be given by \eqref{wlambda},  the nilpotent element corresponding to this element in the Weyl group is
\[
\begin{aligned}
gl_n:\qquad f_{\lambda}&=\sum_{b=1}^s\frac1{\lambda_b} \sum_{i=1}^{\lambda_b-1} e^{bb}_{i, i+1}\\
&=
\sum_{b=1}^s\frac1{\lambda_b}\left(t e^{bb}_{\lambda_b, 1} +\sum_{i=1}^{\lambda_b-1} e^{bb}_{i, i+1}\right)\Big|_{t=0}
.
\end{aligned}
\]
This means that $f_{\lambda}$ has the Jordan normal form decomposition with $s$ Jordan blocks of size $\lambda_1$, $\lambda_2$,  $\ldots, \ \lambda_s$.

Let $w_{(\lambda,\mu)}$ be given by \eqref{wlambdamu},  where in the case of $so_{2n}$ $r$ is even, 
 the nilpotent element corresponding to this element in the Weyl group is 
\[
\begin{aligned}
so_{2n+1}:&\qquad
f_{(\lambda,\mu)}=\sum_{b=1}^s \frac1{\lambda_b}\sum_{i=1}^{\lambda_b-1}e_{i,i+1}^{+b,-b}
 -\sum_{c=1}^{\frac{r}2}\big(
\frac1{\mu_{2c-1} }\big(\frac{\gamma^{-1}_r}{\sqrt 2}e_{12}^{+(s+2c-1),-(s+2c-1)}+\\
&+\frac{1}{\sqrt 2}e_{\mu_{2c-1}\mu_{2c}}^{+(s+2c-1),-(s+2c)}
+\frac{1}{\sqrt 2}e_{\mu_{2c-1}\mu_{2c}}^{+(s+2c-1),+(s+2c)}+
\sum_{j=2}^{\mu_{2c-1}-1} e_{j,j+1}^{+(s+2c-1),-(s+2c-1)}
\big)+\\
&+\frac1{\mu_{2c-1} }\big(
\frac{\sqrt{-1}}{\sqrt 2}e_{\mu_{2c}-1,\mu_{2c}}^{+(s+2c),-(s+2c)}
-\frac{\sqrt{-1}}{\sqrt 2}e_{\mu_{2c}-1,\mu_{2c}}^{+(s+2c),+(s+2c)}+
\sum_{j=1}^{\mu_{2c-1}-2} e_{j,j+1}^{+(s+2c),-(s+2c)}
\big)
\big)-
\\
&-\frac{\eta}{\mu_r} \big(
e_{\mu_r}^{+(s+r)}+
\sum_{j=1}^{\mu_{r}-1} e_{j,j+1}^{+(s+r),-(s+r)}
\big);
\\
sp_{2n}:&\qquad
f_{(\lambda,\mu)}=\sum_{b=1}^s \frac1{\lambda_b}\sum_{i=1}^{\lambda_b-1}e_{i,i+1}^{+b,-b}
 +\sum_{c=1}^r \frac1{\mu_c} (e^{+(s+c),+(s+c)}_{\mu_c\mu_c}+\sum_{i=1}^{\mu_c-1}
e^{+(s+c),-(s+c)}_{i,i+1});
\end{aligned}
\]
\[
\begin{aligned}
so_{2n}:&\qquad
f_{(\lambda,\mu)}=\sum_{b=1}^s \frac1{\lambda_b}\sum_{i=1}^{\lambda_b-1}e_{i,i+1}^{+b,-b}
 -\sum_{c=1}^{\frac{r}2}\big(
\frac1{\mu_{2c-1} }\big(\frac{\gamma^{-1}_r}{\sqrt 2}e_{12}^{+(s+2c-1),-(s+2c-1)}+\\
&+\frac{1}{\sqrt 2}e_{\mu_{2c-1}\mu_{2c}}^{+(s+2c-1),-(s+2c)}
+\frac{1}{\sqrt 2}e_{\mu_{2c-1}\mu_{2c}}^{+(s+2c-1),+(s+2c)}+
\sum_{j=2}^{\mu_{2c-1}-1} e_{j,j+1}^{+(s+2c-1),-(s+2c-1)}
\big)+\\
&+\frac1{\mu_{2c-1} }\big(
\frac{\sqrt{-1}}{\sqrt 2}e_{\mu_{2c}-1,\mu_{2c}}^{+(s+2c),-(s+2c)}
-\frac{\sqrt{-1}}{\sqrt 2}e_{\mu_{2c}-1,\mu_{2c}}^{+(s+2c),+(s+2c)}+
\sum_{j=2}^{\mu_{2c-1}-2} e_{j,j+1}^{+(s+2c),-(s+2c)}
\big)
\big).
\end{aligned}
\]
In the case of $sp_n$,  the nilpotent element $f_{(\lambda,\mu)}$ has   $2s+r$ Jordan blocks of size
$\lambda_1, \lambda_1,\lambda_2, \lambda_2,\ldots,\lambda_s, \lambda_s,$  $2\mu_1, \ 2\mu_2,
\ldots,2\mu_r$.  For $so_{2n}$ we again have  $2s+r$ Jordan blocks, but now of size
$\lambda_1, \lambda_1,\lambda_2, \lambda_2,\ldots,\lambda_s, \lambda_s,$  $2\mu_1+1 ,2\mu_2-1,2\mu_3+1,  2\mu_4-1, 
\ldots,2\mu_r-1$.  For $so_{2n+1}$ there are $2s+r+1$ Jordan blocks of size
$\lambda_1, \lambda_1,\lambda_2, \lambda_2,\ldots,\lambda_s, \lambda_s,$  $2\mu_1+1 ,2\mu_2-1,2\mu_3+1,  2\mu_4-1, 
\ldots,2\mu_r-1,1$, if $r$ is even and $2s+r$ if $r$ is odd, but then of size $\lambda_1, \lambda_1,\lambda_2, \lambda_2,\ldots,\lambda_s, \lambda_s,$  $2\mu_1+1 ,2\mu_2-1,2\mu_3+1,  2\mu_4-1,
\ldots,2\mu_r+1$.
\subsection{A slightly different map}
Unfortunately the map defined in the previous subsection for $so_n$  is not injective if we restrict to the set of all elliptic conjugacy classes, i.e.  all conjugacy classes with $\lambda=\emptyset$.   For instance  in $so_{28}$,
both $\mu=(5,5,3,1)$ and $\mu= (5,4,4,1)$ are mapped onto a nilpotent element with Jordan blocks of size 11,  9,   7,   1.   We can however correct this for $so_{2n}$ by using the following observation:
\begin{remark}
The construction in Section \ref{sec5}  for $so_n$  also holds if we permute  the parts of $\mu$.  This can be seen as follows.
Recall,  that in order to obtain the construction of $\tilde w_{(\lambda,\mu)}$,  we combine each time two blocks of the size $2\mu_{2a-1}$ and $2\mu_{2a}$
(two consecutive parts of $\mu$).  In principle one can combine any two parts of $\mu$.  Hence we can shuffle the parts of $\mu$. This can be achieved  by a permutation 
$\pi$ of $1, 2,\ldots r$, such that $\pi(\mu)=(\mu_{\pi(1)}, \mu_{\pi(2)}, \ldots ,\mu_{\pi(r)})$,  then instead of $f_{(\lambda,\mu)}$, we can consider $f_{(\lambda,\pi(\mu))}$.  Note, that the element in the conjugacy class of the Weyl group of $so_n$ does not change,  but the nilpotent element changes. 
It now has Jordan blocks of size 
$\lambda_1, \lambda_1,\lambda_2, \lambda_2,\ldots,\lambda_s, \lambda_s,$  $2\mu_{\pi(1)}+1 ,2\mu_{\pi(2)}-1,2\mu_{\pi(3)}+1,  2\mu_{\pi(4)}-1, 
\ldots,2\mu_{\pi(r)}-1$,  for $so_{2n}$    (in this case $r$ is always even).  For $so_{2n+1}$, one thus gets
$\lambda_1, \lambda_1,\lambda_2, \lambda_2,\ldots,\lambda_s, \lambda_s,$   $2\mu_{\pi(1)}+1 ,2\mu_{\pi(2)}-1,2\mu_{\pi(3)}+1,  2\mu_{\pi(4)}-1, 
\ldots,2\mu_{\pi(r)}-1,1 $,   if $r$ is even,  and $\lambda_1, \lambda_1,\lambda_2, \lambda_2,\ldots,\lambda_s, \lambda_s,$  $2\mu_{\pi(1)}+1 ,2\mu_{\pi(2)}-1,2\mu_{\pi(3)}+1,  2\mu_{\pi(4)}-1, 
\ldots,2\mu_{\pi(r)}+1$,     if $r$  is odd.
\end{remark}

We now use the above remark and choose the permutation $\pi$ as follows:
\[
\pi(2i-1)=i,\ \mbox{for } i=1,\ldots ,\left[ \frac{r+1}2\right] \ \mbox{and }
\pi(2i)=\left[ \frac{r+1}2\right]+i,\ \mbox{for } i=\left[ \frac{r+3}2\right],\ldots ,r. 
\]
Then the pair $(\lambda, \mu)$ gets mapped to
\begin{equation}
\label{MMMu}
\lambda_1, \lambda_1,\lambda_2, \lambda_2,\ldots,\lambda_s, \lambda_s,2\mu_{1}+1 ,2\mu_{2}+1,\ldots 
2\mu_{\left[ \frac{r+1}2\right]}+1, 
2\mu_{\left[ \frac{r+3}2\right]}-1,
2\mu_{\left[ \frac{r+5}2\right]}-1,\ldots
2\mu_{r}-1,
\end{equation}
in the case of $so_{2n}$,  and $so_{2n+1}$  with $r$ odd,  and to
\[
\lambda_1, \lambda_1,\lambda_2, \lambda_2,\ldots,\lambda_s, \lambda_s,2\mu_{1}+1 ,2\mu_{2}+1,\ldots 
2\mu_{\left[ \frac{r+1}2\right]}+1, 
2\mu_{\left[ \frac{r+3}2\right]}-1,
2\mu_{\left[ \frac{r+5}2\right]}-1,\ldots
2\mu_{r}-1,1,
\]
for $so_{2n+1}$ with $r$ even.
This map is surjective,  and  its restriction to the elliptic conjugacy classes is injective for  $so_{2n}$.
Surjectivity can be shown as follows.  Suppose  we have a Jordan block decomposition 
$n_1\ge n_2\ge \cdots, \ge n_k$
of $n$. Then blocks of even size always come with even multiplicity.  We construct the pair $(\lambda, \mu)$ for which $\lambda$ has the most parts that gets mapped to this Jordan normal form,  so the corresponding $w$, when acting on the Cartan subalgebra, $\mathfrak h$   has the largest subspace of fixed points $\mathfrak h^w$.  For this we first make pairs of Jordan blocks which have the same size,  say
$n_j$ and $n_{j+1}$.  This pair then gives one part $\lambda_i$ of $\lambda$ such that $\lambda_i=n_j$.
So assume that we have removed all pairs from this Jordan normal form, then there are only Jordan blocks left of odd sizes and their sizes  are all different.  Assume that  the blocks have size $m_1> m_2>\cdots >m_\ell$.
Then $\ell$ is even (resp odd)  for $n$ even (resp.  odd).  This is how we construct $\mu$:
\begin{equation}
\label{MMu}
\mu_j=\frac{m_j-1}2\ \mbox{for }j=1, \ldots, \left[\frac{\ell+1}2\right],\ \mbox{and }
\mu_j=\frac{m_j+1}2\ \mbox{for }j=\left[\frac{\ell+3}2\right], \ldots, \ell
 .
\end{equation}
Then this pair $(\lambda,\mu)$ is mapped to the Jordan normal form with Jordan blocks
$n_1\ge n_2\ge \cdots, \ge n_k$.

To prove injectivity for $so_{2n}$ on the elliptic conjugacy classes,  so all classes $(\lambda=\emptyset, \mu)$, we do the following.  Let $\mu^1\ne \mu^2$ be two partitions of $n$, then 
cleary when the number of parts of both partitions are different we also obtain different Jordan normal forms.  So assume 
both $\mu$'s must have the same number of parts and since we have $so_{2n}$ the number of parts must be even, say $r=2k$.  The map which gives the Jordan blocks,  given in \eqref{MMMu} produces 
blocks of sizes
\begin{equation}
\label{MMMuu}
2\mu_{1}+1 \ge 2\mu_{2}+1\ge\ldots\ge
2\mu_{k}+1>
2\mu_{k+1}-1\ge
2\mu_{k+2}-1\ge\cdots\ge
2\mu_{r}-1.
\end{equation}
Now clearly $\mu^1$ and $\mu^2$,  when different, produce different Jordan blocks \eqref{MMMuu}.

Unfortunately, this new map is not injective for $so_{2n+1}$, with $n>3$.  For instance in $so_9$ both $\mu=(2,2)$  and $\mu= (2,1,1)$ give Jordan normal form with blocks of sizes 5, 3, 1.

Note that the above construction in the case of $so_{2n}$  also gives that   in each fibre of the map, given in \eqref{MMMu},     there is a unique conjugacy class $[w]$.,    which has the smallest fixed subspace $\mathfrak h^w$.  One constructs the to $[w]$ corresponding pair $(\lambda,\mu)$ in a similar way as in the proof of surjectivity.  The difference is that now we want to have the element $w$ with smallest subspace $\mathfrak h^w$.  This means that we want to have the number of parts of $\lambda$ as small as possible.  So we produce $\lambda$ only from the even sized  Jordan blocks.  If we remove these we only have an even number of odd sized Jordan blocks  left,   say $m_1\ge m_2\ge\cdots \ge m_\ell$.  From this we construct $\mu$ as in \eqref{MMu}. Since the map is injective on the elliptic conjugacy classes of $so_{m_1+\cdots+m_\ell}$,  this pair $(\lambda, \mu)$, constructed in this way,
is unique and has the desired property.

Unfortunately we do not know how to modify
the construction of Subsection \ref{S6.1} such that it becomes injective when restricted to the set of elliptic conjugacy classes of $W$ for $so_{2n+1}$, $n>3$.

A surjective map of the set of conjugacy classes of $W$ to the set of nilpotent orbits, which is injective on the set of elliptic conjugacy classes,  for any simple Lie algebra $\mathfrak g$ was constructed by Lusztig, see \cite{Yun}. It coincides with our map for $\mathfrak g=sp_n$,  but not for $\mathfrak g= so_n$.
\\
\
\\

\section{Hierarchies of KP type  related to $x_\infty$ and $\hat{\mathfrak g}$ reductions}
\label{S7}
\subsection{$\hat{gl}_n$ and the $\lambda$-KdV  hierarchy}
\label{SS7.1}
Instead of considering the group orbit of $GL_\infty$, we want to describe the orbit of $\hat{SL}_n$, which is the central
extension of the loop group of type $SL_n$ that is a subgroup of $A_\infty$,  leaves the space $F^{(m)}$ invariant. We find that an $\hat{SL}_n$ tau-function satisfies more Hirota bilinear equations,  besides \eqref{BosKP}.
Using Lemma 2.4 of \cite{KvdLB}, an element $g\in A_\infty$, which  has the following action on the generators in the Clifford algebra $C\ell$:
\[
g\psi_j^+g^{-1}=\sum_i a_{ij}\psi_i^+, \ g\psi_k^-g^{-1}=\sum_i b_{ik}\psi_i^-,\ \mbox{with } \sum_i  a_{ij}b_{-ik}=\delta_{j,-k}.
\]
If 
$g\in \hat{GL}_n$, then the $a_{ij}$ and $b_{ij}$ satisfy $a_{i+n,j+n}=a_{ij}$ and  $b_{i+n,j+n}=b_{ij}$. Using this we can show that $g\otimes g$ commtes with $ \sum_i \psi_i^+ \otimes \psi_{pn-i}^-$ for $p=0,1,2,\ldots$:
\[
\begin{aligned}
\sum_i g \psi_i^+ \otimes g \psi_{pn-i}^-&=\sum_i g \psi_i^+ g^{-1}g \otimes g\psi_{pn-i}^-g^{-1}g\\
&=\sum_{i, j, k} a_{ij}\psi^+_jg \otimes b_{pn-i,k}\psi_{k} g\\
&=\sum_{i, j, k} a_{ij} b_{-i,k-pn}\psi^+_jg \otimes\psi_{k} g\\
&=\sum_{j} \psi^+_jg \otimes\psi_{pn-j} g.\\
\end{aligned}
\]
 Since $\sum_i  \psi_i^+|0\rangle  \otimes  \psi_{pn-i}^- |0\rangle=0$, we have that for all $p=0,1,2,\ldots$ ($p=0$ is the $s$-component KP hierarchy \eqref{BosKP}):
\[
\sum_{i\in\frac12+\mathbb Z}  \psi_i^+g|0\rangle  \otimes  \psi_{pn-i}^- g |0\rangle=0,
\]
which turns into the following equation in the $\lambda$-reduction
\[
\sum_{j=1}^s\sum_{i\in\frac{1}{2}+\mathbb{Z}
} \psi_i^{+j}\tau \otimes \psi_{p\lambda_j-i}^{-j}\tau=0,\quad p=0,1,2,\ldots,
\]
and,  after using the isomorphism $\sigma$, into
\begin{equation}
\label{BosKdV}
\begin{aligned}
{\rm Res}_{z=0}\sum_{j=1}^s &
(-1)^{|\underline k+\underline \ell|_{j-1}} z^{ p\lambda_j+k_j-\ell_j-2+2\delta_{js}}
e^{z\cdot (t^{(j)}-\overline t^{(j)}) }
e^{z^{-1}\cdot (\tilde\partial_{\overline t^{(j)} }-t^{(j)}-\tilde\partial_{t^{(j)}})}
\tau_{\underline k+\underline e_s-\underline e_j}(t)\times\\
&\tau_{\underline \ell+\underline e_j-\underline e_s}(\overline t)
 dz=0.
\end{aligned}
\end{equation}
This is the Hirota bilinear equation for the $\lambda$-KdV. The special case when $s=1$ and $\lambda_1=n$  is the  $n$-Gelfand-Dickey hierarchy, and $n=2$ is the Korteweg-de Vries  (KdV) hierarchy.
\begin{proposition}
\label{prop8.1}
Equation \eqref{BosKdV}  for $p=0,1,\ldots $  is equivalent to Equation \eqref{BosKP}, which is equal to \eqref{BosKdV} for $p=0$,  and 
\begin{equation}
\label{BosKdV2}
\begin{aligned}
{\rm Res}_{z=0}\sum_{j=1}^s &
(-1)^{|\underline k+\underline \ell|_{j-1}} z^{  k_j-\ell_j-2+2\delta_{js}}
e^{z\cdot (t^{(j)}-\overline t^{(j)}) }
e^{z^{-1}\cdot (\tilde\partial_{\overline t^{(j)} }-t^{(j)}-\tilde\partial_{t^{(j)}})}
\left(\sum_{i=1}^s \frac{\partial\tau_{\underline k+\underline e_s-\underline e_j}(t)}{\partial t_{p\lambda_i}^{(i)}}\right)\times
\\
&\tau_{\underline \ell+\underline e_j-\underline e_s}(\overline t)
 dz=0.
\end{aligned}
\end{equation}
\end{proposition}
{\bf Proof}
Differentiate   Equation \eqref{BosKdV} for $p=0$  by 
$\sum_{j=1}^s \frac{\partial }{\partial t_{q\lambda_j}^{(j)}}$,  for $q>0$.  We thus get that the sum of the left-hand side of \eqref{BosKdV}  for $p=q$ and the left-hand side of \eqref{BosKdV2}   for $p=q$ is  equal to zero.  Hence for all $p>0$, \eqref{BosKdV} implies \eqref{BosKdV2} and vice versa.\hfill $\square$\break
\
\\
Note that if equation  \eqref{BosKdV} holds for $p=0$,  then  the equations
\begin{equation}
 \label{BosKdV3}
\sum_{i=1}^s \frac{\partial\tau_{m}(t)}{\partial t_{p\lambda_i}^{(i)}}
=C_p \tau_{m}(t),\ \mbox{with } C_p\in\mathbb C,\qquad p=1,2,\ldots . 
\end{equation}
imply  \eqref{BosKdV2} and hence  \eqref{BosKdV} for $p>1$.   In fact one can prove 
\begin{proposition}
\label{prop20}
Equation  \eqref{BosKdV}  is equivalent to  equation \eqref{BosKdV} for $p=0$ together with  Equation  \eqref{BosKdV3}.
 \end{proposition}

Equation \eqref{BosKdV} turns into ($p=0,1,\ldots$):
\begin{equation}
\label{BosKdV6}
{\rm Res}_{z=0}
\sum_{j=1}^s
V^+(\underline \alpha; x,t,z)z^{p\lambda_j}E_{jj}V^-(\underline \beta;\overline x, \overline t,z)^t dz=0.
\end{equation}
Using the fundamental lemma (see \cite{KLmult},  Lemma 4.1),  we deduce the following expression for the matrix pseudo-differential operators
\begin{equation}
\label{VV2b}
\sum_{j=1}^s
(P^+(\underline \alpha;x,t,\partial) R^+(\underline \alpha;\partial)\partial^{p\lambda_j}E_{jj}R^-(\underline\beta;x,t,\partial)^*
P^-(\underline \beta; x,t,\partial)^*)_-=0.
\end{equation}
This gives that
\begin{equation}
\label{sato2b}
\sum_{j=1}^s (P^+(\underline \beta; x,t,\partial)\partial^{p\lambda_j}E_{jj}P^+(\underline \beta; x,t,\partial)^{-1})_-=0,
\end{equation}
and thus that
\begin{equation}
\label{lax222}
 \sum_{j=1}^s (L^{p\lambda_j}C^j)_-=0.
\end{equation}
From this it is obvious that ${\cal L}= \sum_{j=1}^s L^{\lambda_j}C^j$ is a differential operator that also satisfies
\begin{equation}
\frac{\partial \cal L}{\partial t_i^j}=[(L^iC^j)_+,\cal L].
\end{equation}

Let $D_p=\sum_{j=1}^s\frac{\partial}{\partial  t^{(j)}_{p\lambda_j}}$,
then 
\begin{equation}
\label{Dp}
D_p(P^+)=D_p(L)=D_p(C^j)=D_p({\cal L})=0.
\end{equation}
We now have all the ingredients to prove  Proposition \ref{prop20}.\\
\ 
\\
{\bf Proof of Proposition \ref{prop20}} We only have to show that \eqref{BosKdV2} implies \eqref{BosKdV3}.
So assume that \eqref{BosKdV2} holds. 
We first take $\underline k=\underline\ell$,  and $t=\overline t$, this gives, see \eqref{Dp}, that
\begin{equation}
\label{prop20a}
\frac{\partial}{\partial t_1^{(j)}} \left(\frac{D_p(\tau_{\underline\alpha}(x,t))}{\tau_{\underline\alpha}(x,t)}\right)=0.
\end{equation}
Next take $j=s$, $\underline\ell=\underline k+\underline e_i-\underline e_j$,  and $t=\overline t$ in \eqref{BosKdV2},  this gives
\[
 \frac{D_p(\tau_{\underline\alpha}(x,t))}{\tau_{\underline\alpha}(x,t)}=\frac{D_p(\tau_{\underline\alpha+\underline e_i-\underline e_j}(x,t))}{\tau_{\underline\alpha+\underline e_i-\underline e_j}(x,t)},
\]
which means that
\begin{equation}
\label{prop20b}
 \frac{D_p(\tau_{\underline\alpha}(x,t))}{\tau_{\underline\alpha}(x,t)}=\frac{D_p(\tau_{\underline\beta}(x,t))}{\tau_{\underline\beta}(x,t)} \quad \mbox{ for all } \underline\alpha\ \mbox{and }\underline \beta.
\end{equation}
Now take \eqref{BosKdV2} for $\underline k=\underline\alpha-\underline e_s+\underline e_i$ and $\underline\ell=\underline\alpha+\underline e_s-\underline e_m$ and divide by $\tau(\underline \alpha; x,t)
\tau(\underline \alpha;\overline  x,\overline t)$ and use \eqref{prop20b},  to obtain 
\begin{equation}
\label{BosKdV4}
\begin{aligned}
{\rm Res}_{z=0}\sum_{j=1}^s &
\exp\left(-\sum_{i=1}^\infty \frac\partial{\partial \overline  t^{(j)}_i }\frac{z^{-j}}{j}\right)\left(
\frac{D_p(\tau_{\underline\beta}(x,t))}{\tau_{\underline\beta}(x,t)}
\right)V^+(\underline\alpha; x,t)E_{jj} V^-(\underline\alpha; \overline x,\overline  t)^t 
 dz=0.
\end{aligned}
\end{equation}
Subtract a multiple of \eqref{BosKdV6} for $p=0$ and $\underline\beta=\underline \alpha$,   this gives 
\begin{equation}
\label{BosKdV5}
\begin{aligned}
{\rm Res}_{z=0}\sum_{j=1}^s &\left\{
\exp\left(-\sum_{i=1}^\infty \frac\partial{\partial \overline  t^{(j)}_i }\frac{z^{-j}}{j}\right)-1\right\}\left(
\frac{D_p(\tau_{\underline\beta}(x,t))}{\tau_{\underline\beta}(x,t)}
\right)V^+(\underline\alpha; x,t)E_{jj} V^-(\underline\alpha; \overline x,\overline  t)^t 
 dz=0.
\end{aligned}
\end{equation}
Introduce the pseudo-differential operator $S(\underline \beta; x,t,\partial)$ by
\[
S(\underline \beta; x,t,z)=\sum_{j=1}^s \left\{
\exp\left(-\sum_{i=1}^\infty \frac\partial{\partial \overline  t^{(j)}_i }\frac{z^{-j}}{j}\right)-1\right\}\left(
\frac{D_p(\tau_{\underline\beta}(x,t))}{\tau_{\underline\beta}(x,t)}
\right)E_{jj}.
\]
Since \eqref{prop20a} holds $\partial S(\underline \beta; x,t,\partial)=S(\underline \beta; x,t,\partial)\partial$.
Hence, using the fundamental Lemma, i.e.  Lemma 4.1 of \cite{KLmult}, we deduce that
\[
\begin{aligned}
0&=
(P^+(\underline \alpha;x,t,\partial)R^+(\underline\alpha;\partial )S(\underline \beta; x,t,\partial)
R^+(\underline\alpha;\partial )^{-1}P^+(\underline \alpha;x,t,\partial)^{-1})_-\\
&=(P^+(\underline \alpha;x,t,\partial) S(\underline \beta; x,t,\partial)
 P^+(\underline \alpha;x,t,\partial)^{-1})_-\\
&=P^+(\underline \alpha;x,t,\partial) S(\underline \beta; x,t,\partial)
 P^+(\underline \alpha;x,t,\partial)^{-1}.
\end{aligned}
\]
Thus $S(\underline \beta; x,t,\partial)=0$,  which gives that 
\[
\frac{\partial}{\partial t_i^{(j)}}
\left(
\frac{D_p(\tau_{\underline\beta}(x,t))}{\tau_{\underline\beta}(x,t)}
\right)=0, \quad 1\le j\le s,\quad i=1,2,\ldots .
\]
Hence,
\[
\frac{D_p(\tau_{\underline\beta}(x,t))}{\tau_{\underline\beta}(x,t)}=C_p\in\mathbb C
\]
and because of \eqref{prop20b},   $C_p$ must be the same for all $\underline \beta$.  Thus \eqref{BosKdV3} holds.
\hfill$\square$
\subsection{$\hat{so}_{2n}$ and the $(\lambda,\mu)$-reduced DKP hierarchy}
\label{subs7.2}
The group $\hat{SO}_{2n}$ leaves the module  $F_d^{\overline a}$ invariant. It consists of all  elements $g\in D_\infty$
 that satisfy $g\phi_ig^{-1}=\sum_{i} a_{ji}\phi_j$, with  $\sum_{k} a_{ki}a_{-k+1,j}=\delta_{i},$ and hence  $\sum_{k} a_{ik}a_{j,-k+1}=\delta_{i+j,1}$, and $a_{i+2n,j+2n}=a_{ij}$. Since for $p=0,1,2,\ldots$
\[
\begin{aligned}
(g\otimes g) 
\sum_{i\in\mathbb{Z}
} \phi_i \otimes \phi_{2pn-i+1}
&=\sum_{i\in\mathbb{Z}
}g \phi_i g^{-1}g\otimes g \phi_{2pn-i+1}g^{-1} g\\
&=\sum_{i,j,k\in \mathbb{Z}} a_{ji} a_{k,2pn-i+1}\phi_j g\otimes \phi_k g\\
&=\sum_{i,j,k\in\mathbb{Z}} a_{ji} a_{k-2pn,-i+1}\phi_j g\otimes \phi_k g\\
 &=\sum_{j\in\mathbb{Z}} \phi_jg\otimes \phi_{2pn-j+1}g\\
&=\sum_{j\in\mathbb{Z}
} \phi_j\otimes \phi_{2pn-j+1} (g\otimes g) 
,
\end{aligned}
\]
and 
\[
 \sum_{i\in\mathbb{Z}
} \phi_i |\overline a\rangle\otimes \phi_{2pn-i+1}|\overline a\rangle =0,
\]
  we have that elements $\tau_a\in F_d^{\overline a}$, which belong to the  $\hat{SO}_{2n}$-group orbit of $|\overline a\rangle$ satisfy
\[
\sum_{i\in\mathbb{Z}
} \phi_i \tau_a \otimes \phi_{2pn-i+1}\tau_a=0,\quad\mbox{for }p=0,1,2,\ldots .
\]
This equation can be rewritten, with respect to the relabeling with respect to $(\lambda,\mu)$ to:
\begin{equation}
\label{SD2n}
\left(\sum_{b=1}^s\sum_{i\in\frac{1}{2}+\mathbb{Z}}\left( \psi_i^{+b} \otimes \psi_{p\lambda_b-i}^{-b}
+ \psi_i^{-b} \otimes \psi_{p\lambda_b-i}^{+b}\right)+
\sum_{c=s+1}^{r+s}\sum_{i\in\mathbb{Z}}(-1)^i\tilde\phi^c_i\otimes \tilde\phi^c_{2p\mu_c-i}\right)(\tau_a\otimes \tau_a)=0,
\end{equation}
{\bf Note that $r$ is even in this case.}
In terms of the fields we obtain:
\begin{equation}
\label{SD2nfield}
\begin{aligned}
{\rm Res}_{z=0}& 
\left(\sum_{b=1}^s  z^{p\lambda_b}\left( \psi^{+b}(z) \otimes \psi^{-b}(z)
+   \psi^{-b}(z) \otimes \psi^{+b}(z)\right)+\right.
\\
&\qquad+\left.
\sum_{c=s+1}^{r+s}z^{2p\mu_c-1} \tilde\phi^c(z)\otimes \tilde\phi^c(-z)\right)(\tau_a\otimes \tau_a)dz=0,\quad \mbox{for }p=0,1,2,\ldots .
\end{aligned}
\end{equation}
Now using $\sigma$ and substituting the vertex operators \eqref{intro1.18} and \eqref{intro1.18b} we get the following hierarchy of differential equations.
We write
\[
\sigma(\tau_a)=\tau^a(q,t,\theta) =\sum_{\underline k\in \mathbb Z^s,\underline \ell\in \mathbb Z_2^{\frac{r}2},\, |\underline k|+|\underline\ell|=a\, {\rm mod}\,  2}\tau_{\underline k,\underline \ell}(t)q^{\underline k}\theta^{\underline\ell},
\]
where $q^{\underline k}=q_1^{k_1}q_2^{k_2}\cdots q_s^{k_s}$ and  $\theta^{\underline\ell}=\theta_1^{\ell_1}\theta_2^{\ell_2}\cdots\theta_{\frac{r}2}^{\ell_{\frac{r}2}}$. Then \eqref{SD2nfield} turns into the following set of equations for all $\underline k,\overline{\underline k}\in \mathbb Z^s$, $\underline \ell,\overline{\underline\ell}\in \mathbb Z_2^{\frac{r}2}$, such that $|\underline k|+|\underline\ell|+1=|\overline{\underline k}|+|\overline{\underline\ell}|+1=a\mod 2$ and 
   $p=0,1,2,\ldots$,   
\begin{equation}
\label{BosDKPred}
\begin{aligned}
&{\rm Res}_{z=0}\big(\sum_{b=1}^s \big(
(-1)^{|\underline k+\overline{\underline k}|_{b-1}} z^{ p\lambda_b+k_b-\overline k_b-2+2\delta_{bs}}
e^{z\cdot t^{(b)}-z\cdot \overline t^{(b)}}e^{z^{-1}\cdot \tilde\partial_{\overline t^{(b)}}-z^{-1}\cdot \tilde\partial_{t^{(b)}}}
\tau_{\underline k+\underline e_s-\underline e_b,\underline\ell}(t)\tau_{\overline{\underline k}+\underline e_b-\underline e_s,\overline{\underline \ell}}(\overline t)+
\\
&\qquad\qquad-(-1)^{|\underline k+\overline{\underline k}|_{b-1}} (-z)^{ p\lambda_b-k_b+\overline k_b-2-2\delta_{bs}}
e^{(-z)\cdot \overline t^{(b)}-(-z)\cdot t^{(b)}}
\times\\  &\qquad \qquad \qquad \qquad \qquad 
e^{(-z)^{-1}\cdot \tilde\partial_{t^{(b)}}-(-z)^{-1}\cdot \tilde\partial_{\overline t^{(b)}}}
\tau_{\underline k+\underline e_s+\underline e_b,\underline\ell}(t)\tau_{\overline{\underline k}-\underline e_b-\underline e_s,\overline{\underline\ell}}(\overline t)\big)+\\
&+\frac12
\sum_{c=1}^{r} (-1)^{{|\underline k+\overline{\underline k}|}+|\underline\ell+\overline{\underline\ell}|_{\left[\frac{c-1}2\right]}+(c-1)(\ell_{\left[\frac{c+1}2\right]}+\overline\ell_{\left[\frac{c+1}2\right]}+1)}z^{2p\mu_c-1}
e^{z\circ t^{(s+c)}-z\circ \overline t^{(s+c)}}
\times\\  &\qquad \qquad \qquad \qquad \qquad 
e^{2(z^{-1}\circ (\tilde\partial_{\overline t^{(s+c)}}-\circ \tilde\partial_{t^{(s+c)}})}
%
\tau_{\underline k+\underline e_s ,\underline\ell+\underline e_{\left[\frac{c+1}2\right]}}(t)
\tau_{\overline{\underline k} -\underline e_s,\overline{\underline\ell}+\underline e_{\left[\frac{c+1}2\right]}}(\overline t)
\big) dz=0.
\end{aligned}
\end{equation}
Note, that all the equations for $p=0$ form the DKP hierarchy.
\\
\
\\
{\bf The case $r=0$}\\
Assume that $\mu=0$ thus $r=0$. The wave operators $V^{\pm}(\underline \alpha: x,t, z)$ also satisfy
($p=0,1,\ldots$):
\begin{equation}
\label{BosDKPred6b}
{\rm Res}_{z=0}
\sum_{j=1}^s
V^+(\underline \alpha; x,t,z)
\left(z^{p\lambda_j}E_{jj}
-(-z)^{p\lambda_j}E_{s+j,s+j}
\right)
V^-(\underline \beta;\overline x, \overline t,z)^t dz=0.
\end{equation}
Using the fundamental lemma (see \cite{KLmult},  Lemma 4.1),  we deduce the following expression for the matrix pseudo-differential operators
\begin{equation}
\label{DDD2}
\sum_{j=1}^s
(P^+(\underline \alpha;x,t,\partial) R^+(\underline \alpha;\partial)\left(\partial^{p\lambda_j}E_{jj}+(-\partial)^{p\lambda_j}E_{s+j,s+j}\right)R^-(\underline\beta;x,t,\partial)^*
P^+(\underline \beta; x,t,\partial)^{-1})_-=0.
\end{equation}
From this it is obvious that ${\cal L}= \sum_{j=1}^s L^{\lambda_j}E^j$ is a differential operator that satisfies
\[
\frac{\partial \cal L}{\partial t_i^j}=[(L^iD^j)_+,\cal L].
\]
If we define ${\cal M}= \sum_{j=1}^s L^{\lambda_j}D^j$,  then ${\cal L}^2={\cal M}^2$.\\
%
%
%
%
\ \\
{\bf The case $r>0$ (first approach)}
Recall that $r$ is always even.  Following the construction of Section \ref{section4},  we set  all $t^{s+c}_j=\overline t^{s+c}_j=0$, for $1\le c\le r$ in 
Equation \eqref{BosDKPred} and replace all other $t^a_1$ (resp. $\overline t^a_1$) by $t^a_1+x$ (resp. $\overline t^a_1=\overline x$). Let $V^{\pm}(\alpha; x,t,z)$ be given as before (but of course with the substitution $t^{s+c}_j =0$) and $W^{\pm}(\alpha; x,t,z)$ as in \eqref{WW1}, then \eqref{BosDKPred} turns into
\begin{equation}
\label{BosDKPred5}
\begin{aligned}
{\rm Res}_{z=0}&
\sum_{j=1}^s
V^+(\underline \alpha; x,t,z)
\left(z^{p\lambda_j}E_{jj}-(-z)^{p\lambda_j}E_{s+j,s+j}\right)
V^-(\underline \beta;\overline x, \overline t,z)^t dz=\\ &\qquad{\rm Res}_{z=0}\frac12 \sum_{i=1}^r W^+(\underline \alpha; x,t,z) z^{2p\mu_i-1}E_{ii}
W^-(\underline \beta;\overline x, \overline t,z)^{t}dz.
\end{aligned}
\end{equation}
Applying the fundamental lemma  gives
\begin{equation}
\begin{aligned}
\label{DD22}
\sum_{j=1}^s
&(P^+(\underline \alpha;x,t,\partial) R^+(\underline \alpha;\partial)\left(\partial^{p\lambda_j}E_{jj}-(-\partial)^{p\lambda_j}E_{s+j,s+j}\right)R^-(\underline\beta;x,t,\partial)^*
P^-(\underline \beta; x,t,\partial)^*)_-=\\
&(P^+(\underline \alpha;x,t,\partial) R^+(\underline \alpha;\partial)\left(\partial^{p\lambda_j}E_{jj}+(-\partial)^{p\lambda_j}E_{s+j,s+j}\right)R^-(\underline\beta;x,t,\partial)^*
P^+(\underline \beta; x,t,\partial)^{-1}J)_-=\\
&\qquad \frac12 
\sum_{i=1}^r 
\sum_{n=0}^{2p\mu_i} W^+(\underline \alpha; x,t)^n E_{ii}\partial^{-1}
\left(W^-(\underline \beta;x, t)^{2p\mu_i-n  }\right)^t.
\end{aligned}
\end{equation}
Note that if $p=0$ and  $\underline\alpha=\underline \beta$, the right-hand side of \eqref{DD22} is equal to 0.
Set $\underline\alpha=\underline \beta$ in \eqref{DD22},  and define $L$ and $D^j$ as before,   then   ${\cal L}= \sum_{j=1}^s L^{\lambda_j}E^j$  satisfies the Lax equation 
$\frac{\partial \cal L}{\partial t^a_j}=[ (L^j D^a)_+,{\cal L}]$   and 
is equal to
\[
{\cal L}(\underline\alpha;x,t)={\cal L}(\underline\alpha;x,t)_++\frac12 \sum_{i=1}^r 
\sum_{n=0}^{2\mu_i} W^+(\underline \alpha; x,t)^n E_{ii}\partial^{-1}
\left(JW^-(\underline \alpha;x, t)^{2\mu_i-n  }\right)^t.
\]
\
\\
{\bf The case $r>0$ (second approach)}
We start with the following observation.
\begin{remark}
\label{mur}
Note that in principle we can permute the elements of $\mu$ in such a way that $\mu_r$ is no longer the smallest part of $\mu$.  We will do this allowing $\mu_r$ to have any value greater than zero.  We will assume that all  the other $\mu_i$ are still in decreasing order,  hence that $(\mu_1,\mu_2,\ldots ,\mu_{r-1})$ is a genuine partition.  From now on always whenever we use the second approach we will assume this, viz. that $\mu_r$ is no longer the smallest part of $\mu$.
\end{remark}
As before we assume that $r\ge 1$, in this case $s$ can be 0 and  $r=2p$         
We follow the approach of  {\bf The DKP hierarchy (second approach)}   and define the $(2s+1)\times(2s+1)$-matrix valued wave functions by
\eqref{VPMB2},  where the coefficients  $P^\pm (\underline \alpha;x,t,\pm z)$ are given by \eqref{VVV} and \eqref{vvv} and the other matrices by \eqref{VVVV}. The $(2s+1)\times (r-1)$ is 
$W^{\pm}(\underline\alpha;x,t,z) $ is again defined as  in \eqref{WW2} and \eqref{WW3} and $T^{\pm}\underline\alpha;x,t)=0$ in this case, because $r=2p$.
Equation \eqref{BosDKPred} now translates to
Then we obtain the following bilinear identity:
\begin{equation}
\label{BosBKPHIR2Dred}
\begin{aligned}
{\rm Res}_{z=0}&
    \tilde V^+(\underline \alpha; x,t,z) 
   J_k(z)
\tilde V^-(\underline \beta;\overline x, \overline t,z)^t dz=\\ 
 &{\rm Res}_{z=0}\frac12 W^+(\underline \alpha; x,t,,z) M_k(z) W^-(\underline \beta;\overline x, \overline t,z)^{t} dz
,\quad k=0,1,2,\ldots ,
\end{aligned}
\end{equation}
where   
\begin{equation}
\label{J_k(z)}
J_k(z)=z^{2\mu_r k-1}E_{2s+1,2s+1}+\sum_{j=1}^s z^{\lambda_j k}E_{jj}-(-z)^{\lambda_j k}E_{s+j,s+j},
\quad
M_k(z)=\sum_{j=1}^{r-1}z^{2\mu_j k-1}E_{j,j}.
\end{equation}
Note that $J(\partial)=J_0(\partial)$.
Using the fundamental lemma,  we obtain 
\begin{equation}
\label{WW}
\begin{aligned}
\left(P^{+}(\underline\alpha;x,t,\partial)J_k(\partial)P^{-}(\underline\alpha;x,t,\partial)^*\right)_-\\
=\frac12
\sum_{a=1}^{r-1}\sum_{j=0}^{2k\mu_a}
W^+(\underline \alpha; x,t)^j E_{aa}\partial^{-1} \left(W^-(\underline \alpha; x,  t)^{2k\mu_a-j}\right)^{t},
\end{aligned}
\end{equation}
Then for $k=0$ we again obtain \eqref{sato1B},  hence we define $\hat V^{\pm }$ and $\hat P^{\pm}$ as in \eqref{hatV},
then we obtain \eqref{SATO1B}, \eqref{SATOB2} and \eqref{SATOB3}.  
Hence we can define  the Lax operators as in \eqref{BLax} and they clearly satisfy
\eqref{BLax2}.  Define now 
$\hat W^\pm(\underline \alpha; x,t,z)=(P^\pm(\underline \alpha; x,t,z)^0)^{-1}W^\pm(\underline \alpha; x,t,z)$,
then \eqref{WW}
gives
 \begin{equation}
\label{WW22}
\begin{aligned}
\left(\hat P^{+}(\underline\alpha;x,t,\partial)J_k(\partial)J(\partial)^{-1}\hat P^{-}(\underline\alpha;x,t,\partial)^{-1}J(\partial)\right)_- J(\partial)^{-1}=
\\
=\frac12
\sum_{a=1}^{r-1}\sum_{j=0}^{2k\mu_a}
\hat W^+(\underline \alpha; x,t)^j E_{aa}\partial^{-1} \left( \hat  W^-(\underline \alpha; x,  t)^{2k\mu_a-j}\right)^{t}J(\partial)^{-1}
\end{aligned}
\end{equation}
and  
\[
{\cal L}(\underline \alpha; x,t,\partial)=L(\underline \alpha; x,t,\partial)^{2\mu_r} D^{s+r}(\underline \alpha; x,t,\partial)+\sum_{a=1}^s
L(\underline \alpha; x,t,\partial)^{\lambda_a} E^{a}(\underline \alpha; x,t,\partial).
\]
Then clearly ${\cal L} $  satisfies the same Lax equations and
\[
{\cal L}(\underline \alpha; x,t,\partial)={\cal L}(\underline \alpha; x,t,\partial)_{\ge}+\frac12 \sum_{a=1}^{r-1}\sum_{j=0}^{2\mu_a}
\hat W^+(\underline \alpha; x,t)^j E_{aa}\partial^{-1} \left(\hat W^-(\underline \alpha; x,  t)^{2\mu_a-j}\right)^{t}J(\partial)^{-1}.
\]

\subsection{$\hat{so}_{2n+1}$ and the $(\lambda,\mu)$-reduced DKP hierarchy}
\label{SecDred}
The group $\hat{SO}_{2n+1}$  acs on both $F_b$ and $F_d$ and leaves the module  $F_d^{\overline a}$ invariant \cite{KvdLB}. It consists of all  elements  $g\in B_\infty$ (resp. $D_\infty$)
 that satisfy $g\tilde\phi_jg^{-1}=\sum_{i} a_{ji}\tilde\phi_i$ (N.B.  in the $D_\infty$-case the tildes have to be removed, since in that case this is a genuine field), with  $\sum_{k} a_{ik}a_{j,-k}=\delta_{i,-j}$ and $a_{i+2n+1,j+2n+1}=a_{ij}$.  Let $\epsilon=0$ for $F_b$ and $ \frac 12$ for $F_d$. Since for $p=0,1,2,\ldots$
\[
\begin{aligned}
(g\otimes g) 
\sum_{i\in\epsilon +\mathbb{Z}
} \tilde\phi_i \otimes \tilde\phi_{p(2n+1)-i}
&=\sum_{i\in\epsilon +\mathbb{Z}
}g \tilde\phi_i g^{-1}g\otimes g \tilde\phi_{p(2n+1)-i}g^{-1} g\\
&=\sum_{i,j,k\in\epsilon+\mathbb{Z}} a_{ji} a_{k,p(2n+1)-i}\tilde\phi_j g\otimes \tilde\phi_k g\\
&=\sum_{i,j,k\in\epsilon +\mathbb{Z}} a_{ji} a_{k-p(2n+1)-i}\tilde\phi_j g\otimes \tilde\phi_k g\\
 &=\sum_{j\in\epsilon +\mathbb{Z}} \tilde\phi_jg\otimes \tilde\phi_{p(2n+1)-j}g\\
&=\sum_{i\in\epsilon+\mathbb{Z}
} \tilde\phi_i \otimes \tilde\phi_{p(2n+1)-i} (g\otimes g) 
\end{aligned}
\]
and 
\[
 \sum_{i\in \mathbb{Z}
} \tilde\phi_i |0\rangle\otimes \tilde\phi_{p(2n+1)-i}|0\rangle =\frac{\delta_{p0}}2 |0\rangle\otimes|0\rangle,\quad 
 \sum_{i\in\frac12 +\mathbb{Z}
} \phi_i |\overline a\rangle\otimes\phi_{p(2n+1)-i}|\overline a\rangle =0,
\]
  we have that elements $\tau\in F_b$ (resp. $\tau_a\in F_d^{\overline a}$), which belong to the  $\hat{SO}_{2n+1}$-group orbit of $|0\rangle$ (resp. $|\overline a\rangle$) satisfy
\begin{equation}
\label{BDREDUCTION}
\sum_{i\in \mathbb{Z}
} \phi_i T \otimes \phi_{p(2n+1)-i}\tau=\frac{\delta_{p0}}2 \tau\otimes \tau\quad(\mbox{resp. }
\sum_{i\in\frac12 +\mathbb{Z}
} \tilde\phi_i \tau_a \otimes \tilde\phi_{p(2n+1)-i}\tau_a=0),\quad\mbox{for }p=0,1,\ldots .
\end{equation}
In this section we want to obtain this as a reduction of the DKP hierarchy,  hence we assume that the second equation of \eqref{BDREDUCTION} holds. 
In terms of the fermions related to $(\lambda, \mu)$, this is 
\begin{equation}
\label{SD}
\begin{aligned}
&\sum_{b=1}^s\sum_{i\in\frac{1}{2}+\mathbb{Z}}\left( \psi_i^{+b}\tau_a \otimes \psi_{p\lambda_b-i}^{-b}\tau_a+ \psi_i^{-b}\tau_a \otimes \psi_{p\lambda_b-i}^{+b}\tau_a\right)+\\
&+\sum_{c=s+1}^{r+s}\sum_{i\in\mathbb{Z}}(-1)^i\tilde\phi^c_i\tau_a\otimes \tilde\phi^c_{2p \mu_c-i}\tau_a
+\sum_{j\in \delta+\mathbb Z} \sigma_j\tau_a\otimes\sigma_{p-j}\tau_a=0.
\end{aligned}
\end{equation}
For the value of $\delta$,  see the two  possibilities below \eqref{FockB}.  $\delta=0$ if $r=2r'-1$ is odd and $\delta=-\frac12$ if $r=2r'$ is even.  This equation for $p=0$ is in fact the  $(s,R)$-component DKP Equation \eqref{SD2evenfield} when $r=2r'-1$  and $R=r+1=2r'$ and Equation \eqref{SD2oddfield} with $p$ replaced by $r'+1$ if $r=2r'$ and $R=r+1=2r'+1$.  In that case, we have the following relation,   for $R=r+1$:
\begin{equation}
\label{psisigma}
\begin{aligned}
r=2r'-1:&\quad \sigma_j
=\begin{cases}
\tilde\phi_j^{R}&\mbox{if } j\ \mbox{is even},\\
\sqrt{-1}\tilde\phi_j^{R}&\mbox{if } j\ \mbox{is odd},
\end{cases}\\
r=2r':&\quad \sigma_{j+\frac12}
=\begin{cases}
\frac{\tilde\phi_{0}^{R}-\sqrt{-1}\sigma_0}{\sqrt 2}&\mbox{if } j=-1,\\
\frac{\tilde\phi_{0}^{R}+\sqrt{-1}\sigma_0}{\sqrt 2}&\mbox{if } j=0,\\
\tilde\phi_j^{R}&\mbox{if } 0<j\ \mbox{is even or }j<-1\ \mbox{odd} ,\\
\sqrt{-1}\tilde\phi_j^{R}&\mbox{if } 0<j\ \mbox{is odd or }j<-1\ \mbox{even}.
\end{cases}
\end{aligned}
\end{equation}
We do not bosonize the field $\sigma(z)=\tilde\phi^{R}(z)=\tilde\phi^{r+1}(z)$ with a vertex operator,  because the corresponding Heisenberg algebra,  which is given by $\frac12 :\tilde\phi^{r+1}(z)\tilde\phi^{r+1}(-z):$ does not belong to $\hat{so}_{2n+1}$.   Instead we define $\sigma\sigma(z)\sigma^{-1}= \Xi(z)$, see
\eqref{sigmaXi}.
We first rewrite the Equations \eqref{SD} using the fermionic fields
\begin{equation}
\label{SD2}
\begin{aligned} &{\rm Res}_{z=0}\big\{
\sum_{b=1}^s z^{\lambda_b}\left( \psi^{+b}(z)\tau_a \otimes \psi^{-b}(z)\tau_a+ \psi^{-b}(z)\tau_a \otimes \psi ^{+b}(z)\tau_a\right)
+\\
&+\sum_{c=s+1}^{r+s} z^{2p\mu_c-1}\tilde\phi^c(z)\tau_a\otimes \tilde\phi^c(-z)\tau_a
+z^{p-1-2\delta}\sigma(z)\tau_a\otimes\sigma(z)\tau_a\big\}dz=0,
\end{aligned}
\end{equation}
and then apply the bosonization $\sigma$.  Note that 
\[
\sigma(\tau_a)=\sum_{\underline k\in \mathbb Z^s,\underline \ell\in \mathbb Z_2^{r'}, \alpha\in \mathbb Z_2; \, \underline k+\underline \ell+\alpha= a\, {\rm mod}\; 2} q^{\underline k}\theta^{\underline\ell} \tau_{\underline k,\underline \ell, \alpha}(t, \xi ) .
\]
where $q^{\underline k}=q_1^{k_1}q_2^{k_2}\cdots q_s^{k_s}$, $\theta^{\underline\ell}=\theta_1^{\ell_1}\theta_2^{\ell_2}\cdots\theta_{r'}^{\ell_{r'}}$.
Let $N_\xi (f(t)\xi_{m_1}\xi_{m_2}\cdots \xi_{m_p})=p$,
Then $(-1)^{N_\xi}(\tau_{\underline k,\underline \ell, \alpha}(t, \xi )=(-1)^\alpha (\tau_{\underline k,\underline \ell, \alpha}(t, \xi )$.
This gives for    \eqref{SD2}:
($p=0,1,2,\ldots$)
\begin{equation}
\label{BosDBKPred}
\begin{aligned}
{\rm Res}_{z=0}&\big(\sum_{b=1}^s \big(
(-1)^{|\underline k+\overline{\underline k}|_{b-1}} z^{ p\lambda_b+k_b-\overline k_b-2+2\delta_{bs}}
e^{z(\cdot t^{(b)}- \overline t^{(b)})}e^{z^{-1}\cdot (\tilde\partial_{\overline t^{(b)}}-  \tilde\partial_{t^{(b)}})}
\tau_{\underline k+\underline e_s-\underline e_b,\underline\ell, \alpha}(t,\xi)\tau_{\overline{\underline k}+\underline e_b-\underline e_s,\overline{\underline \ell}, \overline \alpha}(\overline t,\overline \xi)
\\
&+(-1)^{|\underline k+\overline{\underline k}|_{b-1}} z^{ p\lambda_b-k_b+\overline k_b-2-2\delta_{bs}}
e^{(-z)\cdot (t^{(b)}-\overline t^{(b)}}
e^{(-z)^{-1}\cdot( \tilde\partial_{t^{(b)}}-\tilde\partial_{\overline t^{(b)}})}
%
\tau_{\underline k+\underline e_s+\underline e_b,\underline\ell, \alpha}(t,\xi)\tau_{\overline{\underline k}-\underline e_b-\underline e_s,\overline{\underline\ell},\overline\alpha}(\overline t,\overline \xi) \\
&+\frac12
\sum_{c=1}^{r} (-1)^{{|\underline k+\overline{\underline k}|}+|\underline\ell+\overline{\underline\ell}|_{\left[\frac{c-1}2\right]}+(c-1)(\ell_{\left[\frac{c+1}2\right]}+\overline\ell_{\left[\frac{c+1}2\right]}+1)}z^{2p\mu_c-1}
e^{z\circ (t^{(s+c)}- \overline t^{(s+c)})}
\times\\  &\qquad \qquad \qquad \qquad \qquad 
e^{2(z^{-1}\circ (\tilde\partial_{\overline t^{(s+c)}}- \tilde\partial_{t^{(s+c)}}))}
%
\tau_{\underline k+\underline e_s ,\underline\ell+\underline e_{\left[\frac{c+1}2\right]},\alpha}(t,\xi)
\tau_{\overline{\underline k} -\underline e_s,\overline{\underline\ell}+\underline e_{\left[\frac{c+1}2\right]},\overline\alpha}(\overline t,\overline \xi)
\big) dz
\\
&+
(-1)^{{|\underline k+\overline{\underline k}|}+|\underline\ell+\overline{\underline\ell}|}
\big\{
\sum_{j=1}^\infty 
\xi_{j+\delta}\tau_{\underline k+\underline e_s ,\underline\ell,\alpha+1}(t,\xi)\frac{\partial 
\tau_{\overline{\underline k} -\underline e_s,\overline{\underline\ell},\overline\alpha+1 }(\overline t, \overline\xi)}{\partial \overline\xi_{j+p+\delta}} 
\\
&\quad +
\frac{\partial \tau_{\underline k+\underline e_s ,\underline\ell,\alpha+1}(t,\xi)}{\partial \xi_{j+\delta}}\overline\xi_{j+p+\delta}
\tau_{\overline{\underline k} -\underline e_s,\overline{\underline\ell},\overline\alpha+1 }(\overline t, \overline\xi)
%
%
+ \sum_{j=1}^{p-1-2\delta}
\frac{\partial \tau_{\underline k+\underline e_s ,\underline\ell,\alpha+1}(t,\xi)}{\partial \xi_{j+\delta}}
\frac{\partial 
\tau_{\overline{\underline k} -\underline e_s,\overline{\underline\ell} ,\overline\alpha+1}(\overline t, \overline\xi)}{\partial \overline\xi_{p-j-\delta}} 
\\
&\quad +(1+2\delta)(\delta_{p0}-1)\frac{\sqrt{-1}}{\sqrt 2}
\bigg(
(-1)^{\ell_{\frac{r+1}2}}
\tau_{\underline k+\underline e_s ,\underline\ell +\underline e_{\frac{r+1}2},\alpha}(t,\xi)
\frac{\partial 
\tau_{\overline{\underline k} -\underline e_s,\overline{\underline\ell},\overline\alpha+1 }(\overline t, \overline\xi)}{\partial \overline\xi_{p}} 
\\
&  \qquad\qquad\qquad\qquad\qquad\qquad+
(-1)^{\overline\ell_{\frac{r+1}2}}
\frac{\partial \tau_{\underline k+\underline e_s ,\underline\ell ,\alpha+1}(t,\xi)}{\partial \xi_{p}}
\tau_{\overline{\underline k} -\underline e_s,\overline{\underline\ell}+\underline e_{\frac{r+1}2},\overline\alpha}(\overline t,\overline\xi )
\bigg)
\\
&\qquad-\frac{\delta_{p0}}2(1+2\delta) (-1)^{\ell_{\frac{r+1}2}+\underline\ell_{\frac{r+1}2}}
\tau_{\underline k+\underline e_s ,\underline\ell+\underline e_{\frac{r+1}2},\alpha}(t,\xi)
\tau_{\overline{\underline k} -\underline e_s,\overline{\underline\ell}+\underline e_{\frac{r+1}2},\overline\alpha}(\overline t,\overline\xi )
\big\}=0
.
\end{aligned}
\end{equation}
As for $\hat{so}_{2n}$, we will consider several cases.\\
\
\\
{\bf Case 1,  assume that $\mu=\emptyset$, thus that $r=0$.}
In this case $\delta=-\frac12$ and
\[
\sigma(\tau_a)=\sum_{\underline k\in \mathbb Z^s,\nu\in \mathbb Z_2; \, \underline k+\nu= a\, {\rm mod}\; 2} q^{\underline k}  \tau_{\underline k,\nu}(t, \xi ) ,\quad \mbox{with }
(-1)^{N_\xi}(\tau_{\underline k,\nu}(t, \xi ))=(-1)^\nu (\tau_{\underline k,\nu}(t, \xi )),
\]
we find when substituting   $\xi=\overline \xi=0$,  that $\nu=\overline \nu=\overline 0$:
($p=0,1,2,\ldots$)
\begin{equation}
\label{NewBosDBKPred}
\begin{aligned}
&{\rm Res}_{z=0}\big(\sum_{b=1}^s \big(
(-1)^{|\underline k+\overline{\underline k}|_{b-1}} z^{ p\lambda_b+k_b-\overline k_b-2+2\delta_{bs}}
e^{z\cdot (t^{(b)}-\overline t^{(b)})}
\times\\  &\qquad
e^{z^{-1}\cdot (\tilde\partial_{\overline t^{(b)}}- \tilde\partial_{t^{(b)}})}
\tau_{\underline k+\underline e_s-\underline e_b,\underline\ell, \alpha}(t,0)\tau_{\overline{\underline k}+\underline e_b-\underline e_s,\overline{\underline \ell}, \overline \alpha}(\overline t, 0)+
(-1)^{|\underline k+\overline{\underline k}|_{b-1}} z^{ p\lambda_b-k_b+\overline k_b-2-2\delta_{bs}}
\times
\\
&\qquad\qquad
e^{(-z)\cdot (t^{(b)}-\overline t^{(b))}}
e^{(-z)^{-1}\cdot \tilde\partial_{t^{(b)}}-(-z)^{-1}\cdot \tilde\partial_{\overline t^{(b)}}}
%
\tau_{\underline k+\underline e_s+\underline e_b,\overline 0}(t,0)\tau_{\overline{\underline k}-\underline e_b-\underline e_s,\overline 0}(\overline t,0)\big\}
\big) dz
\\
&\qquad\qquad\qquad\quad=-
(-1)^{{|\underline k+\overline{\underline k}|} }
 \sum_{j=1}^{p}
\frac{\partial \tau_{\underline k+\underline e_s ,\overline 1}(t,\xi)}{\partial \xi_{j-\frac12}}\big|_{\xi=0}
\frac{\partial 
\tau_{\overline{\underline k} -\underline e_s ,\overline 1}(\overline t, \overline\xi)}{\partial \overline\xi_{p-j+\frac12}} \big|_{\overline\xi=0}
.
\end{aligned}
\end{equation}
\begin{remark}
Note that for $p=0$, the right hand side is equal to 0.  One obtains  a  similar equation in the DKP case when $r=1$ and one  substitutes $t^{s+1}_j=0$,   for all $j=1, 3,\ldots$, in Equation \eqref{BosDKP}.   In particular, we have that $P^{-*}=JP^{+-1}$ for $J$ as in \eqref{J}.
\end{remark}
Define,  now  for $j=\frac12, \frac32, \frac52,\ldots$,
\begin{equation}
\label{WD}
\begin{aligned}
T^{\pm j}(\underline\alpha;x,t)_i=&\sqrt{-1}\frac{(-1)^{|\underline \alpha|}}{\tau_{\underline\alpha,\overline 0}(x,t,0)}
\frac{\partial \tau_{\underline \alpha\pm\underline e_i,\overline 1}(x,t,\xi)}{\partial \xi_{j}}\big|_{\xi=0},\\
T^{\pm j}(\underline\alpha;x,t)_{s+i}=&\sqrt{-1}\frac{(-1)^{|\underline \alpha|}}{\tau_{\underline\alpha,\overline 0}(x,t,0)}
\frac{\partial \tau_{\underline \alpha\mp\underline e_i,\overline 1}(x,t,\xi)}{\partial \xi_{j}}\big|_{\xi=0}
\end{aligned}
\end{equation}
and $V^\pm (\underline \alpha,;x,t,z)$ as before.  For which one has
\begin{equation}
\label{BosBKPred6}
\begin{aligned}
{\rm Res}_{z=0}& \,
\sum_{j=1}^s
V^+(\underline \alpha; x,t,z)
\left(z^{p\lambda_j}E_{jj}
-(-z)^{p\lambda_j}E_{s+j,s+j}
\right)
V^-(\underline \beta;\overline x, \overline t,z)^t dz=\\
&=
\sum_{j=\frac12}^{p-\frac12}
T^{+ j}(\underline\alpha;x,t)T^{-(p-j)}(\underline\beta;\overline x,\overline t)^t
.
\end{aligned}
\end{equation}
Note that 
\[
T^{\pm j}(\underline\alpha;x,t)_i =T^{\mp j}(\underline\alpha;x,t)_{s+i}.
\]
 Thus 
\begin{equation}
\label{W+-}
P^\pm(\underline \alpha;x,t,\partial) = NP^\mp(\underline \alpha;x,t,\partial) N, \quad NT^{\pm j}(\underline\alpha;x,t)=T^{\mp j}(\underline\alpha;x,t)
\end{equation}
where $N$ is defined in \eqref{N}.
Let $\partial=\partial_x$ and using the fundamental lemma (see \cite{KLmult},  Lemma 4.1),  we deduce the following expression for the matrix pseudo-differential operators

\begin{equation}
\label{BB2}
\begin{aligned}
\sum_{j=1}^s&
(P^+(\underline \alpha;x,t,\partial) R^+(\underline \alpha;\partial)\left(\partial^{p\lambda_j}E_{jj}-(-\partial)^{p\lambda_j}E_{s+j,s+j}\right)R^-(\underline\beta;x,t,\partial)^*
P^-(\underline \beta; x,t,\partial)^*)_-\\
&=\sum_{j=\frac12}^{p-\frac12}
T^{+ j}(\underline\alpha;x,t)\partial^{-1}T^{-(p-j)}(\underline\beta; x, t)^t
.
\end{aligned}
\end{equation}
Take 
$p=0$ and $\underline\alpha=\underline \beta$ and $J= \sum_{j=1}^s E_{jj}-E_{s+j,s+j}=J^{-1}$, as in \eqref{J},  we obtain again  \eqref{sato3}.
However, taking $p\ne 0$,  we get 
\begin{equation}
\label{satoB4}
\begin{aligned}
\sum_{j=1}^s& (P^+(\underline \alpha; x,t,\partial)\left(\partial^{p\lambda_j}E_{jj}+(-\partial)^{p\lambda_j}E_{s+j,s+j}\right)P^+(\underline \alpha; x,t,\partial)^{-1})_-=
\\
&\qquad \sum_{k=\frac12}^{p-\frac12}
T^{+ k}(\underline\alpha;x,t)
\partial^{-1}(J
T^{- (p-k)}(\underline\alpha;x,t))^t
.
\end{aligned}
\end{equation}

Then $P^\pm (\alpha;x,t,\partial)$,  
$L(\alpha;x,t,\partial)$,  $D^j(\alpha;x,t,\partial)$,
$
E^j(\alpha;x,t,\partial)$ 
and
${\cal L}(\alpha;x,t,\partial)$ do not change and their relations also do not change,  except that  ${\cal L}(\alpha;x,t,\partial)_-$ is no longer equal to zero.
More general we have  
\[
\sum_{j=1}^s (L^{p\lambda_j}E^j)_-=\sum_{k=\frac12}^{p-\frac12}
T^{+ k}(\underline\alpha;x,t)
\partial^{-1}
(JT^{- (p-k)}(\underline\alpha;x,t))^t
\]
and in particular 
\[
{\cal L}_-=
T^{+ \frac12}(\underline\alpha;x,t)
\partial^{-1}
(JT^{- \frac12}(\underline\alpha;x,t))^t
.
\]
The adjoint of  $\cal L$ is $${\cal L}^*=K{\cal L}K^{-1},$$   for $K$  as in \eqref{K}.

If we differentiate in this case \eqref{BosBKPred6} for $\underline\alpha=\underline\beta$ and  $p=1$ by $t_i^j$ and use the fundamental Lemma  of \cite{KLmult},  we obtain 
\[
	\left(\frac{\partial P^+}{\partial t_i^j}(P^+)^{-1}{\cal L}+ L^iD^j{\cal L}\right)_-=\frac{\partial T^{+ \frac12}}{\partial t_i^j}\partial^{-1}{T^{- \frac12}}^t  J.
\]
Now substitute  the second equation of quation \eqref{sato3},  this gives
that 
\[
\begin{aligned}
\frac{\partial T^{+ \frac12} }{\partial t_i^j}\partial^{-1}{T^{- \frac12}}^TJ&=
\left(  \left( L^iD^j\right)_+{\cal L}\right)_-\\
&=
\left(  \left( 
L^i D^j
\right)_+
{\cal L}_-
\right)_-
\\
&=
\left(  \left( 
L^i D^j
\right)_+
 T^{+ \frac12} \partial^{-1}{T^{- \frac12}}^t J
\right)_-
\\
&=
\left( L^iD^j\right)_+\left(
T^{+ \frac12}\right) \partial^{-1}{T^{- \frac12}}^tJ.
\end{aligned}
\]
Thus 
\[
\frac{\partial T^{+ \frac12} }{\partial t_i^j}=\left( L^iD^j\right)_+\left(
T^{+ \frac12}\right).
\]
If we differentiate \eqref{BosBKPred6} for $\underline\alpha=\underline\beta$ and  $p=1$ by $\overline t_i^j$, we obtain in a similar way that
\[
\frac{\partial T^{- \frac12}J }{\partial t_i^j}=-\left( L^iD^j\right)^*_+\left(
T^{- \frac12}J\right).
\]
Thus $T^{+\frac12}$,  resp.   $T^{ -\frac12}J$,  is an eigenfunction,  resp. adjoint eigenfunction, for the operators $L$ and $D^j$,    $j=1,2, \ldots ,s$.
 \\
\
\\
{\bf Second case,  $\mu\ne \emptyset$  (first approach).}
In this case,
\[
\sigma(\tau_a)=\!\!\!\!\! \!\!\!\!\! \!\!\!\!\! 
\sum_{\underline k\in \mathbb Z^s,\underline\ell\in\mathbb Z_2^r,\nu\in \mathbb Z_2; \, \underline k+\underline\ell+\nu= a\, {\rm mod}\; 2} q^{\underline k} \theta^{\underline\ell} \tau_{\underline k,\nu}(t, \xi ) ,\quad \mbox{with }
(-1)^{N_\xi}(\tau_{\underline k,\nu}(t, \xi ))=(-1)^\nu (\tau_{\underline k,\nu}(t, \xi )).
\]
We substitute    $\xi=\overline \xi=t_j^c=\overline t_j^c=0$, for $c>s$  in \eqref{BosDBKPred}.
As before, replace all $t_1^a$ by $t_1^a+x$ and define the $2s\times 2s$-matrix  and $V^\pm (\underline \alpha;x,t,z)$ in the usual way,   but clearly only for $\nu=0$.  In this case we also have an $2s\times r$-matrix
$W^\pm (\underline \alpha,x,t,z)$,  defined as in \eqref{WW1} and \eqref{WW2} and the $2s\times 1$-matrixces
\begin{equation}
\label{TTD}
\begin{aligned}
T^{\pm (j+\delta)}(\underline\alpha;x,t)_i=&\sqrt{-1}\frac{(-1)^{|\underline \alpha|}}{\tau_{\underline\alpha,\overline 0}(x,t,0)}
\frac{\partial \tau_{\underline \alpha\pm\underline e_i,\overline 1}(x,t,\xi)}{\partial \xi_{j+\delta}}\big|_{\xi,t^{>s}=0},\\
T^{\pm (j+\delta)}(\underline\alpha;x,t)_{s+i}=&\sqrt{-1}\frac{(-1)^{|\underline \alpha|}}{\tau_{\underline\alpha,\overline 0}(x,t,0)}
\frac{\partial \tau_{\underline \alpha\mp\underline e_i,\overline 1}(x,t,\xi)}{\partial \xi_{j+\delta}}\big|_{\xi,t^{>s}=0} \quad \mbox{for }
j+\delta>0,
\end{aligned}
\end{equation}
and if $\delta=0$, we also have
\begin{equation}
\label{TTDD}\begin{aligned}
T^{\pm 0}(\underline\alpha;x,t)_i=\frac{(-1)^{|\underline \alpha|}
 \tau_{\underline \alpha\pm\underline e_i,\overline 0}(x,t,0)}{\sqrt 2\tau_{\underline\alpha,\overline 0}(x,t,0)}\big|_{t^{>s}=0},\quad
T^{\pm 0}(\underline\alpha;x,t)_{s+i}=&\frac{(-1)^{|\underline \alpha|}
 \tau_{\underline \alpha\mp\underline e_i,\overline 0}(x,t,0)} 
{\sqrt 2\tau_{\underline\alpha,\overline 0}(x,t,0)}
\big|_{t^{>s}=0}.
\end{aligned}
\end{equation}
For $p=0$, we thus obtain 
\begin{equation}
\label{BosDBKPredred}
\begin{aligned}
{\rm Res}_{z=0}
\sum_{j=1}^s
&
V^+(\underline \alpha; x,t,z)
\left( E_{jj}
- E_{s+j,s+j}
\right)
V^-(\underline \beta;\overline x, \overline t,z)^t dz= \\
 &\frac12 W^+(\underline \alpha; x,t)^0
W^-(\underline \beta;\overline x, \overline t)^{0t}
+
(1+2\delta)
T^{+0}(\alpha; x,t) 
T^{-0}(\beta; \overline x,\overline t)^t
,
\end{aligned}
\end{equation}
which is  similar to  \eqref{BosDKPr ne 0}.
Now substituting $\alpha=\beta$, gives that the right-hand side of \eqref{BosDBKPredred} is equal to $0$. And we are exactly in the situation of the first approach of the BKP hierarchy of Subsection \ref{section4}, and we can continue as in that section and defne the Lax operators in the same way as in that subsection.
For $p>0$, we have
\begin{equation}
\label{BosBKPred6a}
\begin{aligned}
&{\rm Res}_{z=0}
\sum_{j=1}^s
V^+(\underline \alpha; x,t,z)
\left(z^{p\lambda_j}E_{jj}
-(-z)^{p\lambda_j}E_{s+j,s+j}
\right)
V^-(\underline \alpha;\overline x, \overline t,z)^t dz=\\
&\frac12
\sum_{a=1}^{r}\sum_{j=0}^{2p\mu_a} 
\hat W^+(\underline \alpha; x,t)^j E_{aa} \left( \hat  W^-(\underline \alpha;\overline x, \overline t)^{2p\mu_a-j}\right)^{t}
+
\sum_{j=-\delta}^{p+\delta}
T^{+ j}(\underline\alpha;x,t)T^{-(p-j)}(\underline\alpha;\overline x,\overline t)^t
.
\end{aligned}
\end{equation}
Which gives that the integral part of the Lax operator 
${\cal L}= \sum_{j=1}^s L^{\lambda_j}E^j$ is equal to
\[
{\cal L}_-=\frac12
\sum_{a=1}^{r}\sum_{j=0}^{2\mu_a} 
\hat W^+(x,t)^j E_{aa} \partial^{-1}\left( \hat  W^-(   x,   t)^{2\mu_a-j}\right)^{t}J
+
\sum_{j=-\delta}^{1+\delta}
T^{+ j}( x,t)\partial^{-1}T^{-(1-j)}(  x,  t)^tJ.
\]
\
\\
 {\bf Third  case,  $\mu\ne \emptyset$  (second  approach).} As before,  we substitute    $\xi=\overline \xi=t_j^c=\overline t_j^c=0$, for $s<c<s+r$  in \eqref{BosBKPred}.   Here also Remark \ref{mur} holds.
Replace all $t_1^a$ by $t_1^a+x$,  for $a=1,2,\ldots,s, s+r$,  and define the $(2s+1)\times (2s+1)$-matrix   $V^\pm (\underline \alpha;x,t,z)$ as in \eqref{VPMB2},  \eqref{VVV}   and    \eqref{vvv},   but now with $\tau_{\underline \alpha}$ replaced by $\tau_{\underline \alpha,\overline 0}$.  In this case we also have a $(2s+1)\times (r-1)$-matrix
$W^\pm (\underline \alpha,x,t,z)$,   defined as in \eqref{WW1}  and  $(2s+1)\times 1$-matrices $T^{\pm  (j+\delta)}(\underline \alpha,x,t,z)$, defined as in \eqref{TTD}, \eqref{TTDD}, except that we do not put $t_j^{s+r}=0$,  and by
\begin{equation}
\label{TTDDD}
\begin{aligned}
T^{\pm (j+\delta)}(\underline\alpha;x,t)_{2s+1}=&\sqrt{-1}\frac{(-1)^{|\underline \alpha|}}{\tau_{\underline\alpha,\overline 0}(x,t,0)}
\frac{\partial \tau_{\underline \alpha ,\overline 1}(x,t,\xi)}{\partial \xi_{j+\delta}}\big|_{\xi,t^{(s<a<s+r)}=0},\\
T^{\pm 0}(\underline\alpha;x,t)_{2s+1}=&\frac{(-1)^{|\underline \alpha|}
 \tau_{\underline \alpha,\overline 0}(x,t,0)}{\sqrt 2\tau_{\underline\alpha,\overline 0}(x,t,0)}\big|_{t^{(s<a<s+r)}=0},\quad \mbox{only for }\delta=0.
\end{aligned}
\end{equation}
This  resembles the second approach of the reduction of the BKP hierarchy (cf. Subsection \ref{section4}).  We obtain thw following  equation for $p=0$, which is similar to Equation \eqref{BosDKPr ne 0 2!!}:
\begin{equation}
\label{BosDKPr ne 0 2!!b}
\begin{aligned}
\left( P^+(\underline \alpha; x,t,\partial)R^+(\underline \alpha-\underline\beta,\partial)J(\partial)
P^-(\underline \beta; x,  t,\partial)^*\right)_-=\\
 \frac12 W^+(\underline \alpha; x,t)^0\partial^{-1}
W^-(\underline \beta; x,  t)^{0t}+ 
(1+2\delta)
T^{+0}(\alpha; x,t) 
\partial^{-1}
T^{-0}(\beta; x,  t)^t
.
\end{aligned}
\end{equation} 
Now take  $\underline \alpha=\underline \beta$.  Again, the right-hand side is equal 
to 
$$
 \frac12 P^+(\underline \alpha; x,t)^0
E_{2s+1,2s+1}\partial^{-1}
P^-(\underline \alpha;x, t)^{0t},$$
which gives Equation
\eqref{sato1B}.  So we can continue in exactly the same way as in the case of the second approach to the BKP hierarchy in Subsection \ref{section4} and we refer the reader to that subsection for the details.

For $p=1$ and $\underline \alpha=\underline \beta$,  we find
\begin{equation}
\label{BosDKPr ne 0 2!!c}
\begin{aligned}
\left( P^+(\underline \alpha; x,t,\partial) J_1(\partial)
P^-(\underline \alpha; x,  t,\partial)^*\right)_-&=
\frac12
\sum_{a=1}^{r-1}\sum_{j=0}^{2\mu_a}
W^+(\underline \alpha; x,t)^j E_{aa}\partial^{-1} \left(W^-(\underline \alpha;x,  t)^{2\mu_a-j}\right)^{t}\\
+
(1+2\delta)(
T^{+0}(\alpha; x,t) 
\partial^{-1}&
T^{-1}(\alpha; x,  t)^t+T^{+1}(\alpha; x,t) 
\partial^{-1}
T^{0}(\alpha; x,  t)^t)\\
&-2\delta
T^{+\frac12}(\alpha; x,t) 
\partial^{-1}
T^{-\frac12}(\alpha; x,  t)^t
,
\end{aligned}
\end{equation} 
where $J_1(\partial)$ is  given by 
\eqref{J_k(z)}.
Note that $J(\partial)=J_0(\partial)$.

 Define as before 
$\hat W^\pm(\underline \alpha; x,t,z)=(P^\pm(\underline \alpha; x,t,z)^0)^{-1}W^\pm(\underline \alpha; x,t,z)$ and
define \break
$\hat T^\pm(\underline \alpha; x,t,z)=(P^\pm(\underline \alpha; x,t,z)^0)^{-1} T^\pm(\underline \alpha; x,t,z)$ 
then \eqref{BosDKPr ne 0 2!!c}
gives
 \begin{equation}
\label{WWW22}
\begin{aligned}
&\left(\hat P^{+}(\underline\alpha;x,t,\partial)J_k(\partial)J(\partial)^{-1}\hat P^{-}(\underline\alpha;x,t,\partial)^{-1} \right)_{< }=
\\
&\qquad\frac12
\sum_{a=1}^{r-1}\sum_{j=0}^{2\mu_a}
\hat W^+(\underline \alpha; x,t)^j E_{aa}\partial^{-1} \left(\hat W^-(\underline \alpha;x,  t)^{2\mu_a-j}\right)^{t}J(\partial)^{-1}\\
&\qquad+
(1+2\delta)(
\hat T^{+0}(\alpha; x,t) 
\partial^{-1}
\hat T^{-1}(\alpha; x,  t)^t+\hat T^{+1}(\alpha; x,t) 
\partial^{-1}
\hat T^{0}(\alpha; x,  t)^t)J(\partial)^{-1}\\
&\qquad\qquad-2\delta
\hat T^{+\frac12}(\alpha; x,t) 
\partial^{-1}
\hat T^{-\frac12}(\alpha; x,  t)^t
J(\partial)^{-1}.
\end{aligned}
\end{equation}
Define
\begin{equation}
\label{calcalLL}
{\cal L}(\underline \alpha; x,t,\partial)=L(\underline \alpha; x,t,\partial)^{2\mu_r} D^{s+r}(\underline \alpha; x,t,\partial)+\sum_{a=1}^s
L(\underline \alpha; x,t,\partial)^{\lambda_a} E^{a}(\underline \alpha; x,t,\partial),
\end{equation}
then clearly ${\cal L} $  satisfies the same Lax equations as $L$\, viz.   \eqref{BLax2} and
\begin{equation}
\label{calcalL}
\begin{aligned}
{\cal L}(\underline \alpha; x,t,\partial)&={\cal L}(\underline \alpha; x,t,\partial)_{\ge}+\frac12 \sum_{a=1}^{r-1}\sum_{j=0}^{2\mu_a}
\hat W^+(\underline \alpha; x,t)^j E_{aa}\partial^{-1} \left(\hat W^-(\underline \alpha;\overline x, \overline t)^{2\mu_a-j}\right)^{t}J(\partial)^{-1}\\
&+
(1+2\delta)(
\hat T^{+0}(\alpha; x,t) 
\partial^{-1}
\hat T^{-1}(\alpha; x,  t)^t+\hat T^{+1}(\alpha; x,t) 
\partial^{-1}
\hat T^{0}(\alpha; x,  t)^t)J(\partial)^{-1}\\
&\qquad -2\delta
\hat T^{+\frac12}(\alpha; x,t) 
\partial^{-1}
\hat T^{-\frac12}(\alpha; x,  t)^t
J(\partial)^{-1}.
\end{aligned}
\end{equation}

\subsection{$\hat{so}_{2n+1}$ and the $(\lambda,\mu)$-reduced BKP   hierarchy}
\label{subs7.4}

The group $\hat{SO}_{2n+1}$    leaves the module  $F_b $ invariant.  In this case one gets that $\tau$ satisfies the first equation of \eqref {BDREDUCTION}, see Section  \ref{SecDred}.
In terms of the fermions related to $(\lambda, \mu)$, this is 
\begin{equation}
\label{SB}
\begin{aligned}
\sum_{b=1}^s\sum_{i\in\frac{1}{2}+\mathbb{Z}}&\left( \psi_i^{+b}\tau \otimes \psi_{p\lambda_b-i}^{-b}\tau+ \psi_i^{-b}\tau\otimes \psi_{p\lambda_b-i}^{+b}\tau\right)+
\sum_{c=s+1}^{r+s}\sum_{i\in\mathbb{Z}}(-1)^i\tilde\phi^c_i\tau\otimes \tilde\phi^c_{2p\mu_c-i}\tau+\\
& +\sum_{j\in \delta+\mathbb Z} \sigma_j\tau\otimes\sigma_{p-j}\tau=\frac{\delta_{p0}}2 \tau\otimes \tau.
\end{aligned}
\end{equation}
For the value of $\delta$ see the two possibilities below \eqref{FockB}.  Again we have a similar  relation between $\psi$,  $\tilde\phi^R_j$ and $\sigma_k$ as in  \eqref {psisigma},  where $R$ is again equal to $R=r+1$, but now the first case appears when $r=2r'$ and the latter when $r=2r'-1$. Hence, for $p=0$ this equation is the same as the BKP Equation \eqref{SB2oddfield} or \eqref{SB2evenfield}.
We can rewrite \eqref{SB} using the fermionic fields:
\begin{equation}
\label{SB2}
\begin{aligned}{\rm Res}_{z=0}\big\{
\sum_{b=1}^s&z^{p\lambda_b} \left( \psi^{+b}(z)\tau \otimes \psi^{-b}(z)\tau+ \psi^{-b}(z)\tau \otimes \psi^{+b}(z)\tau\right)+\\
+&\sum_{c=s+1}^{r+s} z^{2p\mu_c-1} \tilde\phi^c(z)\tau\otimes \tilde\phi^c(-z)\tau
+z^{p-1-2\delta}\sigma(z)\tau\otimes\sigma(z)\tau\big\}dz=\frac{\delta_{p0}}2 \tau\otimes \tau.
\end{aligned}
\end{equation}
We write
\[
\sigma(T)=\sum_{\underline k\in \mathbb Z^s,\underline \ell\in \mathbb Z_2^{r'}} q^{\underline k}\theta^{\underline\ell} \tau_{\underline k,\underline \ell}(t, \xi ) ,
\]
where $q^{\underline k}=q_1^{k_1}q_2^{k_2}\cdots q_s^{k_s}$, $\theta^{\underline\ell}=\theta_1^{\ell_1}\theta_2^{\ell_2}\cdots\theta_{r'}^{\ell_{r'}}$.
Let $N_\xi (f(t)\xi_{m_1}\xi_{m_2}\cdots \xi_{m_p})=p$,
then   \eqref{SB2}, for the embedding in $b_\infty$, which is the highest weight module with weight $\Lambda_n$,  turns into the following equations 
($p=0,1,2,\ldots$)
\begin{equation}
\label{BosBKPred}
\begin{aligned}
&{\rm Res}_{z=0}\big(\sum_{b=1}^s \big(
(-1)^{|\underline k+\overline{\underline k}|_{b-1}} z^{ p\lambda_b+k_b-\overline k_b-2+2\delta_{bs}}
e^{z\cdot (t^{(b)}-\overline t^{(b)})}e^{z^{-1}\cdot (\tilde\partial_{\overline t^{(b)}}- \tilde\partial_{t^{(b)}})}
\tau_{\underline k+\underline e_s-\underline e_b,\underline\ell, \alpha}(t,\xi)\tau_{\overline{\underline k}+\underline e_b-\underline e_s,\overline{\underline \ell}, \overline \alpha}(\overline t,\overline \xi)+
\\
&+(-1)^{|\underline k+\overline{\underline k}|_{b-1}} z^{ p\lambda_b-k_b+\overline k_b-2-2\delta_{bs}}
e^{(-z)\cdot (t^{(b)}-\overline t^{(b)})}
e^{(-z)^{-1}\cdot (\tilde\partial_{t^{(b)}}- \tilde\partial_{\overline t^{(b)}})}
%
\tau_{\underline k+\underline e_s+\underline e_b,\underline\ell, \alpha}(t,\xi)\tau_{\overline{\underline k}-\underline e_b-\underline e_s,\overline{\underline\ell},\overline\alpha}(\overline t,\overline \xi)+\\
&+\frac12
\sum_{c=1}^{r'} (-1)^{{|\underline k+\overline{\underline k}|}+|\underline\ell+\overline{\underline\ell}|_{\left[\frac{c-1}2\right]}+(c-1)(\ell_{\left[\frac{c+1}2\right]}+\overline\ell_{\left[\frac{c+1}2\right]}+1)}z^{2p\mu_c-1}
e^{z\circ (t^{(s+c)}-  \overline t^{(s+c)})}
\times\\  &\qquad \qquad \qquad \qquad \qquad 
e^{2 (z^{-1}\circ (\tilde\partial_{\overline t^{(s+c)}} -\tilde\partial_{t^{(s+c)}}))}
%
\tau_{\underline k+\underline e_s ,\underline\ell+\underline e_{\left[\frac{c+1}2\right]},\alpha}(t,\xi)
\tau_{\overline{\underline k} -\underline e_s,\overline{\underline\ell}+\underline e_{\left[\frac{c+1}2\right]},\overline\alpha}(\overline t,\overline \xi)
\\
&-\delta
 (-1)^{{|\underline k+\overline{\underline k}|}+|\underline\ell+\overline{\underline\ell}|+  N_\xi+N_{\overline \xi}}z^{2p\mu_r-1}
e^{z\circ (t^{(s+r)}-  \overline t^{(s+r)})}
\times\\  &\qquad \qquad \qquad \qquad \qquad 
e^{2(z^{-1}\circ( \tilde\partial_{\overline t^{(s+c)}}- \tilde\partial_{t^{(s+c)}}))}
\tau_{\underline k+\underline e_s ,\underline\ell}(t,\xi)
\tau_{\overline{\underline k} -\underline e_s,\overline{\underline\ell} }(\overline t, \overline\xi)
\big) dz
\\
&+
(-1)^{{|\underline k+\overline{\underline k}|}+|\underline\ell+\overline{\underline\ell}|}
\big\{
\sum_{j=1}^\infty 
\xi_{j+\delta}\tau_{\underline k+\underline e_s ,\underline\ell}(t,\xi)\frac{\partial 
\tau_{\overline{\underline k} -\underline e_s,\overline{\underline\ell} }(\overline t, \overline\xi)}{\partial \overline\xi_{j+p+\delta}} +
\frac{\partial \tau_{\underline k+\underline e_s ,\underline\ell}(t,\xi)}{\partial \xi_{j+\delta}}\overline\xi_{j+p+\delta}
\tau_{\overline{\underline k} -\underline e_s,\overline{\underline\ell} }(\overline t, \overline\xi)
\\
&\qquad \qquad +
 \sum_{j=1}^{p-1-2\delta}
\frac{\partial \tau_{\underline k+\underline e_s ,\underline\ell}(t,\xi)}{\partial \xi_{j+\delta}}
\frac{\partial 
\tau_{\overline{\underline k} -\underline e_s,\overline{\underline\ell} }(\overline t, \overline\xi)}{\partial \overline\xi_{p-j-\delta}} 
\\
&\qquad\qquad +(1-\delta_{p0})\frac{1+2\delta}{\sqrt 2}
\big(
(-1)^{N_\xi}\tau_{\underline k+\underline e_s ,\underline\ell }(t,\xi)
\frac{\partial 
\tau_{\overline{\underline k} -\underline e_s,\overline{\underline\ell} }(\overline t, \overline\xi)}{\partial \overline\xi_{p}} 
\\
& \qquad\qquad\qquad\qquad\qquad\qquad\qquad\qquad\qquad+(-1)^{N_{\overline\xi}}
\frac{\partial \tau_{\underline k+\underline e_s ,\underline\ell}(t,\xi)}{\partial \xi_{p}}
\tau_{\overline{\underline k} -\underline e_s,\overline{\underline\ell}}(\overline t,\overline\xi )
\big)\big\}
=\\
&\qquad={\delta_{p0}}[\frac12 -(\frac12+\delta)(-1)^{{|\underline k+\overline{\underline k}|}+|\underline\ell+\overline{\underline\ell}|+   N_\xi+N_{\overline \xi}}]
\tau_{\underline k+\underline e_s ,\underline\ell }(t,\xi)
\tau_{\overline{\underline k} -\underline e_s,\overline{\underline\ell}}(\overline t,\overline\xi )
.
\end{aligned}
\end{equation}

We will now turn the $(\lambda, \mu)$-reduced BKP hierarchy into a hierarchy of  pseudo-differential operators.  Again,  we will  consider several cases.
\\
\
\\
{\bf Case 1, $\mu=\emptyset$. }
Thus  $r=0$.   
We substitute in equation \eqref{BosBKPred} for all Grassmann variables $\xi_j=0$.
The first equation of \eqref{BosBKPred},  where $\delta=0$ and $r=r'=0$ is then equal to 
($p=0,1,2,\ldots$)
\begin{equation}
\label{BosBKPred2}
\begin{aligned}
&{\rm Res}_{z=0} \sum_{b=1}^s \big(
(-1)^{|\underline k+\overline{\underline k}|_{b-1}} z^{ p\lambda_b+k_b-\overline k_b-2+2\delta_{bs}}
e^{z\cdot (t^{(b)}-  \overline t^{(b)})}e^{z^{-1}\cdot (\tilde\partial_{\overline t^{(b)}}-  \tilde\partial_{t^{(b)}})}
\tau_{\underline k+\underline e_s-\underline e_b,\underline\ell, \alpha}(t,0)\tau_{\overline{\underline k}+\underline e_b-\underline e_s,\overline{\underline \ell}, \overline \alpha}(\overline t,0)+
\\
&+(-1)^{|\underline k+\overline{\underline k}|_{b-1}} z^{ p\lambda_b-k_b+\overline k_b-2-2\delta_{bs}}
e^{(-z)\cdot (t^{(b)}-  \overline t^{(b)})}
e^{(-z)^{-1}\cdot (\tilde\partial_{t^{(b)}}- t \tilde\partial_{\overline t^{(b)}})}
%
\tau_{\underline k+\underline e_s+\underline e_b,\underline\ell, \alpha}(t,0)\tau_{\overline{\underline k}-\underline e_b-\underline e_s,\overline{\underline\ell},\overline\alpha}(\overline t,0)\big)dz
\\
&=-
(-1)^{{|\underline k+\overline{\underline k}|}}
\big\{
 \sum_{j=1}^{p-1}
\frac{\partial \tau_{\underline k+\underline e_s }(t,\xi)}{\partial \xi_{j}}\big|_{\xi=0}
\frac{\partial 
\tau_{\overline{\underline k} -\underline e_s }(\overline t, \overline\xi)}{\partial \overline\xi_{p-j}} \big|_{\overline \xi=0}
\\
&  \qquad+(1-\delta_{p0})\frac{1}{\sqrt 2}
\big(
 \tau_{\underline k+\underline e_s }(t,0)
\frac{\partial 
\tau_{\overline{\underline k} -\underline e_s}(\overline t, \overline\xi)}{\partial \overline\xi_{p}}\big|_{\overline \xi=0}
+
\frac{\partial \tau_{\underline k+\underline e_s }(t,\xi)}{\partial \xi_{p}}\big|_{\xi=0}
\tau_{\overline{\underline k} -\underline e_s}(\overline t,0)
\big)\big\}
\\
&\qquad +{\delta_{p0}}[\frac12 -\frac12(-1)^{{|\underline k+\overline{\underline k}|} }]
\tau_{\underline k+\underline e_s }(t,0)
\tau_{\overline{\underline k} -\underline e_s}(\overline t,0)
.
\end{aligned}
\end{equation}
As before, replace all $t_1^a$ by $t_1^a+x$ and define 
$V^\pm (\underline \alpha; x,t,z)$ as before,  and 
define the column vector $T^{\pm j}(\underline\alpha;x,t)= (W^{\pm j}(\underline\alpha;x,t)_i)_{1\le i\le 2s}$  again by \eqref{TTD} and  \eqref{TTDD}
Note that again \eqref{W+-} holds.
Then
\begin{equation}
\label{BosBKPred62}
\begin{aligned}
{\rm Res}_{z=0}& \,
\sum_{j=1}^s
V^+(\underline \alpha; x,t,z)
\left(z^{p\lambda_j}E_{jj}
-(-z)^{p\lambda_j}E_{s+j,s+j}
\right)
V^-(\underline \beta;\overline x, \overline t,z)^t dz=\\
&=
-
T^{+ 0}(\underline\alpha;x,t)T^{- 0}(\underline\beta;\overline x,\overline t)^t
+ \sum_{j=0}^p
T^{+ j}(\underline\alpha;x,t)T^{-(p-j)}(\underline\beta;\overline x,\overline t)^t.
\end{aligned}
\end{equation}
Let $\partial=\partial_x$ and using the fundamental lemma (see \cite{KLmult},  Lemma 4.1),  we deduce the following expression for the matrix pseudo-differential operators
\begin{equation}
\label{BB22}
\begin{aligned}
\sum_{j=1}^s&
(P^+(\underline \alpha;x,t,\partial) R^+(\underline \alpha;\partial)\left(\partial^{p\lambda_j}E_{jj}-(-\partial)^{p\lambda_j}E_{s+j,s+j}\right)R^-(\underline\beta;x,t,\partial)^*
P^-(\underline \beta; x,t,\partial)^*)_-=\\
&-
T^{+0}(\underline\alpha;x,t)\partial
T^{- 0}(\underline\beta; x,t)^tJ\ 
+ \sum_{j=0}^p
T^{+ j}(\underline\alpha;x,t)
\partial
T^{-(p-j)}(\underline\beta;x,t)^tJ
.
\end{aligned}
\end{equation}
As in the previous section,   if we take take 
$p=0$ and $\underline\alpha=\underline \beta$ and $J= \sum_{j=1}^s E_{jj}-E_{s+j,s+j}=J^{-1}$,  we obtain again  \eqref{sato3}
However, taking $p\ne 0$,  we get 
\begin{equation}
\label{satoB44}
\begin{aligned}
\sum_{j=1}^s& (P^+(\underline \alpha; x,t,\partial)\left(\partial^{p\lambda_j}E_{jj}+(-\partial)^{p\lambda_j}E_{s+j,s+j}\right)P^+(\underline \alpha; x,t,\partial)^{-1})_-=
\\
&\qquad \sum_{k=0}^p
T^{+ k}(\underline\alpha;x,t)
\partial^{-1}
T^{- (p-k)}(\underline\alpha;x,t)^tJ
.
\end{aligned}
\end{equation}
We deduce from  \eqref{satoB4} that
\begin{equation}
\label{laxB22}
\begin{aligned}
 \sum_{j=1}^s (L^{p\lambda_j}E^j)_-=\sum_{k=0}^p
T^{+ k}(\underline\alpha;x,t)
\partial^{-1}
T^{- (p-k)}(\underline\alpha;x,t)^tJ
.
\end{aligned}
\end{equation}
From this it is obvious that the integral part of ${\cal L}= \sum_{j=1}^s L^{\lambda_j}E^j$ is equal to
\begin{equation}
\label{calLLL}
{\cal L}_-=
T^{+ 0}(\underline\alpha;x,t)
\partial^{-1}
T^{- 1}(\underline\alpha;x,t)^t J+T^{+ 1}(\underline\alpha;x,t)
\partial^{-1}
T^{- 0}(\underline\alpha;x,t)^tJ
.
\end{equation}
As before,
\[
  {\cal L}^*=K{\cal L}K^{-1}.
\]
Again we have
\begin{equation}
\label{WB2}
\frac{\partial T^{+ k} }{\partial t_i^j}=\left( L^iD^j\right)_+\left(
T^{+ k}\right), \quad\frac{\partial T^{- k} J}{\partial t_i^j}=-\left( L^iD^j\right)^*_+\left(
T^{- k}J\right),\quad \mbox{for } k=0,1.
\end{equation}
Thus $W^{+k}$,  resp. $W^{ -k}J$,  for $k=0,1$, is an eigenfunction,  resp. adjoint eigenfunction, for the operators $L$ and $D^j$,    $j=1,2, \ldots ,s$.
We  obtain this from
\[
\frac{\partial W^{+0} }{\partial t_i^j}
\partial^{-1}
{W^{- 1}}^t+\frac{\partial W^{+1} }{\partial t_i^j} \partial^{-1}
{W^{- 0}}^t=\left( L^iD^j\right)_+
  (W^{+0} )
\partial^{-1}
{W^{- 1}}^t + \left( L^iD^j\right)_+(W^{+1} )  \partial^{-1}
{W^{- 0}}^t.
\]
Namely, we can eliminate one of the two terms, by letting the matrices of this equation act on a specific  vector of functions of the variables $t$.
Thus obtaining the first equation of \eqref{WB2}.
The second formula of \eqref{WB2} is obtained in a similar fashion.
\\
\
\\
{\bf Case 2, $\mu\ne\emptyset$ (first approach).} This case is similar to the $\mu\ne\emptyset$ (first approach) case for the embedding in 
$d_\infty$.   The only difference is the value of $\delta$.  If $r$ is even (resp.  odd), then in the reduction of the DKP hierarchy, $\delta=-\frac12$ (resp.  $\delta=0$), while in this case,   for the reduction of the BKP hierarchy,  $\delta =0$ (resp.  $\delta=-\frac12$) when $r$ is even (resp.  odd).
\\
\
\\
{\bf Case 3, $\mu\ne\emptyset$ (second approach).}
Here also Remark \ref{mur} holds.  This case is similar to the $\mu\ne\emptyset$ (second approach) case for the embedding in 
$d_\infty$.   In this case,   however,  $\delta =0$ (resp.  $\delta=-\frac12$) when $r$ is even (resp.  odd).  Again
$\cal L$ is given by \eqref{calcalLL}
is  equal to \eqref{calcalL}
and satisfies   the same Lax equations as $L$\, viz.   \eqref{BLax2}.

\subsection{$\hat{sp}_{2n}$ and the $(\lambda,\mu)$-reduced CKP  hierarchy}
\label{SubsecC}

The group $\hat {SP}_{2n}$  acs on  $F_c$ and leaves the module  $F_c^{\overline a}$ invariant.  It consists of all  elements  $g\in C_\infty$
 that satisfy $gb_jg^{-1}=\sum_{i} a_{ij}b_i$, with  $\sum_{k} (-1)^{k-\frac12}a_{ki}a_{-k,j}=(-1)^{i-\frac12}\delta_{i,-j}$,  hence
 $\sum_{k} (-1)^{k+\frac12}a_{ik}a_{j,-k}=(-1)^{i+\frac12}\delta_{i,-j}$,
 and $a_{i+2n,j+2n}=a_{ij}$. Then 
\[
\begin{aligned}
(g\otimes g) 
\sum_{i\in\frac12+\mathbb{Z}
} (-1)^{i+\frac12}b_i \otimes b_{p(2n)-i}
&=\sum_{i\in\frac12  +\mathbb{Z}
}(-1)^{i+\frac12}g b_i g^{-1}g\otimes g b_{p(2n+1)-i}g^{-1} g\\
&=\sum_{i,j,k\in\frac12+\mathbb{Z}}(-1)^{i+\frac12} a_{ji} a_{k,p(2n+1)-i}b_j g\otimes b_k g\\
&=\sum_{i,j,k\in\frac12 +\mathbb{Z}} (-1)^{i+\frac12} a_{ji} a_{k-p(2n+1),-i}b_j g\otimes b_k g\\
 &=\sum_{j\in\delta +\mathbb{Z}}(-1)^{j+\frac12} b_jg\otimes b_{p(2n+1)-j}g\\
&=\sum_{j\in\delta +\mathbb{Z}} (-1)^{j+\frac12}
 b_j\otimes b_{p(2n+1)-j} (g\otimes g) 
.
\end{aligned}
\]
Since $\sum_{i\in\frac12+\mathbb{Z}
} (-1)^{i+\frac12}b_i|0\rangle \otimes b_{p(2n)-i}|0\rangle =0$, we obtain
\begin{equation}
\label{CCKP}
\sum_{i\in\frac12+\mathbb{Z}
} (-1)^{i+\frac12}b_i \tau_0\otimes b_{p(2n)-i}\tau_0=0,\quad p=0,1,2,\ldots
\end{equation}
In a similar way we deduce that 
\[
\sum_{i\in\frac12+\mathbb{Z}
} (-1)^{i+\frac12}b_i \tau_1\otimes b_{p(2n)-i}\tau_1=0,\quad p=1,2,\dots, \quad\mbox{and }S^2_c(\tau_1\otimes \tau_1)=-4T_1\otimes T_1.
\]
In terms of $(\lambda,\mu)$  equation \eqref{CCKP} turns into
\begin{equation}
\label{SC}
\begin{aligned}
\sum_{a=1}^s\sum_{i\in\frac{1}{2}+\mathbb{Z}}&\left( b_i^{-a}\tau_0 \otimes b_{p\lambda_a-i}^{+a}\tau_0- b_i^{+a}\tau_0 \otimes b_{p\lambda_a-i}^{-a}\tau_0\right)+
\sum_{c=s+1}^{r+s}\sum_{i\in\frac12+\mathbb{Z}}(-1)^{i+\frac12}b^c_i\tau_0\otimes b^c_{2p\mu_{c-s}-i}\tau_0 =0.
\end{aligned}
\end{equation}
In terms of the  bosonic fields, this is
\begin{equation}
\label{SC2}
\begin{aligned}
{\rm Res}_{z=0}\big\{
\sum_{a=1}^s &z^{p\lambda_a} \left(b^{-a}(z)\tau_0 \otimes b^{+a}(z)\tau_0- b^{+a}(z)\tau_0 \otimes b^{-a}(z)\tau_0\right)+\\
+&\sum_{c=s+1}^{r+s}z^{2p\mu_{c-s}} b^c(z)\tau_0\otimes b^c (-z)\tau_0 \big\} dz =0.
\end{aligned}
\end{equation}

We can rewrite \eqref{SC} for the wave function introduced in \eqref{VPMC}  ($ p=0,1,\ldots$)
\begin{equation}
\label{CKP5red}
{\rm Res}_{z=0} V(x,t,z) \left(\sum_{a=1}^s \left( z^{p\lambda_a}E_{aa}+ (-z)^{p\lambda_a}E_{s+a,s+a}\right ) 
+
\sum_{i=1}^r z^{2p\mu_i}E_{2s+i,2s+i} \right)V(\overline x,\overline t,-z)^t dz =0 .
\end{equation}
This  equation for $p=1$ is equivalent to the fact that
\begin{equation}
\label{LaxCC}
{\cal L:}=
\sum_{a=1}^s L^{\lambda^a} E^a +
 \sum_{i=1}^r L^{2\mu_i}C^{2s+i}
=({\cal L})_+.
\end{equation}
This ${\cal L}$ also satisfies  the Lax equations 
\[
\frac{\partial {\cal L}}{\partial t_k^a}= [(L^kD^a)_+, {\cal L}]
,\quad
\frac{\partial {\cal L}}{\partial t_k^c}= [(L^k C^{s+c})_+, {\cal L}],\quad a=1,\ldots s,\ c=s+1,\ldots, s+r
.
\]
Using \eqref{LaxCC} \eqref{P-C2}, \eqref{NI} and \eqref{Lax-op-C},  we find that 
\[
{\cal L}^*=(N+I){\cal L}(N+I)
.
\]
\section{Hierarchies of KP type  related to $x_\infty$ and $\hat{\mathfrak g}^{(2)}$ reductions}
The Lie group that corresponds to  $\hat{gl}_n^{(2)}$ (resp. $\hat{so}_{2n}^{(2)}$ ) is the twisted loop group  $\hat{GL}_n^{(2)}$ (resp. $\hat{SO}_{2n}^{(2)}$) that preserves the bilinear form on $\mathbb C[t,t^{-1}]^{n}$ (resp.  $\mathbb C[t,t^{-1}]^{2n}$):
\[
({\bf v}(t),{\bf w}(t)) ={\rm Res}\, {\bf v}(-t)^tJ_n {\bf w}(t)\frac{dt}t, \quad (\mbox{resp.   }({\bf v}(t),{\bf w}(t)) ={\rm Res}\, {\bf v}(-t)^tJ_{2n}O {\bf w}(t)\frac{dt}t).
\]
Hence, an element $G(t)\in \hat{GL}_n^{(2)}$  (resp.  $G(t)\in \hat{SO}_{2n}^{(2)}$ satisfies
\[
J_n G(-t)^tJ_nG(t)=I_n, \quad  ( \mbox{resp.   } OJ_{2n}G(-t)^tJ_{2n}OG(t)=I_{2n}).
\]
Date, Jimbo, Kashiwara and Miwa obtain in \cite {DJKM1}, via  the embedding of $\hat{\mathfrak g}^{(2)}\subset d_\infty $ or $b_\infty$,  so-called  reduced DKP (in \cite{DJKM1} this hierarchy corresponds to the 2-component BKP) and  BKP hierarchies.  
This means in the one component case that one obtains the additional equations 
\begin{equation}
\label{BKPextra}
\sum_{i\in\mathbb{Z}}
 \sigma^k(\tilde\phi_i) \tau\otimes \tilde\phi_{kP-i}\tau =0,\quad (\mbox{resp.  }\sum_{i\in\mathbb{Z}}
\tilde\phi_i \tau\otimes \tilde\phi_{kP-i}\tau =0),\quad 
k=1,2,\dots
\end{equation}
in the BKP case,  with $P=2m+1$ (resp.  $P=2m$) for $\hat{gl}_{2m+1}^{(2)}$   (resp. $\hat{so}_{2m}^{(2)}$),
and 
\begin{equation}
\label{DKPextra}
\sum_{i\in\frac12 +\mathbb{Z}
} \sigma^k(\phi_i) \tau_a\otimes \phi_{2km-i}\tau_a =0,\quad k=1,2,\dots 
\end{equation}
in the DKP case for $\hat{gl}_{2m}^{(2)}$.
In the case of $\hat{so}_{2n}^{(2)}$ this is clear.  Because the matrices in $b_\infty$ that correspond to $\hat{so}_{2n}^{(2)}$  are $2n$ periodic, the same proof as in the untwisted case can be used.  For $\hat{gl}_n^{(2)}$, one can use the following observation.

We consider  $\hat{GL}_{2m}^{(2)}$. This group leaves the module  $F_d^{\overline a}$ invariant.  It consists of all  elements $g\in D_\infty$
 that satisfy $g\phi_ig^{-1}=\sum_{i} a_{ji}\phi_j$, with  $\sum_{k} a_{ki}a_{-k,j}=\delta_{i,-j},$ and hence  $\sum_{k} a_{ik}a_{j,-k}=\delta_{i,-j}$, and $a_{i+2pm,j+2pm}=J^pa_{ij}J^p=\sigma_i^p \sigma_j^pa_{ij}$.  Recall from Remark \ref{Jremark2} that $\sigma_i=\sigma_{-i}$.  Thus,  for $p=1,2,\ldots$, we find that
\[
\begin{aligned}
(g\otimes g) 
\sum_{i\in\frac12+\mathbb{Z}
}  J^p\phi_i \otimes \phi_{2pm-i}
&=\sum_{i\in\frac12+\mathbb{Z}
}\sigma_i^pg \phi_i g^{-1}g\otimes g  \phi_{2pm-i} g^{-1} g   \\
&=\sum_{i,j,k\in \frac12+\mathbb{Z}} \sigma_i^p  a_{ji} a_{k,2pm-i}\phi_j g\otimes \phi_k g\\
&=\sum_{i,j,k\in\frac12+\mathbb{Z}}  \sigma_i^p   \sigma_{k-2pm}^p\sigma_{-i}^pa_{ji}  a_{k-2pm,-i}\phi_j g\otimes \phi_k g\\
&=\sum_{i,j,k\in\frac12+\mathbb{Z}}  \sigma_{k-2pm}^pa_{ji}  a_{k-2pm,-i}\phi_j g\otimes \phi_k g\\
 &=\sum_{j\in\frac12+\mathbb{Z}}\sigma_j^p\phi_jg\otimes \phi_{2pm-j}g\\
&=\sum_{j\in\frac12+\mathbb{Z}
} J^p\phi_j\otimes \phi_{2pm-j} (g\otimes g).
\end{aligned}
\]
Since also
\[
 \sum_{i\in\frac12+\mathbb{Z}
} J^p\phi_i |\overline a\rangle\otimes \phi_{2pm-i}|\overline a\rangle =0,
\]
  we have that elements $\tau_a\in F_d^{\overline a}$, which belong to the  $\hat{GL}_{2m}^{(2)}$-group orbit of $|\overline a\rangle$ satisfy
\[
\sum_{i\in\frac12+\mathbb{Z}
} J^p\phi_i \tau_a \otimes \phi_{2pm-i}\tau_a=0,\quad\mbox{for }p=0, 1,2,\ldots .
\]
Finally,  let $J^p\otimes 1$ act on this equation, this gives \eqref{BKPextra}.
In a similar way we find that $\tau\in F_b$, which belong to the  $\hat{GL}_{2m+1}^{(2)}$-group orbit (resp.     $\hat{SO}_{2n}^{(2)}$-group orbit)
of $|  0\rangle$, satisfies
\eqref{BKP} and the first equation (resp.  second equation) of \eqref{BKPextra}.
\subsection{$\hat{gl}_{2m}^{(2)}$ and the $(\lambda,\mu)$-reduced   DKP hierarchy}
In this case we assume that $\lambda=(\lambda_1, \lambda_2,\ldots, \lambda_s)$,  where all parts $\lambda_i=2\ell_i$ are even,  and 
$\mu=(\mu_1,\mu_2,\ldots,\mu_r)$, where all parts $\mu_i=2m_i+1$  are odd, such that $|\lambda|+|\mu|=2m$. This means that $r$ can only be $r=2R$ even.  
Transforming the equations \eqref{DKP} and \eqref{DKPextra}  to the $(r,s)$ multicomponent description we obtain 
\begin{equation}
\label{SDgl}
\begin{aligned}
{\rm Res}_{z=0}\big\{
&
\sum_{b=1}^s
z^{k\ell_b}\left( \psi^{+b}(z)
 \tau_a \otimes \psi^{-b}(z)\tau_a 
 +
\psi^{-b}(z)
 \tau_a \otimes \psi^{+b}(z)\tau_a 
\right)+\\
&
\qquad+\frac1{z}\sum_{c=s+1}^{2R+s}z^{(2m_{c-s}+1)k} \tilde\phi^c (z)\tau_a\otimes \tilde\phi^c(-z)\tau_a\big\}  dz
 =0,  \quad  k=0,1,\ldots .
\end{aligned}
\end{equation}
This resembles the $\hat{so}_{2n}$ reduction \eqref{SD2nfield} of the DKP hierarchy.  However in this case the factor $z^{(2m_{c-s}+1)k}$ in front of the neutral twisted fields has an odd power  of $z$,  while this power in the $\hat{so}_{2n}$-case is always even.
This means that the same approach works as in subsection \ref{subs7.2}, but then with all $\lambda_b$ and $\mu_i$'s replaced by 
$\ell_b$ and $\frac{2m_{c-s}+1}2$,  respectively. 
We refer the reader to that subsection for the construction of wave function and Lax operators.
\\
In the {\bf first approah to the DKP hierarchy},  we find that the operator 
\[
\begin{aligned}
{\cal L}(\underline\alpha;x,t)&=
\sum_{b=1}^s L(\underline\alpha;x,t)^{\ell_b}E^b(\underline\alpha;x,t)
\\
={\cal L}&(\underline\alpha;x,t)_++\frac12 \sum_{c=s+1}^{s+r }
\sum_{n=0}^{2m_{c-s}+1}    W^+(\underline \alpha; x,t)^n E_{c-s,c-s}\partial^{-1}
\left(JW^-(\underline \alpha;x, t)^{2m_{c-s}-n+1  }\right)^t,
\end{aligned}
\] 
satisfies
$\frac{\partial \cal L}{\partial t^a_j}=[ (L^j D^a)_+,{\cal L}]$.\\
In the {\bf second approach to the DKP hierarchy},  where again Remark \ref{mur} holds,   we find  
that
\[
{\cal L}(\underline \alpha; x,t,\partial)=L(\underline \alpha; x,t,\partial)^{m_r  }D^{s+r}(\underline \alpha; x,t,\partial)+\sum_{b=1}^s
L(\underline \alpha; x,t,\partial)^{\ell_b} E^{b}(\underline \alpha; x,t,\partial),
\]
  satisfies the same Lax equation as $L$, viz. \eqref{BLax2},  and
\[
\begin{aligned}
&{\cal L}(\underline \alpha; x,t,\partial)={\cal L}(\underline \alpha; x,t,\partial)_{\ge}+\\
&\qquad+\frac12 \sum_{c=s+1}^{s+r-1}\sum_{j=0}^{ 2m_{c-s}+1}
\hat W^+(\underline \alpha; x,t)^j E_{c-s,c-s}\partial^{-1} \left(\hat W^-(\underline \alpha; x,  t)^{2m_{c-s}-j+1}\right)^{t}J(\partial)^{-1}.
\end{aligned}
\]

\subsection{$\hat{gl}_{2m+1}^{(2)}$ and      $\hat{so}_{2n}^{(2)}$ 
and the $(\lambda,\mu)$-reduced   BKP hierarchy}
In the case of $\hat{gl}_{2m+1}^{(2)}$ (resp.  
$\hat{so}_{2n}^{(2)}$)
we assume that $\lambda=(\lambda_1, \lambda_2,\ldots, \lambda_s)$,  where all parts $\lambda_i=2\ell_i$ are even,  and 
$\mu=(\mu_1,\mu_2,\ldots,\mu_r)$, where all parts $\mu_i=2m_i+1$  are odd (resp.      $\mu_i=2m_i$ are even),
 such that $|\lambda|+|\mu|=2m+1$ (resp.  $|\lambda|+|\mu|=n$ and  $r=2R+1$  odd).  This means in the  $\hat{gl}_{2m+1}^{(2)}$
that $r$ can only be $r=2R+1$  odd.   

Transforming the equations \eqref{BKP} and \eqref{BKPextra}  to the $(r,s)$ multicomponent description we obtain for $\hat{gl}_{2m+1}^{(2)}$ and  $\hat{so}_{2n}^{(2)}$:
\begin{equation}
\label{SBgl}
\begin{aligned}
{\rm Res}_{z=0}\big\{
&
\sum_{b=1}^s
z^{pn_b}\left( \psi^{+b}(z)
 \tau \otimes \psi^{-b}(z)\tau
 +
\psi^{-b}(z)
 \tau \otimes \psi^{+b}(z)\tau
\right)+\\
&
\qquad+\frac1{z}\sum_{c=s+1}^{2R+s+1}z^{pn_c} \tilde\phi^c (z)\tau\otimes \tilde\phi^c(-z)\tau\big\}  dz
 =\frac{\delta_{p,0} }2 \tau\otimes \tau,  \quad  p=0,1,\ldots ,
\end{aligned}
\end{equation}
where $n_{b}=\ell _b $ and $n_{c}=2m_{c-s}+1$ for $\hat{gl}_{2m+1}^{(2)}$, and  
$n_{b}=\lambda_b$ and  $n_{c}=2\mu_{c-s}$ for 
for $\hat{so}_{2n}^{(2)}$, with $1\le b\le s<c\le r+s$.
This equation resembles 
 \eqref{SB2}, the  $\hat{so}_{2n+1}$ reduction of the BKP hierarchy,  however in this case  we have odd powers of $z$, in front of the neutral twisted fields,  viz.  we have $ z^{(2m_{c-s}+1)k}$,  for 
$\hat{gl}_{2m+1}^{(2)}$ 
while in the $\hat{so}_{2n+1}$ case these powers are even.   In the  $\hat{so}_{2n}^{(2)}$ case,
this resembles also the $\hat{so}_{2n+1}$ reduction   of the BKP hierarchy.  However,   in this case there can only be  an odd number of neutral twisted fields.
\begin{remark}
The automorphism ${\rm Ad}\, O$ which defines the twisted Lie algebra $\hat{so}_{2n}^{(2)}$, has as fixed points the Lie algebra $so_{2n-1}$.  Hence $\hat{so}_{2n-1}$ appears as subalgebra inside  of $\hat{so}_{2n}^{(2)}$, viz.  as
\[
\mathbb C K\oplus \bigoplus_{j\in \mathbb Z} t^{2j} {so}_{2n-1}.
\]
Equation \eqref{SBgl}   is in fact the same as the description of the reduced BKP equation,  when we reduce to this $\hat{so}_{2n-1}$,  with $s$ pairs of charged fermion fields and $2R$ neutral twisted fermion fields. 
The additional field $\sigma(z)$, which is needed in that case,  is proportional to $\tilde\phi^{2R+1}(z)$.     However,  for $\hat{so}_{2n-1}$,  the   corresponding to $\sigma(z)$ Heisenberg algebra does not lie in 
$\hat{so}_{2n-1}$.  This is no problem now,  the modes of  $\alpha^{2R+1}(z)=\frac12 :\tilde\phi^{2R+1}(z)\tilde\phi^{2R+1}(-z):$ lie in $\hat{so}_{2n}^{(2)}$, they correspond to
the elements $t^{2k+1}(e_{nn}-e_{nn})$.
\end{remark}
We again find   the two cases of the BKP hierarchy of Subsection \ref{section4}.  However since $r$ is odd in both cases,  the case $\mu=\emptyset$ does not hold. \\
In the {\bf first approah to the BKP hierarchy},  we find that the operator  
${\cal L}= \sum_{b=1}^s L^{n_b}E^b$  satisfies the Lax equation 
$\frac{\partial \cal L}{\partial t^b_j}=[ (L^j D^b)_+,{\cal L}]$   and 
is equal to
\[
{\cal L}(\underline\alpha;x,t)={\cal L}(\underline\alpha;x,t)_++\frac12 \sum_{c=s+1}^{s+2R+1}
\sum_{j=0}^{n_c} W^+(\underline \alpha; x,t)^j E_{c-s,c-s}\partial^{-1}
\left(JW^-(\underline \alpha;x, t)^{n_c-j  }\right)^t.
\]
In the {\bf second approach to the BKP hierarchy},  where again Remark \ref{mur} holds,   we find that the operator  
is equal to
\[
{\cal L}(\underline \alpha; x,t,\partial)=L(\underline \alpha; x,t,\partial)^{n_{s+2R+1} }D^{s+2R+1}(\underline \alpha; x,t,\partial)+\sum_{b=1}^s
L(\underline \alpha; x,t,\partial)^{n_b} E^{b}(\underline \alpha; x,t,\partial),
\]
Then clearly ${\cal L} $  satisfies the same Lax equation as $L$, viz. \eqref{BLax2},  and
\[
{\cal L}(\underline \alpha; x,t,\partial)={\cal L}(\underline \alpha; x,t,\partial)_{\ge}+\frac12 \sum_{c=s+1}^{s+2R}\sum_{j=0}^{ n_{c}}
\hat W^+(\underline \alpha; x,t)^j E_{c-s,c-s}\partial^{-1} \left(\hat W^-(\underline \alpha; x,  t)^{n_{c}-j}\right)^{t}J(\partial)^{-1}.
\]

\pagebreak


\end{document}